\documentclass[]{aa} 

\include{definitions}
\usepackage{graphicx}
\usepackage{txfonts}
\usepackage[utf8]{inputenc}
\usepackage{dcolumn}
\usepackage{natbib}
\usepackage{enumerate}
\usepackage{color}
\usepackage{grffile}
\usepackage{array,float}
\usepackage{multirow}
\usepackage{lscape}
\usepackage{xcolor}
\usepackage{hyperref} 
\usepackage{pdflscape}
\usepackage{afterpage}
\usepackage{capt-of}
\usepackage{longtable}
\usepackage[export]{adjustbox}
\usepackage{array}
\usepackage{soul,xcolor}
\setstcolor{red}
\usepackage{amsmath,bm}
\newcolumntype{R}[2]{%
    >{\adjustbox{angle=#1,lap=\width-(#2)}\bgroup}%
    l%
    <{\egroup}%
}

\newcommand{\hii}{H~{\sc ii}}

\newcommand{\blobcat}{\texttt{BLOBCAT}}

\newcommand{\glo}{GLOSTAR}

\newcommand{\fstblobs}{3880}
\newcommand{\Rblobs}{3325}
\newcommand{\Spblobs}{555}
\newcommand{\HRC}{1457}  
\newcommand{\LRC}{1866}

\newcommand{\Plike}{904}
\newcommand{\Clike}{455}
\newcommand{\Elike}{100}

\begin{document}

\title{A global view on star formation: The GLOSTAR Galactic plane survey. VI. Radio Source Catalog II: $28\degr < \ell < 36\degr$ and $|b| < 1$\degr, VLA B-configuration.}

\author{ S.\,A.\,Dzib\inst{1,2};\thanks{E-mail: sdzib@mpifr-bonn.mpg.de}
A.\,Y.\,Yang\inst{2}, 
J.\,S.\,Urquhart\inst{3}, 
S.-N.\,X. Medina\inst{2}, 
A. Brunthaler\inst{2}, 
K.\,M.\,Menten\inst{2}, F.\,Wyrowski\inst{2}, W.\,D.\,Cotton\inst{4},  R. Dokara\inst{2}, G.\,N.\,Ortiz-Le\'on\inst{2,5}, M.\,R.\,Rugel\inst{2}, H.\,Nguyen\inst{2}, Y.\,Gong\inst{2}, A. Chakraborty\inst{6}, H.\,Beuther\inst{7}, S.\,J.\,Billington\inst{3}, C.\,Carrasco-Gonzalez\inst{7}, T.\,Csengeri\inst{9}, P.\, Hofner\inst{4,10}, J.\,Ott\inst{4}, J.\,D.\,Pandian \inst{11}, and N.\,Roy\inst{12} \and V. Yanza\inst{8}}

\offprints{S. A. Dzib}

 \institute{IRAM, 300 rue de la piscine, 38406 Saint Martin d'H\`eres, France.
 \and
 Max-Planck-Institut f\"ur Radioastronomie (MPIfR), 
              Auf dem H\"ugel 69, 53121 Bonn, Germany
            \and
           Centre for Astrophysics and Planetary Science, University of Kent, Canterbury, CT2\,7NH, UK
        \and
           National Radio Astronomy Observatory,  520 Edgemont Road, Charlottesville, VA 22903, USA
           \and Instituto de Astronom\'{\i}a, Universidad Nacional Aut\'onoma de M\'exico (UNAM), Apdo Postal 70-264, M\'exico, D.F., Mexico
        \and McGill University, 3600 rue University, Montreal, QC, Canada H3A 2T8
        \and Max Planck Institute for Astronomy, Koenigstuhl 17, 69117 Heidelberg, Germany.
         \and Instituto de Radioastronom\'{i}a y Astrof\'{i}sica (IRyA), Universidad Nacional Aut\'{o}noma de M\'{e}xico  Morelia, 58089, M\'{e}xico.
         \and Laboratoire d'astrophysique de Bordeaux, Univ. Bordeaux, CNRS, B18N, all\'ee Geoffroy Saint-Hilaire, 33615 Pessac, France
         \and Physics Department, New Mexico Tech, 801 Leroy Place, Socorro, NM 87801, USA.
         \and Department of Earth and Space Science, Indian Institute for Space Science and Tecnology, Trivandrum 695547, India.     
         \and Department of Physics, Indian Institute of Science, Bangalore 560012, India
}

\date{Received / Accepted}
\authorrunning{S. A. Dzib et al.}


\abstract
{As part of the Global View on Star Formation (GLOSTAR) survey 
we have used the Karl G. Jansky Very Large Array (VLA) in its 
B-configuration to observe the 
part of the Galactic 
plane between   longitudes of  $28^\circ$ and
$36^\circ$ and  latitudes from $-1^\circ$ to $+1^\circ$ 
at the C-band (4--8 GHz). { To reduce the contamination of
extended sources that are not well recovered by our coverage of the 
($u,\,v$)-plane we discarded short baselines that are sensitive
to emission on angular scales $<4''$.} The resulting 
radio continuum images have an angular resolution of  ${1\rlap{.}''0}$, 
and a sensitivity of {${\sim60\,\mu}$Jy~beam$^{-1}$;}  making it the most
sensitive radio survey covering a large area of the Galactic
plane with this angular resolution. 
An automatic source extraction algorithm was used in combination with  
visual inspection to  
identify a total of { \Rblobs}  radio sources.
A total of {\HRC}  radio sources are  $\geq7\sigma$ and comprise 
our highly 
reliable catalog; {72}  of these are grouped as {22}  fragmented sources,
e.g., multiple components of an extended and resolved source.
To explore the nature
of the catalogued radio sources we searched for counterparts at millimeter 
and infrared wavelengths. Our classification attempts resulted in 
 93 \hii\ region candidates,  104  radio stars,  64  planetary nebulae, while most of the 
remaining radio sources are suggested to be extragalactic sources. 
We investigated the spectral indices ($\alpha$, $S_\nu\propto\nu^{\alpha}$)
of radio sources classified as \hii\ region candidates and found that many
have negative values. This may imply that these radio sources represent young  stellar 
objects that are members of the star clusters around the high mass stars that 
excite the \hii\ regions, but not these \hii\ regions 
themselves. By comparing the {peak flux densities} from the GLOSTAR 
and CORNISH surveys we have identified {49}  variable radio sources, 
most of them with an unknown nature. 
Additionally, we provide the list of {\LRC} radio sources detected within 5 to 7$\sigma$ levels.}


\maketitle

\section{\label{intro}Introduction}

The Global view on star formation (GLOSTAR) survey is presently the most 
sensitive radio survey ($\sim60\,\mu$Jy~beam$^{-1}$) of the northern 
hemisphere of the 
Galactic plane at the C-band (4 to 8 GHz) \citep{brunthaler2021,medina2019}. 
Taking full use of the capabilities of the Karl G. Jansky Very Large Array 
(VLA), a distinction of  GLOSTAR compared to previous  surveys is that it 
simultaneously observes  radio continuum and spectral line emission. 
GLOSTAR is indeed complementary to the wealth of Galactic plane surveys
at infrared and sub-millimeter wavelengths that address star formation in the Galaxy, 
some of which are described in subsection 3.7. 

\begin{figure*}
    \centering
    \includegraphics[width=1.01\textwidth, trim= 20 10 10 0, angle=0] {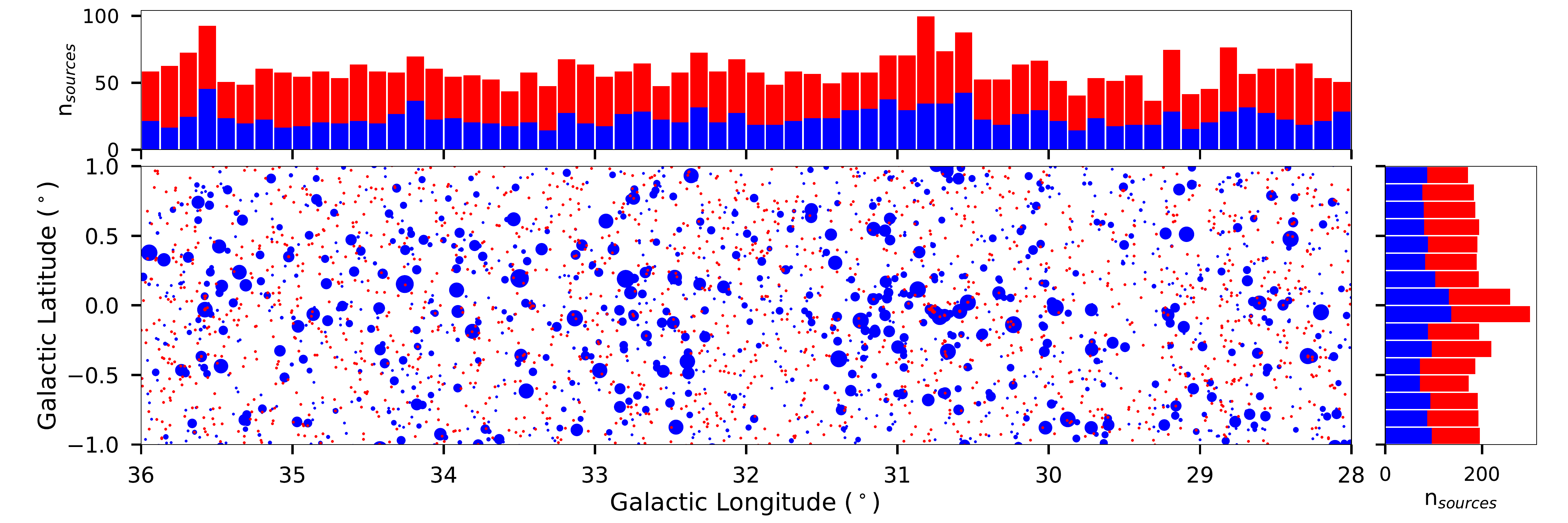}\\ 
    \caption{Spatial distribution of sources extracted from the VLA B-configuration images. Blue circles indicate the positions of highly reliable sources, their sizes are scaled (non-linear) with the flux intensity. Red points indicate the position of sources with 5  to 7$\sigma$ levels. Top and right panel are the histogram of the source distributions in the Galactic longitude and Galactic latitude, respectively.}
    \label{fig:Gspat}
\end{figure*}

The primary goal of the GLOSTAR survey is to localize signposts of massive star 
formation (MSF) activity
\citep[see][for a detailed overview of the survey]{brunthaler2021}.
Towards this goal, in the continuum mode, the survey mainly observes compact, ultra- and  
hyper-compact H{\small II} regions, that trace different early phases 
of MSF activity \citep[e.g.,][]{medina2019,nguyen2021}. Thus, the GLOSTAR 
survey complements previous radio surveys by providing a 
powerful and comprehensive radio-wavelength 
survey of the ionized gas in the Galactic Plane with an unprecedented sensitivity. 
However, the survey area is also populated with radio sources related 
to post-main sequence stars (e.g. Wolf-Rayet stars, pulsars, etc.), planetary nebulae, supernova
remnants and extragalactic radio sources
\citep[][]{medina2019,chakraborty2020,dokara2021}. In spectral line mode,
it traces the radio recombination lines from regions ionized by massive stars and the 
methanol maser line at 6.7~GHz \citep[][]{ortiz2021, nguyen2022}, 
both of which are related to massive star formation.
The formaldehyde absorption line at 4.8~GHz  is also observed. It traces neutral 
molecular gas and its {radial velocity information with respect to the  local standard of rest (LSR) radial}
can help to solve distance ambiguities. 
The final GLOSTAR images will cover the Galactic plane 
between Galactic longitudes, $\ell$, of  $-2^\circ$ and
$60^\circ$ and  latitudes, $b$, from $-1^\circ$ to $+1^\circ$  and the Cygnus X region. 
The final dataset will 
consist of low resolution images ($\sim20''$), using the VLA in the D 
configuration, and high resolution images {($\sim1\rlap{.}''0$)}, using  the B
configuration. {These VLA data sets could also be combined for optimal 
sensitivity of the intermediate spatial ranges, such images will be presented in future 
works.} The VLA observations will be complemented with very low
resolution ($\sim150''$) images of single dish observations from the
Effelsberg radio telescope, to recover the most extended emission 
and solve the ``missing short spacing'' issue affecting interferometer-only images.
{The overview of the full  GLOSTAR capabilities are described in detail by \cite{brunthaler2021}. }

Previously, we have reported the radio source catalog of the low resolution
VLA images  covering the area 
$28^\circ < \ell <  36^\circ$ and $-1^\circ < b < +1^\circ$ \citep{medina2019}.  
Complementary infrared- and sub-millimeter-wavelength data were examined towards the  radio source positions with the goal to elucidate the  nature of the radio sources. In this paper we report
the compact {($\sim1\rlap{.}''0$)}  radio sources from the same region using
the high resolution observations obtained with the VLA in B-configuration.
The interesting part of this new catalog is that we can identify
the most compact, and probably youngest, (hypercompact) HII regions.

\section{Observations}

The Karl G. Jansky Very Large Array (VLA) of the National Radio Astronomy Observatory\footnote{The 
National Radio Astronomy Observatory is a facility of the National 
Science Foundation operated under cooperative agreement by Associated 
Universities, Inc.} was used in its
B-configuration to observe the (4--8 GHz) continuum emission. 
We follow the same instrumental setups and calibration as for  the D-configuration data.
We refer the reader to the overview paper 
\citep{brunthaler2021} for a detailed description 
of the data management that we summarize in the following subsections.

\subsection{Observation Strategy}

The correlator setup consisted of two 1~GHz wide sub-bands, 
centered at 4.7 and 6.9~GHz. 
Each sub-band was divided into eight spectral windows of 128~MHz, 
and each spectral window comprising of 64~channels with a channel width 
of 2~MHz\footnote{Higher frequency resolution correlator windows 
were used to cover  prominent methanol maser emission 
line  at  6.7\,GHz, formaldehyde absorption at 4.8 GHz, and 
seven  radio  recombination  lines. The full results of these line
observations will be reported in forthcoming papers.}.
The chosen setup avoids strong persistent radio frequency 
interference (RFI) seen (most prominently) at 6.3 and 
4.1~GHz and allows estimation of source spectral indices.

For each epoch the total  observing time was five hours. At the beginning
of the observations, the amplitude/bandpass calibrator 3C~286 is 
observed for ten minutes. Then, the phase calibrator, 
J1804+0101, is observed for one minute, followed by pointings 
on target fields for eight minutes, after which the phase calibrator is 
observed for another minute. The observation cycle, consisting of 
phase calibrator--targets scans, is repeated over the full five hours. 
During this time, an area of $2^\circ\times1^\circ$ 
is covered with phase centers for 676 target fields 
in a semi-mosaic mode \citep[see ][ for a detailed overview of the observation strategy]{brunthaler2021}. 
Each pointing was observed twice for 11 seconds which, after 
considering the slewing time, yields a total integration time 
of 15 seconds per field. The theoretical noise level 
in brightness (or peak flux density) from these observations is 90~$\mu$Jy beam$^{-1}$ per field and per sub-band.
The noise is improved after combining the fields and both 
sub-bands by a factor of $\sim2$. We observed a total of eight 
epochs, or a total of 40 hours of telescope time, under project 
ID VLA/13A-334. The total covered area is 16 square degrees. The observations were taken during the period
from 2013 September to 2014 January. 


\begin{table*}
 \begin{center}
 \scriptsize
 \renewcommand{\arraystretch}{1.1}
\setlength{\tabcolsep}{3.2pt}
 \caption{GLOSTAR B-configuration catalog for $28\degr < \ell < 36\degr$ and $|b| < 1$\degr. }
 \label{tbl:glostar_B} \begin{minipage}{\linewidth}
\begin{tabular}{ccccccccccccccccccccc} 
 \hline \hline
 GLOSTAR B-conf.& $\ell$ & $b$&SNR &$S_{\rm peak}$ & $\boldsymbol{\sigma}_\mathbf{S_{\rm peak}}$ &$S_{\rm int}$ &$\mathbf{\boldsymbol{\sigma}_{S_{\rm int}}}$& $Y$ & $R_{\rm eff}$ & $\alpha$  & GLOSTAR D-conf. & \multicolumn{3}{c}{Infrared counterpart} & Sub-mm & Classification \\ 
{name}& (\degr)&(\degr)& &\multicolumn{2}{c}{(mJy\,beam$^{-1}$)} &\multicolumn{2}{c}{(mJy)} & &(\arcsec)&  &name& NIR&MIR&FIR&counterpart& \\ 
(1)&(2)&(3)&(4)&(5)&(6)&(7)&(8)&(9)&(10)&(11)&(12)&(13)&(14)&(15)&(16)&(17)\\ 
 \hline
G028.0014+00.0567 & 28.00144 & +0.05670 & 15.6 & 0.90 & 0.08 & 0.86 & 0.07 & 0.95 & 0.8 & $+0.28\pm0.24$ & G028.002+00.057 &&&&&EgC\\
G028.0050+00.1497 & 28.00503 & +0.14967 & 10.2 & 0.60 & 0.07 & 0.66 & 0.07 & 1.09 & 0.7 & $-0.50\pm0.60$ & G028.005+00.150 &&&&&EgC\\
G028.0065-00.9904 & 28.00648 & -0.99038 & 35.6 & 2.49 & 0.15 & 2.48 & 0.14 & 1.00 & 0.9 & $-0.24\pm0.06$ & G028.007-00.990 &\checkmark &&&&EgC\\
G028.0101-00.3032 & 28.01014 & -0.30323 & 12.6 & 0.69 & 0.07 & 0.72 & 0.07 & 1.04 & 0.8 & $-0.11\pm0.36$ & G028.010-00.303 &&&&&EgC\\
G028.0239+00.5132 & 28.02391 & +0.51318 & 13.8 & 0.92 & 0.08 & 0.97 & 0.08 & 1.05 & 0.8 & $+0.34\pm0.35$ & G028.024+00.514 &&&&&EgC\\
G028.0309+00.6677 & 28.03085 & +0.66769 & 08.2 & 0.54 & 0.07 & 0.62 & 0.07 & 1.13 & 0.7 &       ...      & G028.031+00.669 &&\checkmark &&&EgC\\
G028.0314-00.0726 & 28.03136 & -0.07260 & 31.9 & 2.03 & 0.13 & 2.44 & 0.14 & 1.20 & 1.0 & $-0.79\pm0.14$ &       ...      & &&&&EgC\\
G028.0350+00.3898 & 28.03495 & +0.38978 & 10.2 & 0.69 & 0.08 & 0.70 & 0.08 & 1.01 & 0.7 & $-0.59\pm0.37$ &       ...      & &&&&EgC\\
G028.0403-00.2452 & 28.04027 & -0.24519 & 09.2 & 0.56 & 0.07 & 0.52 & 0.07 & 0.94 & 0.7 &       ...      &       ...      & &&&&EgC\\
G028.0474-00.9892 & 28.04741 & -0.98918 & 23.2 & 1.86 & 0.13 & 2.10 & 0.13 & 1.13 & 0.9 & $-0.33\pm0.25$ & G028.048-00.989 &&&&&EgC\\
G028.0489-00.8699 & 28.04894 & -0.86986 & 16.2 & 1.03 & 0.08 & 1.05 & 0.08 & 1.02 & 0.8 & $+0.45\pm0.25$ & G028.049-00.869 &&&&&EgC\\
G028.0594+00.1197 & 28.05937 & +0.11975 & 11.5 & 0.84 & 0.09 & 0.84 & 0.08 & 1.00 & 0.7 & $+0.36\pm0.46$ & G028.065+00.119 &&&&&EgC\\
G028.0654+00.1184 & 28.06544 & +0.11842 & 25.1 & 1.77 & 0.12 & 2.01 & 0.12 & 1.14 & 1.0 & $-0.71\pm0.26$ &       ...      & &&&&EgC\\
G028.0749-00.2953 & 28.07494 & -0.29530 & 19.1 & 1.28 & 0.10 & 1.42 & 0.10 & 1.11 & 0.9 & $-0.88\pm0.27$ & G028.075-00.296 &\checkmark &&&&EgC\\
G028.0850-00.5583 & 28.08501 & -0.55826 & 28.4 & 1.69 & 0.11 & 1.76 & 0.11 & 1.04 & 0.9 & $-0.34\pm0.16$ & G028.085-00.558 &\checkmark &&&&EgC\\
\hline
 \end{tabular}
  Notes: Only a small portion of the data is provided here, the full table is available in electronic form at the CDS via anonymous ftp to cdsarc.u-strasbg.fr (130.79.125.5) or via http://cdsweb.u-strasbg.fr/cgi-bin/qcat?J/A\&A/.\\
  Classification (17) see also section 3.8: EgC = Extragalctic radio source candidate, HII = HII region candidate, Radio-star, PN = planetary nebula, WR = Wolf-Rayet star, PDR = Photodissociation region, Cataclysmic variable, Pulsar, and Unclear = source with no clear classification.  
 \end{minipage} 
 \end{center}
 \end{table*}

 \begin{table}[!hb]
 \small
 \begin{center}
 \renewcommand{\arraystretch}{1.1}
\setlength{\tabcolsep}{2.8pt}
 \caption{GLOSTAR D-configuration sources that are
 detected as fragmented sources {(e.g., multiple 
 components of an extended and resolved source)} in the B-configuration
 images.}
 \label{tbl:glostar_F} 
\begin{tabular}{ccccccc} 
 \hline \hline
  GLOSTAR B-conf.&GLOSTAR D-conf.& \# & $S_{\rm int}$ & Class \\ 
{name}         &{name}         & frag.&{(mJy)}       & \\ 
(1)&(2)&(3)&(4)&(5)\\ 
 \hline 
G028.2440+000.0143 & G028.245+00.013  & 2 & 01.73 & HII\\
G028.4518+000.0029 & G028.451+00.003 & 3 & 12.98 & HII\\
G028.6520+000.0278 & G028.652+00.027 & 3 & 12.23 & HII\\
G028.6871+000.1771 & G028.688+00.177 & 2 & 10.48 & HII\\
G029.9558-000.0167 & W43 south-center & 5 & 375.51 & HII\\
G030.2529+000.0538 & G030.253+00.053 & 3 & 06.31 & HII\\
G030.9582+000.0868 & G030.955+00.081 & 3 & 06.61 & HII\\
G031.0695+000.0506 & G031.069+00.050 & 2 & 09.11 & HII\\
G031.2435-000.1103 & G031.243-00.110 & 4 & 184.42 & HII\\
G031.2797+000.0623 & G031.279+00.063 & 4 & 10.15 & HII\\
G031.4118+000.3064 & G031.412+00.307 & 3 & 84.14 & HII\\
G032.1501+000.1335 & G032.151+00.133 & 5 & 27.41 & HII\\
G032.2726-000.2257 & G032.272-00.226 & 3 & 12.81 & HII\\
G032.7965+000.1909 & G032.798+00.191 & 2 & 574.36 & HII\\
G032.9272+000.6060 & G032.928+00.606 & 2 & 71.99 & HII\\
G033.9145+000.1104 & G033.914+00.110 & 3 & 106.65 & HII\\
G034.1321+000.4720 & G034.133+00.471 & 8 & 23.45 & HII\\
G034.2571+000.1533 & G034.260+00.125 & 3 & 733.11 & HII\\
G034.8624-000.0630 & G034.862-00.063 & 2 & 34.16 & PN\\
G035.0521-000.5178 & G035.052-00.518 & 2 & 08.62 & HII\\
G035.4666+000.1393 & G035.467+00.139 & 2 & 35.17 & HII\\
G035.5641-000.4909 & G035.564-00.492 & 4 & 07.93 & PN\\

\hline
 \end{tabular}\\
\begin{minipage}{0.950\linewidth}
 Notes: Column (1) gives the name of the brightest fragment.\\ 
  \end{minipage}
 \end{center}
 \end{table}

\subsection{Calibration, Data Reduction and Imaging}

The data were calibrated, edited and imaged using the {\it Obit} 
software environment \citep{cotton2008}, which inter-operates with the
classic Astronomical Image Processing Software package (AIPS) \citep{greisen2003}. We have written 
calibration scripts that handle the GLOSTAR data edition and calibration 
\citep{brunthaler2021}. The calibration 
follows the standard procedures of editing and calibrating interferometric data.
These include 
bandpass, amplitude and phase calibration.

The calibrated data were imaged using the Obit task MFImage.
The CLEANing process from MFImage divides the observed band into 
nine frequency bins, that are narrow enough to perform a spectral 
deconvolution, addressing the effects of variable spectral index
and antenna pattern variations. In the end, we obtained an image representing the data for
the full frequency band and images of each individual frequency bin. The images 
were obtained from the data that had projected baselines with  
{distances in the ${(u,\,v)}$-plane} larger than 50~k$\lambda$;
i.e., discarding all radio {emission on angular scales} larger than {$\approx 4\rlap{.}''13$ at all observed frequencies.}
This choice rejects emission from 
poorly mapped extended structures that introduced artifacts in the images, with a minor impact in the overall sensitivity. 
The images were convolved with a circular beam 
 ${1\rlap{.}''0}$ of size, and with pixel size of $0\rlap{.}''25$. 
The mosaics are constructed to obtain images {of ${35000\times30000}$~pixels, 
containing the  ${1^\circ\times2^{\circ}}$ angular area surveyed by epoch},
following the schemes described by \citet{brunthaler2021}. The mean measured 
noise in the resulting images is {${60\,\mu}$Jy beam$^{-1}$;}  though it can be significantly higher 
in some areas in which  extended emission was not properly recovered or around 
very bright radio sources that produce imperfectly cleaned sidelobe emission.

\section{Catalog construction}

In this section we discuss the procedure used for constructing the
catalogues presented in this work. First, we have extracted the
sources from the images, selected the sources that are real, identified
candidate radio sources, and discarded image artifacts.
We have investigated the astrometry, {flux density} ($S_\nu$), spectral 
indices ($\alpha$; $S_\nu\propto\nu^\alpha$), and have
searched for counterparts at other wavelengths. Based on the  counterpart information,
we attempted a classification of the radio sources.
The final catalog is presented in Table~\ref{tbl:glostar_B}, with the 
reliable radio sources, their properties and the source classification.

\subsection{Source extraction} 

The source extraction was performed following the procedures described
by \cite{medina2019}, and we refer the reader to
that work for the details, while here we will give a brief summary.
Using the SExtractor \citep{bertin1996} tool from the Graphical Astronomy 
and Image Analysis Tool package
(GAIA\footnote{\url{http://starlink.rl.ac.uk/star/docs/sun214.htx/sun214.html}}), 
we first create 
a noise image. Then, we have used the \blobcat\ package \citep{hales2012} 
to extract the sources from each of the final images. Both the
intensity map and the noise map are used as inputs by \blobcat.
This package recognizes islands 
of pixels (blobs)  representing sources with a minimum peak flux above $N$ times 
the noise level in the area.  \citet{purcell2013} noted that with  $N < 4.5$,
large radio images will be dominated by spurious sources. Thus, to diminish the number 
of spurious detections in our catalog, for our extraction 
we have initially defined $N=5$ and require blobs to consist of a minimum 
of {12}  contiguous pixels. The minimum number of pixels was chosen to be the number
of expected pixels in  50\% of the beam area.
The final number of extracted blobs is {\fstblobs} . 
After this first extraction we performed a visual inspection of 
all the blobs 
to identify clear image artifacts, such as sidelobes from very bright radio sources, 
and discarded {\Spblobs}  blobs. The spatial distribution of all the blobs, excluding the artifacts, 
is shown in Figure~\ref{fig:Gspat}. { We detected a total of \HRC blobs with a signal-to-noise ratio (S/N)  $\geq7.0$ and \LRC\, blobs with $5.0\leq {\rm S/N}< 7.0$.}

It has been determined that in large radio surveys such as GLOSTAR the most reliable 
sources are those with peak flux values above $7\sigma_{\rm noise}$ as no spurious sources
are expected at these levels \citep[e.g., ][]{purcell2013,bihr2016,wang2018,medina2019}. 
However, as some blobs with a brightness between 5--7$\sigma_{\rm noise}$ could represent real sources, we have 
searched for counterparts inside a radius of $2''$ in the SIMBAD astronomical 
database\footnote{\url{http://simbad.u-strasbg.fr/simbad/}} for all sources 
(see details below). This comprehensive database is the best option to look for counterparts 
at any wavelength in large areas of the sky but, admittedly, it can miss recent 
catalogs. We have also considered weak blobs (peak flux values $<7\sigma_{\rm noise}$) 
as real radio sources whose positions are consistent 
with the radio sources from our D-configuration catalog by \citet{medina2019}. 
The full number of { weak blobs}  with a known counterpart in the SIMBAD database and
our D-configuration catalog is {142}. The remaining { 1724} 
radio blobs  having a S/N between 5 and 7 and have no known counterpart are referred to as candidate radio source detections. We give the list of 
these { \LRC}  sources with S/N$<7$ in Appendix A (Table \ref{tbl:glostar_can})
labeling those with SIMBAD or D-configuration counterparts, and we do not  
analyze them further.

We consider the remaining {\HRC}  blobs as highly reliable radio sources. They are
represented by the
blue circles in Figure~\ref{fig:Gspat} and are listed in Table~\ref{tbl:glostar_B}. 
The position, 
the SNR, the peak and the integrated flux density values (columns (2) to (8) in 
Table~\ref{tbl:glostar_B}) are taken from the values determined by the 
\blobcat\ software \citep{hales2012}. From now on, this paper will focus on the analysis of these sources. Considering the ratio between 
the integrated flux density (in units of Jy) and the peak flux density (in units of Jy beam$^{-1}$ (here named as the Y-factor; 
$Y = S_{\nu, {\rm Int}}/S_{\nu, {\rm Peak}}$)
we can divide these sources into extended  ($Y >2.0$), compact 
($1.1<Y \leq2.0$) and point-like  ($Y \leq$1.1) sources. 
The $Y$ factor of each source is listed in column (9) of Table~\ref{tbl:glostar_B}. 
Using this classification we obtain { \Elike}  extended, 
{ \Clike}  compact, and {\Plike}  point-like sources. As expected for these images, 
the sources are dominated by compact and point-like sources as we have rejected 
extended structures in our imaging process.

\subsection{Astrometry}\label{sec:astrometry}

For  our previous catalog based on the VLA D-configuration observations we have 
estimated that the accuracy of the astrometry is of the order of $1''$.  
The position errors of the extracted sources with \blobcat\ in 
the VLA B-configuration images have a mean value
of  ${0\rlap{.}''06}$  in both  Galactic longitude and latitude. The ${0\rlap{.}''06}$ 
is a formal statistical error estimate  
from \blobcat\ and does not reflect position errors resulting from imperfect phase calibration. 
As most of the observed radio sources are expected to be background extragalactic
objects, the proper motion of these sources is expected to be zero.
The only other recent Galactic plane survey that observed the same region
at a similar radio frequency
is the Co-Ordinated Radio 'N' Infrared Survey for High-mass star formation 
(CORNISH; \citealt{hoare2012, purcell2013}). The observations of CORNISH were obtained at 5\,GHz
using the VLA in its B-configuration. Because of the similarity of the frequency 
and the use of the same VLA array configuration the angular resolution of CORNISH is ${1\rlap{.}''5}$, {similar}  as that of the \glo\,
survey B array data discussed here ($1\rlap{.}''0$).

\begin{figure}
    \centering
    \includegraphics[width=0.45\textwidth, trim= 20 30 0 0, angle=0] {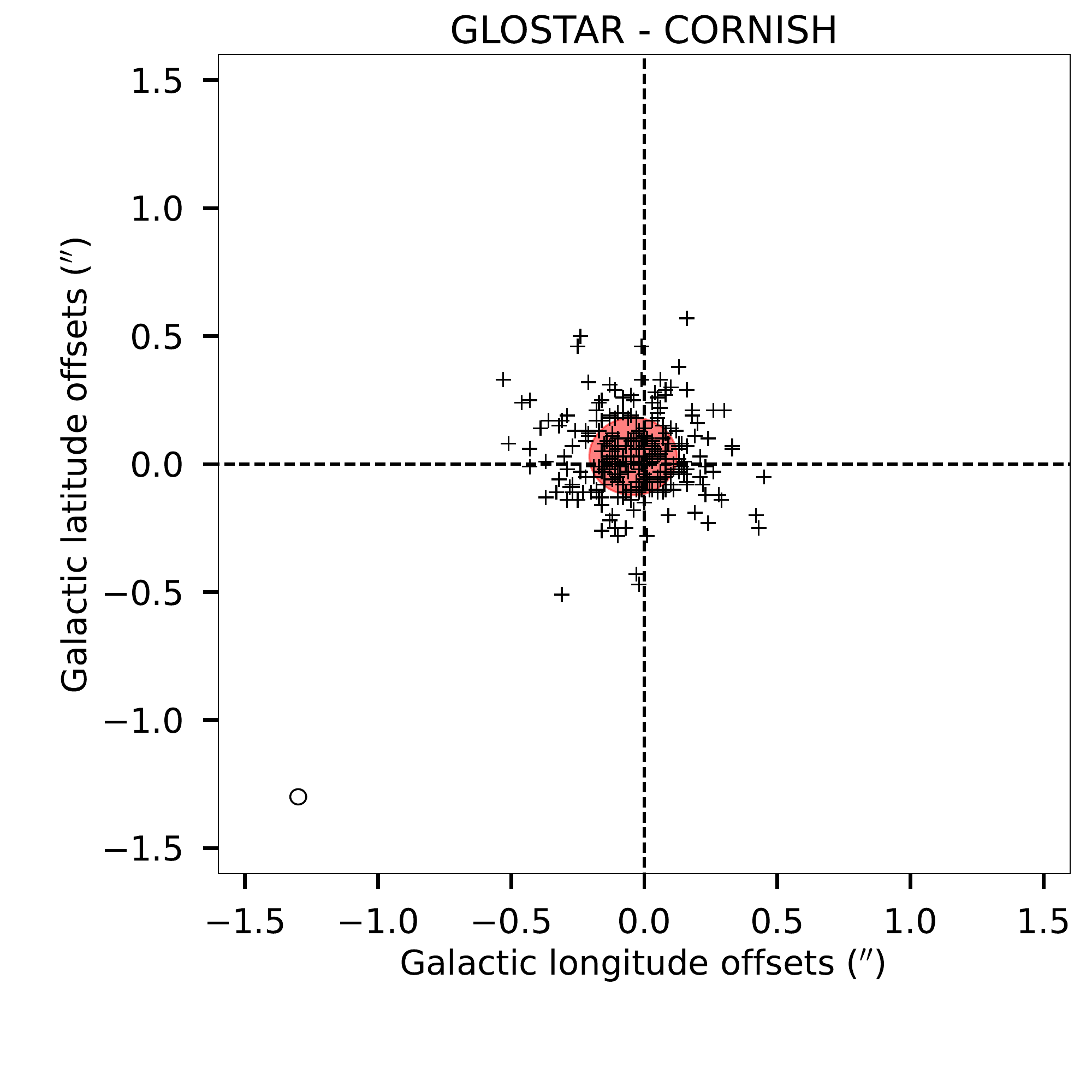}\\ 
    \includegraphics[width=0.45\textwidth, trim= 20 30 0 0, angle=0] {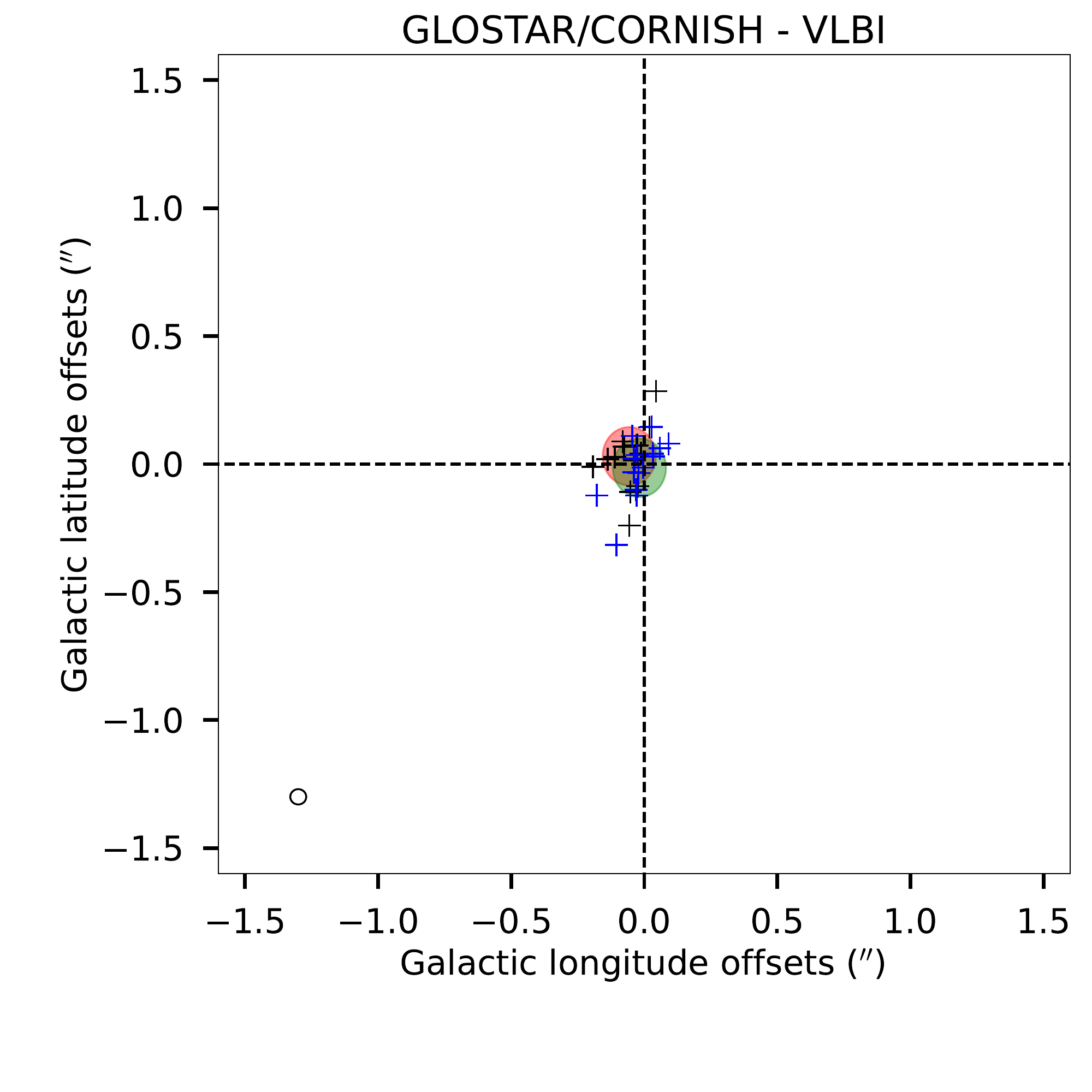}
    \caption{{\it Upper panel:} Position offsets between GLOSTAR compact and point-like sources ($Y\leq\,2$) and CORNISH sources. Filled red ellipse is centered in the mean values of the offsets and the size of the semi-axes are equal to the standard deviations. {\it Lower panel:} Similar as the top panel, but for position offsets of
    GLOSTAR (black crosses, red filled ellipse) and CORNISH (blue crosses, green filled ellipse) sources between the measured positions and the  VLBI positions.  Open black circles in the bottom-left of both panels indicate the mean of position errors of GLOSTAR-B sources.}
    \label{fig:GaVaC}
\end{figure}

A total of {257}  compact and point-like sources are listed in \textit{both} the GLOSTAR-B and CORNISH catalogs \citep{purcell2013} with a maximum angular separation of
$1\rlap{.}''5$, {the CORNISH angular resolution.} In the upper panel of Figure~\ref{fig:GaVaC} 
the measured offsets of these sources are plotted.
We obtain mean and standard deviation for the  position offsets of ${-0\rlap{.}''04\pm0\rlap{.}''01}$ 
and $0\rlap{.}''17$ in Galactic longitude direction, and ${0\rlap{.}''03\pm0\rlap{.}''01}$ and ${0\rlap{.}''15}$ in Galactic latitude direction. {The mean offsets in both directions are smaller than the mean position error
of GLOSTAR radio sources, suggesting a good astrometry.} 

The most accurate positions of radio sources are obtained with the Very Long Baseline
Interferometry (VLBI) technique. The Radio Fundamental catalog of extragalactic radio
sources compiles the positions of $\sim19,000$  sources
that have been measured with VLBI\footnote{The full Radio Fundamental 
catalog can be accessed via the webpage \url{http://astrogeo.org/}.}. We found in 
this catalog 15 sources detected both by 
GLOSTAR and CORNISH. In the lower panel of Figure~\ref{fig:GaVaC} 
we show the offsets between the  GLOSTAR and the VLBI positions as black
crosses, and the offsets between the  CORNISH and VLBI positions as 
blue crosses. The mean GLOSTAR$-$VLBI position offset and its standard deviation is
${-0\rlap{.}''05\pm0\rlap{.}''02}$ and $0\rlap{.}''06$, respectively,  in Galactic longitude direction, 
and $0\rlap{.}''03\pm0\rlap{.}''03$ and $0\rlap{.}''11$, respectively,  in Galactic latitude direction.
On the other hand, the mean CORNISH$-$VLBI position offset and its standard deviation is
$-0\rlap{.}''02\pm0\rlap{.}''02$ and $0\rlap{.}''07$, respectively, in the Galactic longitude direction, and
$-0\rlap{.}''02\pm0\rlap{.}''03$ and $0\rlap{.}''10$, respectively,  in the Galactic latitude direction. { The conclusion from this analysis is that the astrometry of the 
B-configuration GLOSTAR images presented in this paper is accurate to better than $0\rlap{.}''1$.}

\subsection{Comparison with the D-configuration catalog}

\begin{figure}[!h]
    \centering
    \includegraphics[width=0.44\textwidth, trim= 20 30 0 0, angle=0] {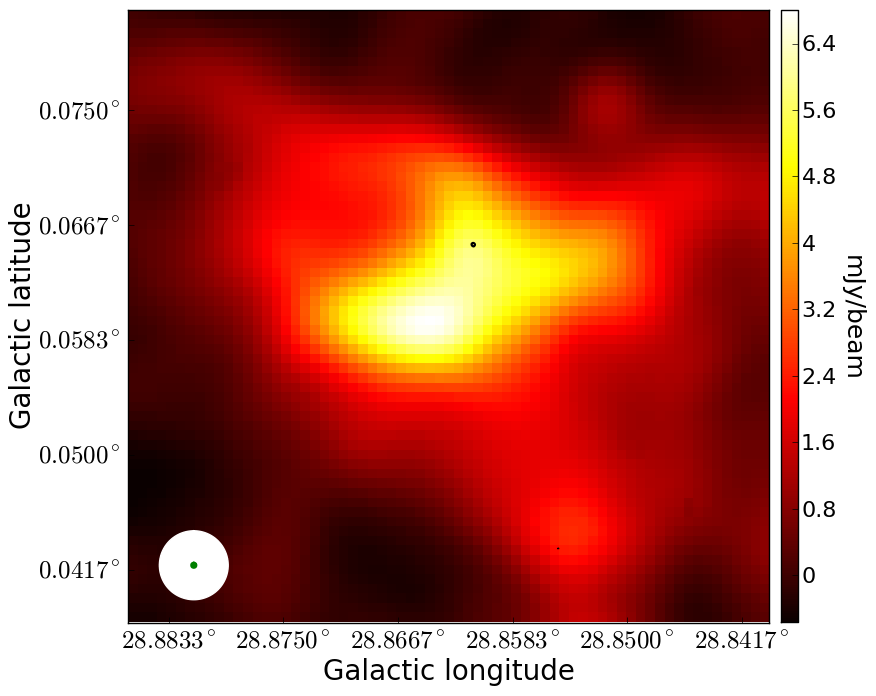} \\
    \includegraphics[width=0.44\textwidth, trim= 20 30 0 0, angle=0] {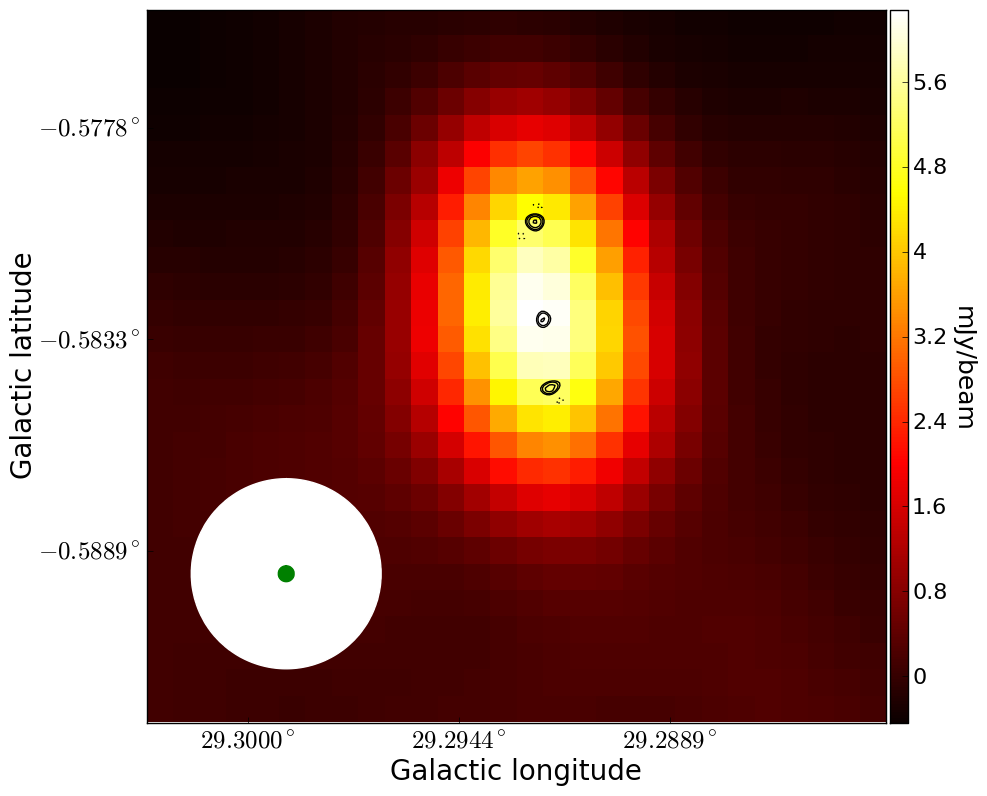} \\
    \includegraphics[width=0.44\textwidth, trim= 20 0 0 0, angle=0] {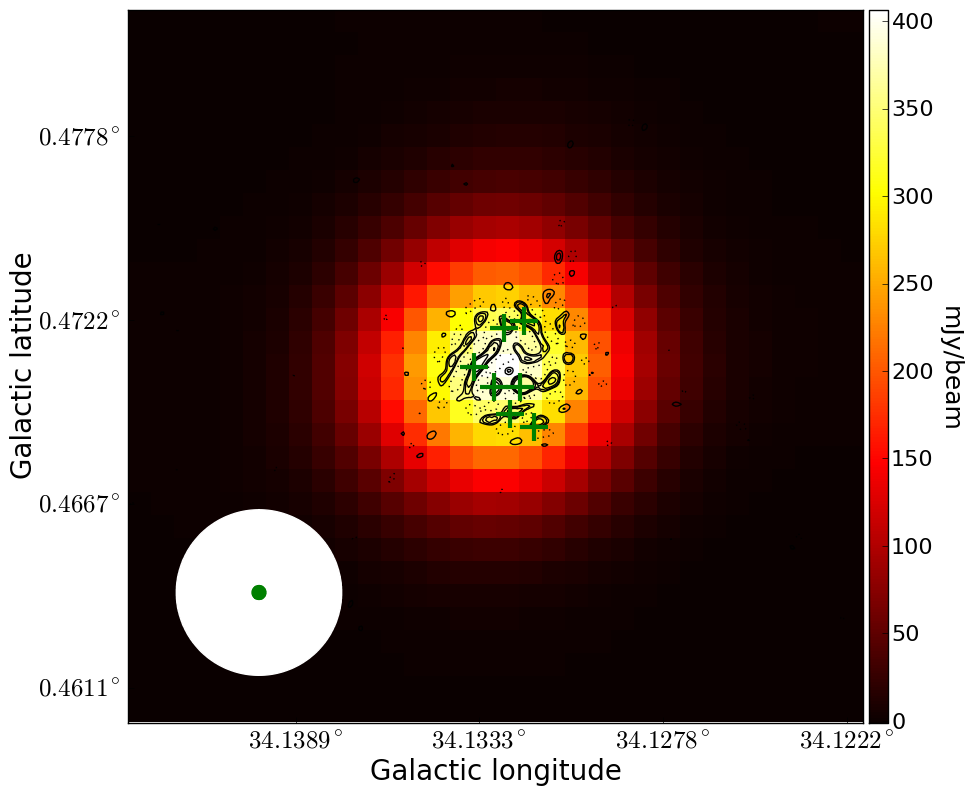}
    \caption{Source examples of the differences between the two sets of VLA images, D-configuration ({\it background}) and 
    B-configuration ({\it contours}). Contour levels are $-5$, 5, 7, 10 and 15 times 60~$\mu$Jy beam$^{-1}$, the mean noise level of the images. Circles at the bottom-left side indicate the beam sizes of D-configuration ($18''$, white) and B-configuration ($1\rlap{.}''5$, green) images. {\it Top: } B-configuration compact radio source seen in projection in the area of a D-configuration extended radio source. {\it Middle: } D-configuration compact source that is resolved in three individual compact radio sources in the B-configuration images. {\it Bottom:} D-configuration compact source that is resolved as a fragmented radio source. The position of fragments considered in the final catalog are indicated with green crosses.  }
    \label{fig:DB}
\end{figure}

In the first GLOSTAR catalog, we have reported a total of 1575 discrete 
sources detected in the same region presented in this paper, but based 
on data obtained with  the VLA in its most compact (D) configuration.
The GLOSTAR images from the VLA D-configuration observations
have an angular resolution of $18''$, i.e., 18 times larger than the 
VLA B-configuration images presented in this paper,  which also excluded 
the shortest baselines to further filter out extended emission, resulting in some 
expected differences. First, given the higher angular resolution observations of
the B-configuration, the extended radio sources reported by \citet{medina2019}
are resolved out and, in some cases, only the brightest peaks of extended sources are
detected. It is worth noting that 
some compact sources detected in the B-configuration images can be seen 
projected on the area of the extended 
sources, although they do not represent their direct counterpart. Their 
possible relation must be  studied further in future (e.g., upper-panel 
of Figure~\ref{fig:DB}). Second, some multiple
component radio sources that are unresolved or slightly resolved in the
D-configuration images will be resolved in the B-configuration images 
(middle-panel of Figure~\ref{fig:DB})
and the {integrated flux densities} of the individual components can be estimated. Third, 
some individual radio sources detected in the D-configuration images 
can be resolved and appear as fragmented radio sources in the B-configuration
images (lower-panel of Figure~\ref{fig:DB}). 
In total, we have found that { 95}  sources in the
D-configuration images are resolved into { 224}  B-configuration 
sources (see column~(12) of Table~\ref{tbl:glostar_B}). 

The components of fragmented radio sources are grouped and treated as 
a single source. However, the information of each single component is given 
in Table~\ref{tbl:glostar_B} and the fragmented sources are listed 
in Table~\ref{tbl:glostar_F}. The { integrated flux density} reported in Table~\ref{tbl:glostar_F} 
is obtained by adding the { integrated flux densities} of the individual fragments. 
In total, 72 sources recovered from \blobcat\ are grouped into 22 fragmented sources.

Other sources were detected as single compact sources 
in both D- and B-configuration images and can be considered as direct 
counterparts.
The mean position error of D-configuration sources is $1\rlap{.}''2$ 
\citep{medina2019} and the beam size of the B-configuration images is 
${1\rlap{.}''0}$. Thus, by adding these values in quadrature we used
a maximum angular separation of $2''$ between sources in both catalogs
to consider them as direct counterparts. The number of matching sources 
between both catalogs  with this criterion is {372}. {
A further 312 matching sources are found using an angular separation of $9''$, 
half of the angular resolution of the D-configuration images, for which 
the association must be investigated further.}
Given the differences described above, 
the matching of sources between both catalogs is not expected to be 
one-to-one.

\begin{figure*}
    \centering
    \includegraphics[width=1.01\textwidth, trim= 20 10 10 0, angle=0] {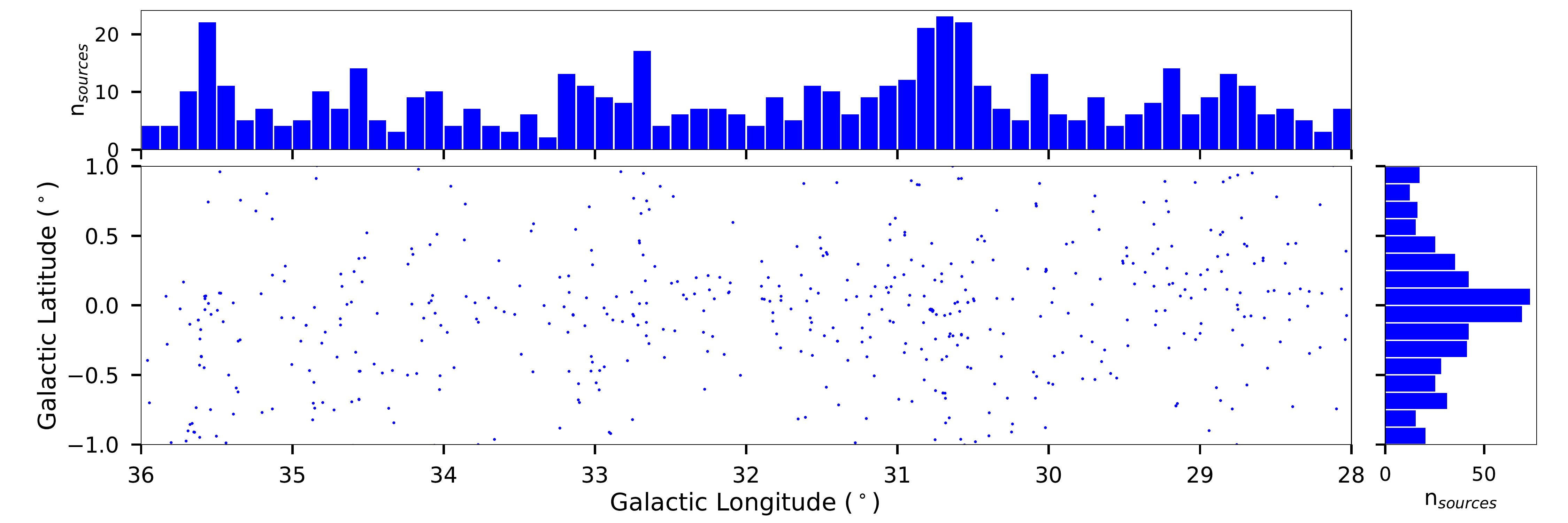}\\ 
    \caption{Spatial distribution of radio sources detected in the B-configuration images that have no counterparts in the D-configuration images. }
    \label{fig:Gspat_D}
\end{figure*}

In column (12) of Table~\ref{tbl:glostar_B} we list the GLOSTAR 
D-configuration name to which the B-configuration source is related.
With the D-configuration name we have also labeled those sources that
are related to two or more B-configuration sources and if they are considered
as individual (I) or fragmented (F) sources.
In total, { 908}  B-configuration sources were related to {780}  D-configuration 
sources. The remaining { 551}  B-configuration sources have no counterpart in
the D-configuration catalog. Most of these sources are located in the inner
parts of the Galactic plane (see Figure~\ref{fig:Gspat_D}) where the noise 
level is higher in the D-configuration images because of the bright and 
extended radio sources  \citep[see the lower panel of Figure 1 in][]{medina2019}.
As the noise levels could be as high as $500~\mu$Jy~beam$^{-1}$, it explains 
why most of the sources were not detected in the D-configuration images, 
but are detected in the B-configuration images where the noise level is 
about 10 times lower.

\subsection{Source sizes}

\begin{figure}
    \centering
    \includegraphics[width=0.45\textwidth, trim= 10 0 0 0, angle=0] {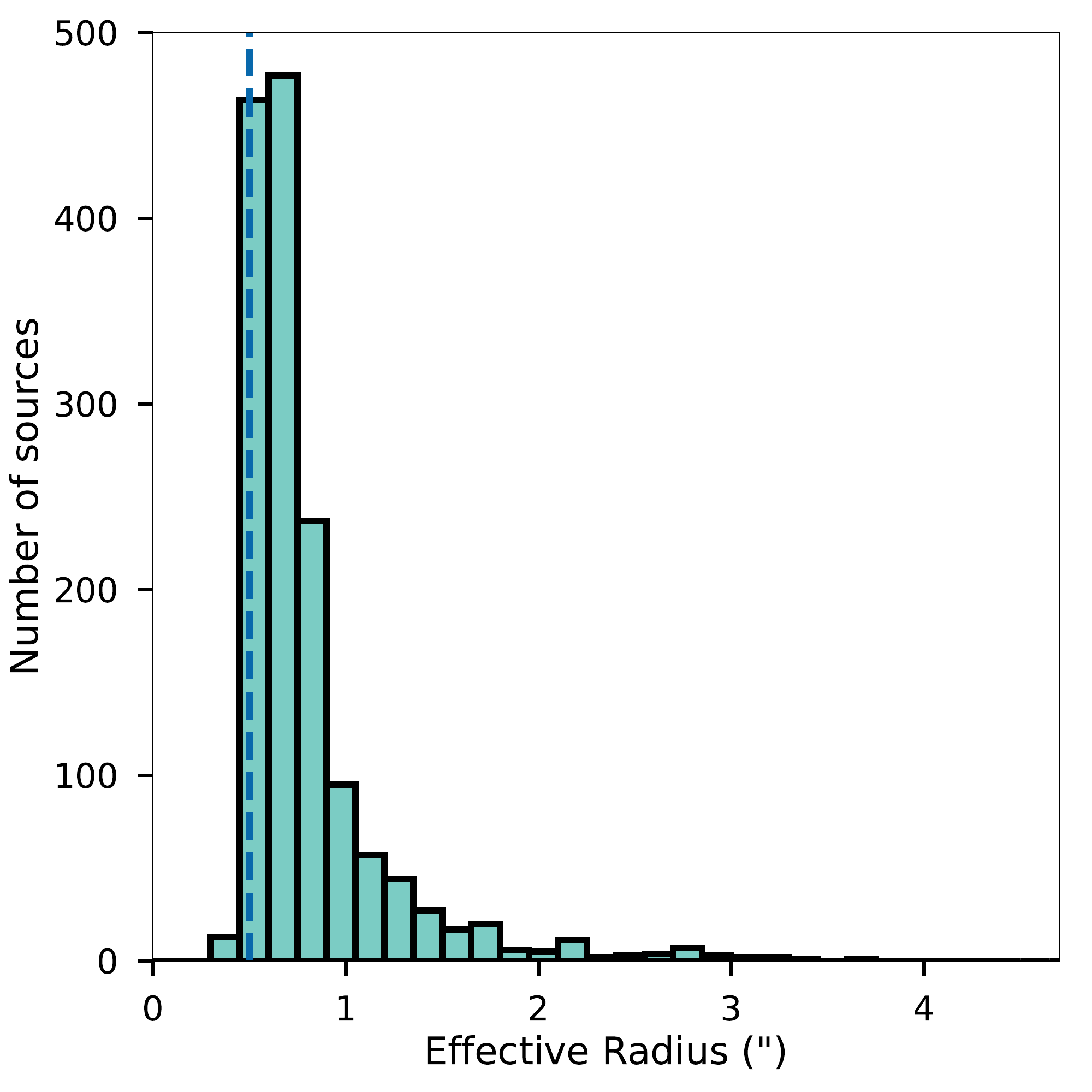}
    \caption{Distribution of the source effective radius. The dashed-dotted line indicates the resolution of the observations (half beam width). { Note that the images are only sensitive to angular scales ${\leq4\rlap{.}''1}$. } The bin size used is $0\rlap{.}''15$. }
    \label{fig:erd}
\end{figure}

The source sizes are obtained following \citet{medina2019}, who determined 
the source effective radius. \blobcat\, determined the number of pixels 
comprising each source, which can be used to estimate the area ($A$) covered by 
the source using the pixel size of $0\rlap{.}''25\times0\rlap{.}''25$.
Then the effective radius can be determined using 
$$R_{\rm eff}=\sqrt{\frac{A}{\pi}}.$$
\noindent The effective radius distribution is shown in Figure~\ref{fig:erd}, 
and the value for each source is listed in column (10) of Table~\ref{tbl:glostar_B}.

\subsection{Flux densities}

To estimate the reliability of the radio source flux densities determined from the \glo\ images, we compare the results from the \blobcat\ extraction with those in the CORNISH catalog.
As most of the sources are expected to be extragalactic sources whose variability
is expected to be low (typically only of a few percent over timescales of several years) these provide a good
point of comparison. We compare sources that are point like, and thus used 
the { peak flux density}. Moreover, we have only compared sources 
whose { peak flux densities} are above 2.7 mJy in both catalogs, as this is the 7$\sigma$ base 
point in the high reliability source catalog of CORNISH  \citep{purcell2013}. 
We found { 207}  sources that meet these criteria.
A difference between both catalogs is the mean observed frequency. CORNISH 
observed at 5~GHz, and GLOSTAR observed a wider bandwidth centered at 5.8~GHz.
The measured {flux density} thus is expected to differ slightly for this observational mismatch 
due to the spectral index.
Assuming that the extragalactic objects have a mean spectral index of 
$-0.7$ \citep{condon1984}, the CORNISH flux values will be on average 10\% 
higher than in GLOSTAR.

\begin{figure}[!h]
    \centering
    \quad \includegraphics[width=0.5\textwidth, trim= 10 0 0 0,clip, angle=0] {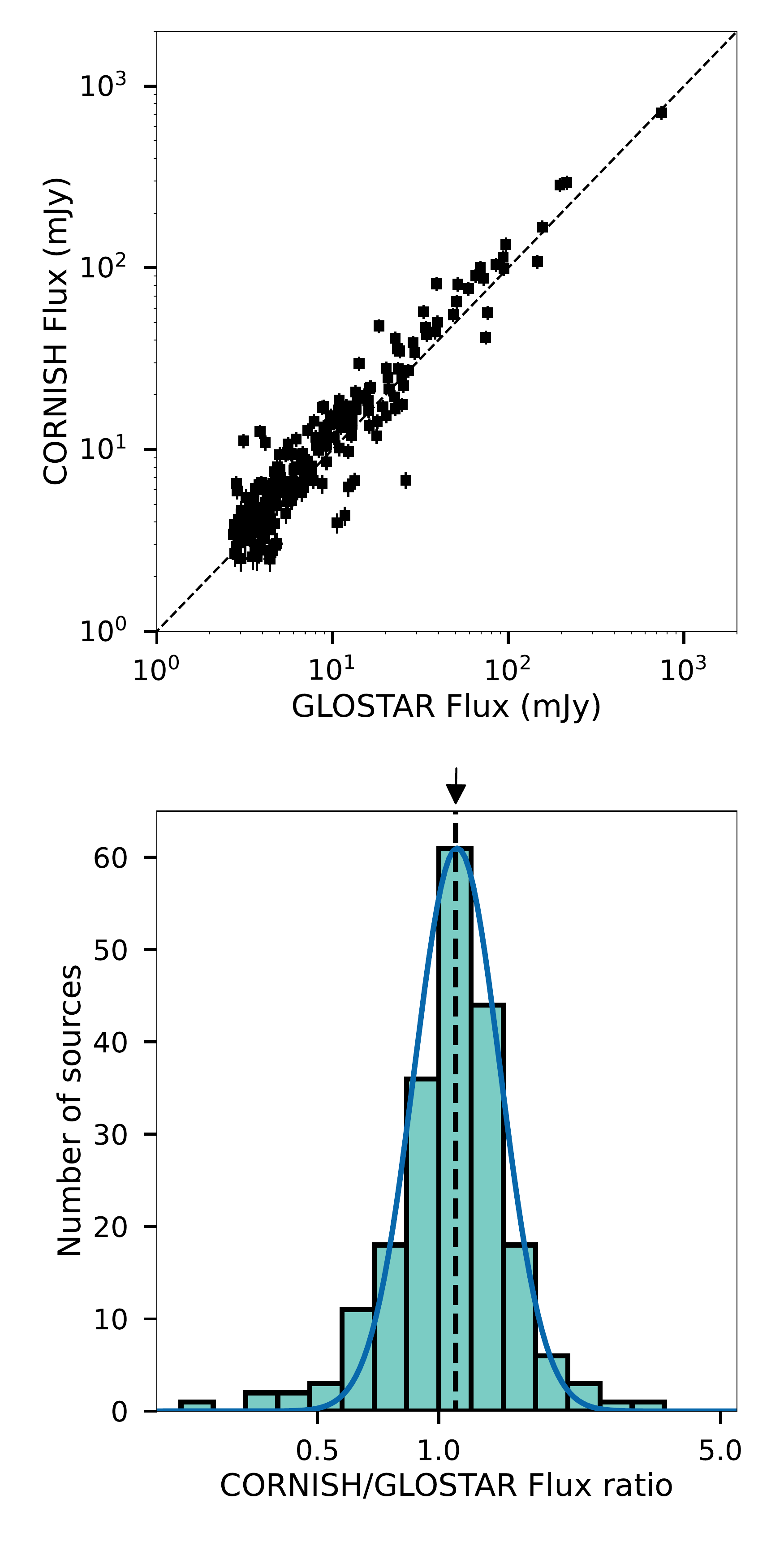}
    \caption{{\it Upper panel:} Comparison of the {peak flux densities} of radio sources detected in GLOSTAR-B and CORNISH. 
    {\it Lower panel:} Histogram of the ratio between the {peak flux densities} of CORNISH and GLOSTAR. The width of the bars is 0.08 dex. The blue line indicates the model from the Gaussian fit to the histogram. {The dashed line indicates the mean value of the distribution. The small arrow at the top of the plot show the expected CORNISH/GLOSTAR flux density ratio for the different observed mean frequency. }}
    \label{fig:FGaC}
\end{figure}

In the upper panel of Figure~\ref{fig:FGaC} we plot the {peak flux densities} measured by 
the CORNISH survey as a function of the {peak flux densities} measured in this work. 
The dashed line indicates the equality line, and most of the sources are around 
this line. The lower panel of this figure shows the distribution of the {peak flux density} 
ratio between the results from both catalogs. A Gaussian fit to the distribution
indicates that the mean value is {${1.11\pm0.03}$} with a standard deviation of { 1.28}. 
Considering the expected higher values in the CORNISH catalog, we conclude that
the { integrated and peak flux densities} of the GLOSTAR-B radio sources are accurate to within 10\%.

\subsection{Spectral indices}

\begin{figure}
    \centering
    \includegraphics[width=0.45\textwidth, trim= 0 0 0 0, angle=0] {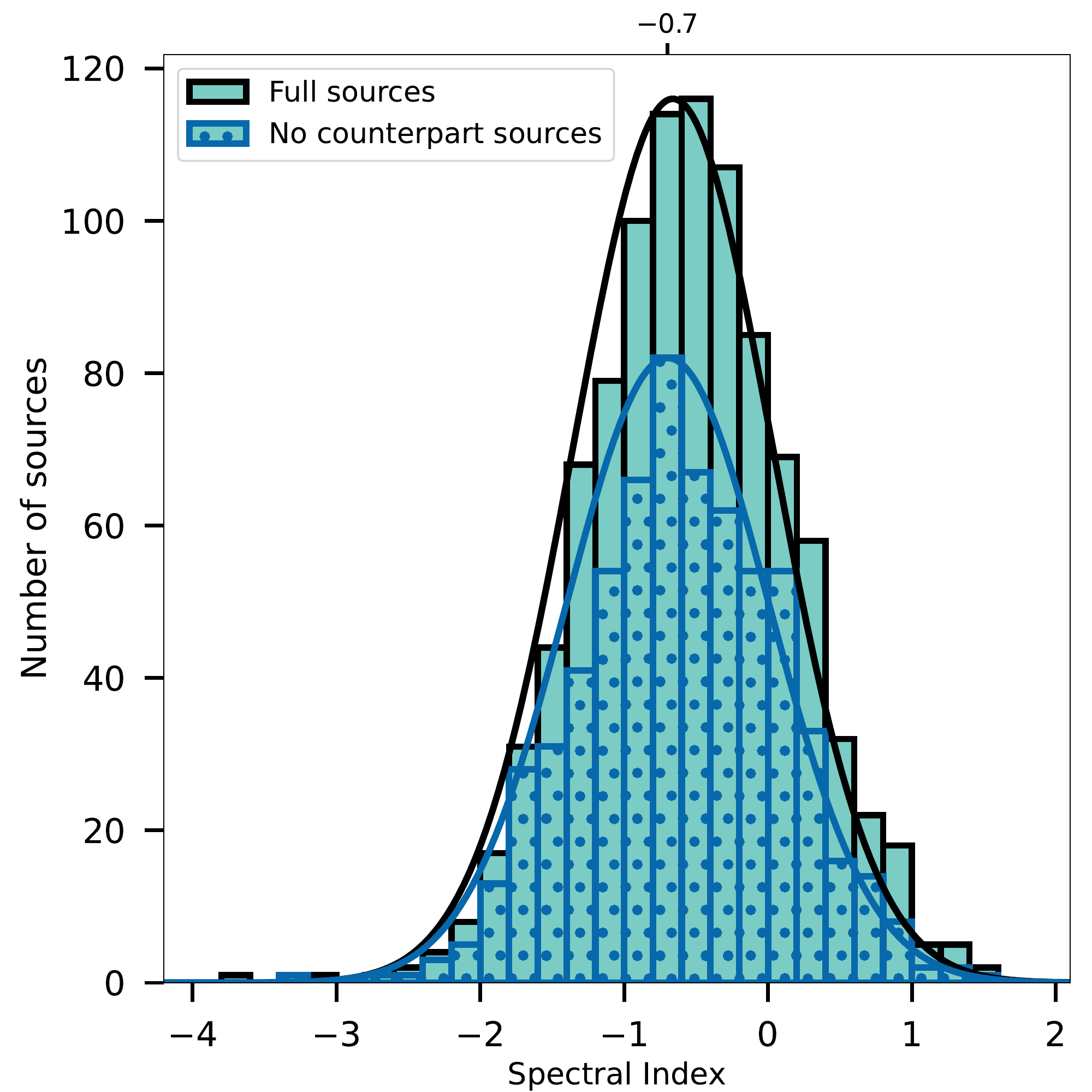}
    \caption{Distribution of the in-band determined spectral indices. Blue dotted bars
    show the spectral index distribution of sources with {no counterparts} at any other
    wavelength.}
    \label{fig:hsi}
\end{figure}

The spectral index of a radio 
source gives us information on the dominant emission mechanism.
{Using the  wide frequency coverage} of 
our observations we can estimate the spectral index within the observed band.
We measure the peak {flux density} in each of the imaged frequency bins for 
compact and point like sources that have S/N$>10$ in our main source extraction, 
a total of {988} sources. The constraint of the S/N was 
chosen to consider that the noise level in each of the imaged frequency bins will 
be $\sim3\times$ higher, and may thus  
not be detected in the individual frequency 
bin images or their determined values will be affected 
by noise. On the other hand, over the area covered by an extended source, flux 
density variations that will depend on its structure may be observed at 
each frequency, and are hence not considered for this analysis.  
{To obtain the spectral index we assume that in the observed frequency range
the flux density is described by the linear equation:}
$${\log{S_\nu}=\alpha\cdot\log{\nu}+C}.$$
{A weighted least-square fitting is made to the measured flux densities.}
Independent {spectral index} values for
each source are listed in column (11) in Table~\ref{tbl:glostar_B}. The distribution
of the determined spectral indices is shown in Figure~\ref{fig:hsi}.
{The distribution of spectral indices has a mean value of $-0.66\pm0.02$.}

\begin{figure*}
    \centering
    \includegraphics[width=1.0\textwidth, trim= 0 0 0 0, angle=0] {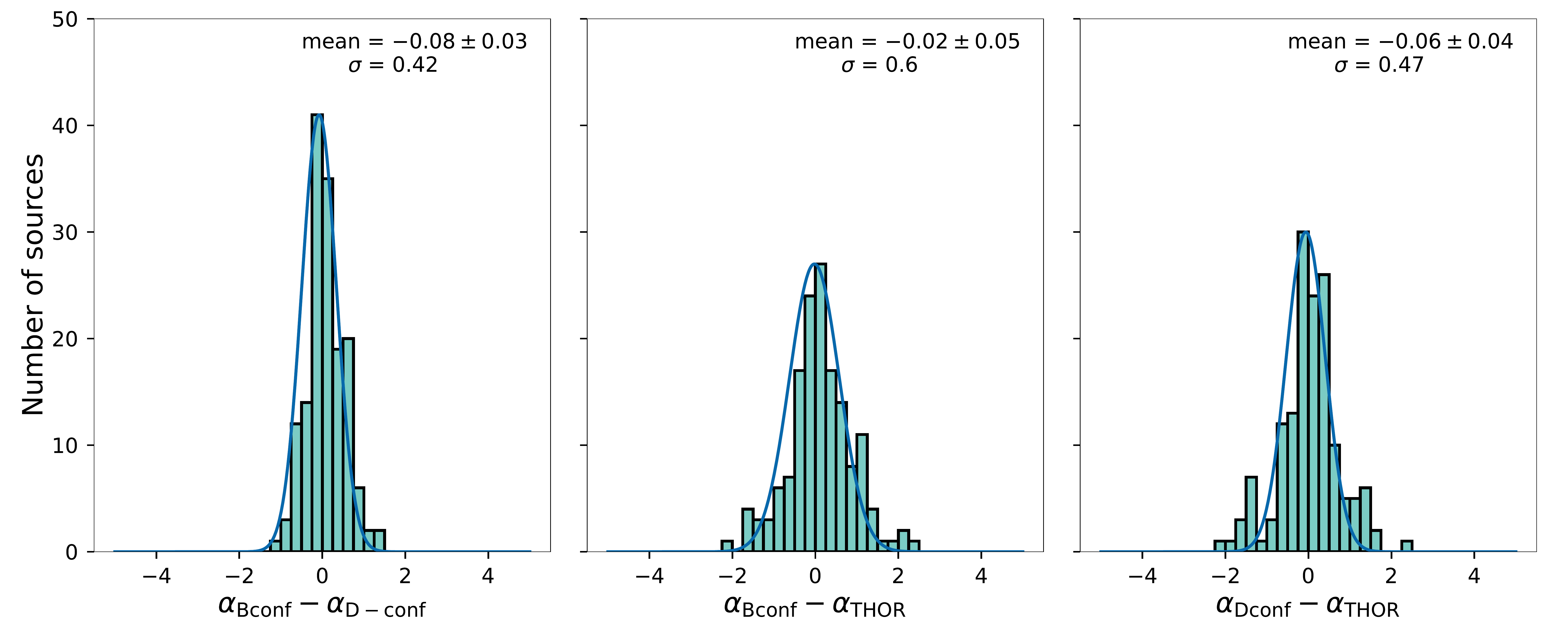}
    \caption{Spectral index difference distributions. From left to right: GLOSTAR B-configuration and GLOSTAR D-configuration, GLOSTAR B-configuration and THOR, and GLOSTAR D-configuration and THOR. } 
    \label{fig:hsico}
\end{figure*}

{ Spectral indices at radio frequencies in the same 
area of the presented images have been measured previously from the 
GLOSTAR D-configuration \citep{medina2019}  and in the THOR
survey \citep[][; described in the next section]{bihr2016}.
To compare the different results, we have selected the
sources that have spectral indices determined and have point
like structure ($Y\leq1.1$) in all three sets of results. 
The last constraint is imposed to diminish effects of possible 
scale structure differences as GLOSTAR D-configuration and THOR
have angular resolutions of $18''$. 
Figure~\ref{fig:hsico}  shows the histogram distributions
of the spectral index differences of the three sets. We found 
that the measured  spectral indices are consistent among the 
three data sets, given that the mean of the differences is
consistent with zero. We thus conclude that the spectral indices 
measured on GLOSTAR B-configuration images are reliable. }

In star forming regions, three
main mechanisms are known to produce compact radio continuum emission, 
and are related to different
astrophysical phenomena \citep{rodriguez2012}. The majority 
of radio sources show thermal free-free radio
from  ionized gas (e.g., HII regions, externally ionized globules, 
proplyds, jets) that has a
spectral index ranging from $2.0$ (optically thick) at low 
frequencies to $-0.1$ (optically thin) at high frequencies. 
Magnetically 
active low-mass stars, may show nonthermal gyrosynchrotron with 
spectral indices ranging from $-2.0$ to $+2.0$. Nonthermal synchrotron 
emission arising from colliding winds in high mass binaries as well 
from jets ejected by high mass stars interacting with the ambient 
{interstellar medium (ISM)} have a typical spectral index of $-0.7$.
However, other phenomena not related to star formation also can produce 
thermal radio emission, namely
gas ionized in planetary nebulae (PNe),
while synchrotron emission is observed from extragalactic sources. 
Background {active galactic nuclei (AGN)} will mostly 
emit optically thin synchrotron emission with spectral indices 
$\sim-0.7$. {On the other hand, a fraction (up to 20\%) of extragalactic 
background radio sources show a flat or positive spectral indices \citep[e.g., ][]{callin2017}. 
These represent a population of star forming galaxies, and progenitors of AGNs
\citep[i.e., high frequency peakers ][]{dallacasa2000,dallacasa2003}. 
} Given the diversity of the radio sources, to better 
understand their nature, information at other
wavelengths is required. This will be discussed in the following subsection.

\subsection{Counterparts at other wavelengths}

To gain  more insight on the nature of the radio sources, we have searched for
counterparts at shorter wavelengths. The search
was focused on catalogs that could give us evidence for
ionized gas, dense cold gas and dust, which are indicators of massive young 
stars and the regions they form in. We now briefly describe the catalogs used 
in our search.

The APEX Telescope Large Area Survey of the Galaxy (ATLASGAL) 
observed the galactic plane at a wavelength of 870~$\mu$m (345\,GHz), 
and angular resolution $\sim20''$ 
\citep{shuller2009}. The emission at this wavelength is dominated by dense 
cool gas and dust. Several ATLASGAL source catalogs have been released
by \citet{contreras2013}, \citet{csengeri2014}, and \citet{urquhart2014},
who list $>10,000$ dense clumps. We will compare our results with the 
compact source catalog presented by \citet{urquhart2014}. 
The differences between the angular
resolution of ATLASGAL and the observations presented here  have 
to be taken into account. Thus, we have used an offset of $18''$ 
(roughly the beam size of ATLASGAL images) to consider an ATLASGAL source to be a 
potential counterpart for the compact radio source.
We found {143 radio sources matching the position of 83 submillimeter sources. } 

The Herschel infrared Galactic plane survey (Hi-GAL) observed the inner 
Galaxy in 5 bands distributed in the wavelength range between 70~$\mu$m
and 500~$\mu$m \citep{molinari2010,molinari2016}. Notably, data taken at these 
wavelengths allow determinations of
the peak of the spectral energy distribution of cold dust and thus 
the source temperatures. The
Hi-GAL observations have angular resolutions from $10''$ down to $35''$ 
from the shortest to the longest wavelength.  The median uncertainty of Hi-GAL sources is $\sim1\rlap{.}''2$.  We have used a
$2''$ offset between a Hi-GAL source and a GLOSTAR source to consider them as
counterparts, and found 98 sources that matched this criterion.

The Wide-field Infrared Survey Explorer (WISE)  mapped the entire sky in 
four infrared bands centered at  3.4,  4.6,  12.0,  and  22.0\,$\mu$m. 
The WISE ALL-sky Release Source Catalog contains astrometry and photometry 
for over half a billion objects \citep{wright2010}. The angular
resolution of the observations is $6''$ at the shortest wavelength
and the position errors are around $0\rlap{.}''2$. To consider a WISE source as 
the counterpart to a GLOSTAR radio source we have used  a maximum offset of 
$2''$. A total of { 125} sources match this criterion.

The Galactic Legacy Infrared Mid-Plane Survey Extraordinaire 
\citep[GLIMPSE;][]{benjamin2003, churchwell2009} mapped a large fraction of the Galactic plane
with the Spitzer space telescope. It observed in four near-infrared sub-bands 
covering the range from 3.6 to 8.0~$\mu$m, and with angular resolutions $\sim2''$.
While the filter widths of the 3.6 and 4.5 $\mu$m bands are similar to those of WISE, 
the GLIMPSE survey has focused on Galactic plane observations and is more sensitive
than the former. The interstellar medium emission at these wavelengths comes mainly from 
warm and dusty embedded sources. Considering the angular resolution of our
observations
and the position uncertainties in the GLIMPSE survey we have used 
an offset of $2''$ for the counterpart searching, leading to {251} matching sources.

The United Kingdom Infrared Deep Sky Survey (UKIDSS) is a suite of 
five public surveys at near infrared wavelengths (NIR) of varying depth 
and area coverage \citep{lucas2008}\footnote{The UKIDSS project is defined in \citet{lawrence2007}. UKIDSS uses the UKIRT Wide Field Camera \citep[WFCAM;][]{casali2007}. The photometric system is described in \citet{hewett2006}, and the calibration is described in \citet{hodgkin2009}. The pipeline processing and science archive are described in \citet{hambly2008}.}. 
Particularly, the UKIDSS Galactic plane survey (UKIDS-GPS) covered a total
of 1878~deg$^2$ of the northern hemisphere Galactic plane. The main observed
bands are the so called J (1.25$\,\mu$m), H (1.65$\,\mu$m) and K (2.20$\,\mu$m)
bands. The spatial resolution of UKIDSS-GPS observation is typically better 
than $0\rlap{.}''8$. For the counterpart search of GLOSTAR radio sources 
in the UKIDSS-GPS catalog we have used this value of $0\rlap{.}''8$, finding {389} matches.  

After our careful search of counterparts at other wavelengths, it is
noticeable that {640 radio sources}  have no counterparts at any other wavelength. 
The spectral index distribution of the sources 
in this category shows that they have preferably negative spectral indices
(see Fig~\ref{fig:hsi}). We will further discuss these sources in the next section.

\subsection{Classification of radio sources}

Using the counterparts of the radio sources in the catalogues described in the previous section,
a robust classification can be carried out. A single classification was performed 
for fragmented sources listed in Table~\ref{tbl:glostar_F} instead of separate
classifications of their fragments, and
thus the classification was done to {1409}  sources. 
The classification criteria are based on our findings of the emission properties and/or counterparts of near-infrared (NIR; UKIDSS),
mid-infrared (MIR; GLIMPSE and WISE), far-infrared (FIR; Hi-Gal), and 
submillimeter (SMM; ATLASGAL). 

Images from the above mentioned infrared surveys are plotted at the position 
of the radio sources, some examples can be seen in Appendix B.
{For SMM and FIR, we show the emission properties of each GLOSTAR source. 
For NIR and MIR, we did a visual inspection of the three-color images for the UKIDSS (red K-band, green H-band and blue J-band), the GLIMPSE (red 8.0\,$\mu$m, green 4.5\,$\mu$m, and blue 3.6\,$\mu$m), and the WISE (red 22.0\,$\mu$m, green 12.0\,$\mu$m  and blue 4.6\,$\mu$m) surveys.}
The sources have been classified into 5 groups,
using the following criteria:

\begin{itemize}
    \item {\it HII region candidates:} Radio sources with emission at SMM and FIR
    wavelengths and with weak or no emission at MIR and NIR 
\citep{hoare2012,anderson2012,urquhart2013,yang2021}.

\item{\it Radio star: } Compact radio sources showing blue compact emission 
in the three color images of  NIR and MIR wavelengths. Weak or no emission at FIR wavelenghts and no-emission at SMM wavelengths 
\citep{hoare2012, lucas2008}. 

\item{\it Planetary nebula (PN):} Show red emission in the three color 
images of NIR and MIR wavelengths 
where it is also seen as an isolated point source. Weak or no emission at 
far infrared (FIR) and
submillimeter (SMM) wavelengths \citep{hoare2012,anderson2012,phillips2011}.

\item{\it Photo dissociated region (PDR): } Ionized gas seen as extended
emission at MIR, and showing only weak or no compact emission 
at FIR and SMM wavelengths
\citep{hoare2012}.

\item{\it Extragalactic candidate (EgC):} Radio sources that have no-counterpart
at any other wavelength, or are only seen as a point source at NIR wavelengths
\citep{hoare2012,lucas2008,marleau2008}.

\item{\it Other sources.} Sources that could not be classified in any of the previous categories. 
\end{itemize}

\noindent The number of radio sources in these groups are {93}  HII region candidates, 
{4}  PDRs, {83}  radio stars, {65}  PNs, {1163} EgCs, and 2 other sources.  
Examples of sources in these classes are shown in Figures~\ref{fig:eHII}, 
\ref{fig:eRS},
\ref{fig:ePN}, and \ref{fig:eEgC}. The individual source classification 
is given in column (17) in Table~\ref{tbl:glostar_B} and in column (5) of
Table~\ref{tbl:glostar_F} for fragmented sources.

\subsection{Sources previously classified}

Classification of the radio sources is a main part of the catalog construction.
However, some of the sources could have been previously classified. A 
search for previous classifications of the radio sources has been performed 
using the SIMBAD astronomical database \citep{wenger2000}, within a radius of $2''$.
We found that {269}  radio sources have counterparts in the SIMBAD database. 
Most of these sources, {138}, are only classified as radio sources, and are thus of an
unknown nature. The classification of the remaining 107 sources suggests
that these are Galactic sources. In Table~\ref{tab:Simb} we have separated 
these sources in seven types of sources, and have listed the source class in SIMBAD considered for these types and the number of sources of each type. 


\begin{table}[!h]
    \centering
    \renewcommand{\arraystretch}{1.4}
    \caption{Source classification from SIMBAD database.}
    \begin{tabular}{p{0.17\textwidth}p{0.22\textwidth}c}
    \hline\hline
    Type & SIMBAD class$^{a}$ & \#\\
    \hline
     Evolved stars & WR*, PN, PN?, AB?, CV*   &  38 \\
     Pulsars       & Psr             &   3 \\
     Stars          & *, Cl*          &   7 \\

     Young stellar objects & Y*O, Y*?&  30 \\
     HII regions   & HII             &  28 \\
     Gamma-Ray source  & gam             & 1 \\
     Extragalactic objects & QSO         & 1 \\
     Other objects & 
                     Mas, mm, smm, IR, NIR, MIR, MoC, EmO, PoC, cor  &  23 \\

     \hline
    \end{tabular}\label{tab:Simb}
    
   \noindent {\footnotesize $^{a}$ The SIMBAD object classification is described in}\\ {\scriptsize \url{http://simbad.u-strasbg.fr/simbad/sim-display?data=otypes}.}
    
\end{table}

\section{Discussion}

\subsection{Overview of the final classification}

The final radio source classification is a combination of the method
described in Section 3.8, and the use of known information of 
individual sources recovered from the SIMBAD database. 
For our final catalog we additionally consider
the SIMBAD classification for sources for which we find no counterpart
in the searched IR and sub-mm catalogs. This was, for example, the case
for pulsars that are usually not detected at infrared wavelength and are non-thermal radio
emitters. Based on the criteria described in the previous section they were 
first classified as EgCs and after consulting the SIMBAD database they were
finally classified as pulsars. A similar situation occurred with a 
Wolf-Rayet star and a {cataclysmic variable}  star,
wherein they were initially classified as 
a PN and as a Radio-star, respectively. We use the classification recovered from 
SIMBAD in these cases. 

In Table~\ref{tab:FCN} we give the number of sources found in each of the classifications. Interesting sources are the HII region candidates, as they are related to massive star formation, that we will discuss later.

\begin{table}[]
    \centering
    \renewcommand{\arraystretch}{1.4}
    \caption{Final source classification of radio sources.}
    \begin{tabular}{lcc}
    \hline\hline
    Classification     & \# \\
    \hline
     HII region candidates      &  93\\ 
     WR                &  2\\
     Pulsar            &  3\\
     Radio stars       &  81\\
     PNe               &  64\\
     Cataclysmic Variable             &  1\\
     PDRs              &  4\\
     Other           & 2\\
     EgCs              &  1157\\
     \hline
    \end{tabular}\label{tab:FCN}
\end{table}

\subsection{Comparison of classifications}

Classifications of some of the detected radio sources have been performed
in previous radio surveys and here we will compare the results of 
these classifications. We will focus on a comparison based on the low resolution GLOSTAR D-configuration image presented by \citet{medina2019}, 
and with
the classification based on the CORNISH survey  which has a similar angular
resolution observation as our B-configuration data \citep{purcell2013}.

\subsubsection{GLOSTAR D-configuration}
Some differences between the present catalog and that derived from the D-configuration
data are expected, since for the latter, the search for counterparts 
was done using larger angular separations than in this work on account of the lower
resolution.  54 of the HII 
region candidates identified by us were also detected in D-configuration
images, and were also identified as HII 
regions. { Four}  other HII region candidates had a D-configuration counterpart,
but were unclassified. The remaining {32}  HII region candidates were not
detected in the D-configuration images. Additionally, { six} sources classified as
HII region candidates in the D-configuration catalog have been classified as PNe in 
the present work as their positions are no longer coincident with ATLASGAL sources.

We have  detected { 32}  radio sources that were classified as PNe in the
D-configuration catalog, of which {30}  are classified as PNe in this work, one is 
classified as a WR star and another as an EgC because it is no longer positionally
coincident with IR emission. We also detect { 13}  sources identified as radio stars 
in the D-configuration, of which {10}  were related to a radio source also classified as 
radio stars. The remaining source, G028.098--00.781, was resolved 
into three different sources, two of them were classified as extragalactic sources 
and the other as PN. A further {749}  sources detected in D-configuration 
images were also detected in the B-configuration images. Among these, {688}  were 
classified as EgCs, consistent with the suggestion by \citet{medina2019} that most 
of their unclassified radio sources are background extragalactic radio sources.
The remaining {61} unclassified sources in D-configuration have now been classified, 
{43} as radio-stars, {14} as PNe, and {4} as HII region candidates. A comparison between 
the method used in the D-configuration data and in the B-configuration data shows 
a consistency better than 90\% in the resulting classes. 

\subsubsection{CORNISH Survey}
The CORNISH survey has a similar angular resolution, and it also
used IR information to classify  detected radio sources. In their 
classification effort, they consider UCHII regions and dark HII regions;
we detected radio emission from {40}  CORNISH sources classified as UCHII and from one
dark HII region, which are classified as HII region candidates 
by us. 
We also have detected {25}  CORNISH radio sources classified as PNe,
that are also classified as PNe in our work. Seventeen radio-stars in the 
CORNISH survey have counterparts in our catalog; we classified fifteen of them 
also as radio-stars and two of them as EgCs because we did not find IR 
counterparts. We have detected 
171 CORNISH radio sources classified as IR-Quiet (not detected at IR
wavelengths); we classify 167 of them as EgCs, 4 as radio-stars as IR emission
is  now reported in the position of these sources, and one as a pulsar based on the 
SIMBAD database. Finally, out of the 29 radio-Galaxy sources in CORNISH, 28 are 
classified as EgCs in our work, and the remaining source is classified as 
radio-star given its IR properties. The consistency in classification between CORNISH
and GLOSTAR B-configuration is high, especially for HII regions 
and PNe where we found a 100\% agreement.

\subsubsection{THOR survey}

The HI/OH/Recombination line survey of the Milky Way 
\citep[THOR;][]{bihr2015,beuther2016,wang2020} observed the northern hemisphere of the  Galactic 
plane with the VLA in C-configuration and using its L-band receivers (1 to 2 GHz). 
It covers Galactic longitudes from 14\rlap{.}$^\circ$5 to 67\rlap{.}$^\circ$4 
and Galactic latitudes from $-1\rlap{.}^\circ$25 to 1\rlap{.}$^\circ$25, i.e., 
including the area of the maps presented in this paper. The angular resolution
of the THOR radio continuum images is $25''$ and they have noise levels from 0.3 to 
1.0 mJy beam$^{-1}$ \citep{wang2018}. { While a detailed comparison of the source 
classification by THOR and GLOSTAR B-configuration is limited given the different 
image angular scales, it is still, however, useful.  }


To compare GLOSTAR B-configuration sources with THOR sources, we used a maximum
separation of $2\rlap{.}''5$, which is the position accuracy of the THOR survey
\citep{wang2018}. We found {554}  sources matching in position between these surveys. 
From these sources, the THOR survey identified {34}  HII regions ({26}  of which 
are identified as HII region candidates) and 9 as PNe (also identified as PNe with 
our classification criteria). Our classification for the remaining eight sources
identified as HII regions by the THOR survey is four EgCs and four PNe. 
Differences between the classification of these sources can be caused by the 
different angular scales used for comparison with IR surveys. In fact, the median 
match radius of IR HII regions \citep[identified with the WISE survey; ][]{anderson2014}
with the THOR survey is $\sim60''$ \citep{wang2018}, almost two orders of magnitude
larger than our angular resolution.


\subsection{HII region candidates}

\begin{figure*}
    \centering
\includegraphics[width=0.35\textwidth, trim= 0 0 0 0,clip, angle=0] {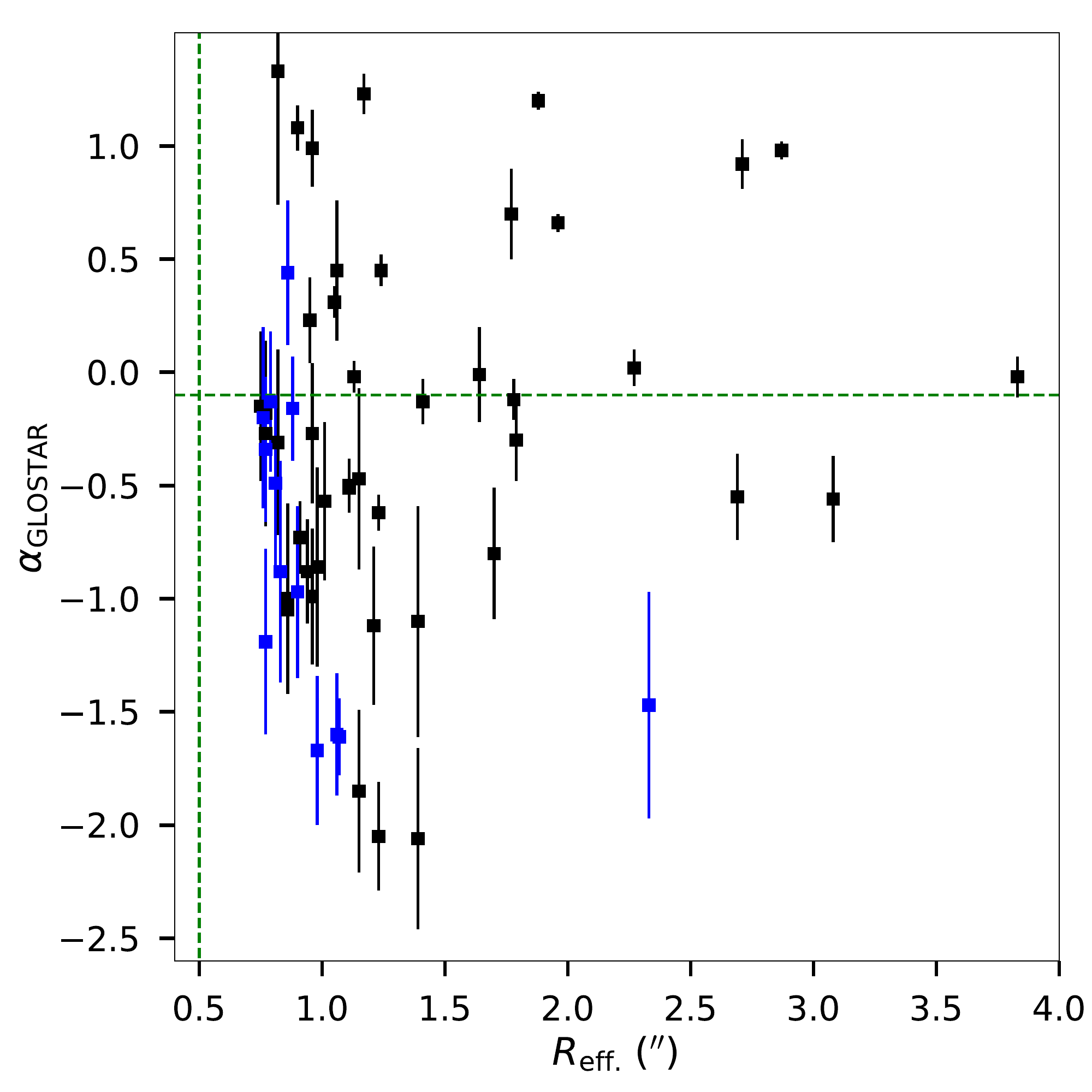}\hspace{0.15cm} 
\includegraphics[width=0.30\textwidth, trim= 85 0 0 0,clip, angle=0] {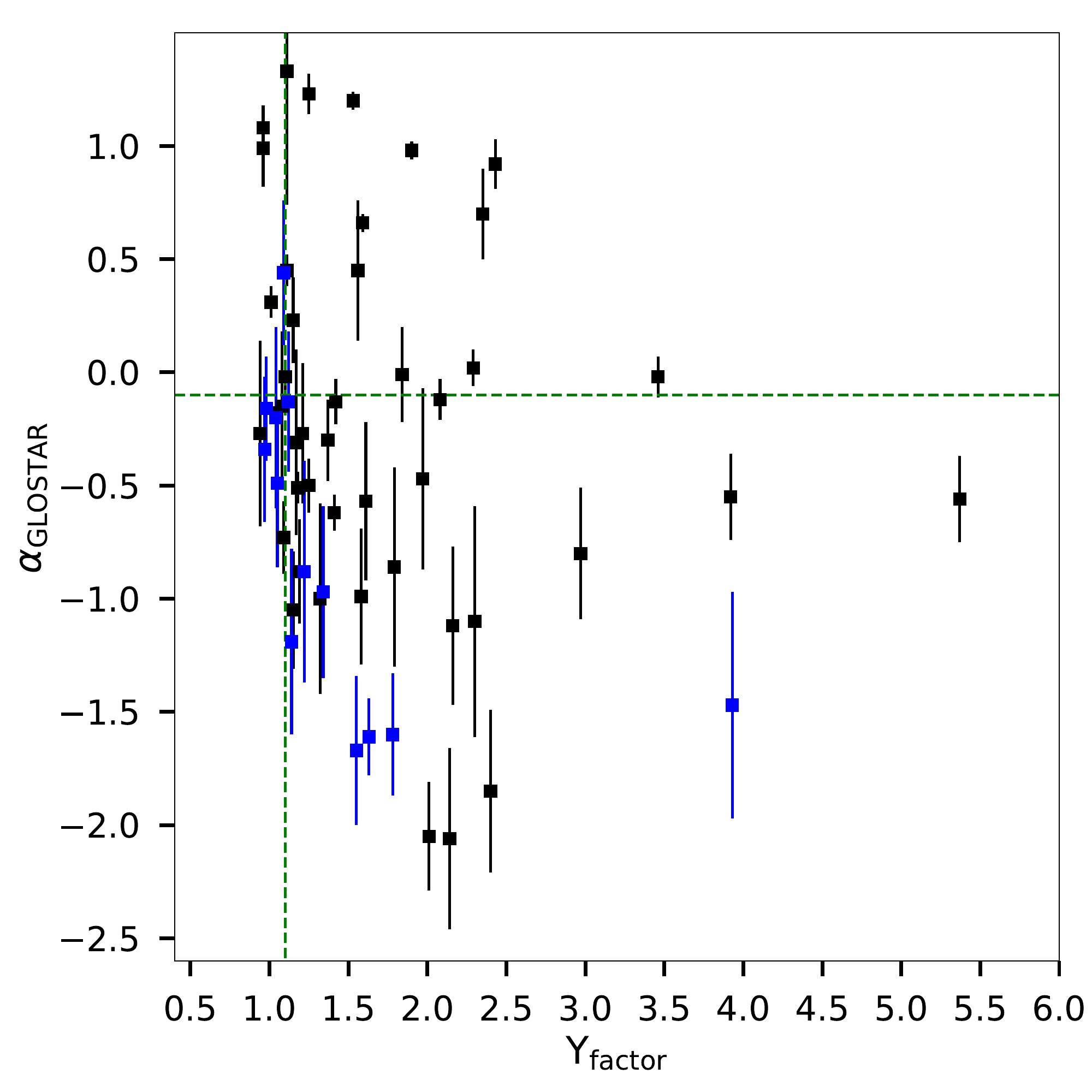}\hspace{0.15cm} 
\includegraphics[width=0.30\textwidth, trim= 85 0 0 0,clip, angle=0] {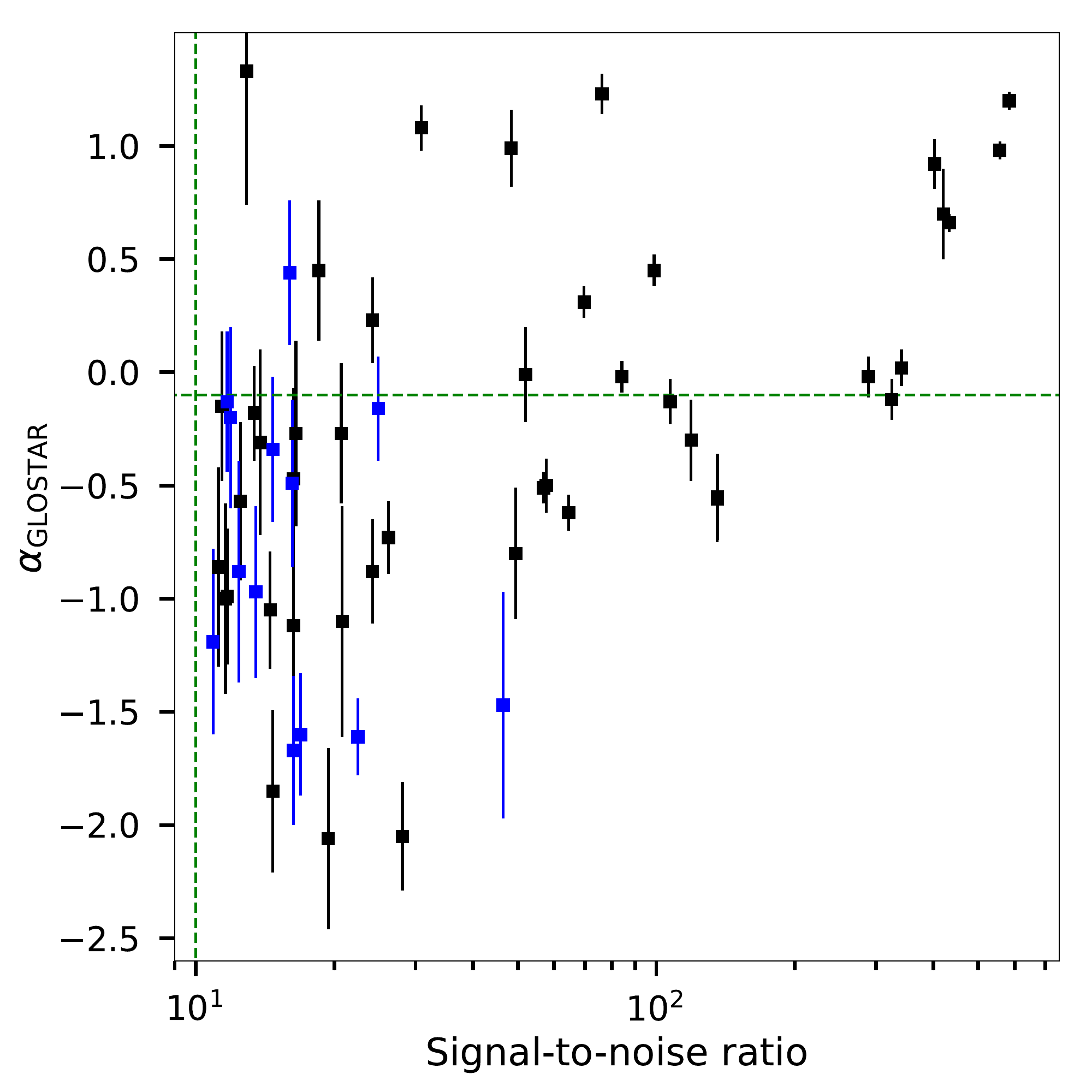}
    \caption{Spectral index of identified HII region candidates as a function
    of the effective radius {(left), Y-factor (center) and S/N (right). } Black symbols are sources that are related to previously 
    identified HII regions, and blue symbols are the new identified HII region candidates.}
    \label{fig:YvsA}
\end{figure*}

We have identified  {93} HII region candidates\footnote{To show the MIR emission around the position
of these sources, a series of images are presented in Appendix~\ref{app:IRHII}. }. 
Out of these sources, {71} were previously related to HII regions from our analysis 
of the D-configuration images {or from CORNISH or from THOR } or from the SIMBAD database and {22} are new 
detections from this work. A characteristic of
the radio emission from HII regions is that the spectral index values ranges
from $-0.1$ (optically thin free-free radio emission) to 2.0 (optically thick
free-free radio emission). Observationally, the spectral index distribution 
of hundreds of young HII regions at similar frequencies  show a mean value of 0.6 \citep{yang2019,yang2021}.  We have determined the in-band spectral index  
(see Section~3.6) for { 57} HII region candidates, {13}  of which are new. 
In Figure~\ref{fig:YvsA}, we plot the spectral index of these HII regions 
as a function of the $R_{\rm eff.}$ { the Y-factor and the S/N}.
Surprisingly, { some} of the determined spectral indices are negative and well below 
the lowest value expected for free-free radio emission of $\alpha=-0.1$, indicating
that the radio emission nature is non-thermal. The results are not affected 
by the size of the source, as most of them are slightly resolved sources 
($R_{\rm eff}\simeq\theta_{resolution}/2={0\rlap{.}''50}$).

{ \citet{kalcheva2018} obtained the spectral index to known ultra-compact 
HII regions and found that about 18\% of their sample were consistent with negative
spectral indices. They  discuss their results and conclude that different 
interferometer array configurations and time variability could explain these 
negative values. These effects, however, are not present in the GLOSTAR observations 
presented here.} Hence, these radio sources with negative spectral indices could 
be related to HII regions but they are not the HII regions themselves  \citep[e.g.][]{purser2016}.

Compact radio sources have been found related to HII regions. 
The most known and nearby case is the Orion Nebula Cluster (ONC) where around 600 compact radio
sources are found in the HII region ionized by the Trapezium 
\citep{forbrich2016,vargas2021}. As first suggested by \citet{Garay1987a, Garay1987b}, most of these fall into two broad categories: first, sources with thermal radio emission from circumstellar matter, often protoplanetary disks (``Proplyds''), that are photo-evaporated in the intense UV field of the brightest Trapezium star $\theta^1$~Ori~C (O7Vp). Second, nonthermal sources associated with the coronal activity of magnetically active low mass members of the ONC many of which also show X-ray  emission and are (highly) variable \citep{forbrich2021,dzib2021}. Other example regions are NGC~6334 
\citep{medina2018} and the M17  \citep{rodriguez2012} star forming regions (SFRs)
where several radio sources are found close to prominent HII regions. 
Also, in these SFRs a significant fraction of the compact 
radio sources have been found to produce non-thermal emission. 
That high number of such radio sources can be detected on the ONC is a result of the cluster's close distance, $D$, of just $\approx 400$~pc \citep{menten2007,kounkel2017}. Extrapolating the Orion case
to a distance of a few kilo-parsecs, most of these magnetically active stars would not be detected,
except for the two strongest  sources with measured {integrated flux densities} of a few tens of mJy. We note that, NGC~6334 and M17, mentioned above, are also relatively nearby, at $D$ = 1.34$^{+0.15}_{-0.12}$   and $1.98_{-0.12}^{+0.14}$~kpc, respectively \citep{reid2014,wu2014}. Time variable weak radio sources around the well-studied ultracompact HII region W3(OH) ($D = 2.0$ kpc) are other examples \citep{Wilner1999}.

For more distant SFRs, thermal radio emission is only  detected from the HII region itself that is excited by the central high mass star(s) of a cluster.

It should be noted that 
other phenomena can produce compact non-thermal radio emission 
in massive star forming regions that are at work in Orion and can be 
observed in regions at distances of a few kpc, such as non-thermal 
radio jets \citep[e.g.][]{carrasco2010, purser2016} and wind collision regions in massive 
binary stars \citep[e.g.][]{dzib2013}. 
Non-thermal radio jets from YSOs, however, are rare and 
only a handful of cases are known. Similarly, non-thermal radio emission from
wind collision regions are not common and are usually associated with evolved 
massive stars that have powerful winds. 

On the other hand, the radio sources with positive spectral index could be true 
HII regions.  More extensive multi-wavelength radio analysis is needed to 
characterize their emission and determine their turnover 
frequency, emission measure, etc. In general, all the radio sources that 
are associated with HII regions discussed in this work deserve further attention
and more detailed studies.

\subsection{Variable radio sources}

By comparing the {peak flux densities} measured in the GLOSTAR images with 
those reported in CORNISH \citep{purcell2013}, we have looked for 
sources with variable radio emission. The search was restricted to sources
with $Y<2.0$, i.e., with compact radio emission.
Variability is identified when the source is detected in 
both catalogs, GLOSTAR and CORNISH, but the flux ratio is larger than 2.0.
We also identify variability when we detect a source in the GLOSTAR catalog
with a peak flux density level $>2.7$\,mJy\,beam$^{-1}$ (the CORNISH detection 
limit) and it is not reported in the CORNISH
catalog. Finally, a source detected as unresolved in the CORNISH catalog 
that is not in the GLOSTAR high reliability catalog or in the 
catalog of sources with 5 to 7$\sigma$, is also considered as a
variable radio source.

\begin{table*}[!h]
 \begin{center}
 \footnotesize
 \renewcommand{\arraystretch}{1.1}
\setlength{\tabcolsep}{3.2pt}
 \caption{Variable radio sources.}
 \label{tbl:HV} 
\begin{tabular}{@{\extracolsep{4pt}}cccccccc@{}} 
 \hline \hline
  & \multicolumn{3}{c}{GLOSTAR} & \multicolumn{3}{c}{CORNISH}& \\ \cline{2-4} \cline{5-7}
Source &$S_{\rm peak}$ & $\boldsymbol{\sigma}_\mathbf{S_{\rm peak}}$ & & $S_{\rm peak}$ & $\boldsymbol{\sigma}_\mathbf{S_{\rm peak}}$& \\ 
{name} &\multicolumn{2}{c}{(mJy\,beam$^{-1}$)}& Classification &\multicolumn{2}{c}{(mJy\,beam$^{-1}$)}& Classification  \\ 
(1)&(2)&(3)&(4)&(5)&(6)&(7)\\ 
 \hline 
G028.1875+00.5047&...&...&...&2.61&0.39&IR-Quiet\\ 
G028.3660--00.9640&...&...&...&2.92&0.48&IR-Quiet\\ 
G028.4012+00.4776&...&...&...&5.25&0.60&Radio-Galaxy\\ 
G028.6043--00.6530&2.79&0.17&EgC&...&...& ... \\
G028.9064+00.2548&...&...&...&3.47&0.53&IR-Quiet\\ 
G028.9224--00.6589&5.07&0.29&EgC&...&...&...\\ 
G029.1640--00.7922&3.05&0.18&EgC&...&...&...\\ 
G029.2620+00.2916&...&...&...&2.76&0.45&IR-Quiet\\ 
G029.3096+00.5124&...&...&...&2.84&0.43&IR-Quiet\\ 
G029.4302--00.9967&...&...&...&2.83&0.48&IR-Quiet\\ 
G029.4404--00.3199&...&...&...&2.81&0.47&IR-Quiet\\ 
G029.4959--00.3000&5.43&0.30&EgC&...&...&...\\ 
G029.5184+00.9478&...&...&...&2.43&0.40&IR-Quiet\\ 
G029.5780--00.2686&12.31&0.69&PN&6.22&0.72&PN\\ 
G029.5893+00.5789&1.73&0.12&EgC&4.17&0.50&IR-Quiet\\
G030.2193+00.6501&...&...&...&2.80&0.42&IR-Quiet\\
G030.3704+00.4824&2.83&0.16&HII& ...&...&...\\ 
G030.6328--00.7232&...&...&...&2.57&0.40&IR-Quiet\\
G030.6517--00.0605&3.08&0.21&PN&...&...&...\\
G030.7859--00.0298&3.72&0.35&Other&...&...&...\\
G030.9704--00.7436&...&...&...&2.47&0.38&IR-Quiet\\ 
G031.0025--00.6330& 	2.78 &	0.17 	&EgC 	&... 	&... 	&...\\
G031.0450--00.0949&...&...&...&2.63&0.40&IR-Quiet\\ 
G031.0777+00.1703&13.37&0.74&EgC&6.75&0.72&IR-Quiet\\ 
G031.2989--00.4929&3.43&0.21&Radio-star&...&...&...\\ 
G031.3444--00.4625&...&...&...&2.78&0.37&IR-Quiet\\ 
G031.3917+01.0265&...&...&...&2.39&0.38&Radio-Star\\ 
G031.3993--00.0813&5.31&0.30&EgC&...&...&...\\ 
G031.5694+00.6870&26.11&1.44&EgC&6.79&0.71&IR-Quiet\\ 
G031.9943+00.5156& 	3.14 &	0.18 	&EgC 	&... 	&... 	&...\\
G032.2739--00.0358& 	2.91 &	0.17 	&EgC 	&... 	&... 	&...\\ 	
G032.3843--00.3397&3.95&0.23&EgC&...&...&...\\ 
G032.4536+00.3679&1.42&0.10&EgC&2.74&0.46&IR-Quiet\\ 
G032.5898--00.4469& 	3.14 &	0.18 	&EgC 	&... 	&... 	&...\\ 	
G032.5996+00.8265&2.87&0.17&EgC&5.92&0.62&IR-Quiet\\
G032.7396+00.8344& 	2.78 &	0.16 	&EgC 	&... 	&... 	&...\\ 	
G032.7568+00.0959& 	2.76 &0.18 	&EgC 	&... 	&... 	&...\\ 	
G033.1860+00.9528& 	2.86 &	0.17 	&EgC 	&... 	&... 	&...\\
G033.3513+00.4056&...&...&...&2.62&0.40&IR-Quiet\\
G033.4100--00.4775& 	3.05 &	0.18 	&EgC 	&... 	&... 	&...\\ 	
G033.4584+00.0163&3.82&0.22&EgC&...&...&...\\ 
G033.8964+00.3251& 	2.87 &	0.17 	&Radio-star 	&... 	&... 	&...\\
G033.9622--00.4966&...&...&...&2.81&0.42&IR-Quiet\\ 
G034.2171--00.6886&...&...&...&2.16&0.34&IR-Quiet\\ 
G034.4794--00.1683&2.78 &	0.16 	&EgC 	&... 	&... 	&...\\
G034.7242+00.6660& 2.8 &	0.17 	&EgC 	&... 	&... 	&...\\ 	
G035.0605+00.6208&...&...&...&2.76&0.43&IR-Quiet\\ 
G035.2618+00.1079&...&...&...&2.36&0.39&IR-Quiet\\
G035.9038--00.4810&2.26&0.14&EgC&4.94&0.55&IR-Quiet\\
\hline
 \end{tabular}\\
  Notes: {Names} for sources not detected in GLOSTAR images come from the CORNISH catalog \citep{purcell2013}. \\ 
 \end{center}
 \end{table*}

With the above criteria we have identified {49}  variable sources. 
They are listed in Table~\ref{tbl:HV} together with the peak flux densities 
in the compared catalogs. The GLOSTAR and CORNISH classification are 
listed when available. Most of the 
identified variable
radio sources are EgC or IR-quiet sources, and have only been detected at radio
wavelengths. Since extragalactic background sources are not expected in general to show pronounced
variability, thus some of these sources could be interesting Galactic 
radio sources whose nature has to be explored. 


\subsection{Extragalactic Objects}

\begin{figure*}[!ht]
\centering
\includegraphics[width=0.85\textwidth, trim= 5 5 5 5,clip, angle=0]{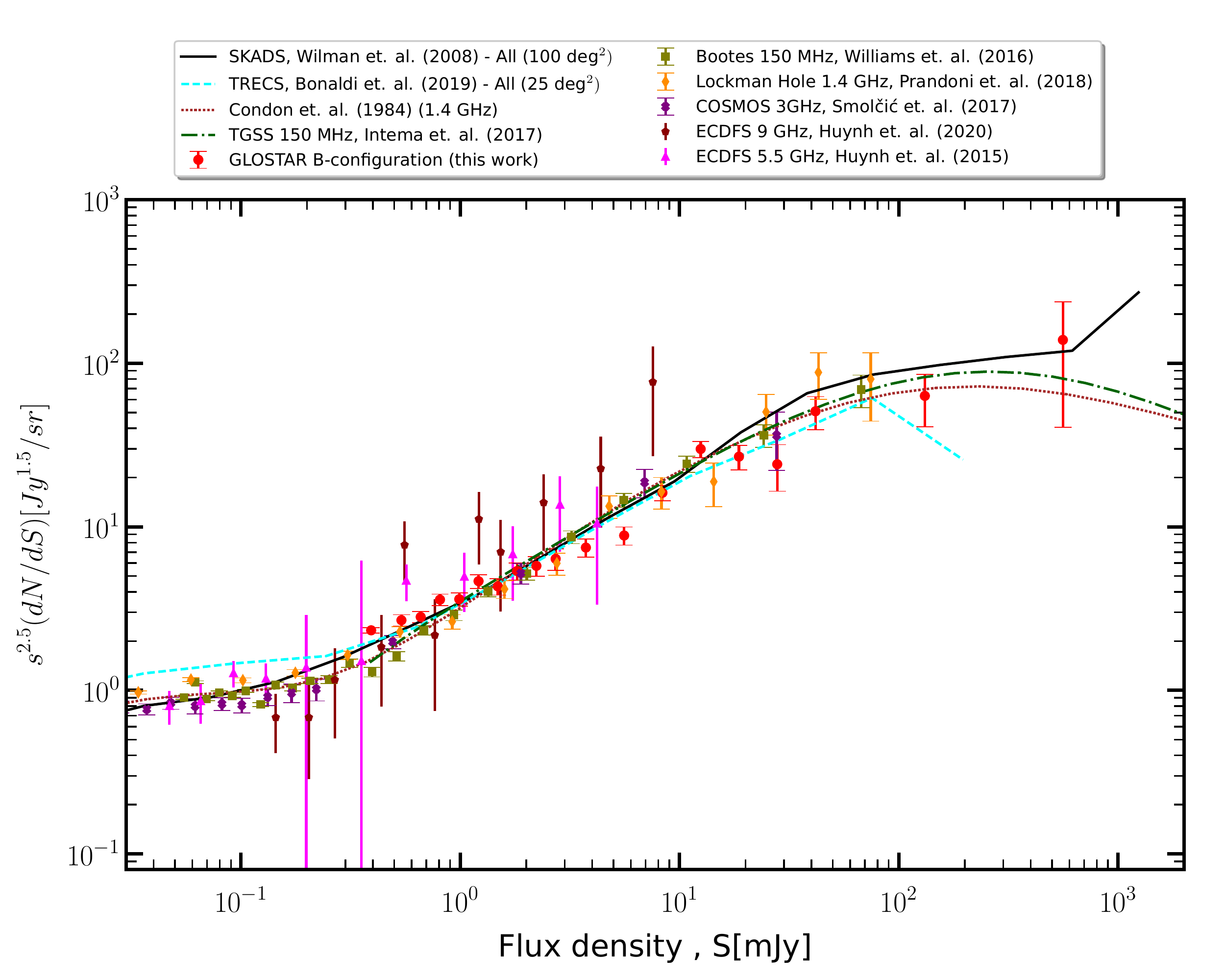} 
\caption{The Euclidean-normalized differential source counts of the point sources classified as of extragalatic origin in the GLOSTAR survey. We have compared the source counts with the  simulated radio sky  and previously observed source populations. For details of simulated catalogues and different observed source populations see text.}
\label{dn_ds}
\end{figure*}

As  has been shown in our previous works, most of the detected radio sources are 
expected to be background extragalactic objects \citep{medina2019,chakraborty2020}.
Following the formulation by \citet{fomalont1991}, and using an observed area of
57,600 arcmin$^2$, assuming a nominal noise level of 
$\sigma=\,$65~$\mu$Jy~beam$^{-1}$ and 
a threshold 7$\times$ the noise level, we estimate that the number of expected
background extragalactic sources in our image is $1138\pm664$. This number 
suggests that most of the detected radio sources above {$455\,\mu$Jy~beam$^{-1}$ } 
(7$\sigma$) are of 
extragalactic origin.

From our classification criteria we have compiled a list 
of {1159} sources that have most probably an extragalactic origin. They
are labeled as EgC in Table~\ref{tbl:glostar_B}. Their 
extragalactic nature is also supported with the negative 
spectral index of most of them. {The spectral index is determined 
for 777 of these sources and 157 (20\%) have a flat or positive spectral index, consistent with 
the expected number of extragalactic radio objects that are expected to have positive spectral indices at our frequencies.}

Similar to our previous work \citep{chakraborty2020}, we have also studied the Euclidean-normalized differential source counts of the point sources characterized as being extragalactic in origin. We have binned the source { integrated flux densities} in logarithmic space and divided the raw counts in each bin by the fraction of image area over which a source with a given {integrated  flux density} value can be detected, known as the visibility area \citep{Windhorst1985ApJ...289..494W}.  The differential source counts have been calculated by dividing the visibility area weighted source counts in each bin,  by the total image area ($\Omega$ in steradians) and bin width ($\Delta S$ in Jy). These differential source counts have been normalized by multiplying  with $S^{2.5}$, where $S$ is the mean {integrated flux density} of sources in each bin \citep{Windhorst1985ApJ...289..494W}.  The normalized differential source counts is shown in  Figure \ref{dn_ds}, where the error bars are Poissonian. We have compared our findings with two simulated catalogues, the SKA Design Study simulations (SKADS, \citealt{Wilman2008MNRAS.388.1335W}) and the Tiered Radio Extragalactic Continuum Simulations (T-RECS, \citealt{Bonaldi2019MNRAS.482....2B}). We have also compared our findings  with the observed extragalactic source populations at low-frequency as well as high-frequency, which include: the TIFR GMRT Sky
Survey  at 150 MHz (TGSS-ADR1; \cite{Intema2017A&A...598A..78I}),  BOOTES field at 150 MHz using LOFAR \citep{Williams2016MNRAS.460.2385W}, Lockman Hole field at 1.4 GHz with the LOFAR \citep{Prandoni2018MNRAS.481.4548P}, COSMOS field at 3 GHz with VLA \citep{Smolcic2017A&A...602A...2S}, ECDFS field at  5 GHz \citep{Huynh2015MNRAS.454..952H} and 9 GHz \citep{Huynh2020MNRAS.491.3395H} with ATCA and the 1.4 GHz source counts based on observations with VLA by \citet{condon1984}.  In all cases we have scaled the source counts to 5.8 GHz using a spectral index, $\alpha = -0.7$. We have found that the source count of these sources classified as extragalactic in the GLOSTAR survey is statistically similar and consistent with the previously observed extragalactic source population as well as with the simulated catalogs.  This shows that the majority of these sources are indeed of extragalactic origin.

\subsection{Perspective on the search of non-thermal galactic sources}

Figure \ref{fig:hsi} shows that the  peak of the spectral index distribution 
is  {${-0.66\pm0.02}$.} As discussed in the previous subsection most of the unidentified 
radio sources will be of extragalactic origin, however some of these will 
be interesting Galactic non-thermal radio sources. These sources are only
detected at radio frequencies and it will be hard to distinguish them. 

Very Long Baseline Interferometry (VLBI) observations to all unidentified 
compact radio sources could help to distinguish the Galactic from the
extragalactic radio sources. Position measurements on the scale of several 
months to years could measure the proper motions {and trigonometric parallaxes} of these
sources. As the
extragalactic background sources are not expected to exhibit proper motions 
{or trigonometric parallaxes}, 
these observations can help to distinguish between the two classes of radio
sources. A wide field VLBI survey of the Galactic Plane is now 
feasible thanks to the DiFX software correlator \citep{deller2011} which 
allows multiple-phase center correlation inside primary beam of the interferometer.

Radio continuum observations can be used to search for pulsar candidates 
\citep[e.g.][]{maan2018}. The advantage of these observations compared
to conventional pulsar searches is that they require less computational power.
The observed frequency by the GLOSTAR survey also has the advantage that
the radio emission is less affected by scattering than the lower frequencies 
used for conventional pulsar searches. As an interesting example, the radio 
emission from PSR J1813--1749, one of the most energetic known pulsars, was 
first detected in radio continuum images at 5~GHz \citep{dzib2010,dzib2018} 
while searches for the pulsed radio emission at lower frequencies failed 
\citep{helfand2007,halpern2012,dzib2018}. It turns out that interstellar scattering 
in the direction of this pulsar is very high, and PSR J1813--1749 is the 
most heavily scattered known pulsar \citep[][]{camilo2021}. Pulsar radio continuum
emission is characterized by point-like structure and steep radio spectrum
\citep[$\alpha=-1.4\pm1.0$;][]{bates2013}. These are criteria that are shared with 
EgCs in our classification scheme, making them hard to differentiate. 
Dedicated observations to search for pulsed emission can distinguish them, with
the advantage that the observations will be target intended instead of a blind
survey.

\section{\label{conclusions} Conclusions}

\noindent As part of the GLOSTAR survey, the VLA in its B- and D-configuration  was used 
to observe a large portion of the Galactic plane in the  C-band. In this paper
we present the B-array observations covering the area within 
$28^\circ\leq \ell<36^\circ$ and $|b|<1^\circ$, which we  previously investigated using data obtained in the D-configuration  \citep{medina2019}.
Using a combination of automatic source extraction with \blobcat\ \citep{hales2012}
and visual inspection we have identified {\Rblobs} radio sources. 

The catalog of these radio sources is divided in two parts. The catalog of highly reliable radio 
sources contains {\HRC}  entries.  Detailed properties of these sources are given, 
such as the positions, the signal-to-noise-ratio, integrated and peak fluxes, the ratio 
between these two values (also known as 
the Y-factor), the effective radii, and the spectral indices. The weak source catalog 
lists {\LRC}  sources with a signal-to-noise ratio between 5 to 7. Only their basic properties 
such as positions, the signal-to-noise-ratio, integrated and peak fluxes are given.

The highly reliable radio sources were further investigated. The positions of these sources 
were compared  with the positions from the CORNISH catalog \citep{purcell2013} and 
radio sources detected with the VLBI from the Radio Fundamental catalog. We found 
that the positions of GLOSTAR sources are { in agreement with those in these catalogs to better than $0\rlap{.}''1$.} We have also compared our 
{integrated flux densities} with those in the CORNISH catalog and conclude that 
the GLOSTAR {integrated flux densities} are accurate to within 
10\%, apart from clearly variable sources. From a comparison with the GLOSTAR D-configuration catalog, we find that {908} of them 
are related to {780} sources detected in the D-configuration images. In particular, {22}  
D-configuration sources are partially resolved and appear as fragmented sources in 
the new high resolution images. A total of {72}  highly reliable B-configuration sources 
comprise these {22}  fragmented sources.

To further investigate the nature of the highly reliable radio sources, we have used
information from surveys at infrared and sub-millimeter wavelengths as well as consulting
the SIMBAD database. The classification of the radio sources resulted in { 93}  HII region
candidates, {64}  PNe, {81} radio stars, and most of the remaining sources as EgCs. 
We compared our classification with the classification done to the D-configuration radio 
sources, and to the sources from the CORNISH survey. We find that the classification
from the catalogs agree in more than 90\% of the sources. An interesting result is 
that 
many sources classified as HII region candidates, however, have a negative in-band 
spectral indices suggesting that the radio emission is predominantly non-thermal and,
thus, they are not the HII regions themselves, but likely related to nearby YSOs. These sources could be other 
radio emitter objects related to star formation. They deserve a further study using
deeper and radio multi-wavelength observations to better characterize their radio 
emission nature, and the nature of the objects themselves. Finally, by comparing the 
{integrated flux densities} from GLOSTAR and CORNISH, whose observations are separated by 7 years, 
we have identified {49} variable sources, whose nature has to be explored for most of
these sources.

\section*{Acknowledgments}
This research was partially funded by the ERC Advanced Investigator
Grant GLOSTAR (247078). AY would like to thank the help of Philip Lucas and Read Mike when using the data of the UKIDSS survey. RD and HN are members of the International Max Planck Research School (IMPRS) for Astronomy and Astrophysics at the Universities of Bonn and Cologne. HB acknowledges support from the European Research Council under the Horizon 2020 Framework Program via the ERC Consolidator Grant CSF-648505. HB also acknowledges support from the DFG in the Collaborative Research Center SFB 881 - Project-ID 138713538 - “The Milky Way System” (subproject B1). VY acknowledge the financial support of CONACyT, M\'exico. This work (partially) uses information from the GLOSTAR database at \url{http://glostar.mpifr-bonn.mpg.de} supported by the MPIfR, Bonn. It also made use of information from the ATLASGAL
database at 
\url{http://atlasgal.mpifr-bonn.mpg.de/cgi-bin/ATLASGAL_DATABASE.cgi}
supported by the MPIfR, Bonn, as well as information from the CORNISH
database at \url{http://cornish.leeds.ac.uk/public/index.php} which
was constructed with support from the Science and Technology 
Facilities Council of the UK.
This work is based in part on observations made with the Spitzer Space 
Telescope, which is operated by the Jet Propulsion Laboratory, California 
Institute of Technology under a contract with NASA.
This publication also makes use of data 
products from the Wide-field Infrared Survey Explorer, which is a 
joint project of the University of California, Los Angeles, and the 
Jet Propulsion Laboratory/California Institute of Technology, funded 
by the National Aeronautics and Space Administration. 
Herschel is an 
ESA space observatory with science instruments provided by European-led 
Principal Investigator consortia and with important participation 
from NASA. This research 
has made use of the SIMBAD database, operated at CDS, Strasbourg, 
France. We have used the collaborative tool Overleaf available at: 
\url{https://www.overleaf.com/}.

\bibliographystyle{aa}
\bibliography{Biblio}

\begin{appendix}

\section{Low signal-to-noise ratio radio sources}

We have identified 1578 radio sources that have signal-to-noise ratio
between 5 and 7$\sigma$. In Tab.~\ref{tbl:glostar_can} we list these
sources with the measured basic properties observed in the radio maps.

\begin{table*}
 \begin{center}
 \small
 \renewcommand{\arraystretch}{1.1}
\setlength{\tabcolsep}{4pt}
\caption{Catalog of sources with signal-to-noise ratio in the range between 5 to 7$\sigma.$}
 \label{tbl:glostar_can} \begin{minipage}{0.7\linewidth}
\begin{tabular}{cccccccccc} 
 \hline\hline
 GLOSTAR name& $\ell$ & $b$&SNR &$S_{\rm peak}$ & $\boldsymbol{\sigma}_\mathbf{S_{\rm peak}}$ &$S_{\rm int}$ &$\boldsymbol{\sigma}_\mathbf{S_{\rm int}}$ & SIMBAD & D-conf. \\ 
{}& (\degr)&(\degr)& & \multicolumn{2}{c}{(mJy\,beam$^{-1}$)} &\multicolumn{2}{c}{(mJy)}&class & name \\ 
(1)&(2)&(3)&(4)&(5)&(6)&(7)&(8)&(9)&(10)\\ \hline 
G028.0142$+00.5778$&28.01418&$ +0.57782$&5.2&0.20&0.07&0.34&0.07&...&...\\ 
G028.0181$+01.0039$&28.01809&$ +1.00390$&5.1&0.34&0.11&0.54&0.11&...&...\\ 
G028.0210$+00.8308$&28.02103&$ +0.83082$&5.3&0.18&0.07&0.37&0.07&...&...\\ 
G028.0217$+00.3191$&28.02170&$ +0.31913$&5.1&0.18&0.07&0.35&0.07&...&...\\ 
G028.0252$-00.1458$&28.02525&$ -0.14580$&5.4&0.20&0.07&0.35&0.07&...&...\\ 
G028.0288$+00.7505$&28.02882&$ +0.75047$&5.0&0.20&0.07&0.33&0.07&...&...\\ 
G028.0295$+00.1057$&28.02948&$ +0.10573$&5.3&0.20&0.07&0.39&0.08&...&...\\ 
G028.0431$+00.5314$&28.04308&$ +0.53136$&5.8&0.32&0.07&0.39&0.07&...&...\\ 
G028.0514$-00.7841$&28.05139&$ -0.78409$&5.1&0.18&0.06&0.32&0.07&...&...\\ 
G028.0573$+00.2387$&28.05733&$ +0.23871$&5.8&0.21&0.07&0.40&0.07&...&...\\ 
G028.0576$+00.8389$&28.05756&$ +0.83895$&5.1&0.16&0.07&0.33&0.07&...&...\\ 
G028.0790$-00.7437$&28.07896&$ -0.74374$&5.0&0.18&0.06&0.32&0.07&...&...\\ 
G028.0894$+00.5602$&28.08940&$ +0.56018$&5.3&0.20&0.07&0.36&0.07&...&...\\ 
G028.0980$-00.7435$&28.09799&$ -0.74353$&6.7&0.25&0.06&0.42&0.07&...&...\\ 
G028.1029$+00.8823$&28.10291&$ +0.88229$&5.2&0.19&0.07&0.33&0.07&...&...\\ 
G028.1052$+00.3603$&28.10524&$ +0.36031$&5.3&0.18&0.07&0.35&0.07&...&...\\ 
\hline
 \end{tabular}\\
 Notes: Only a small portion of the data is provided here, the full table is available in electronic form at the CDS via anonymous ftp to cdsarc.u-strasbg.fr (130.79.125.5) or via http://cdsweb.u-strasbg.fr/cgi-bin/qcat?J/A\&A/.\\ 
 \end{minipage}
\end{center}
\end{table*}

\section{Sample of the multi-wavelength classification}

\begin{figure*}[!h]
    \centering
     \begin{tabular}{cc}
    \includegraphics[width=0.60\textwidth, trim= 0 0 0 0,clip, angle=0, valign=c] {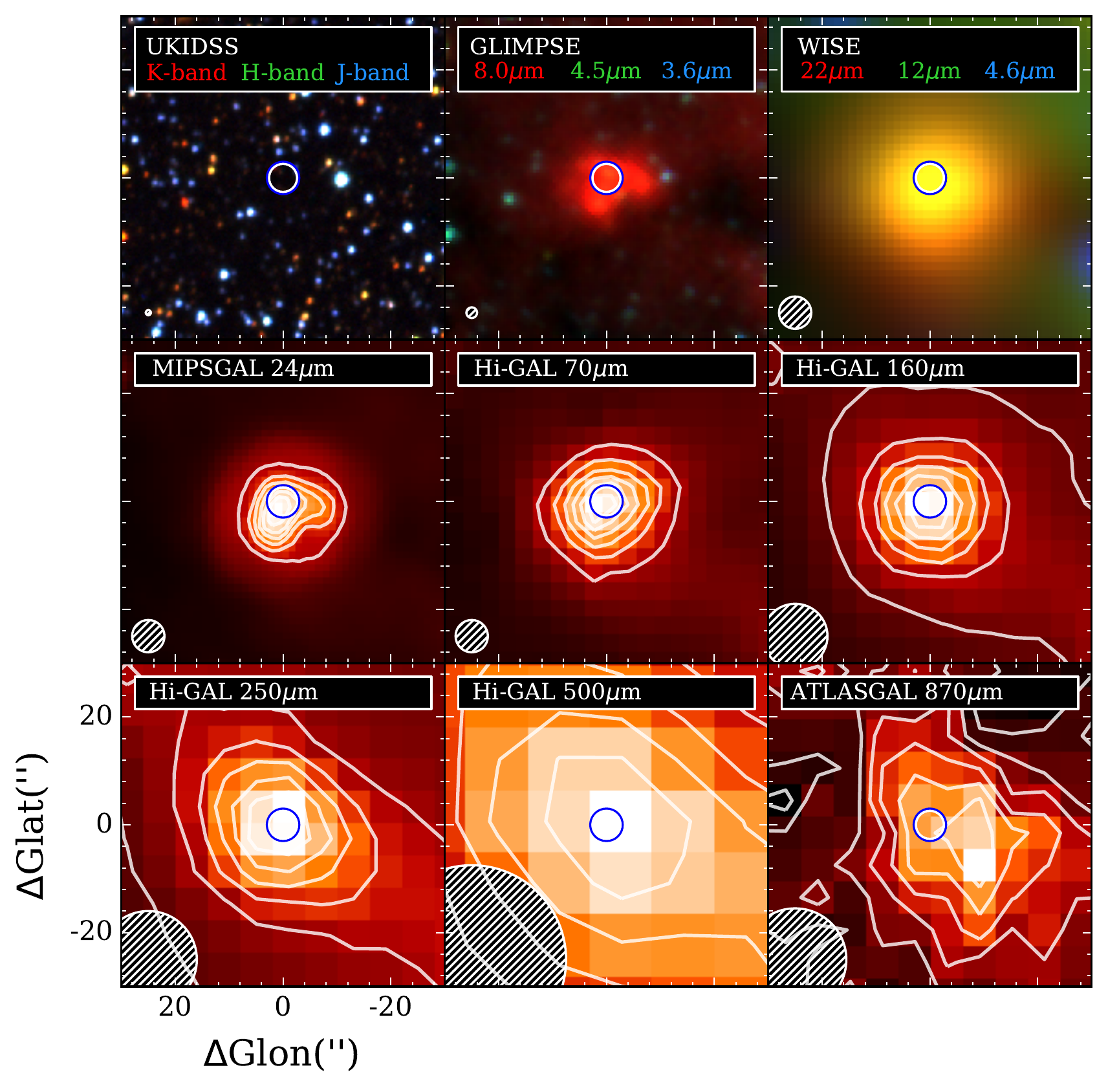}
    & \includegraphics[width=0.35\textwidth, trim= 0 0 0 0,clip, valign=c] {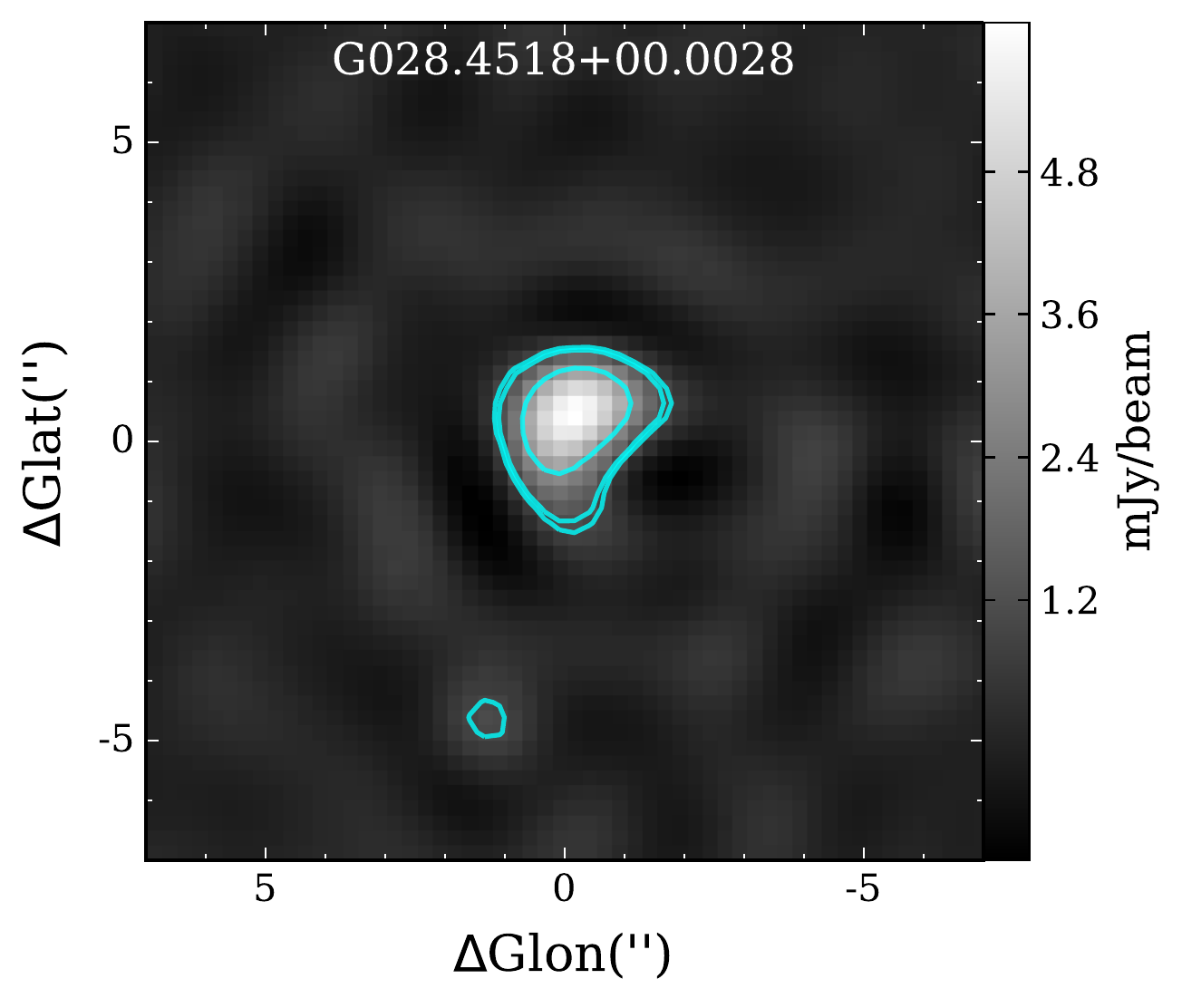}
    \end{tabular}
    \caption{Radio source classified as HII region. {\it Left:} Panels are from top-left to
    bottom right: RGB image of UKIDSS three observed bands, RGB image of GLIMPSE 
    three observed bands, WISE 12~$\mu$m, Hi-GAL 70~$\mu$m, Hi-GAL 160~$\mu$m,
    Hi-GAL 250~$\mu$m, Hi-GAL 500~$\mu$m, and ATLASGAL 870~$\mu$m. {\it Right:}
    Radio source detected in the GLOSTAR B-configuration image.}
    \label{fig:eHII}
\end{figure*}

\begin{figure*}[!h]
    \centering
     \begin{tabular}{cc}
    \includegraphics[width=0.60\textwidth, trim= 0 0 0 0,clip, angle=0, valign=c] {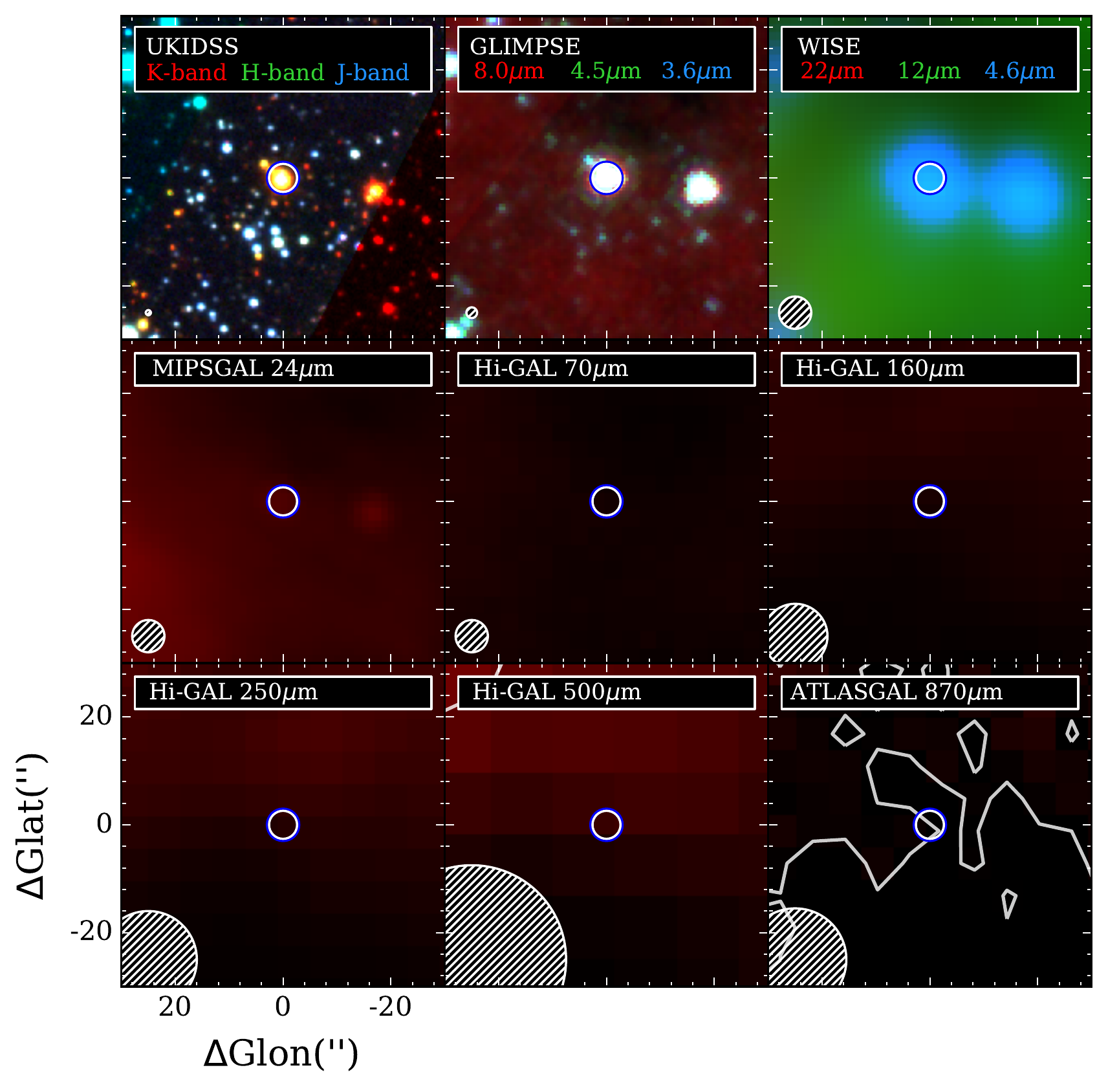}
        & \includegraphics[width=0.35\textwidth, trim= 0 0 0 0,clip, valign=c] {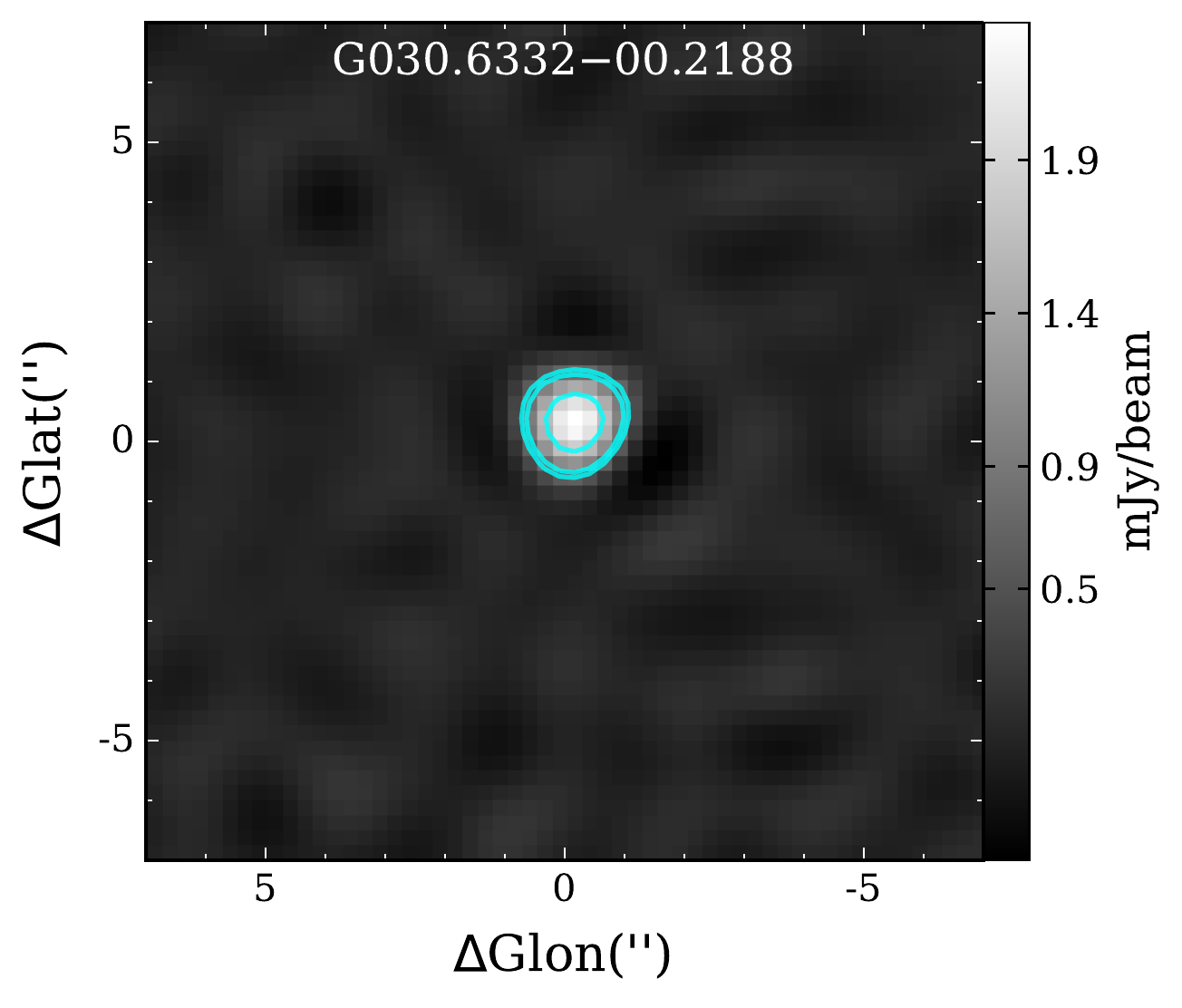}
    \end{tabular}
    \caption{Same as Figure~\ref{fig:eHII}, but for radio source classified as radio star.}
    \label{fig:eRS}
\end{figure*}

\begin{figure*}[!h]
    \centering
    \begin{tabular}{cc}
    \includegraphics[width=0.60\textwidth, trim= 0 0 0 0,clip, angle=0, valign=c] {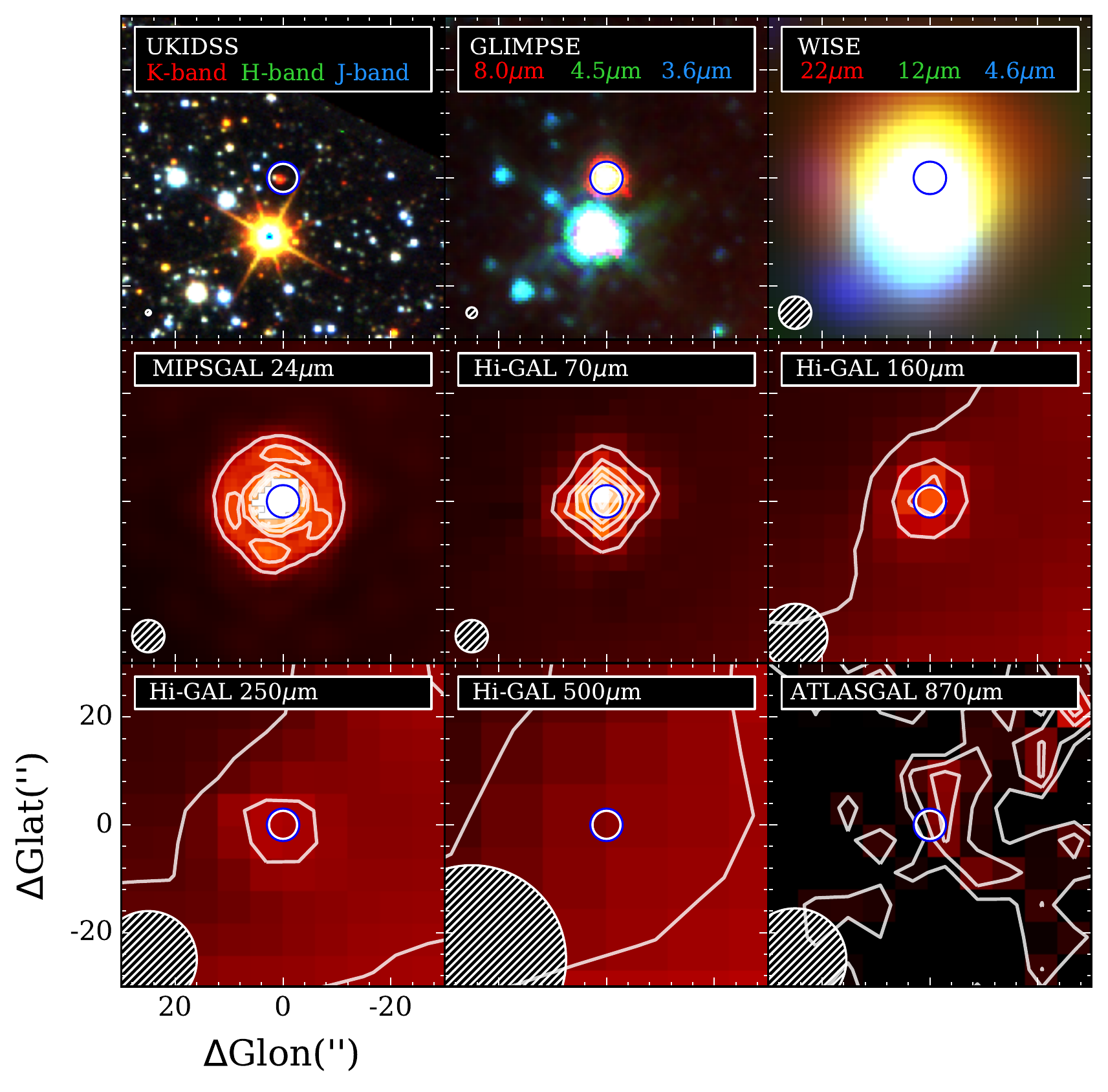}
    & \includegraphics[width=0.35\textwidth, trim= 0 0 0 0,clip, valign=c] {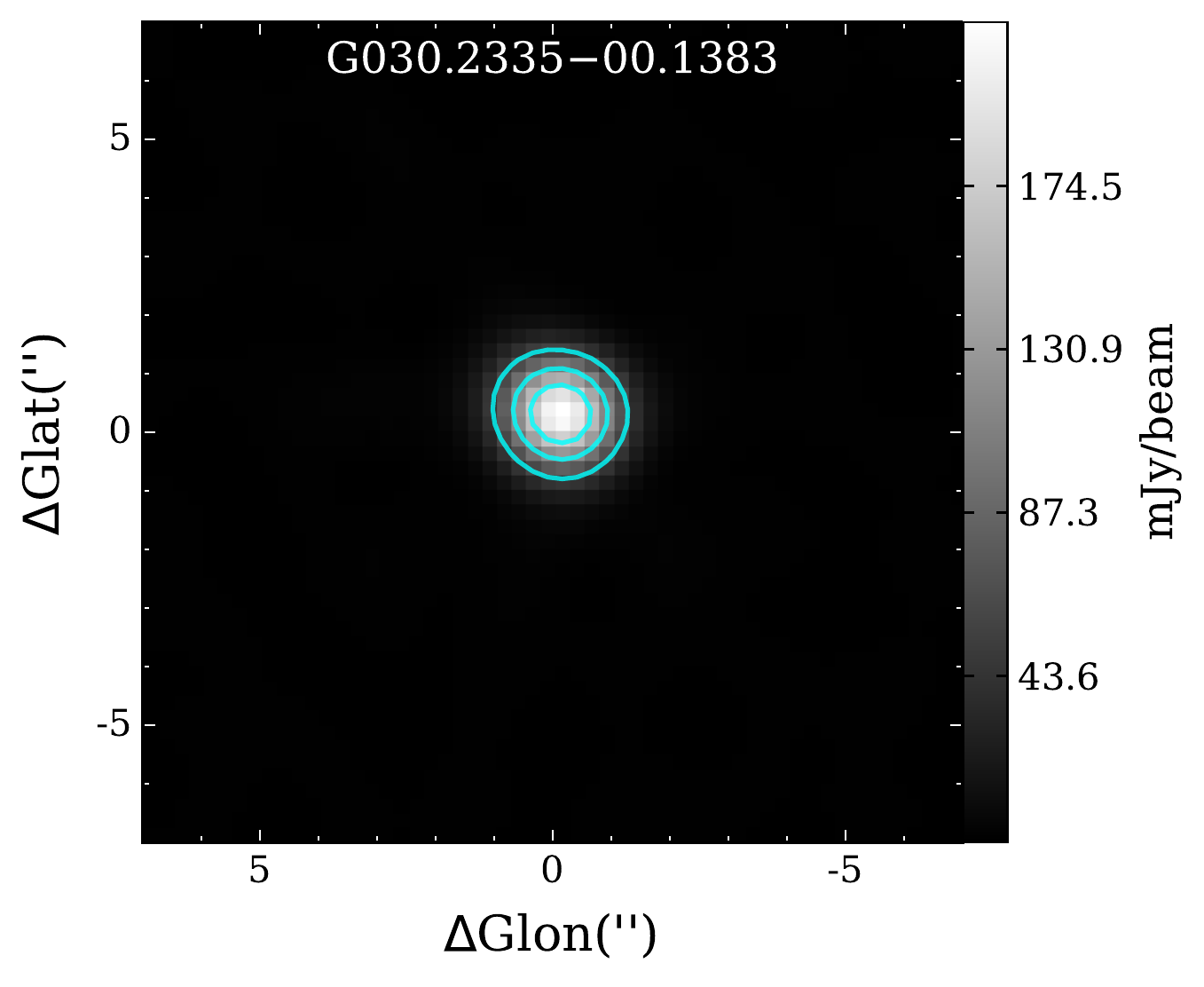}
    \end{tabular}
    \caption{Same as Figure~\ref{fig:eHII}, but for radio source classified as PN.}
    \label{fig:ePN}
\end{figure*}

\begin{figure*}[!h]
    \centering
     \begin{tabular}{cc}
    \includegraphics[width=0.60\textwidth, trim= 0 0 0 0,clip, angle=0, valign=c] {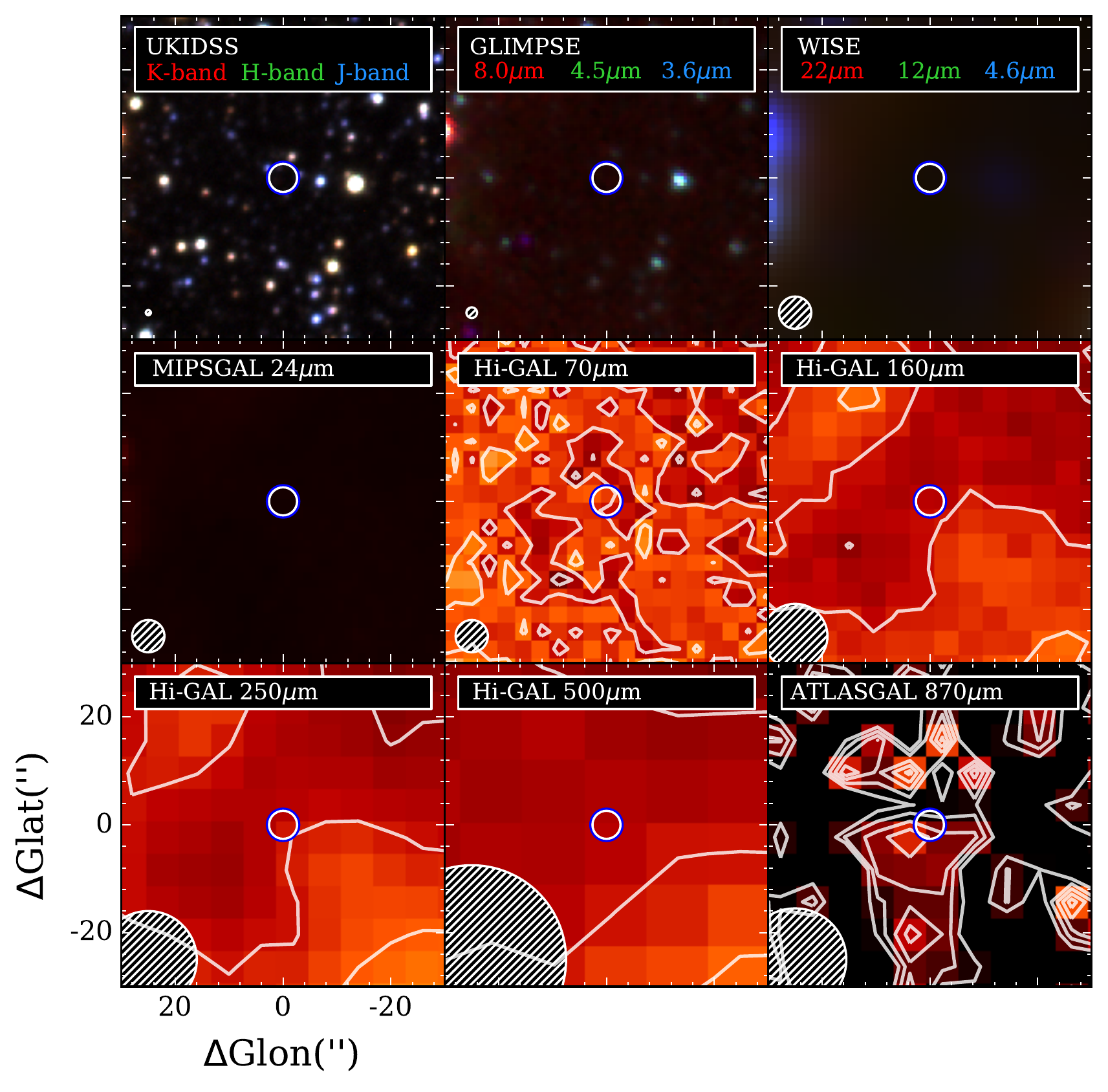}
        & \includegraphics[width=0.35\textwidth, trim= 0 0 0 0,clip, valign=c] {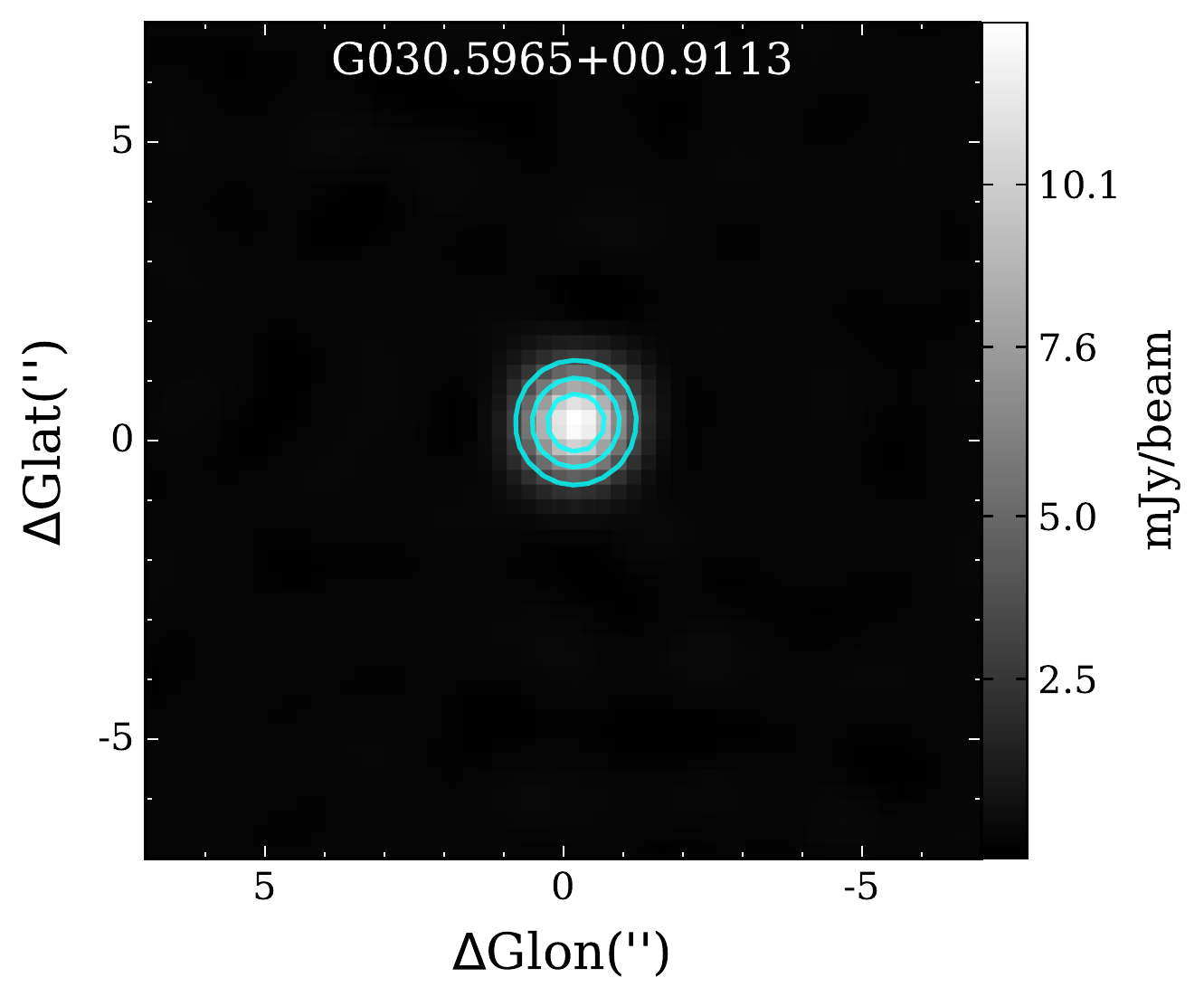}
    \end{tabular}
    \caption{Same as Figure~\ref{fig:eHII}, but for radio source classified as EgC.}
    \label{fig:eEgC}
\end{figure*}

We have searched the prominent IR surveys  to do a visual 
inspection in the position of the highly reliable radio sources.
Using this visual inspection we have done the classification described 
in section 3.8. In Figures~\ref{fig:eHII}, \ref{fig:eRS}, \ref{fig:ePN},
and \ref{fig:eEgC}, we show examples of the images used for the classification
as \hii\ region candidates, radio stars, PNe, and EgCs, respectively. 
In addition to the images of the surveys described in Section 3.8, we have 
included the images of the 24~$\mu$m
emission from the Multiband Imaging Photometer Spitzer Galactic Plane Survey
\citep[MIPSGAL][]{carey2009}.

\section{MIR emission around identified HII region candidates}\label{app:IRHII}

In this appendix, we present a series of images to show the MIR emission as seen by the GLIMPSE 
survey centered at the position of the identified HII region candidates. The images are boxes 
with size of $90''\times90''$, and the false colors are defined to be red, green and blue for 8.0,
4.5 and 3.6$\,\mu$m.

\begin{figure*}[!h]
\centering
\includegraphics[width=0.23\textwidth, trim= 0 0 0 0,clip]{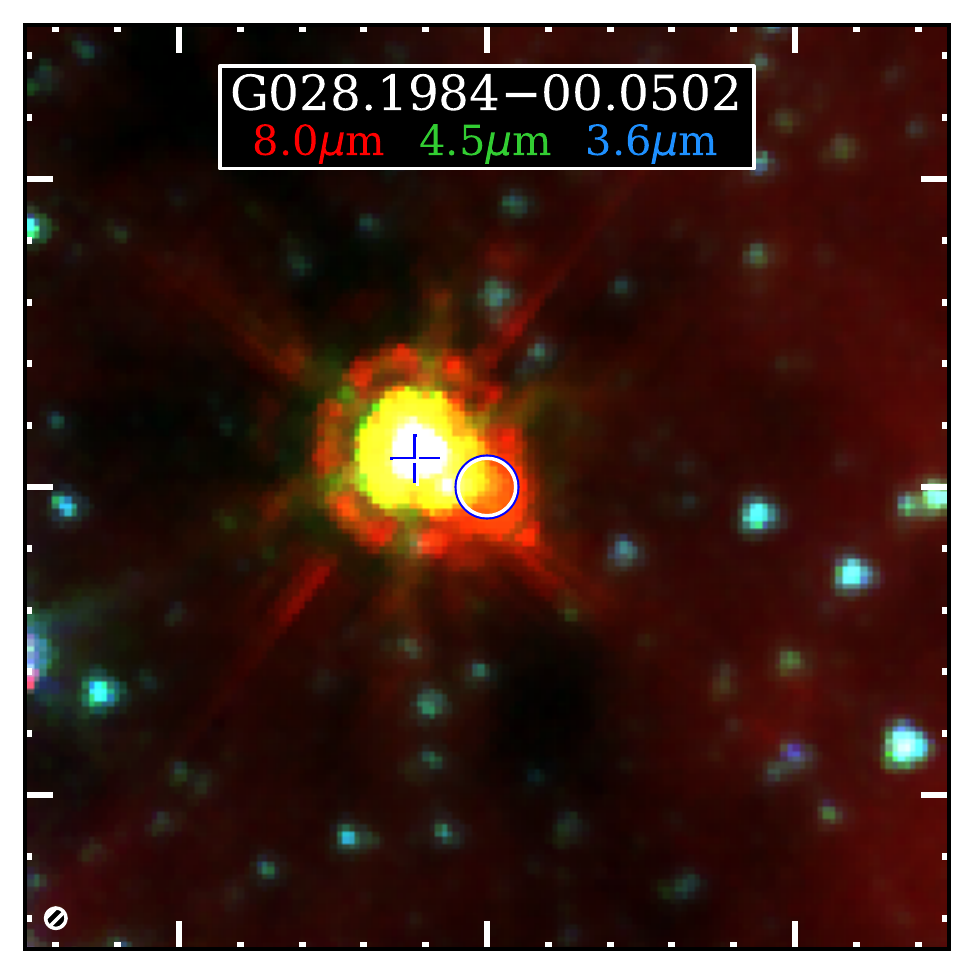}
\includegraphics[width=0.23\textwidth, trim= 0 0 0 0,clip]{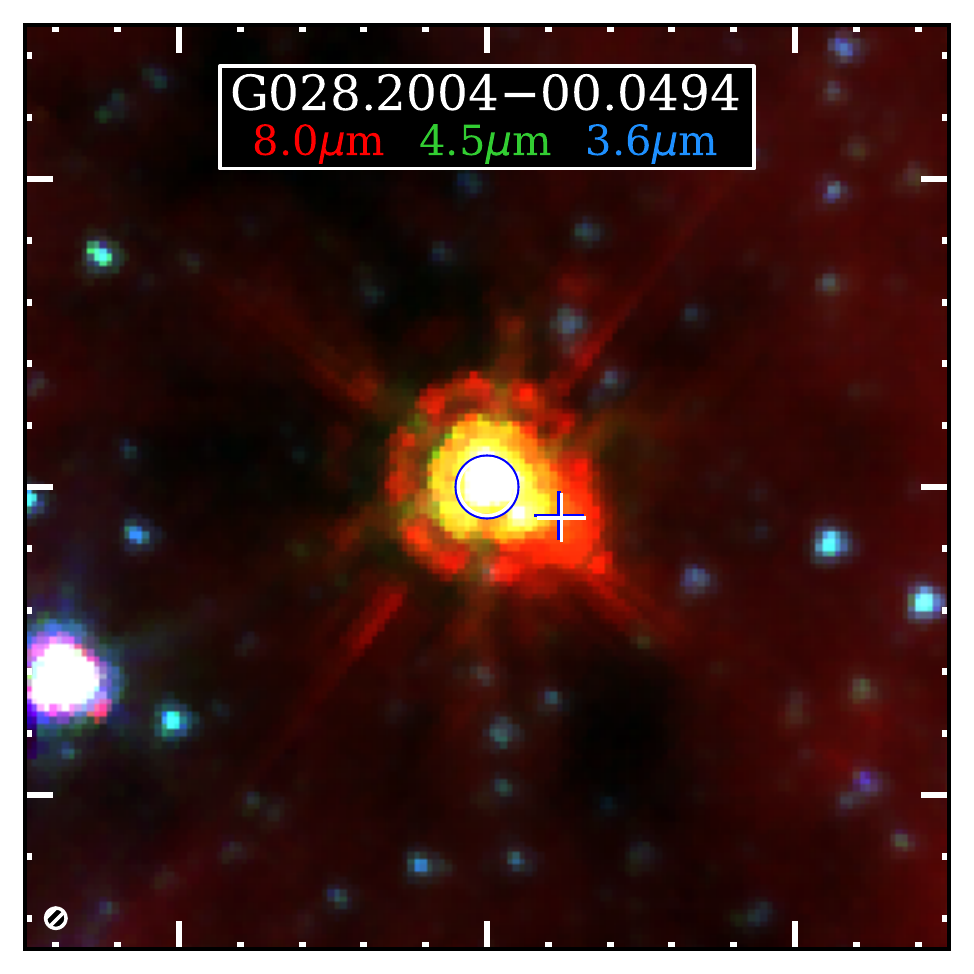}
\includegraphics[width=0.23\textwidth, trim= 0 0 0 0,clip]{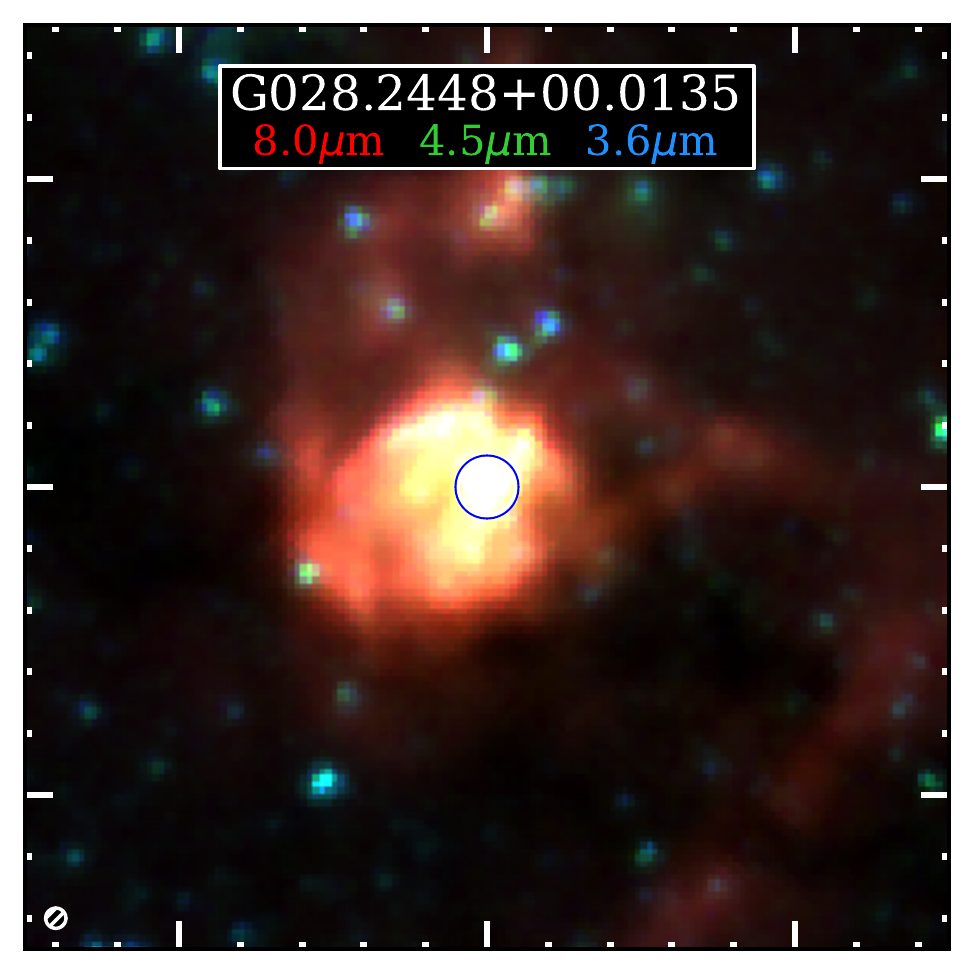}
\includegraphics[width=0.23\textwidth, trim= 0 0 0 0,clip]{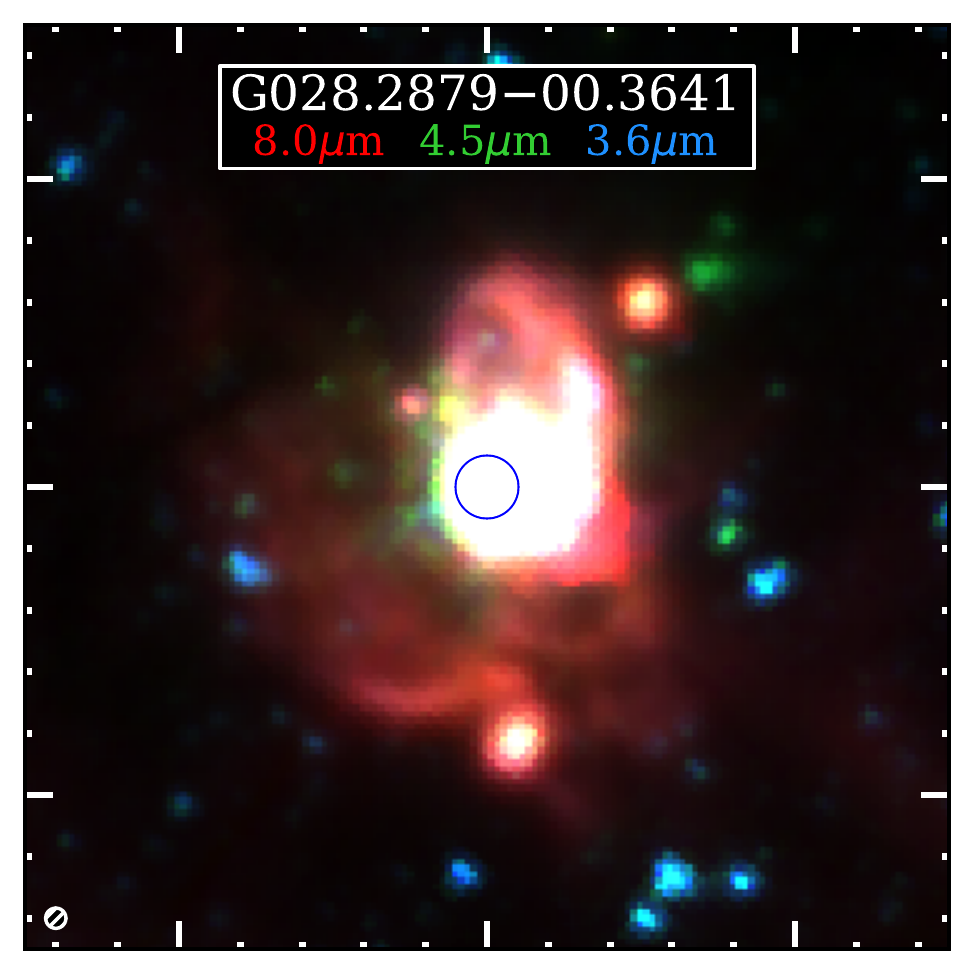}\\ 
\includegraphics[width=0.23\textwidth, trim= 0 0 0 0,clip]{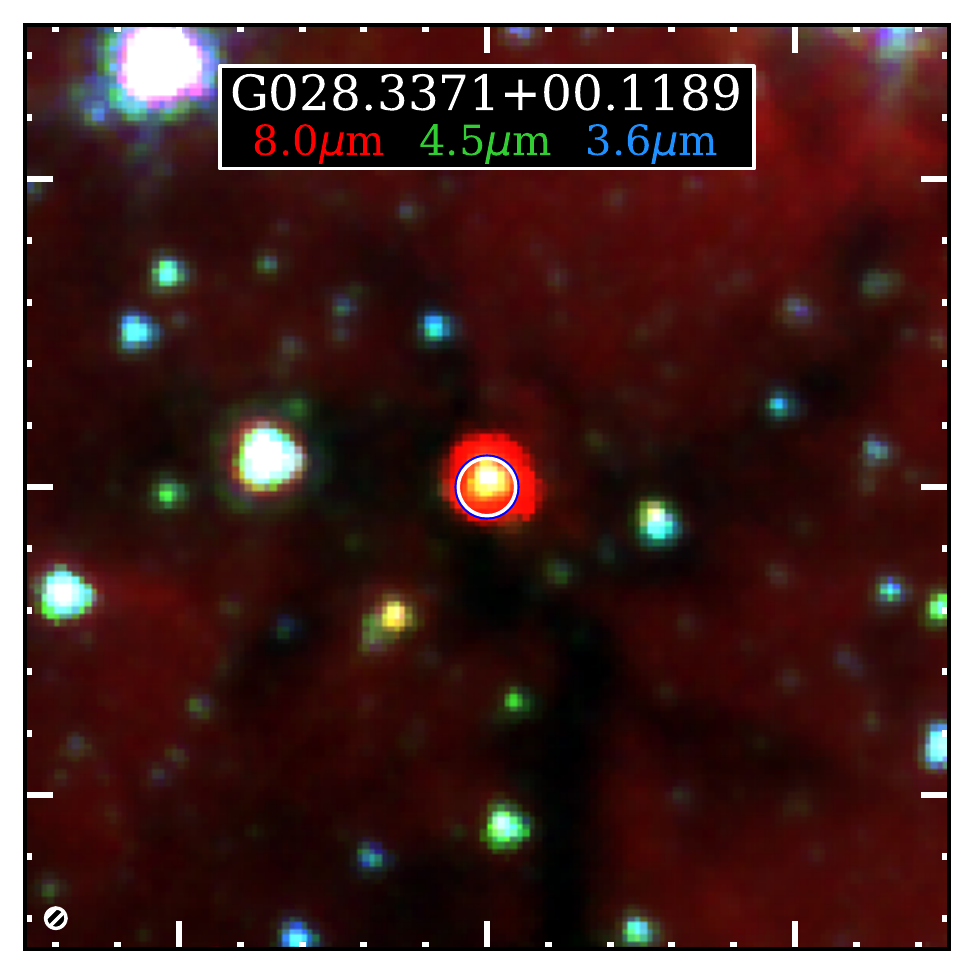}
\includegraphics[width=0.23\textwidth, trim= 0 0 0 0,clip]{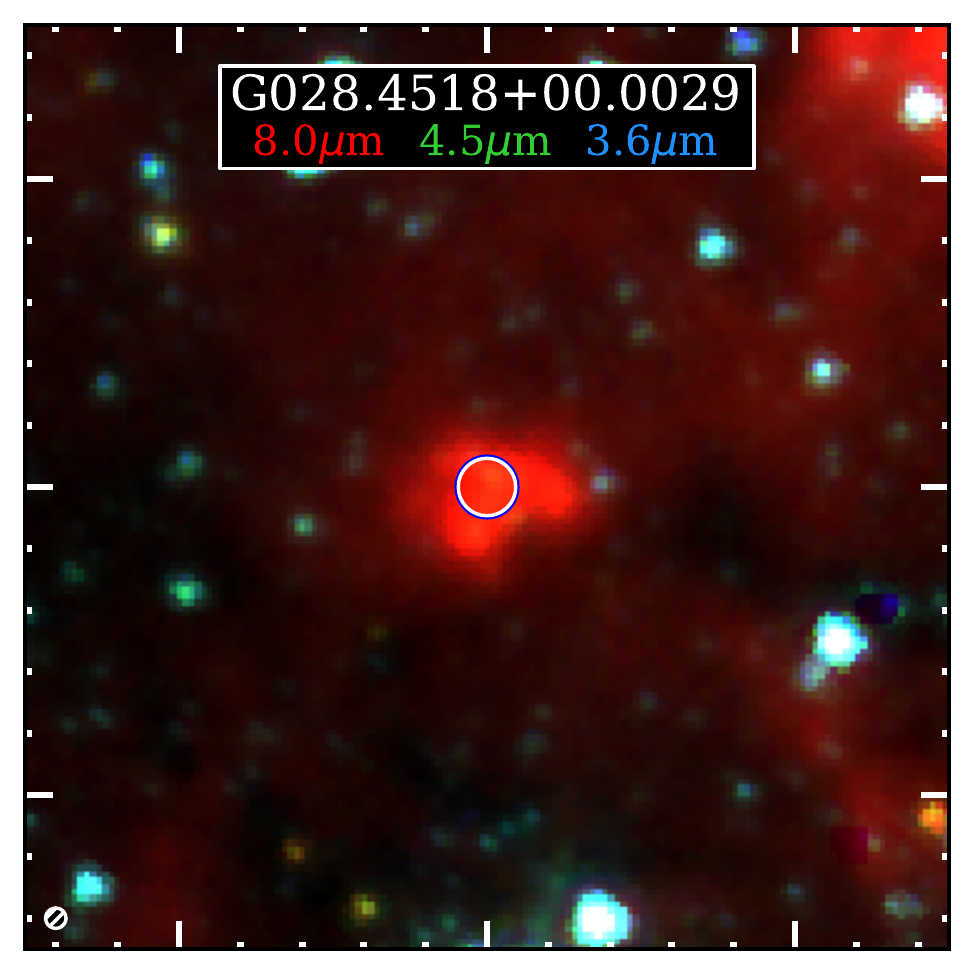}
\includegraphics[width=0.23\textwidth, trim= 0 0 0 0,clip]{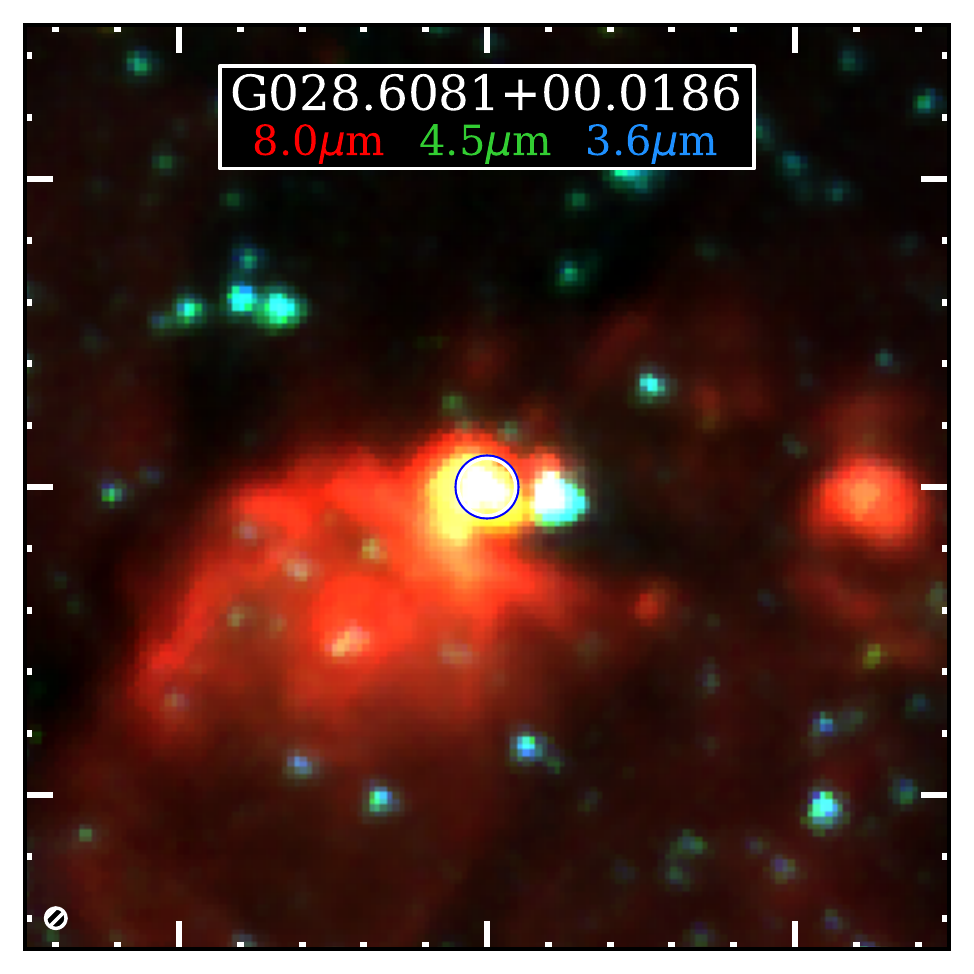}
\includegraphics[width=0.23\textwidth, trim= 0 0 0 0,clip]{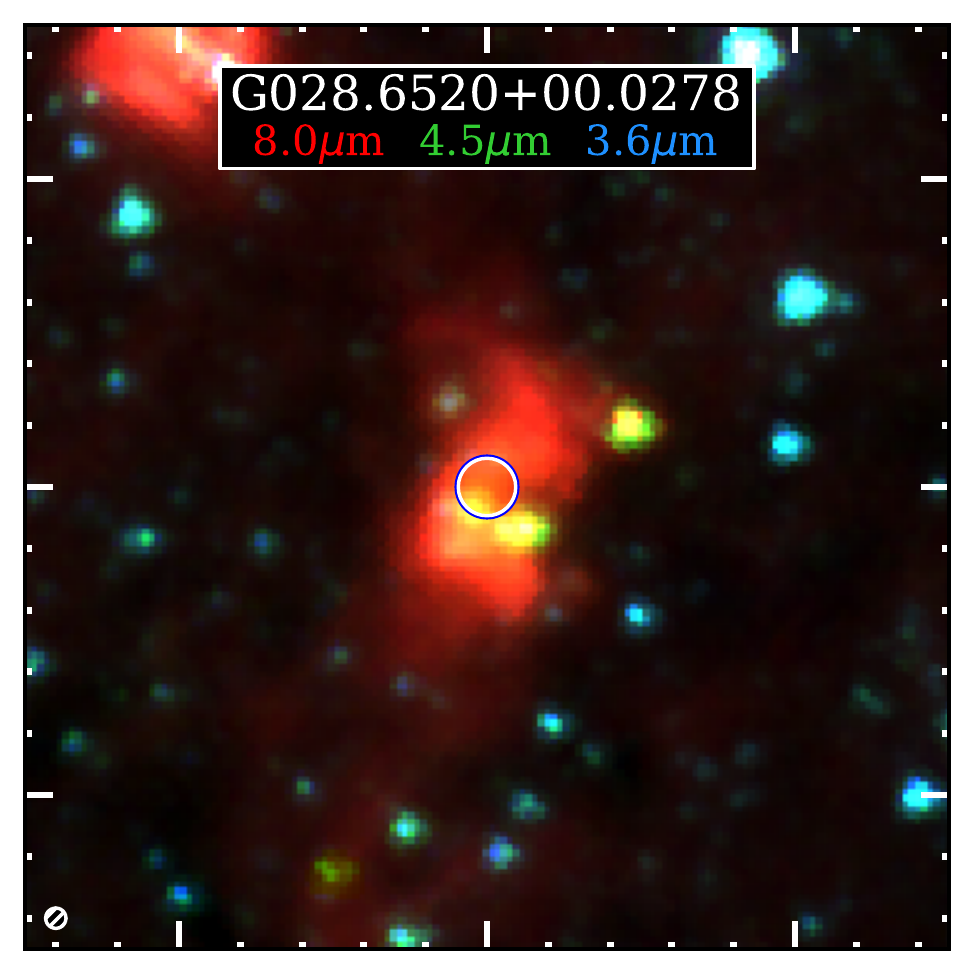}\\ 
\includegraphics[width=0.23\textwidth, trim= 0 0 0 0,clip]{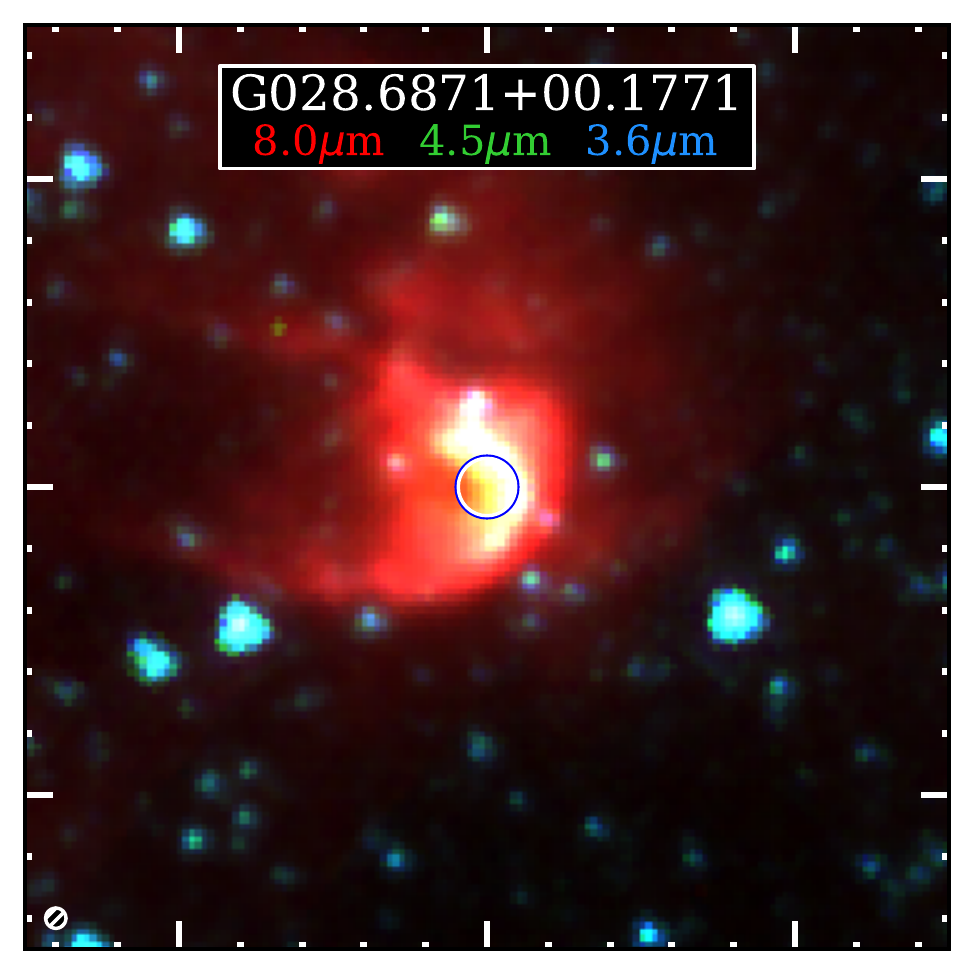}
\includegraphics[width=0.23\textwidth, trim= 0 0 0 0,clip]{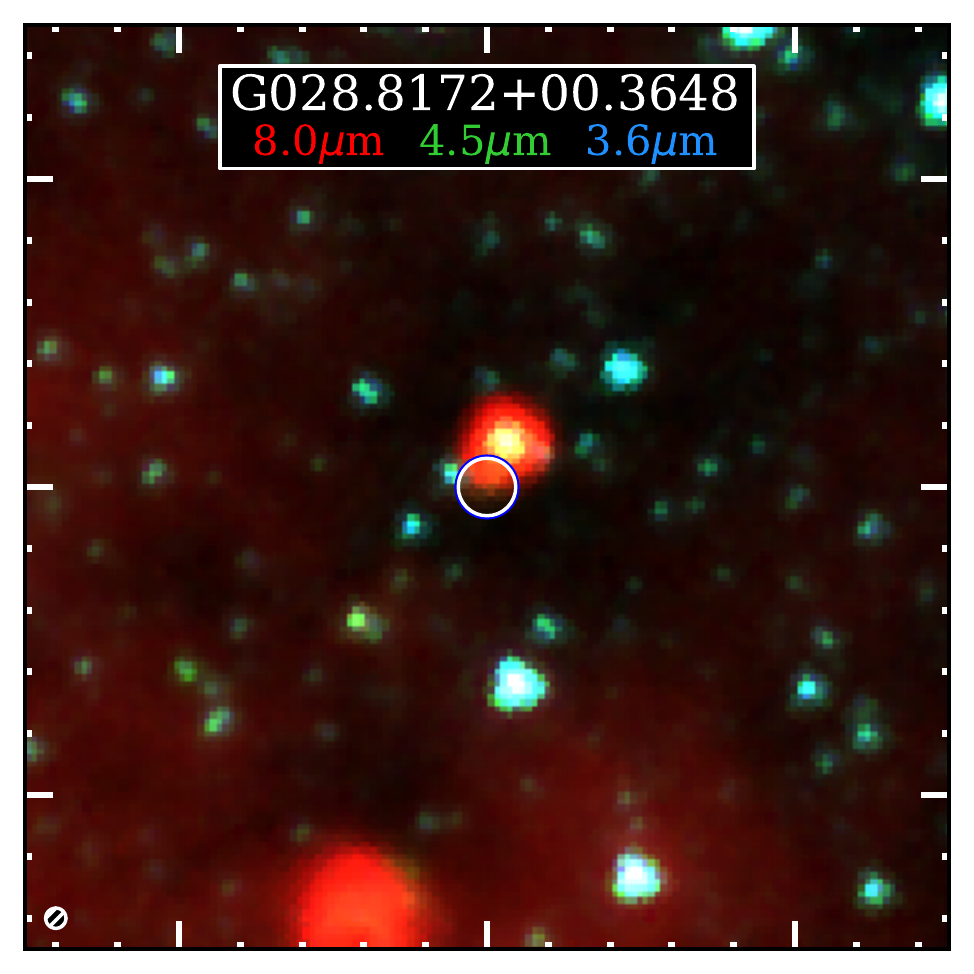}
\includegraphics[width=0.23\textwidth, trim= 0 0 0 0,clip]{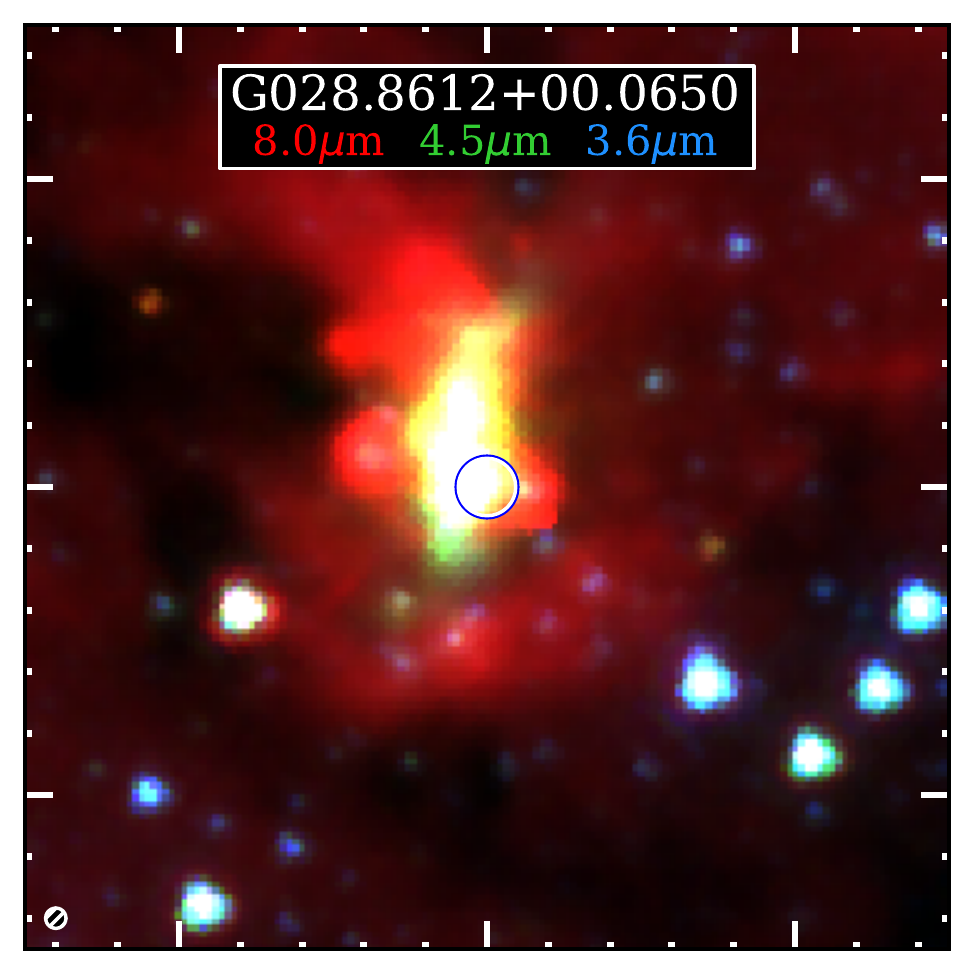}
\includegraphics[width=0.23\textwidth, trim= 0 0 0 0,clip]{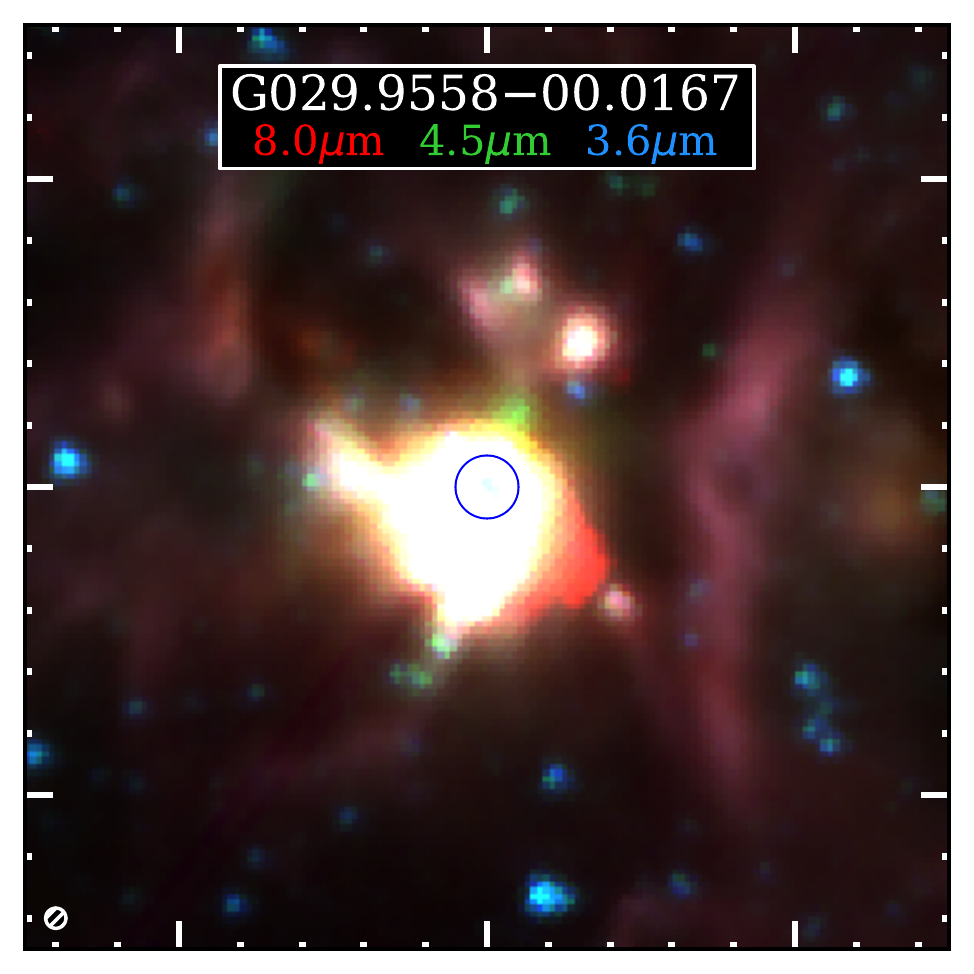}\\ 
\includegraphics[width=0.23\textwidth, trim= 0 0 0 0,clip]{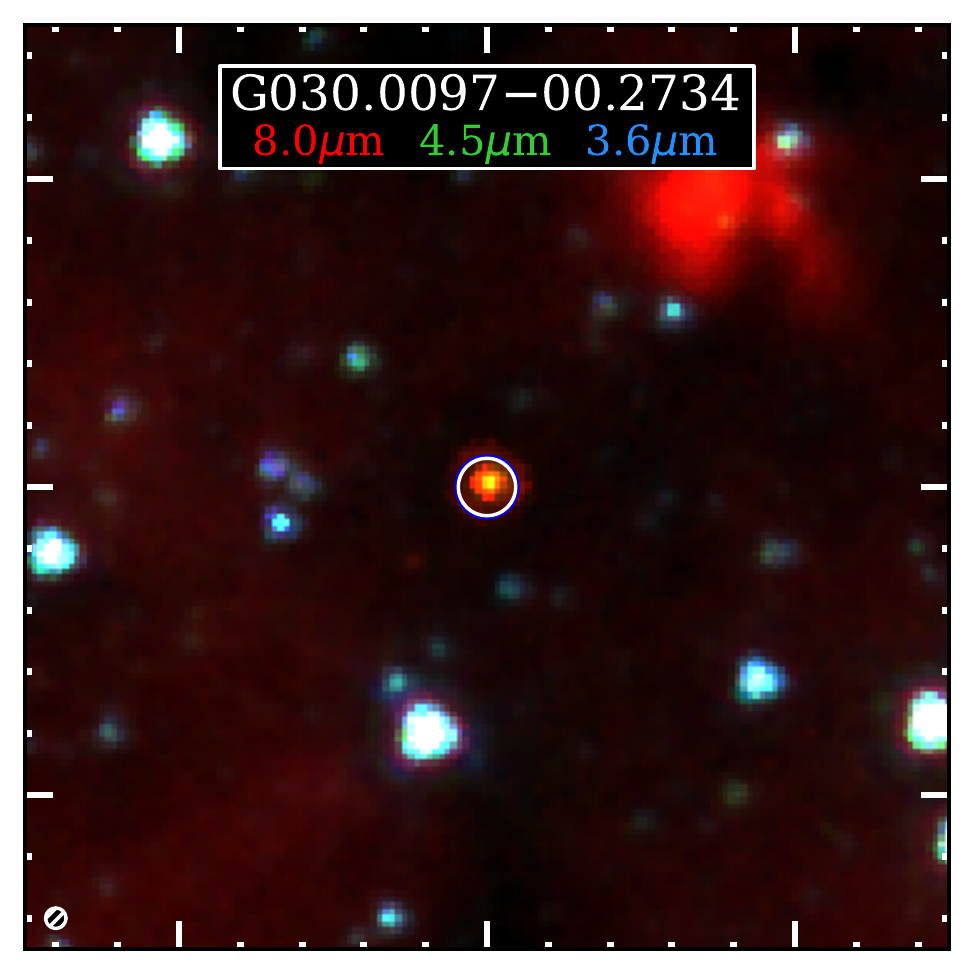}
\includegraphics[width=0.23\textwidth, trim= 0 0 0 0,clip]{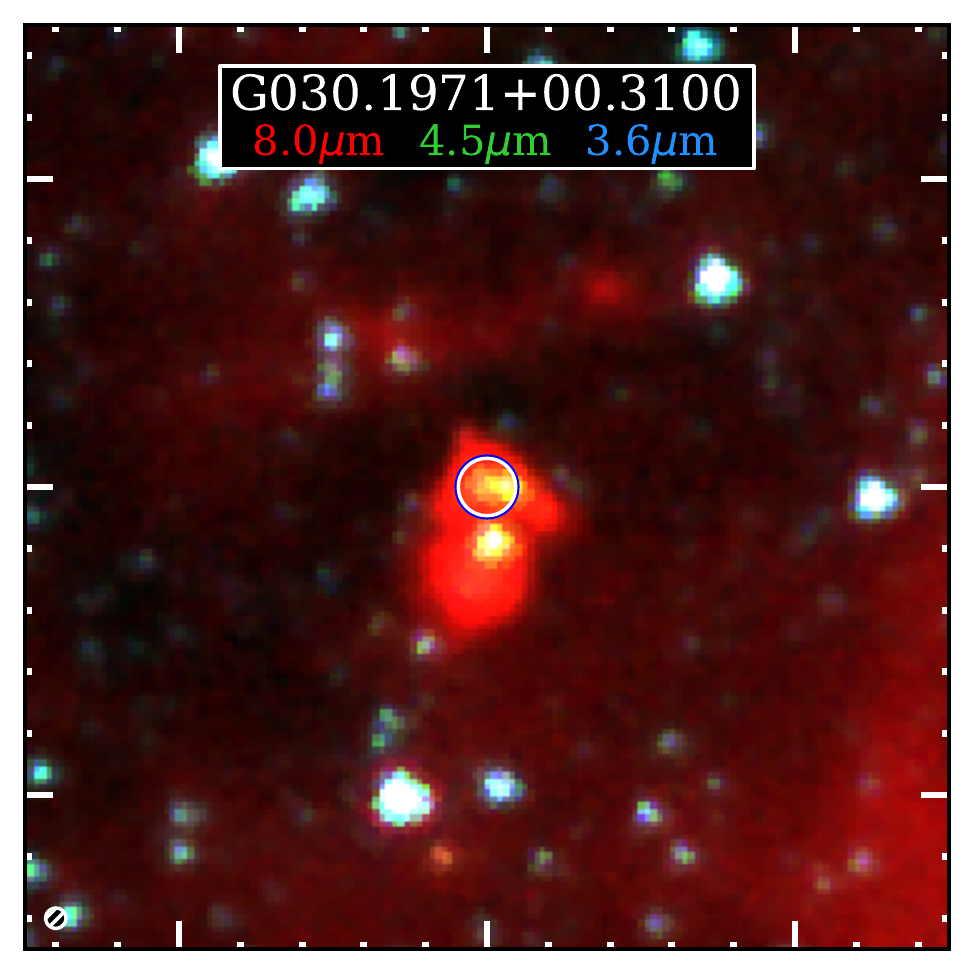}
\includegraphics[width=0.23\textwidth, trim= 0 0 0 0,clip]{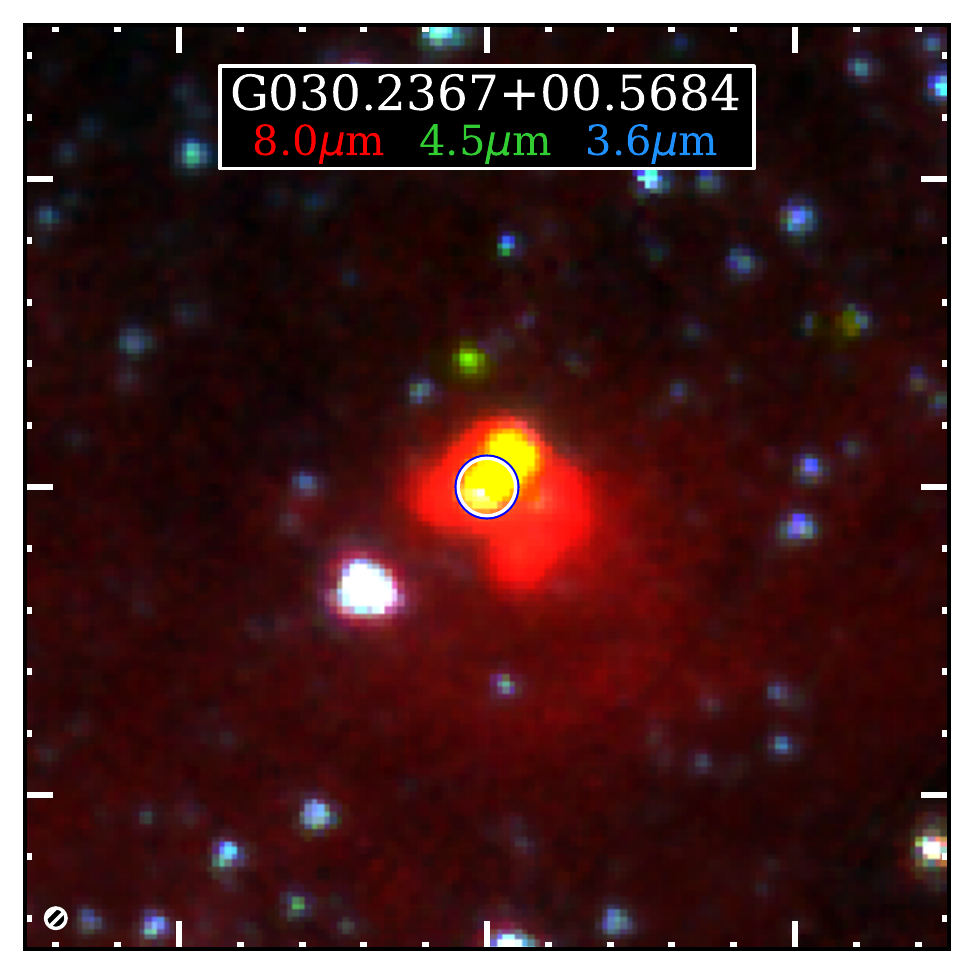}
\includegraphics[width=0.23\textwidth, trim= 0 0 0 0,clip]{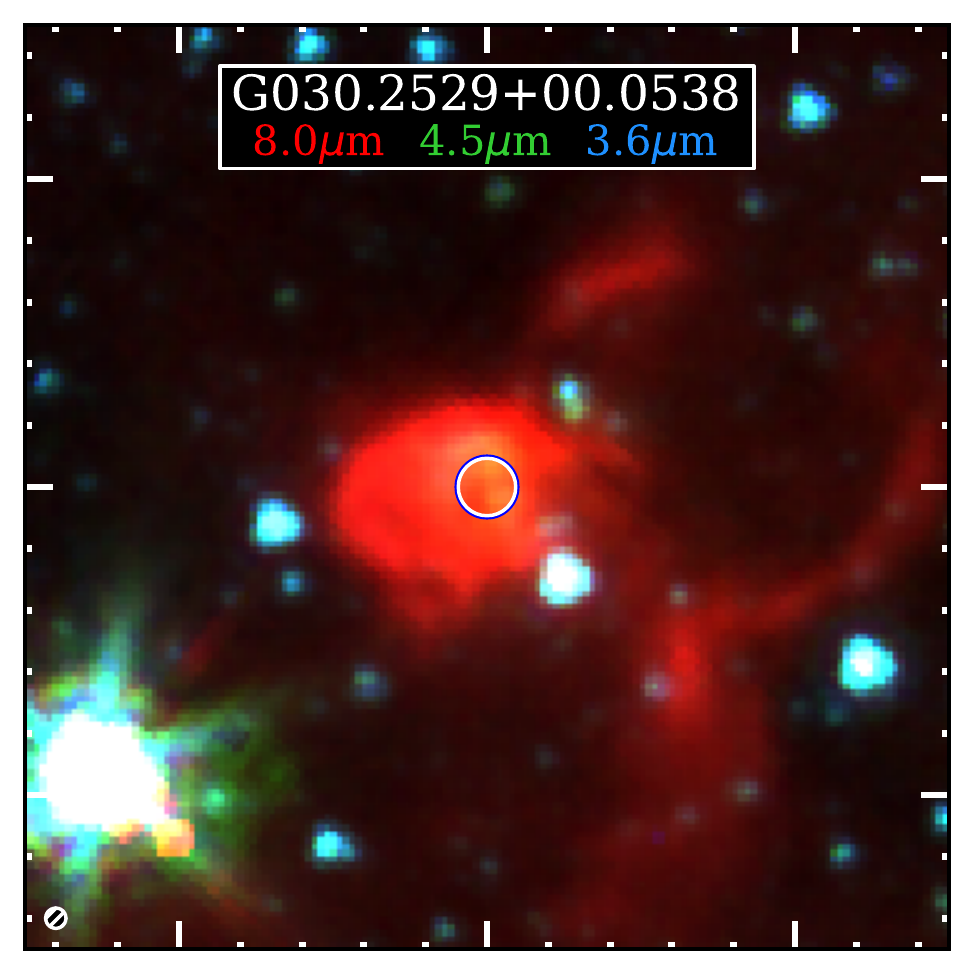}\\ 
\hspace{-0.85cm}\includegraphics[width=0.28\textwidth, trim= 0 0 0 0,clip,valign=t]{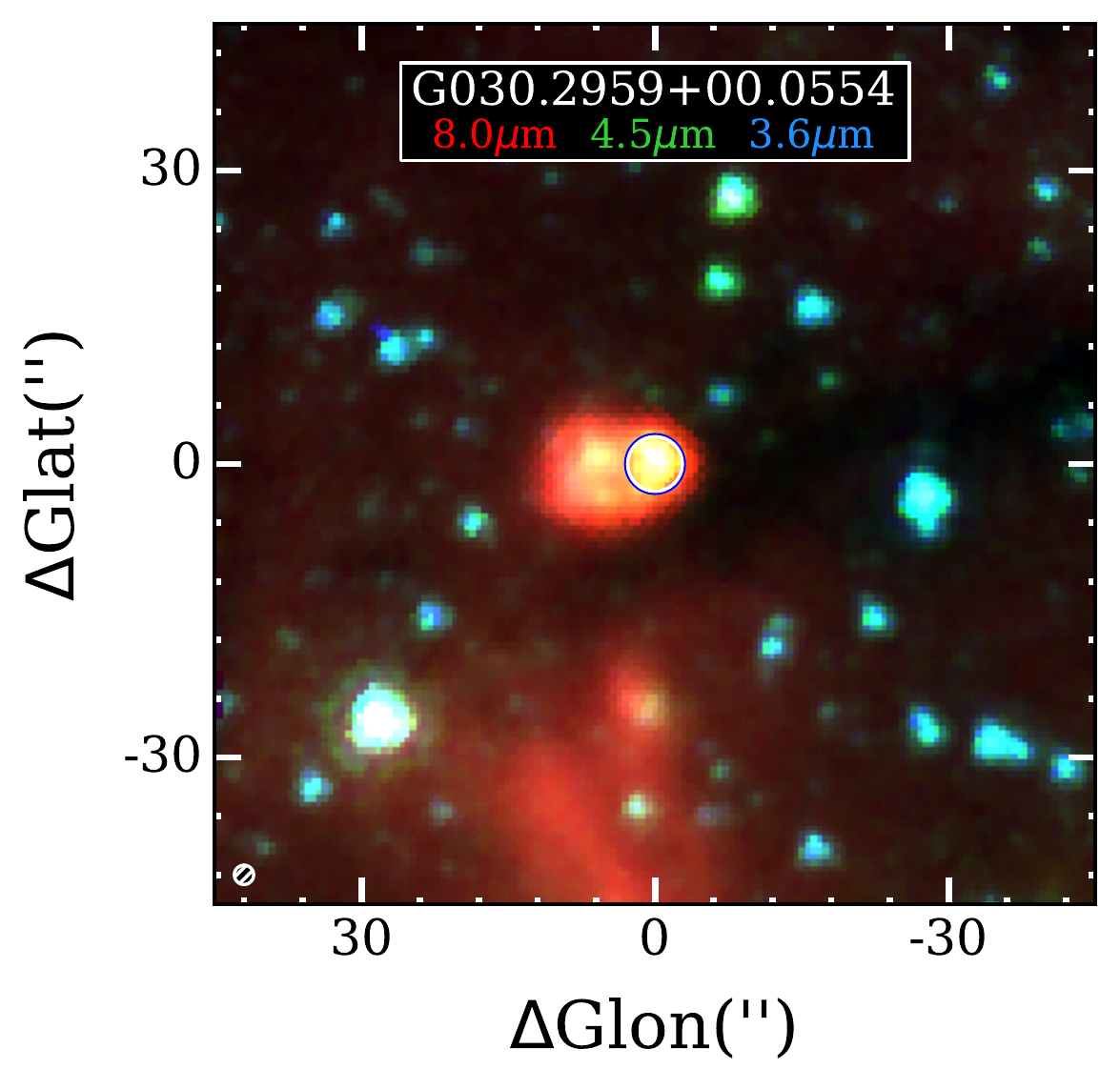}
\includegraphics[width=0.23\textwidth, trim= 0 0 0 0,clip,valign=t]{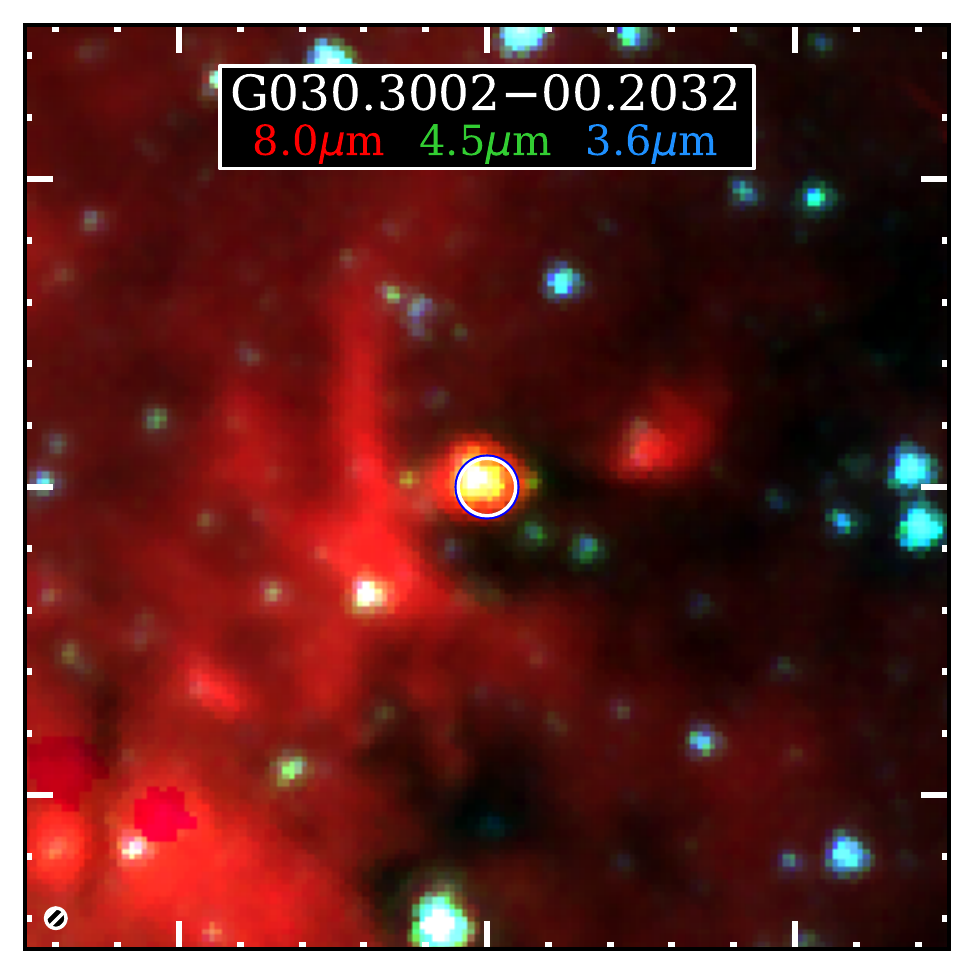}
\includegraphics[width=0.23\textwidth, trim= 0 0 0 0,clip,valign=t]{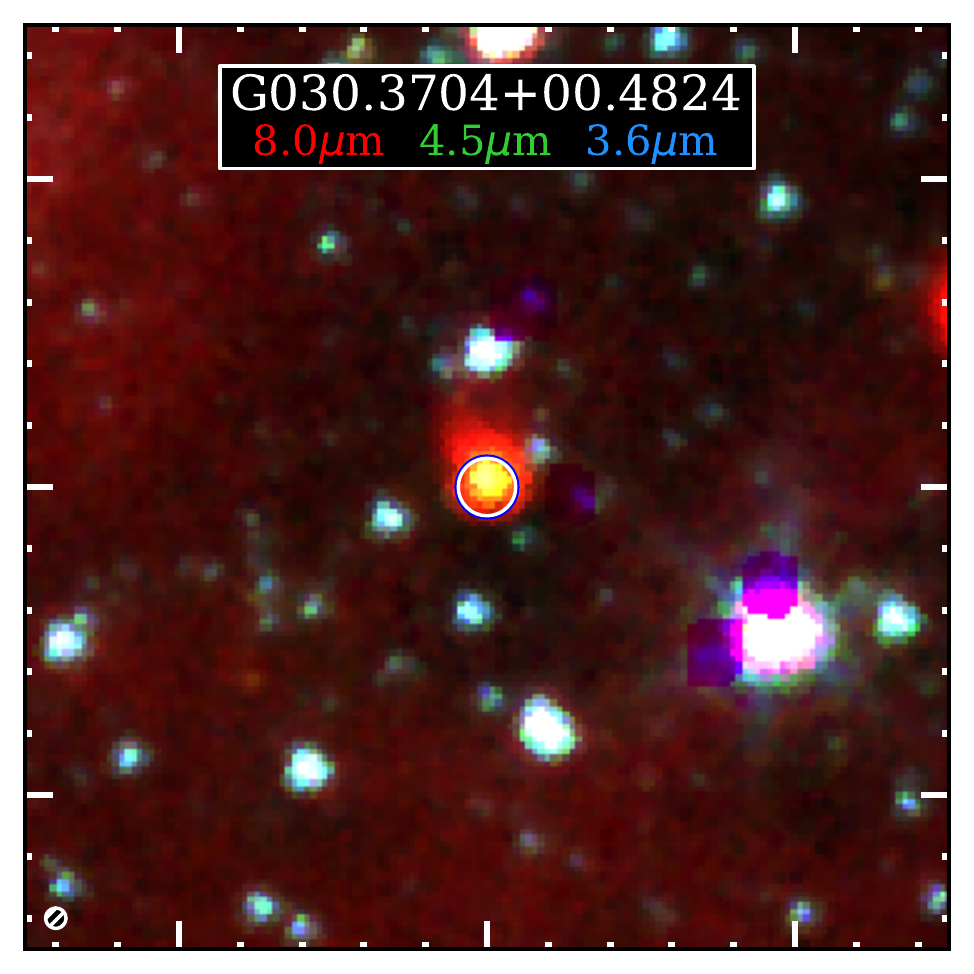}
\includegraphics[width=0.23\textwidth, trim= 0 0 0 0,clip,valign=t]{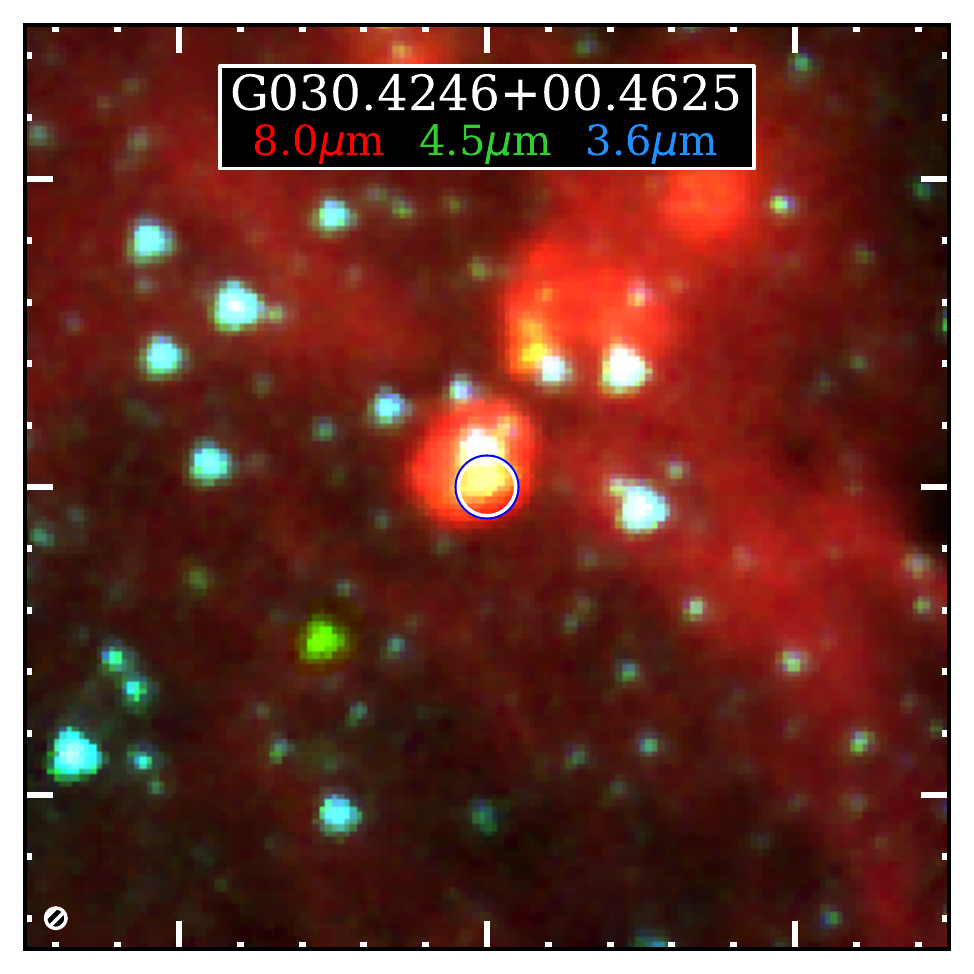}\\
\caption{Three color GLIMPSE images at the position of the HII region candidates identified in the GLOSTAR B-configuration images. In the case of the fragmented HII regions, listed in Table~\ref{tbl:glostar_F}, the position of additional fragment sources are shown with lime-white crosses. }
\end{figure*}

\setcounter{figure}{0}\begin{figure*}[!h]
\centering
\includegraphics[width=0.23\textwidth, trim= 0 0 0 0,clip]{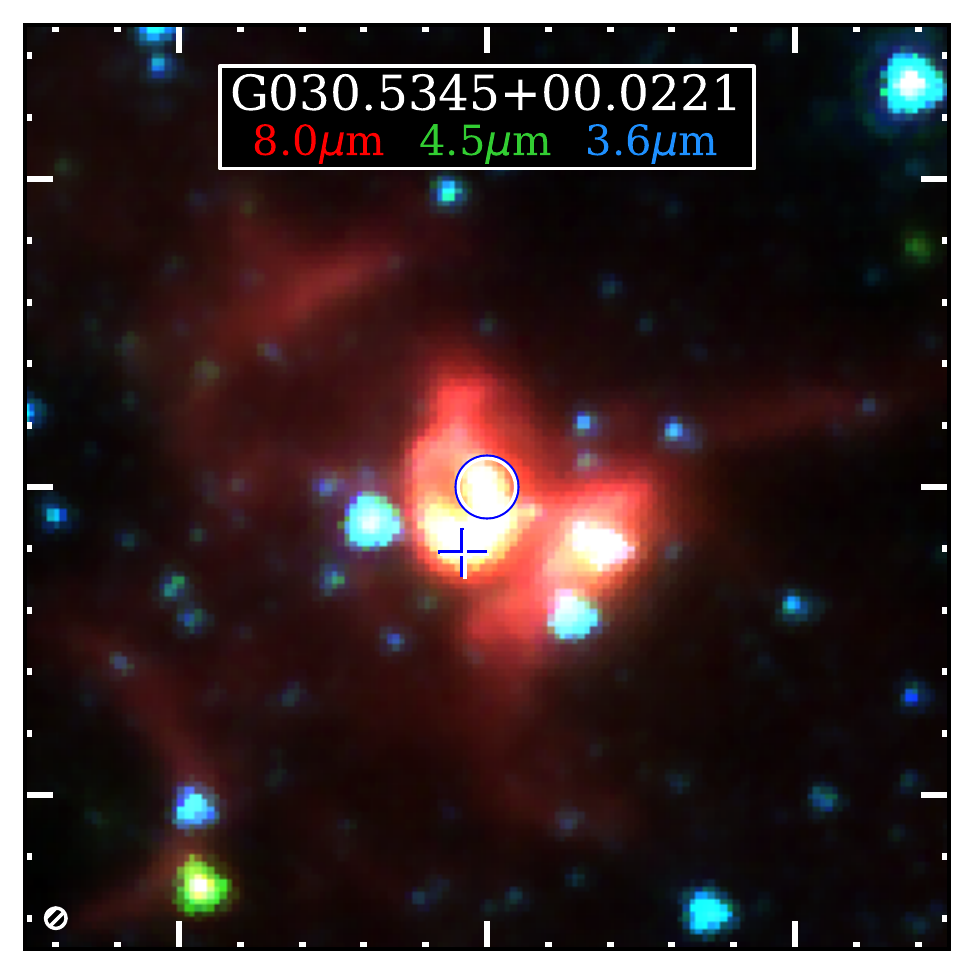}
\includegraphics[width=0.23\textwidth, trim= 0 0 0 0,clip]{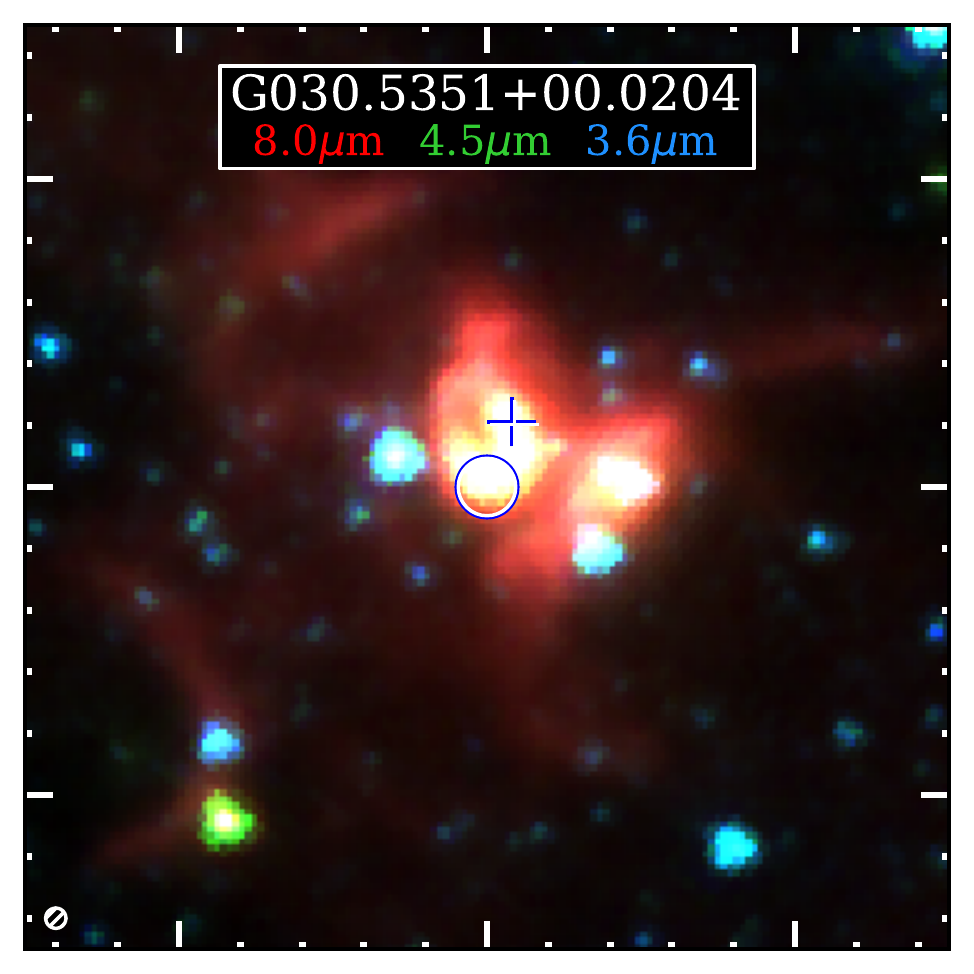}
\includegraphics[width=0.23\textwidth, trim= 0 0 0 0,clip]{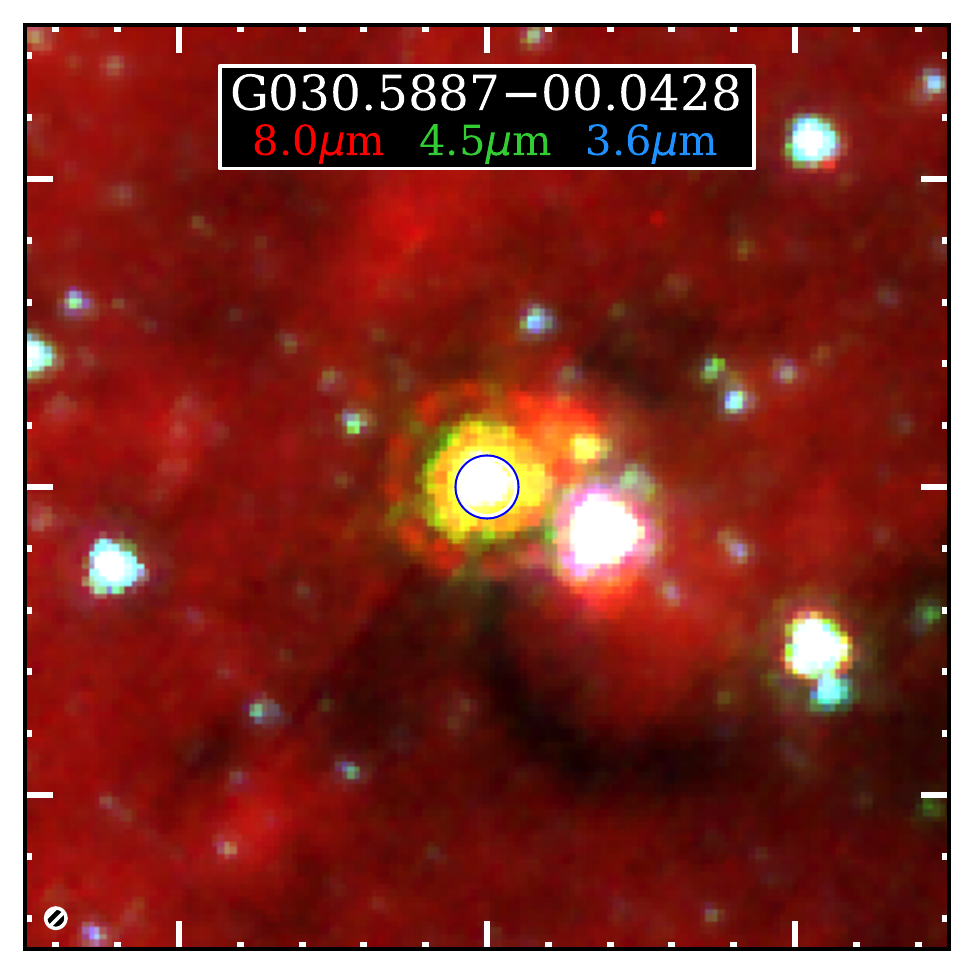}
\includegraphics[width=0.23\textwidth, trim= 0 0 0 0,clip]{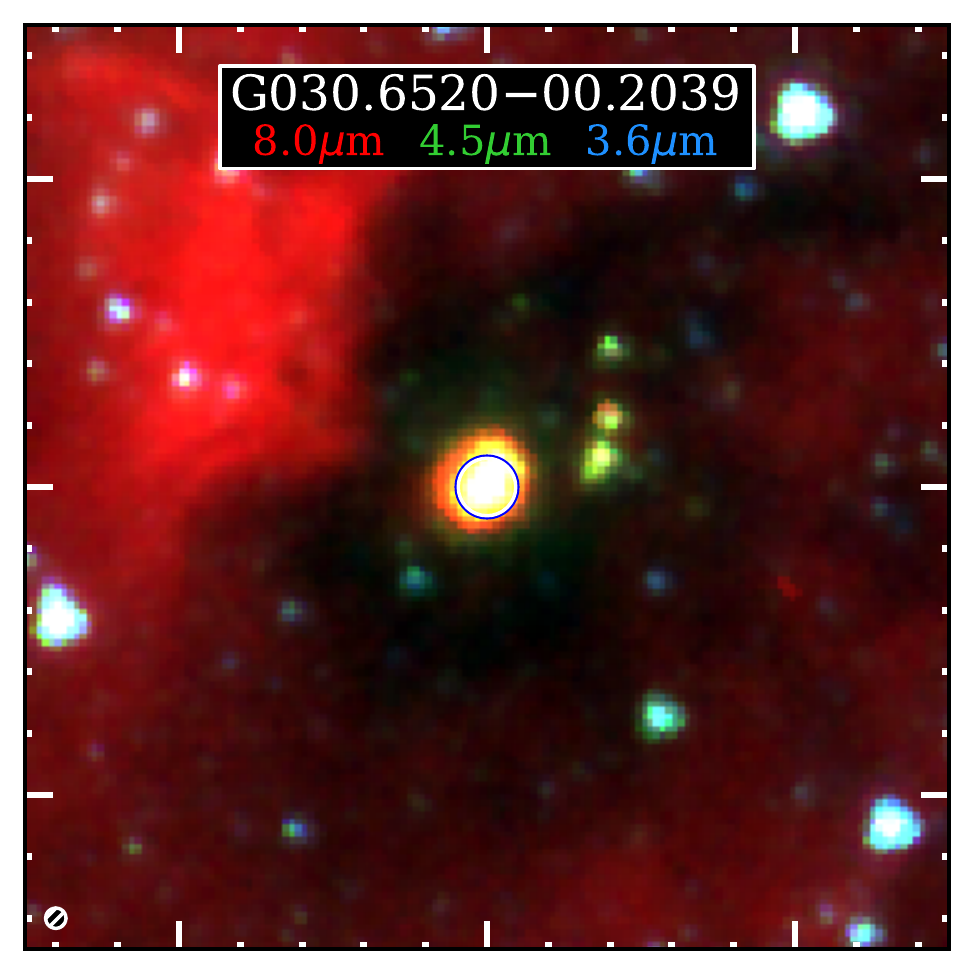}\\ 
\includegraphics[width=0.23\textwidth, trim= 0 0 0 0,clip]{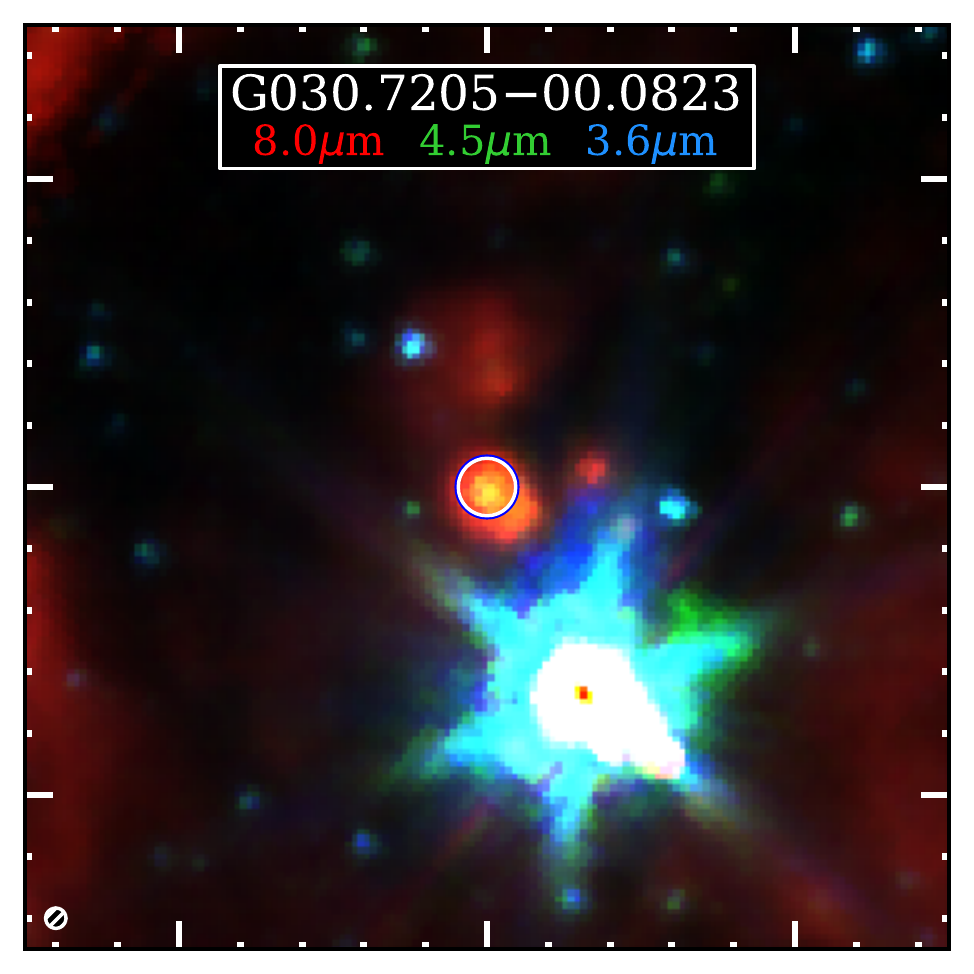}
\includegraphics[width=0.23\textwidth, trim= 0 0 0 0,clip]{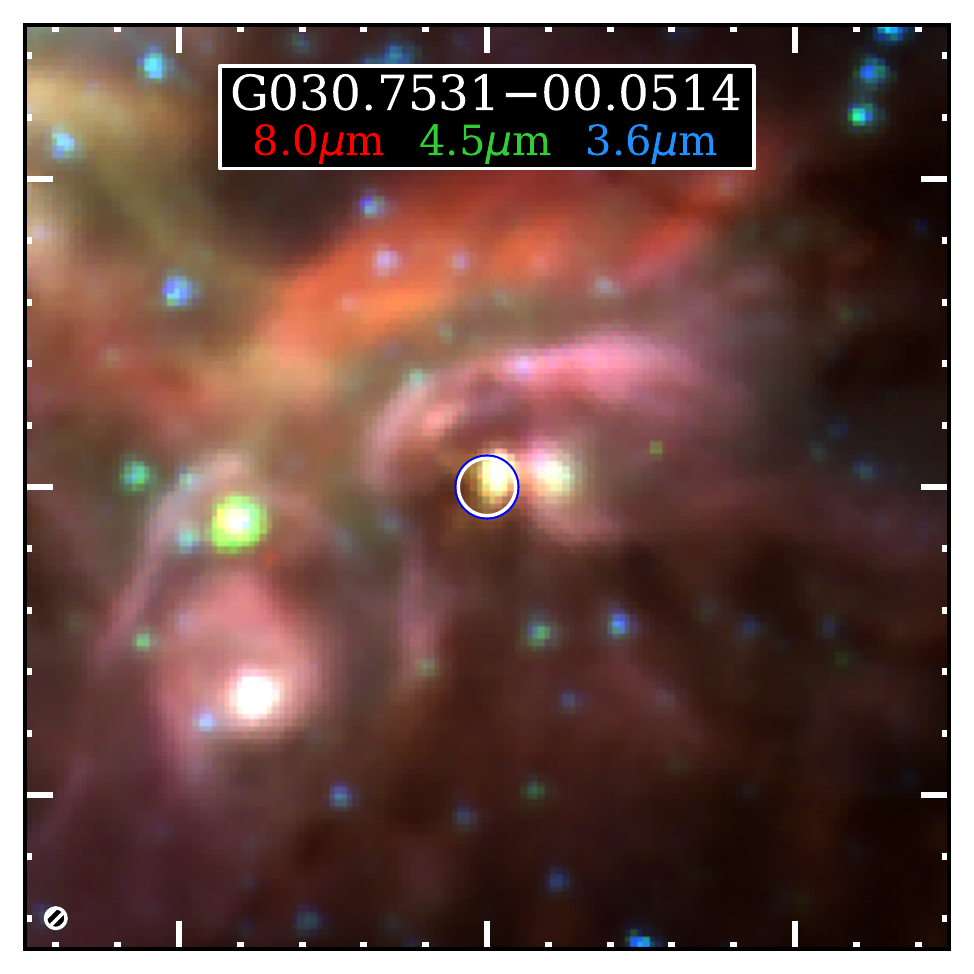}
\includegraphics[width=0.23\textwidth, trim= 0 0 0 0,clip]{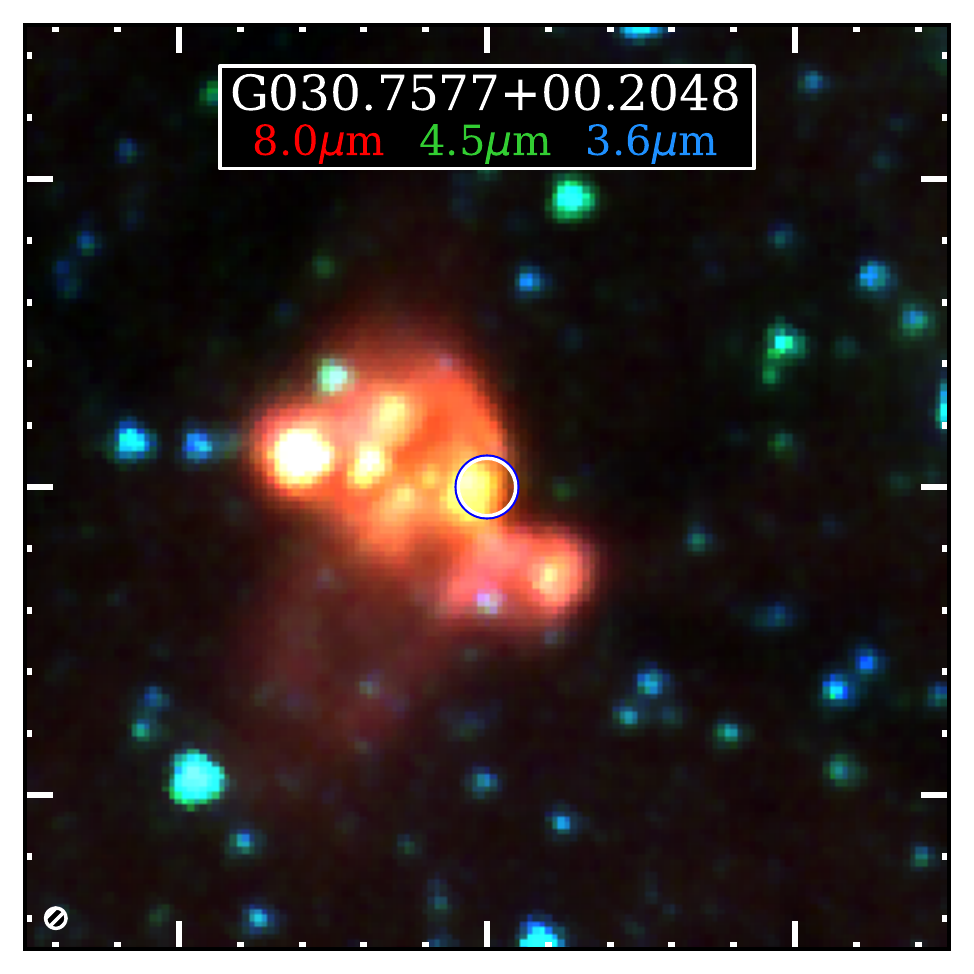}
\includegraphics[width=0.23\textwidth, trim= 0 0 0 0,clip]{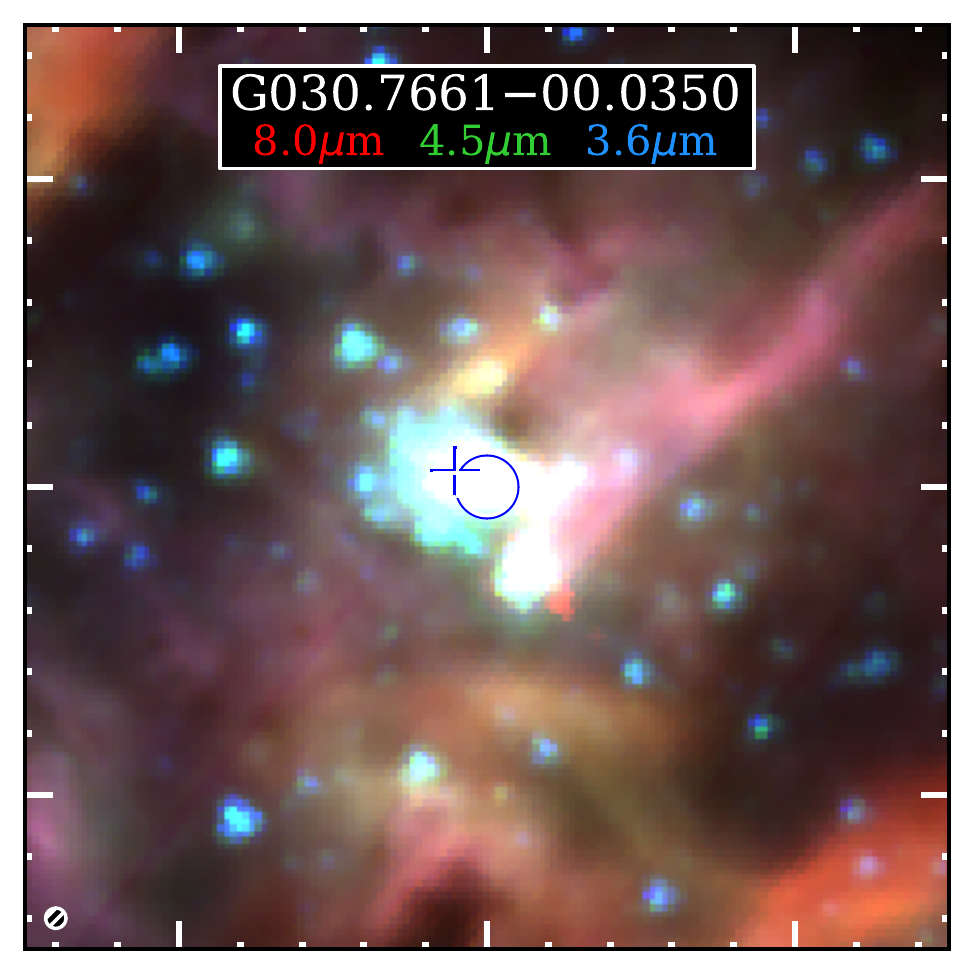}\\ 
\includegraphics[width=0.23\textwidth, trim= 0 0 0 0,clip]{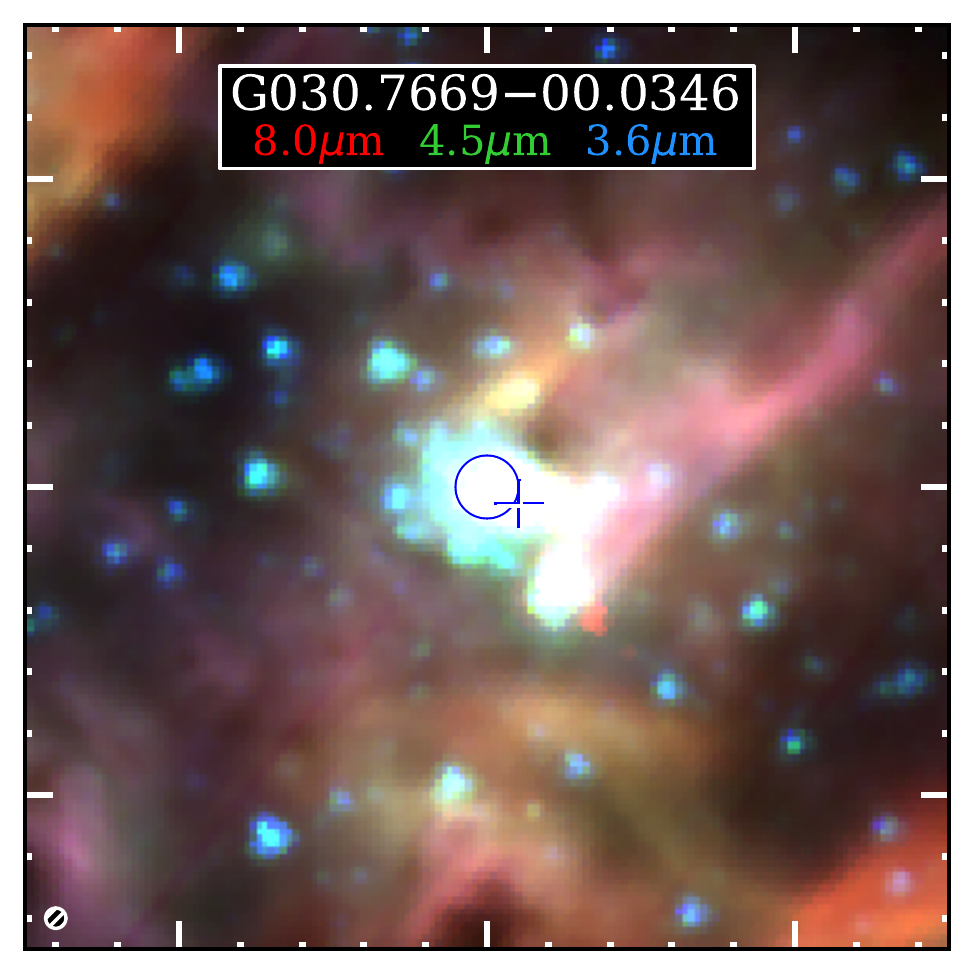}
\includegraphics[width=0.23\textwidth, trim= 0 0 0 0,clip]{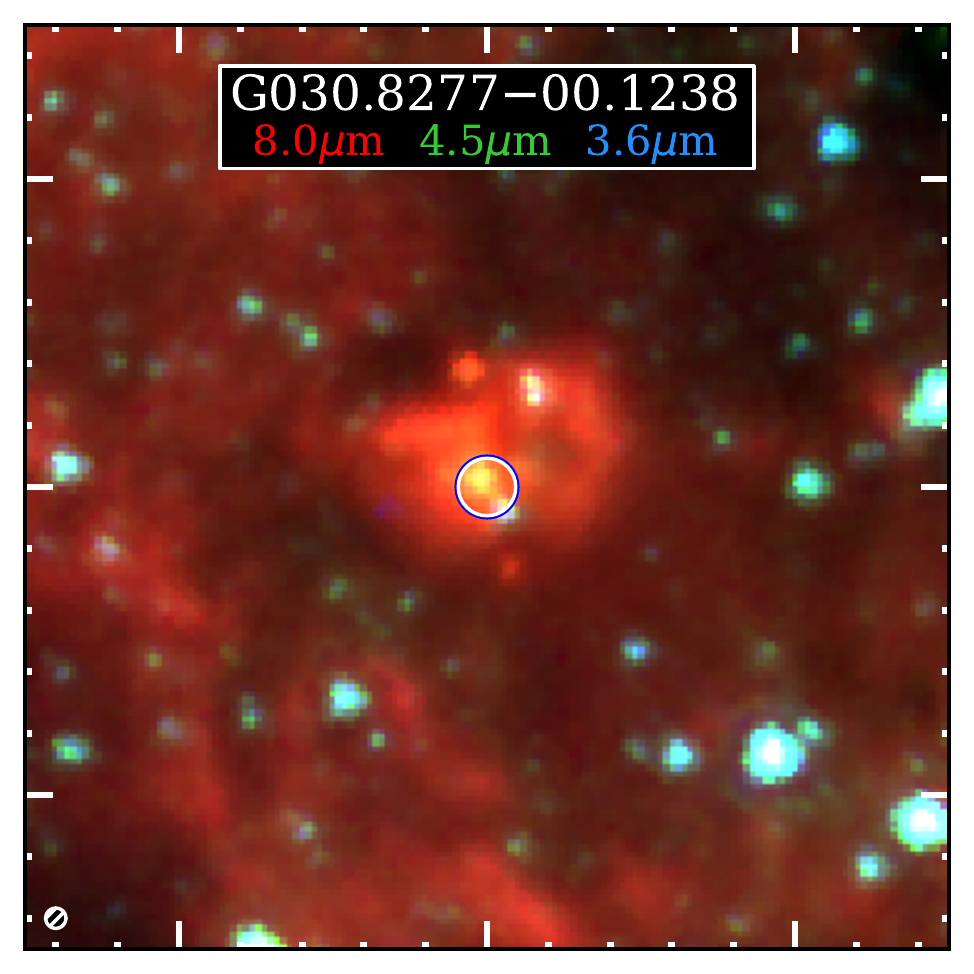}
\includegraphics[width=0.23\textwidth, trim= 0 0 0 0,clip]{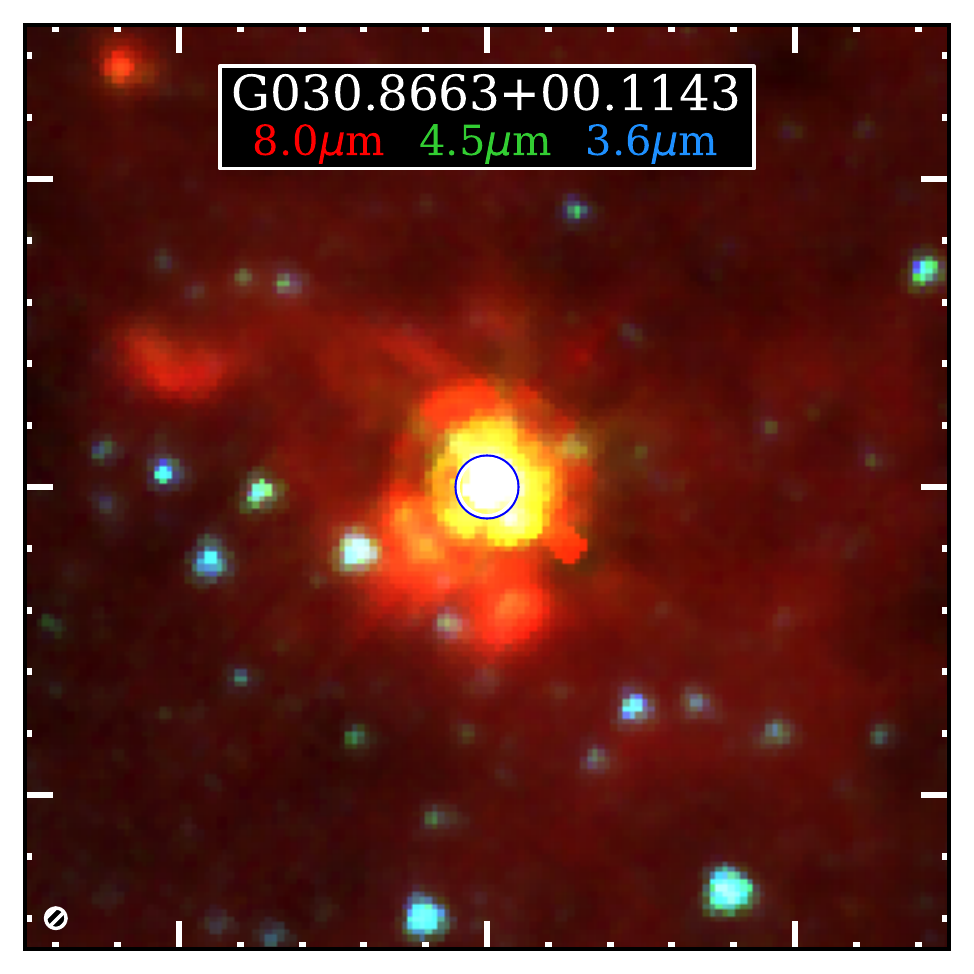}
\includegraphics[width=0.23\textwidth, trim= 0 0 0 0,clip]{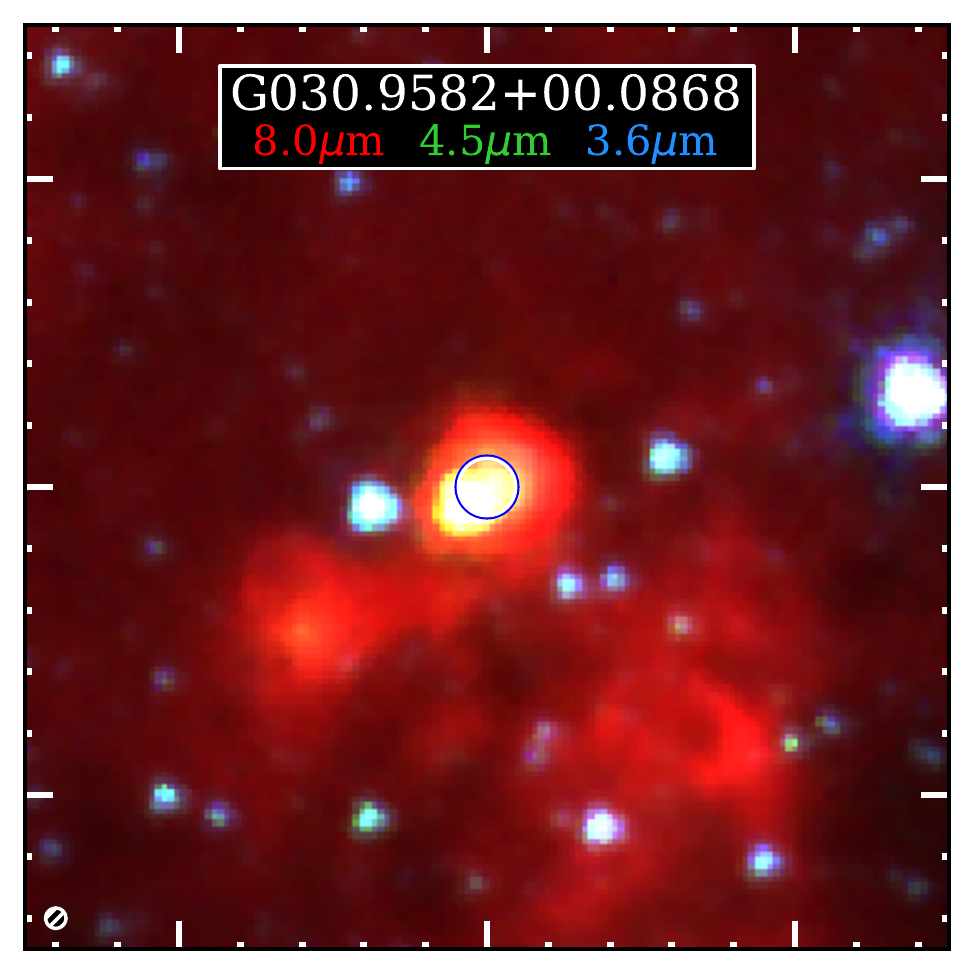}\\ 
\includegraphics[width=0.23\textwidth, trim= 0 0 0 0,clip]{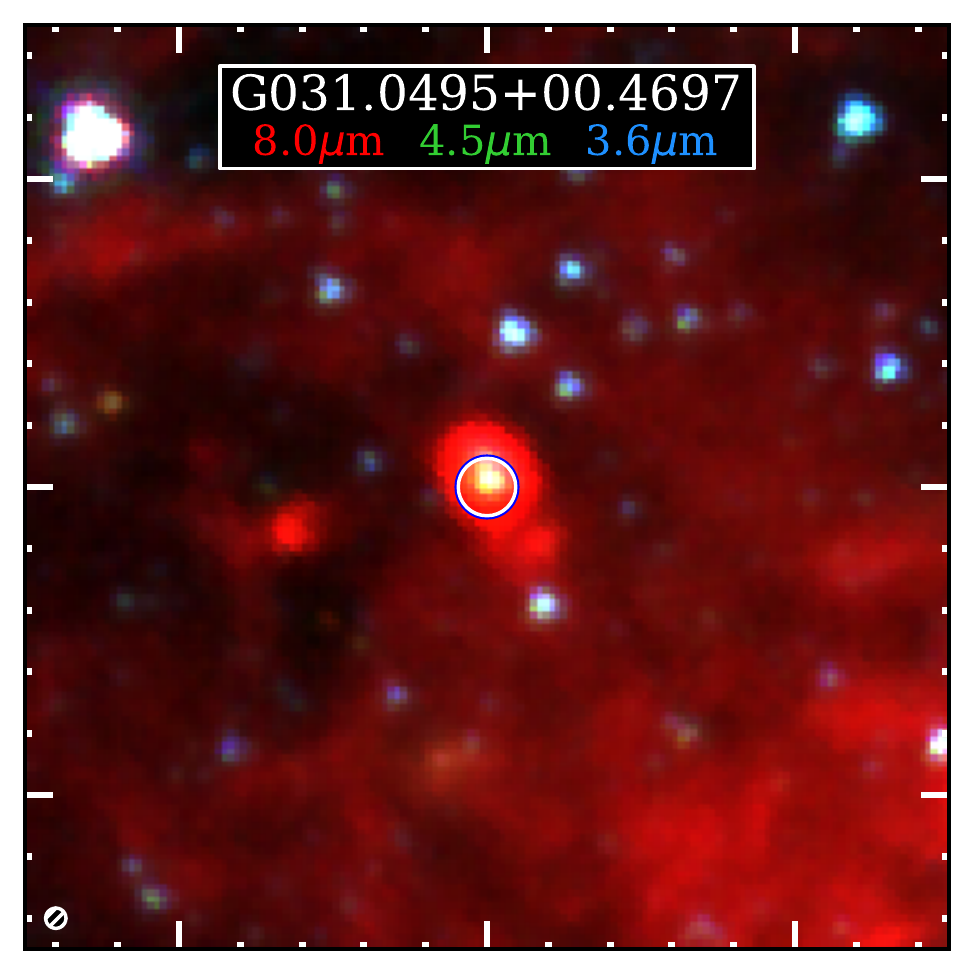}
\includegraphics[width=0.23\textwidth, trim= 0 0 0 0,clip]{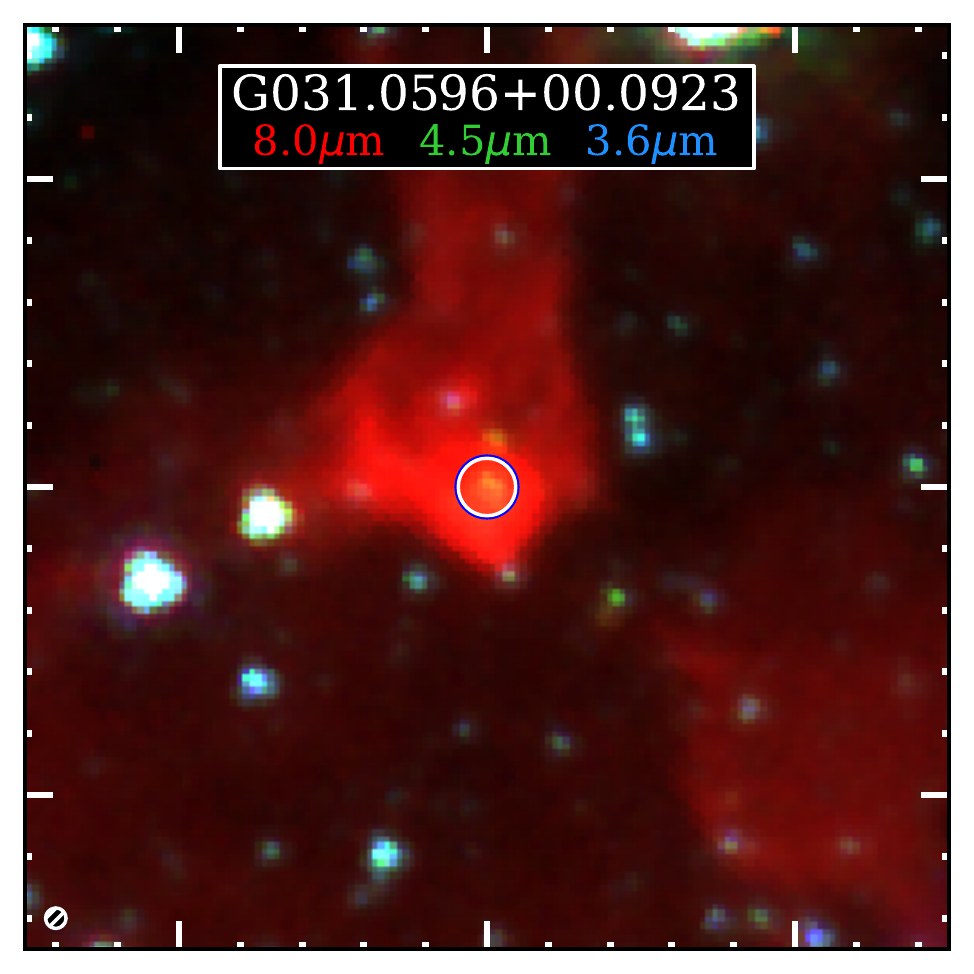}
\includegraphics[width=0.23\textwidth, trim= 0 0 0 0,clip]{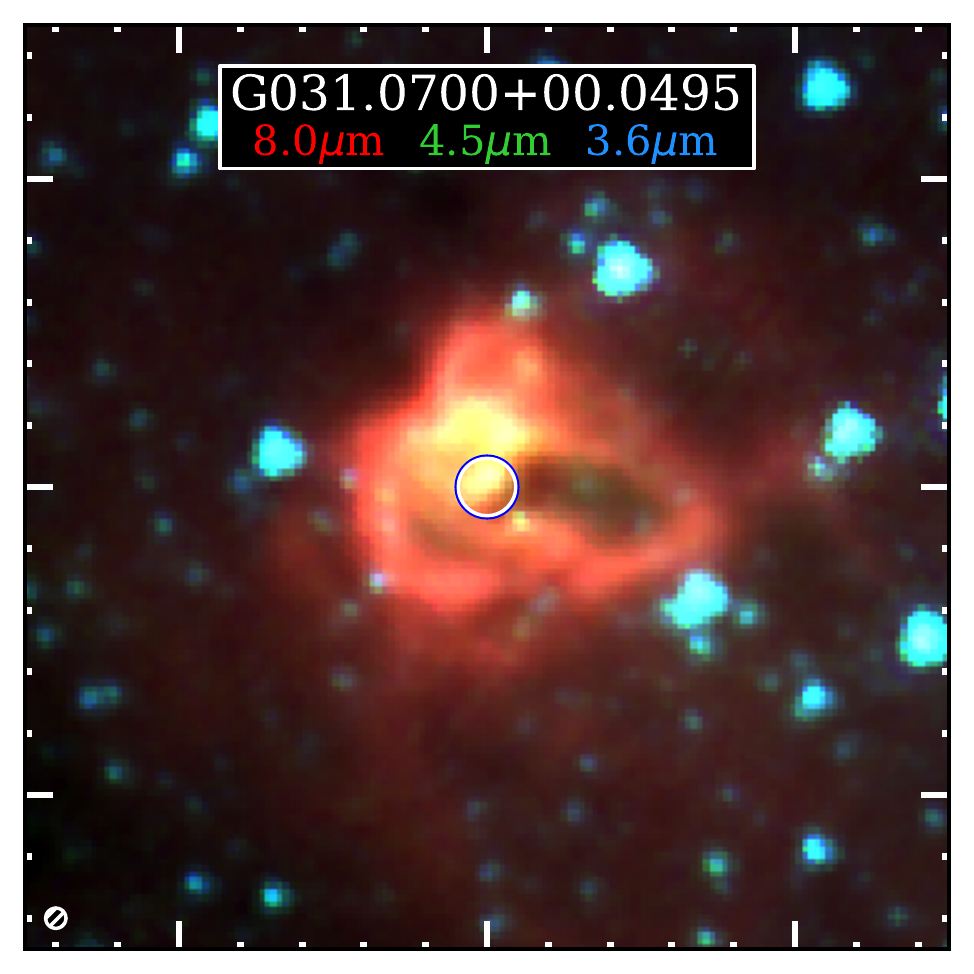}
\includegraphics[width=0.23\textwidth, trim= 0 0 0 0,clip]{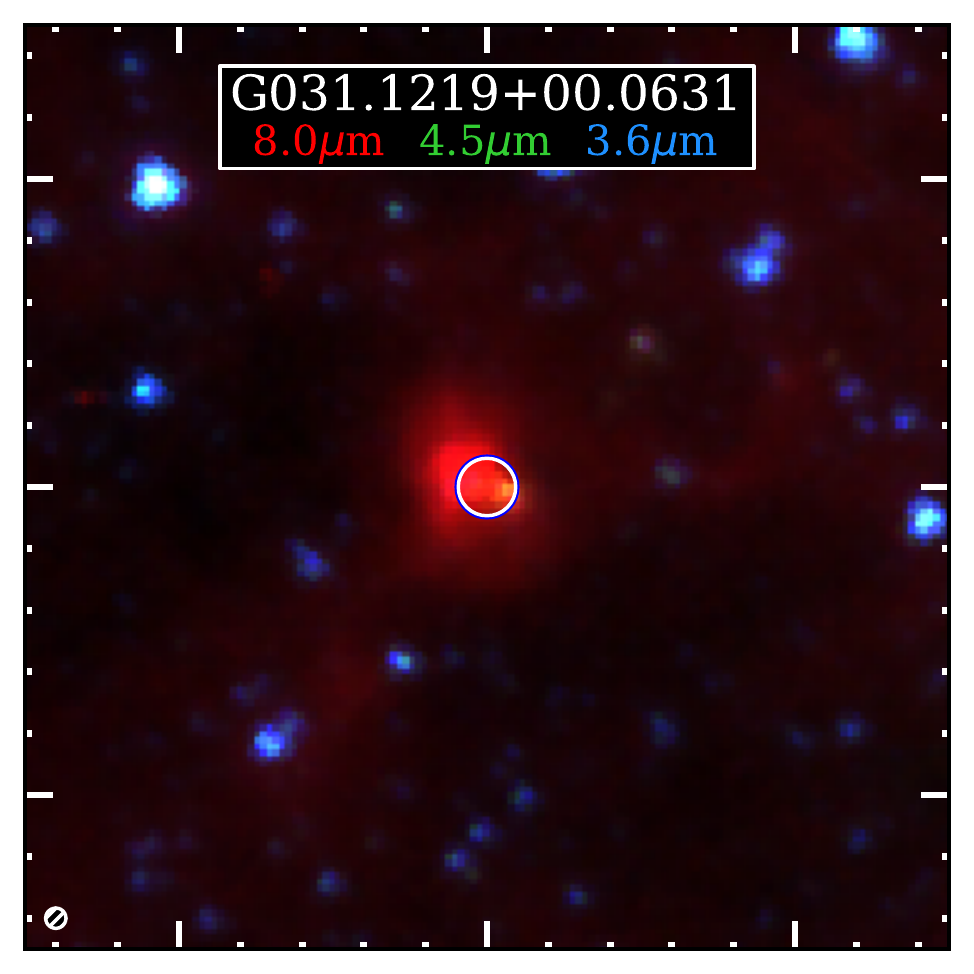}\\ 
\hspace{-0.85cm}\includegraphics[width=0.28\textwidth, trim= 0 0 0 0,clip,valign=t]{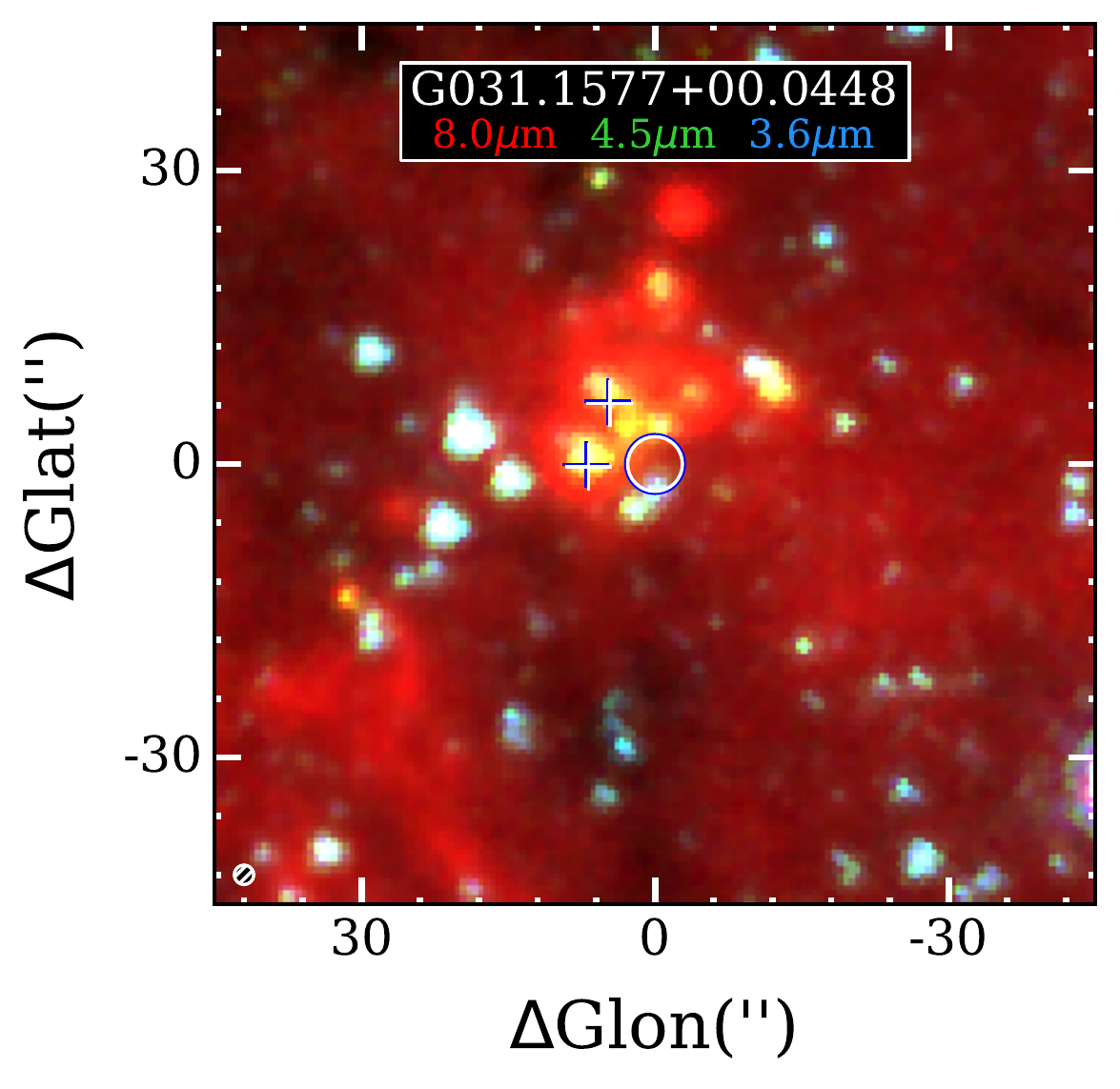}
\includegraphics[width=0.23\textwidth, trim= 0 0 0 0,clip,valign=t]{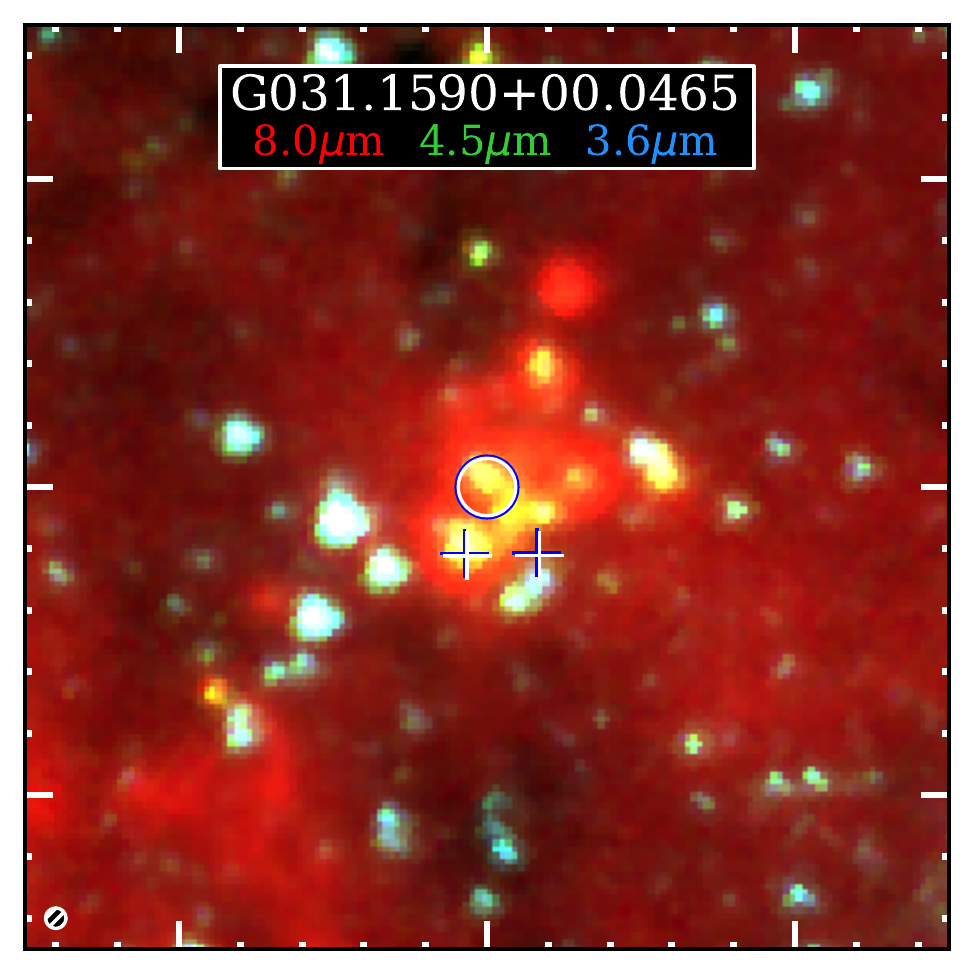}
\includegraphics[width=0.23\textwidth, trim= 0 0 0 0,clip,valign=t]{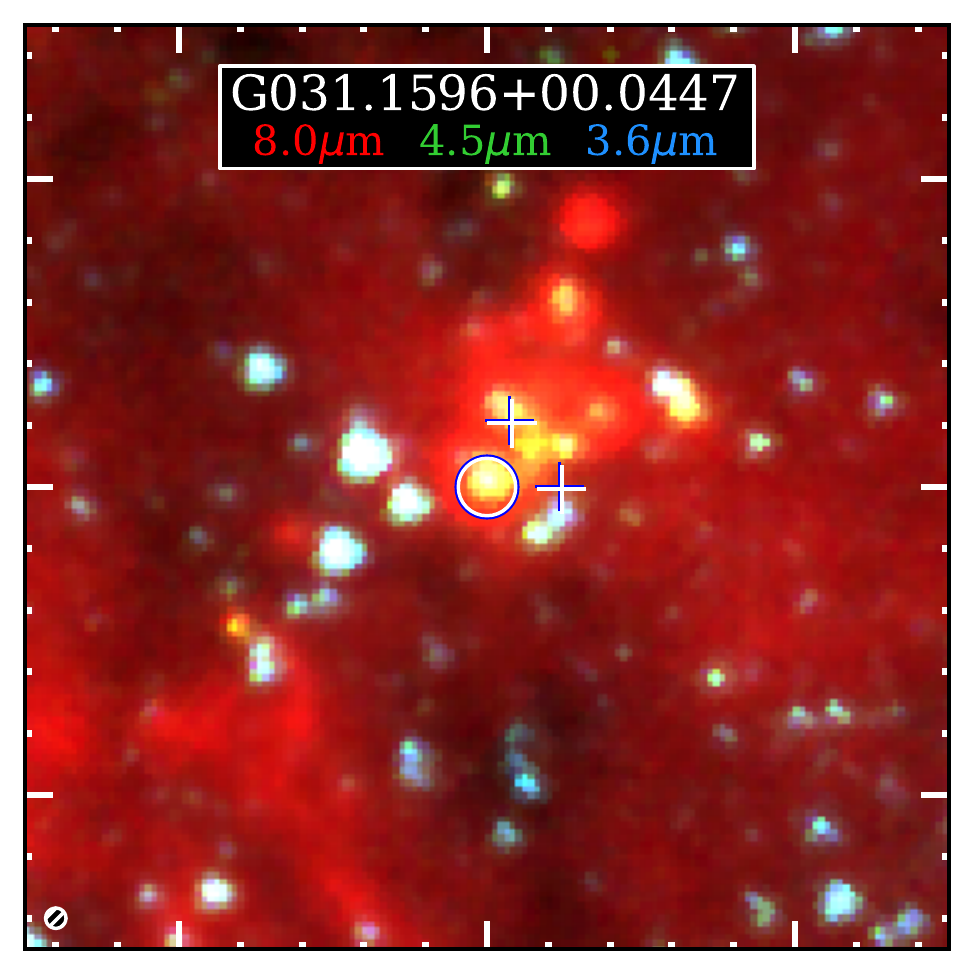}
\includegraphics[width=0.23\textwidth, trim= 0 0 0 0,clip,valign=t]{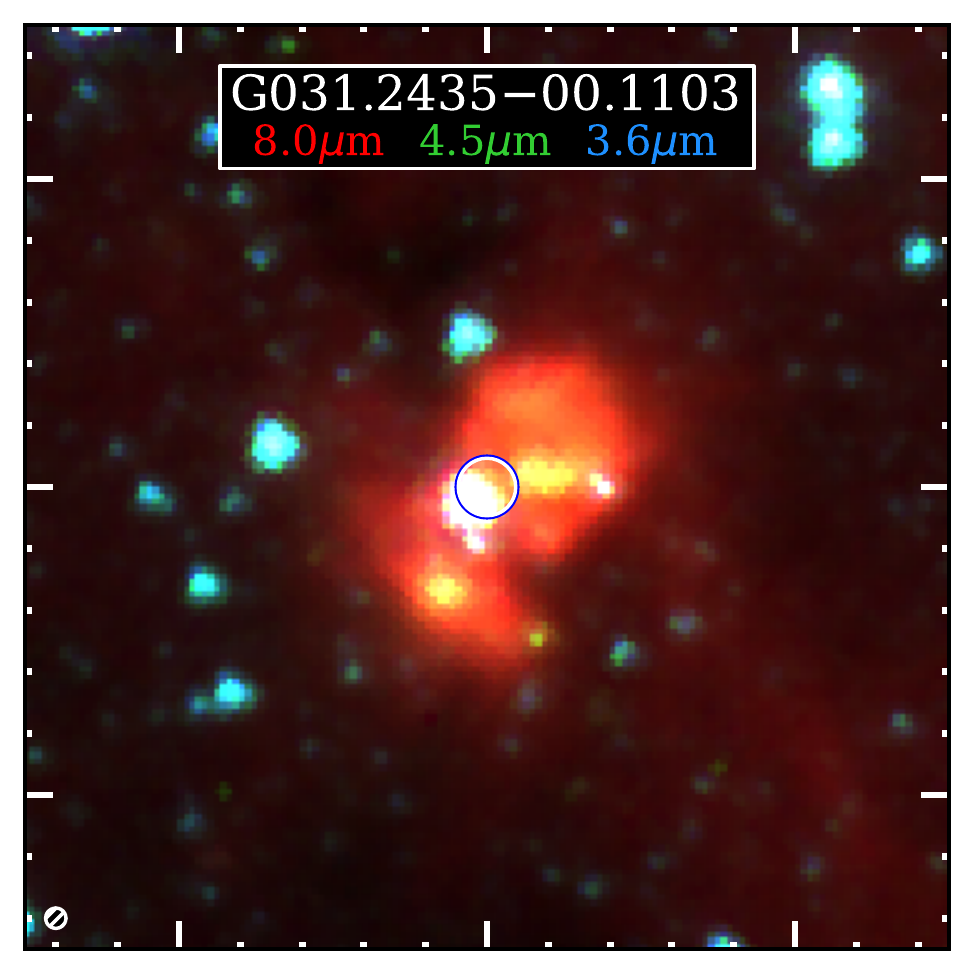}\\    
\caption{Continuation.}
\end{figure*}

\setcounter{figure}{0}\begin{figure*}[!h]
\centering
\includegraphics[width=0.23\textwidth, trim= 0 0 0 0,clip]{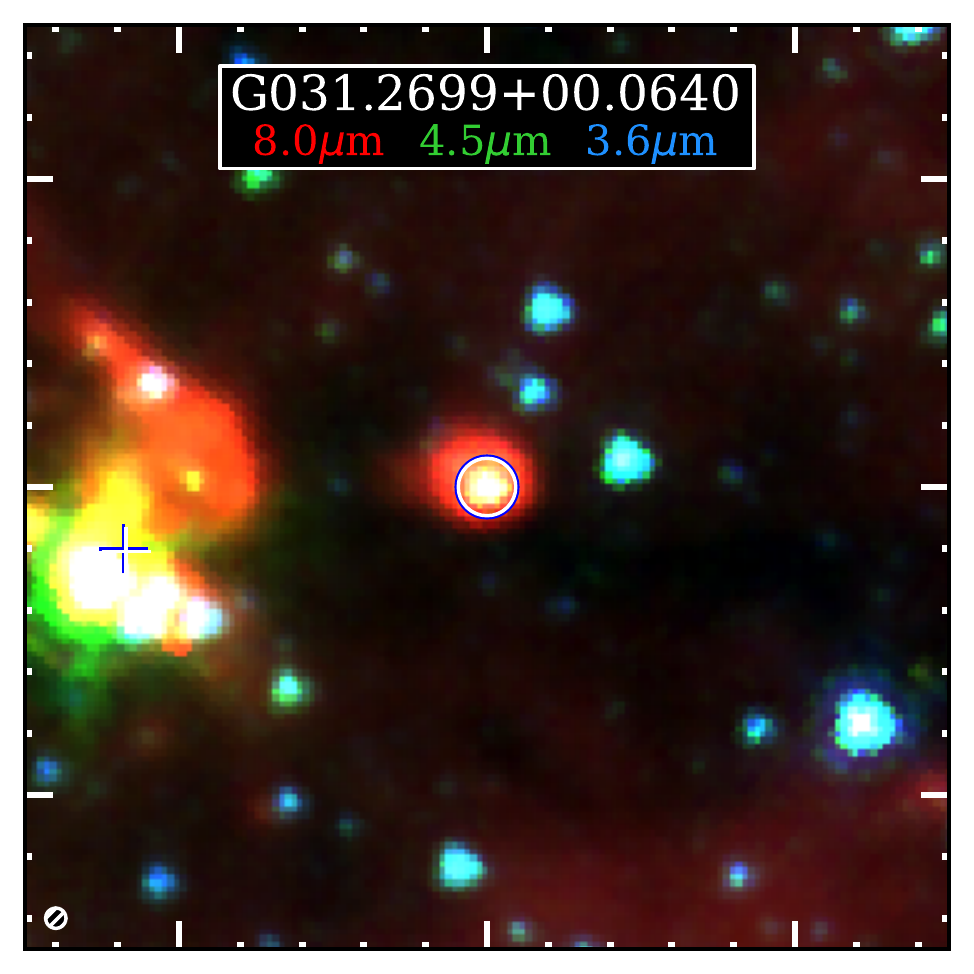}
\includegraphics[width=0.23\textwidth, trim= 0 0 0 0,clip]{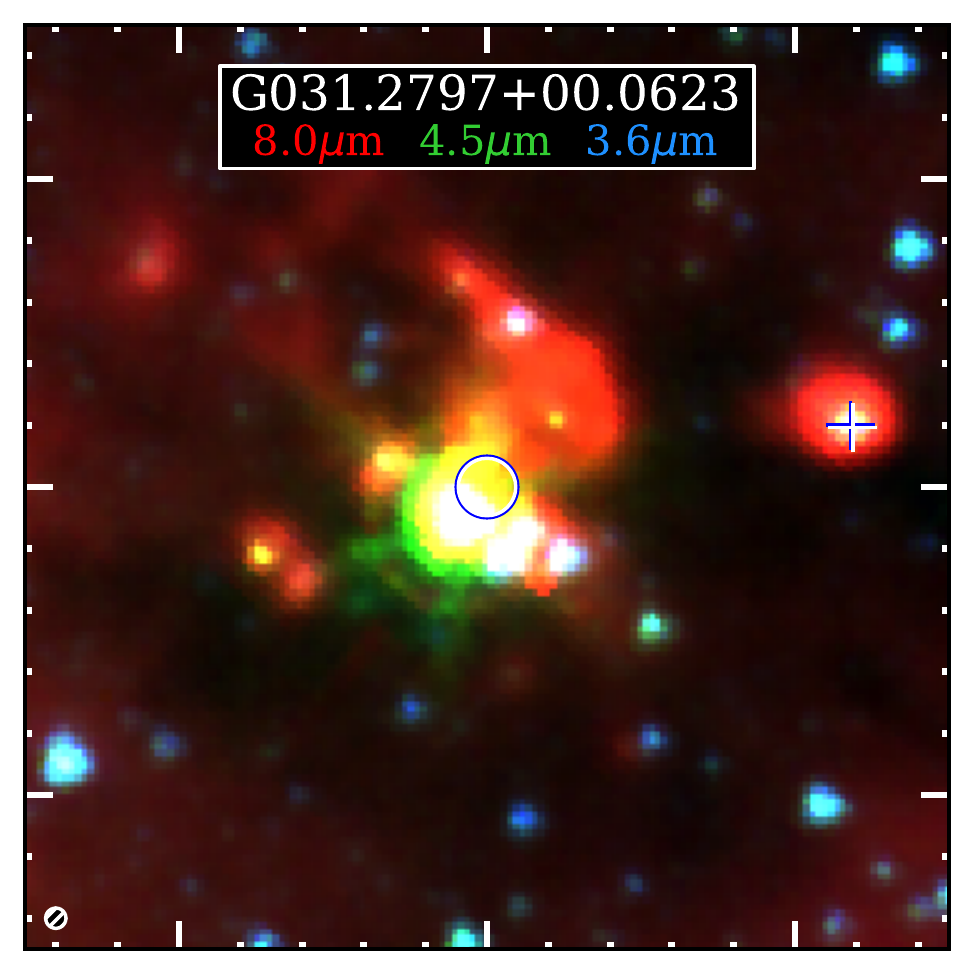}
\includegraphics[width=0.23\textwidth, trim= 0 0 0 0,clip]{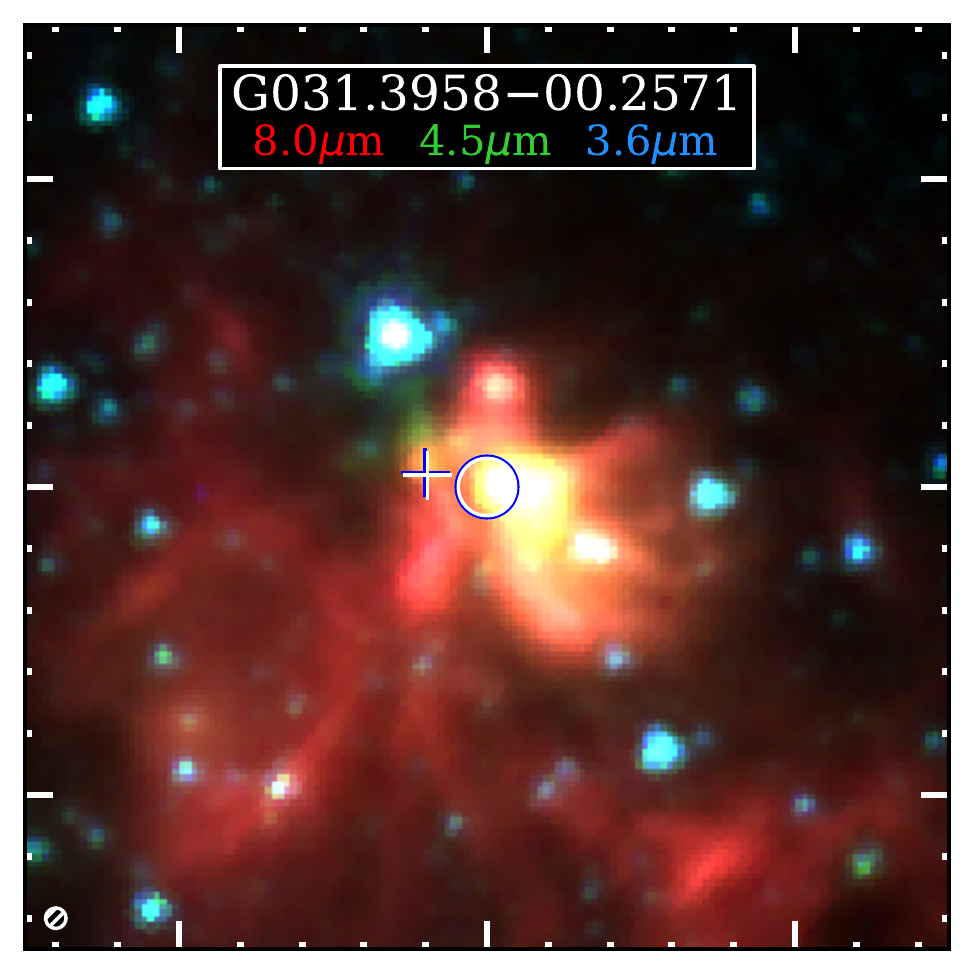}
\includegraphics[width=0.23\textwidth, trim= 0 0 0 0,clip]{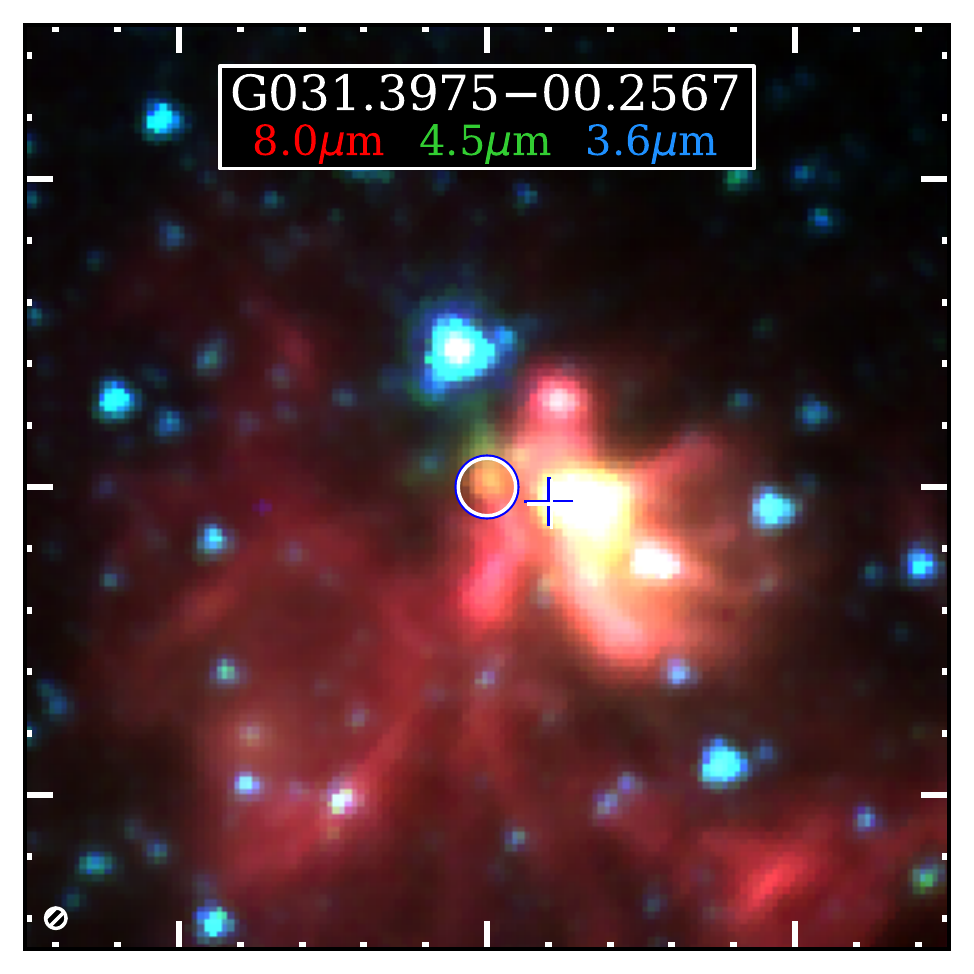}\\ 
\includegraphics[width=0.23\textwidth, trim= 0 0 0 0,clip]{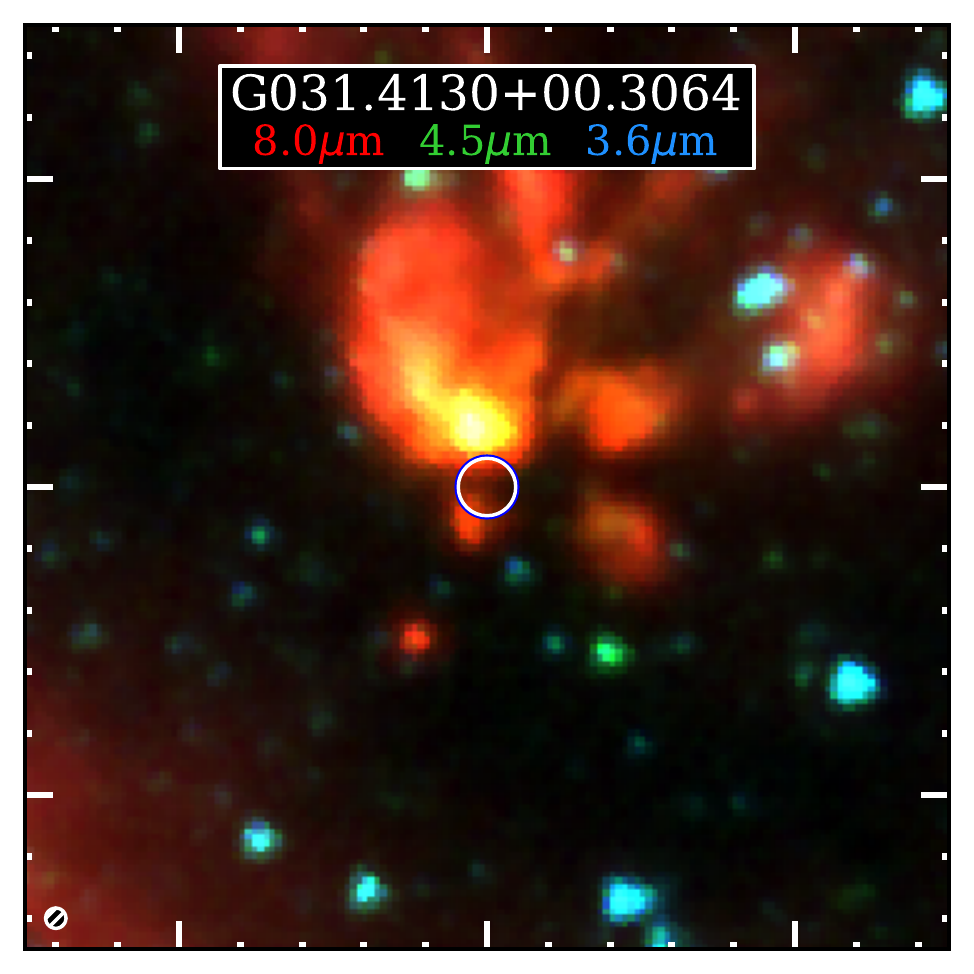}
\includegraphics[width=0.23\textwidth, trim= 0 0 0 0,clip]{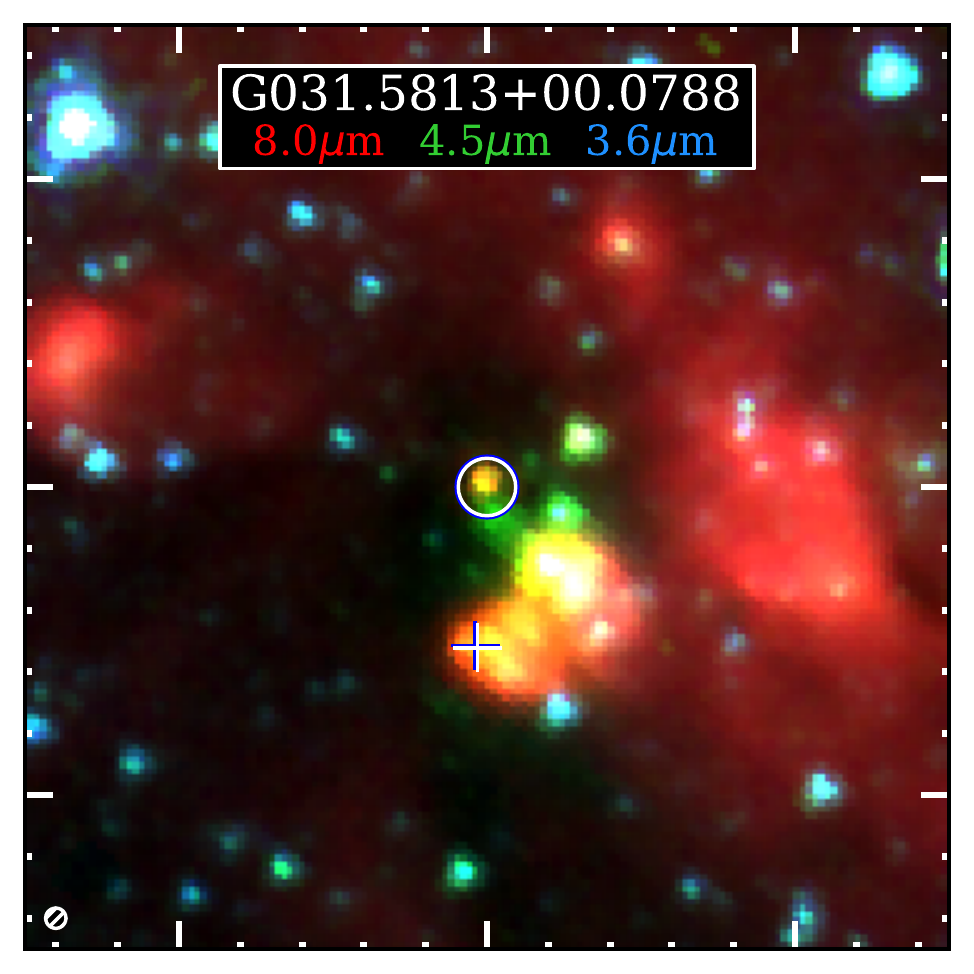}
\includegraphics[width=0.23\textwidth, trim= 0 0 0 0,clip]{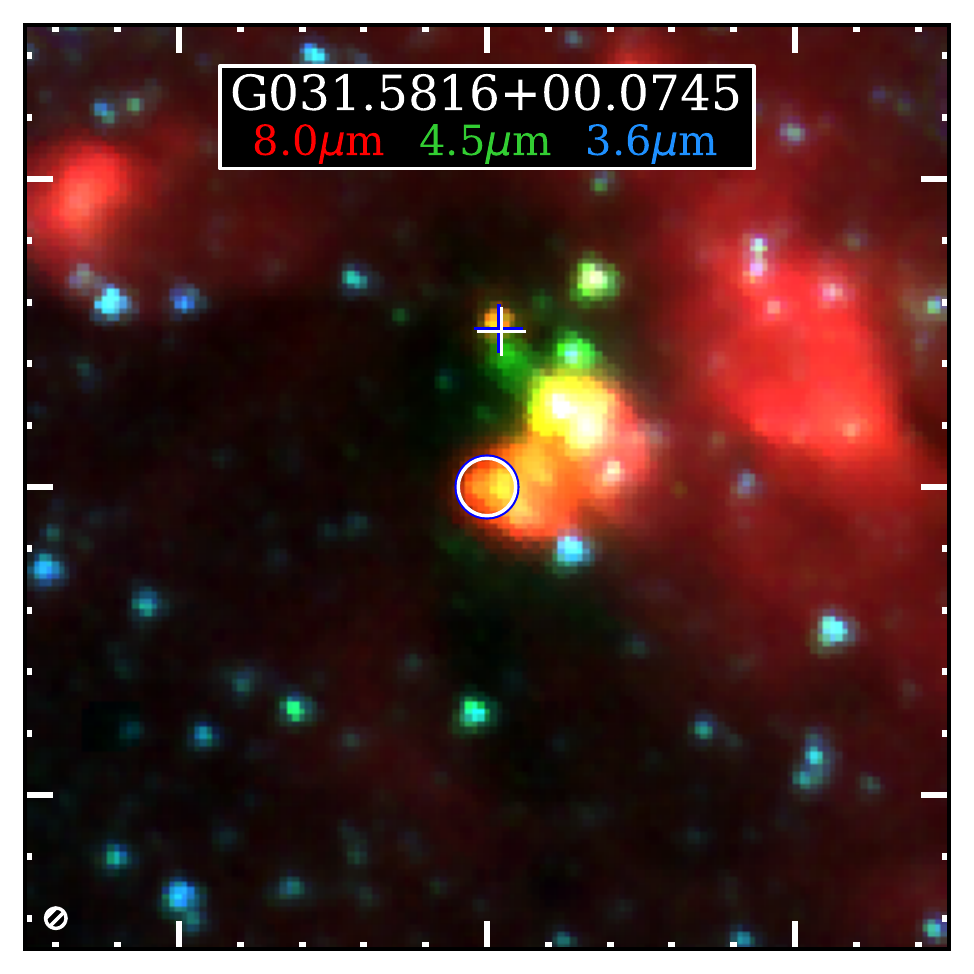}
\includegraphics[width=0.23\textwidth, trim= 0 0 0 0,clip]{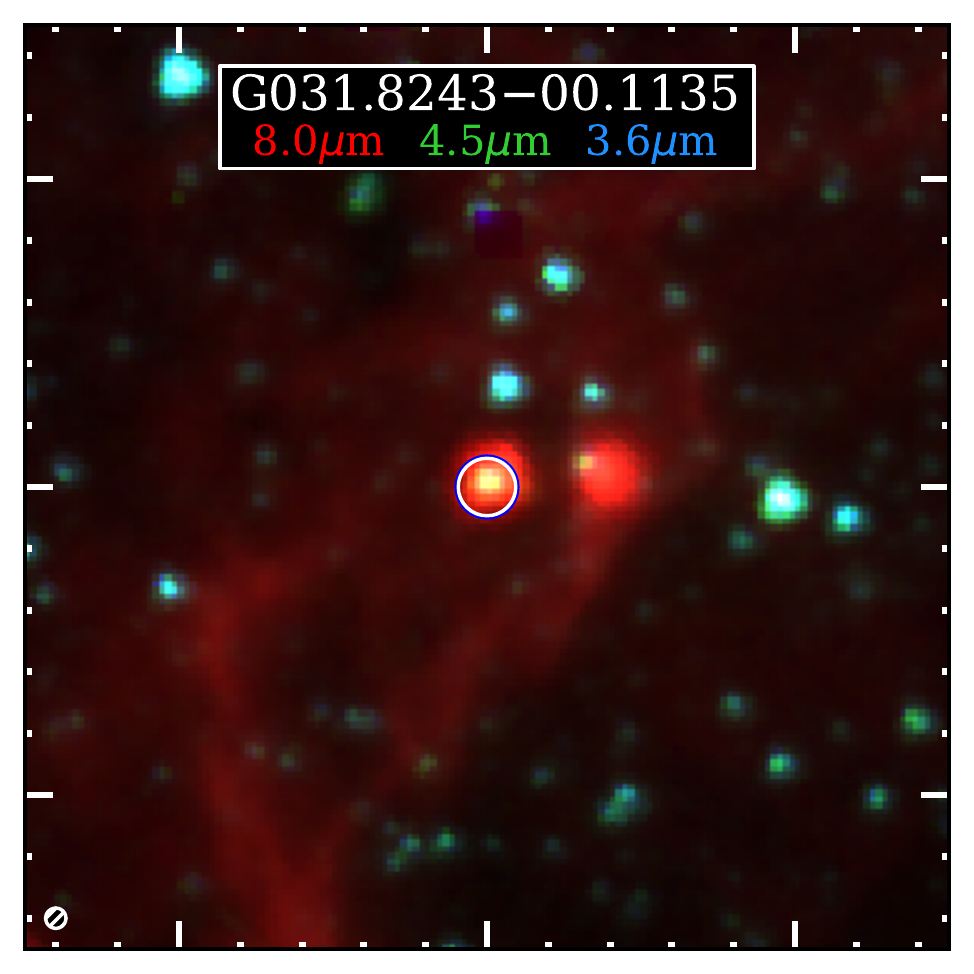}\\ 
\includegraphics[width=0.23\textwidth, trim= 0 0 0 0,clip]{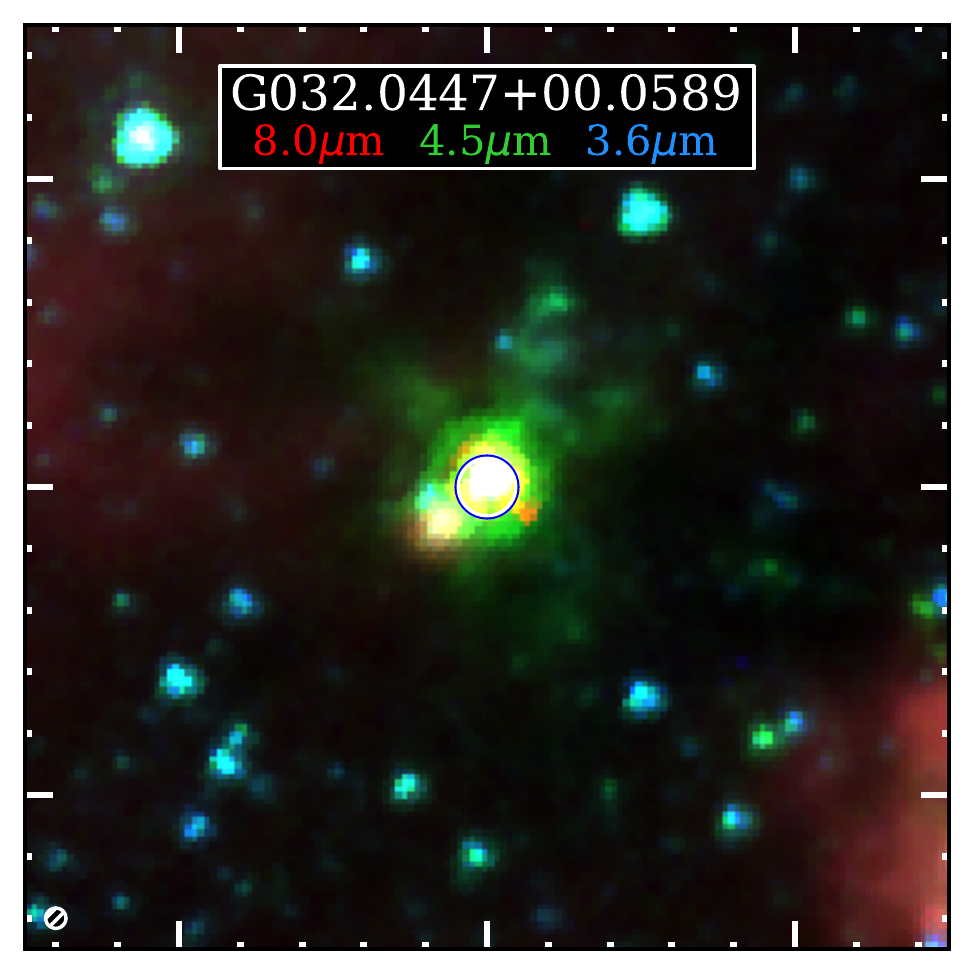}
\includegraphics[width=0.23\textwidth, trim= 0 0 0 0,clip]{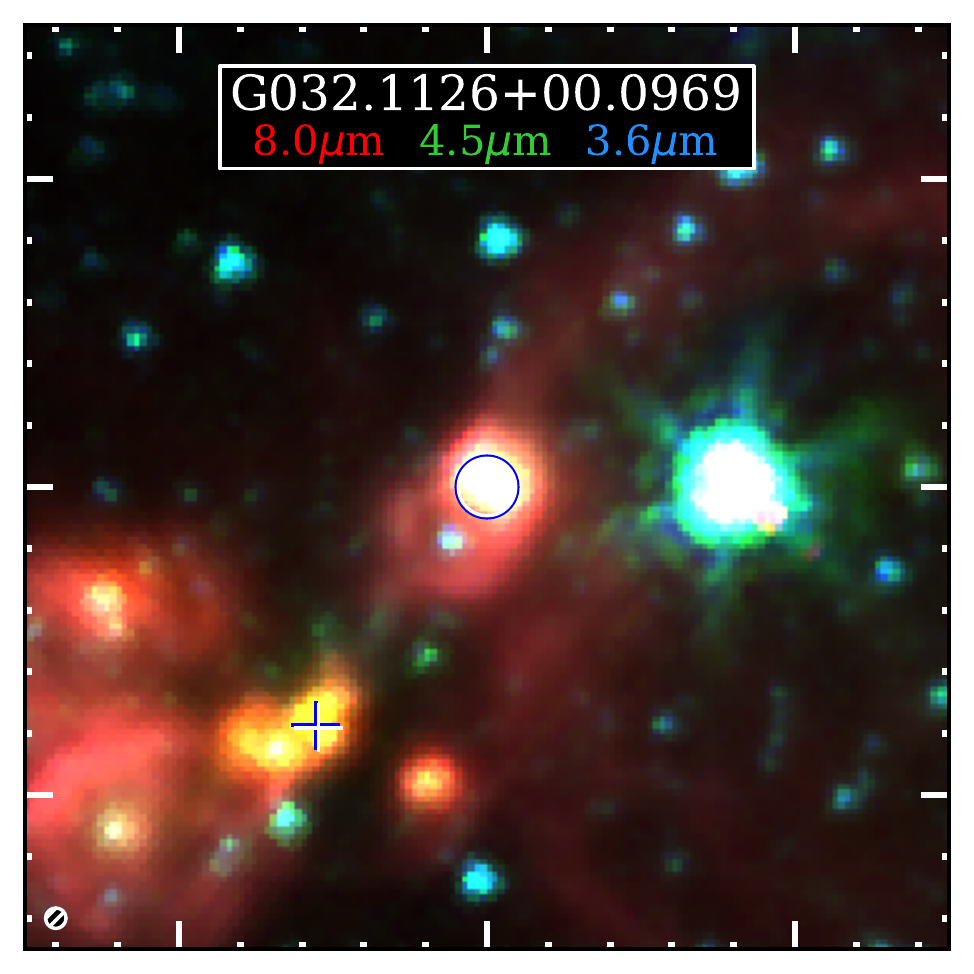}
\includegraphics[width=0.23\textwidth, trim= 0 0 0 0,clip]{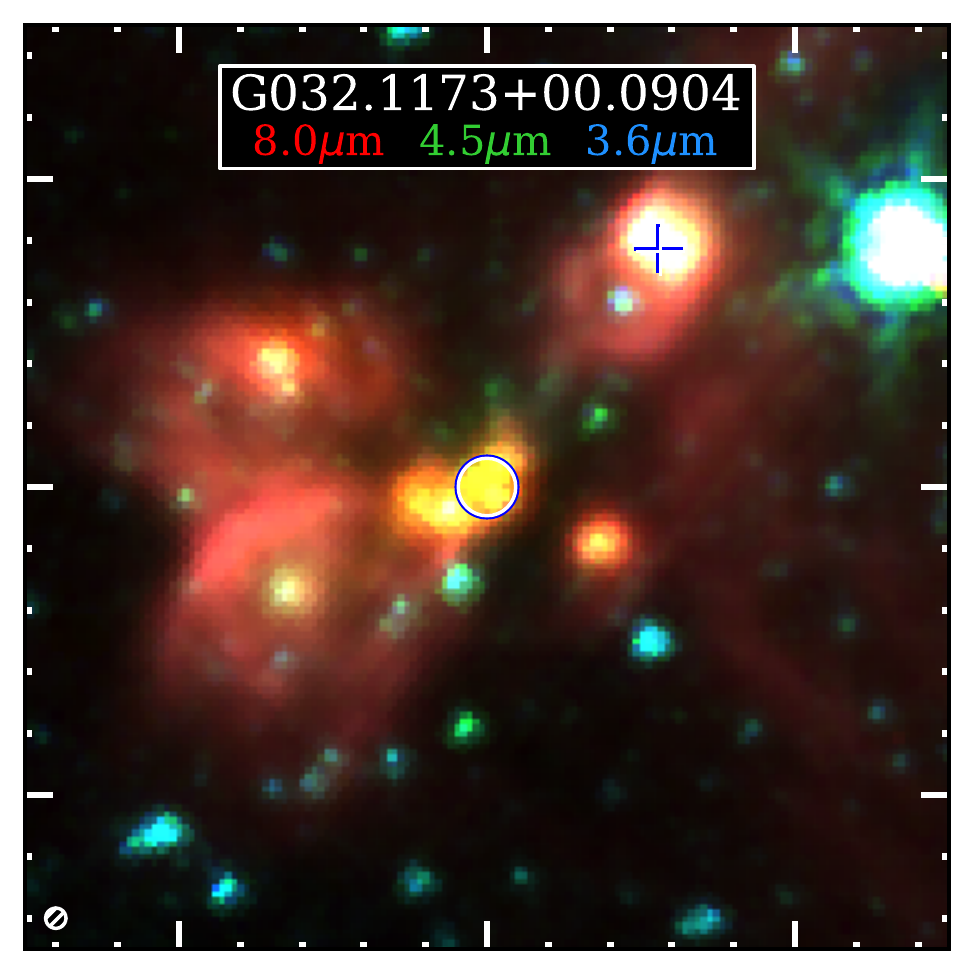}
\includegraphics[width=0.23\textwidth, trim= 0 0 0 0,clip]{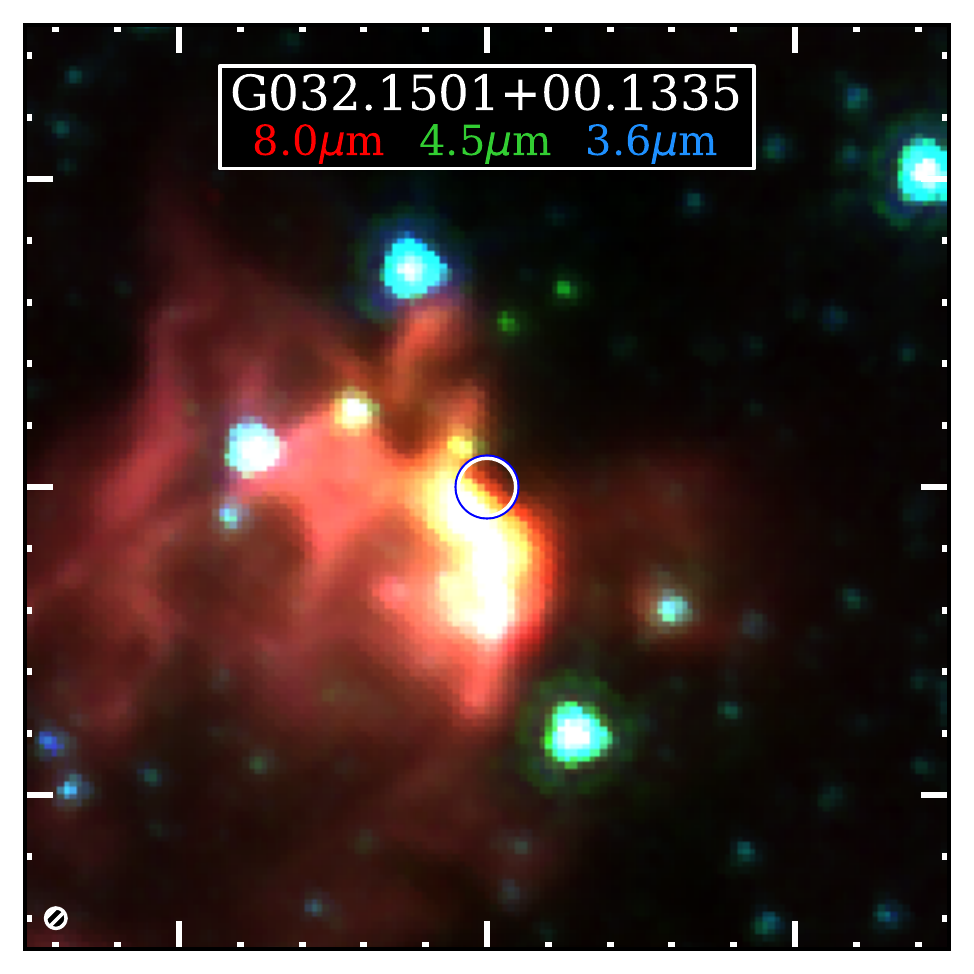}\\ 
\includegraphics[width=0.23\textwidth, trim= 0 0 0 0,clip]{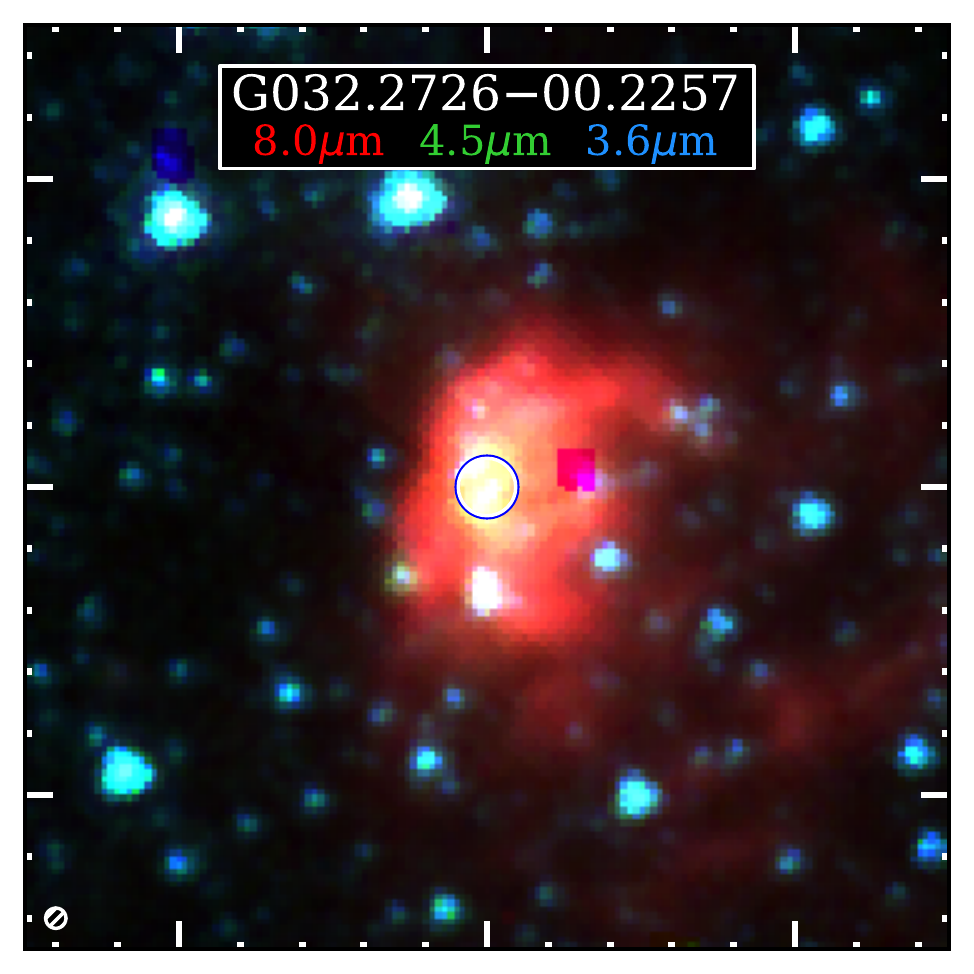}
\includegraphics[width=0.23\textwidth, trim= 0 0 0 0,clip]{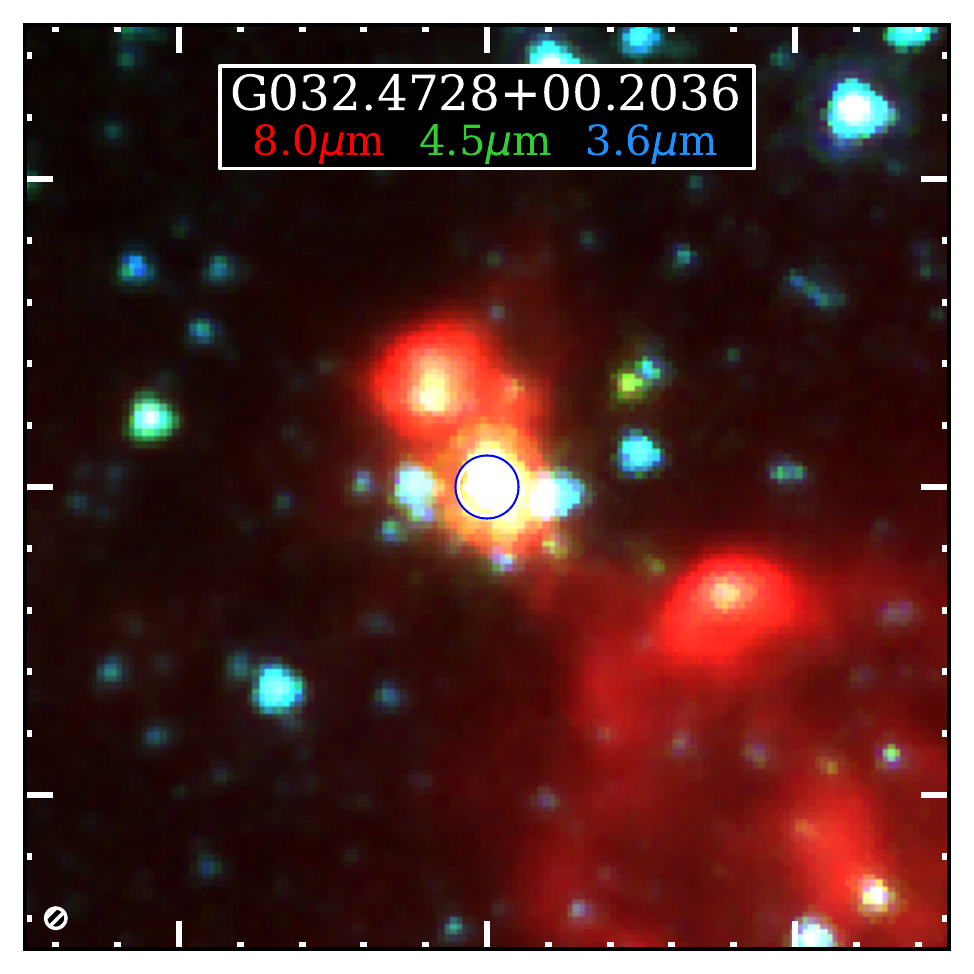}
\includegraphics[width=0.23\textwidth, trim= 0 0 0 0,clip]{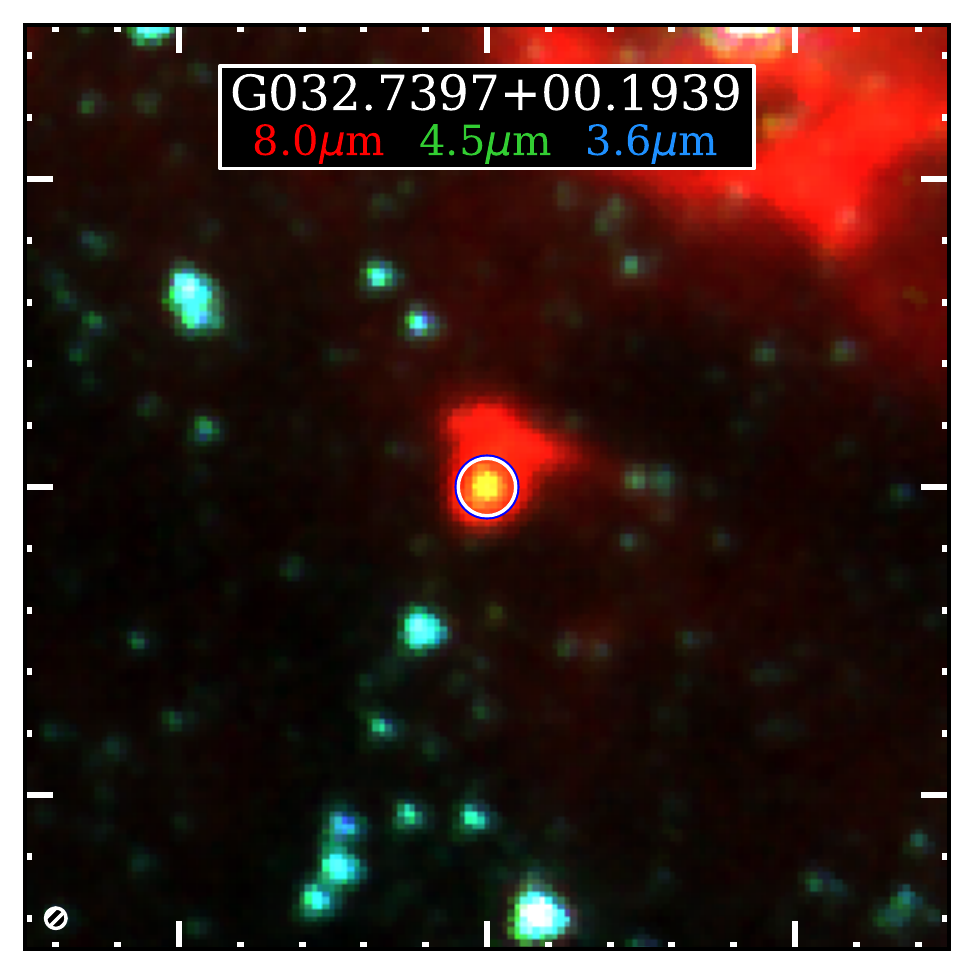}
\includegraphics[width=0.23\textwidth, trim= 0 0 0 0,clip]{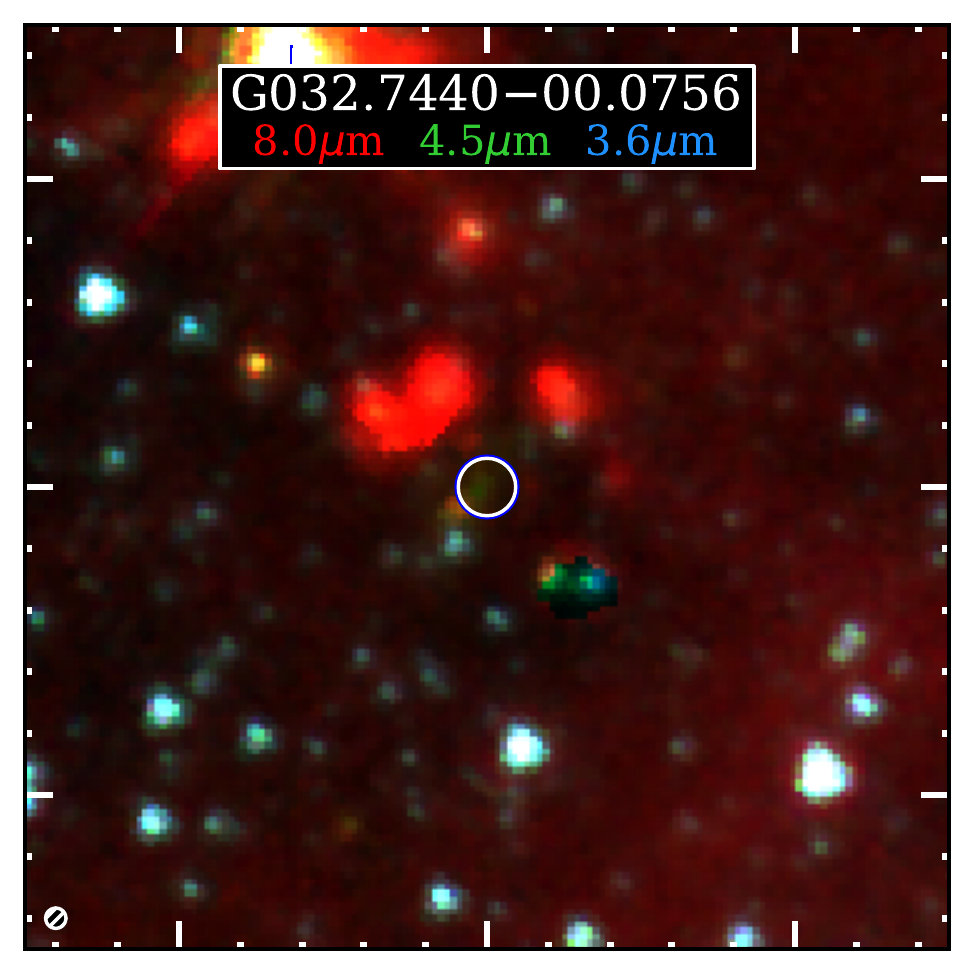}\\ 
\hspace{-0.85cm}\includegraphics[width=0.28\textwidth, trim= 0 0 0 0,clip,valign=t]{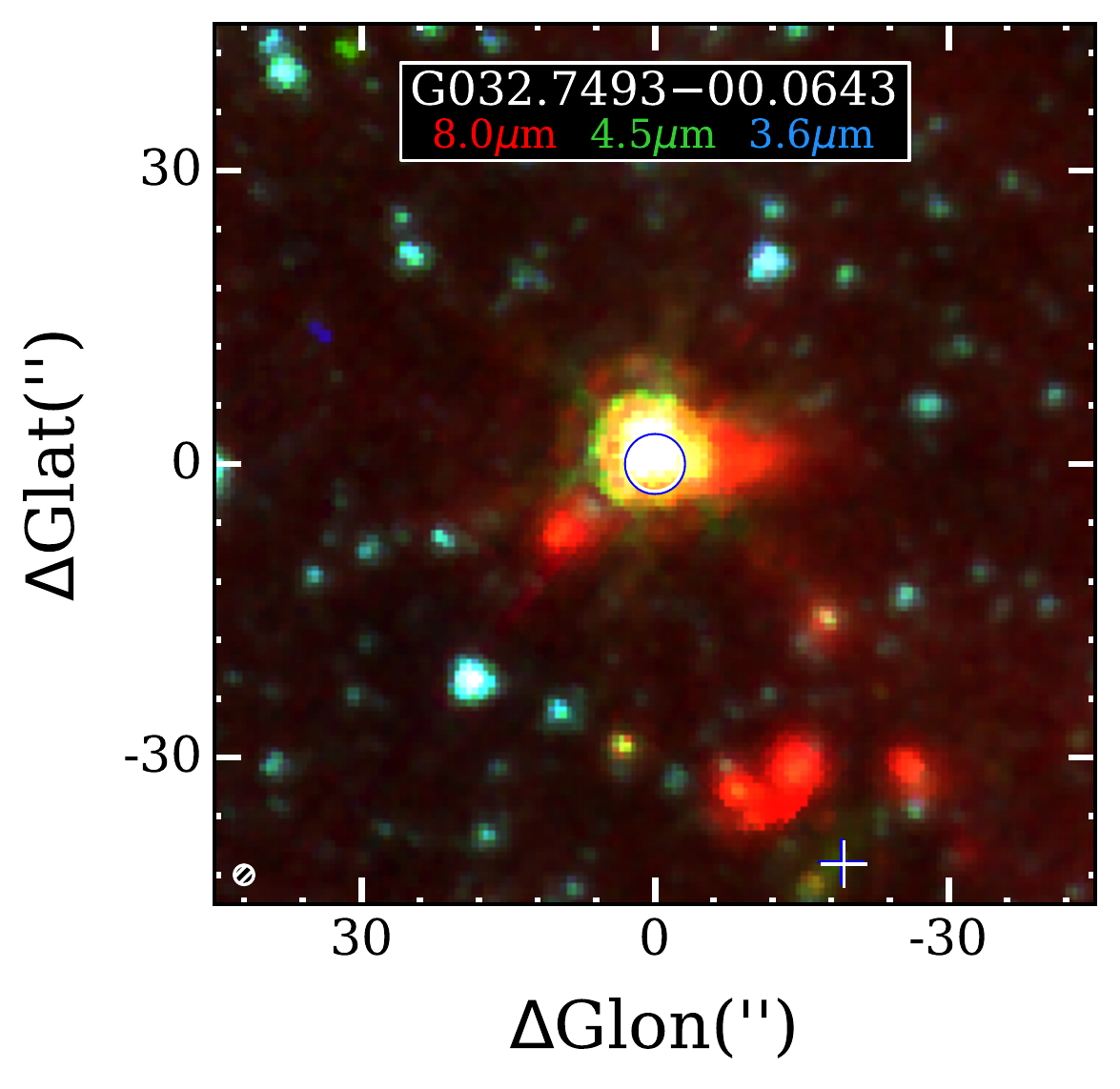}
\includegraphics[width=0.23\textwidth, trim= 0 0 0 0,clip,valign=t]{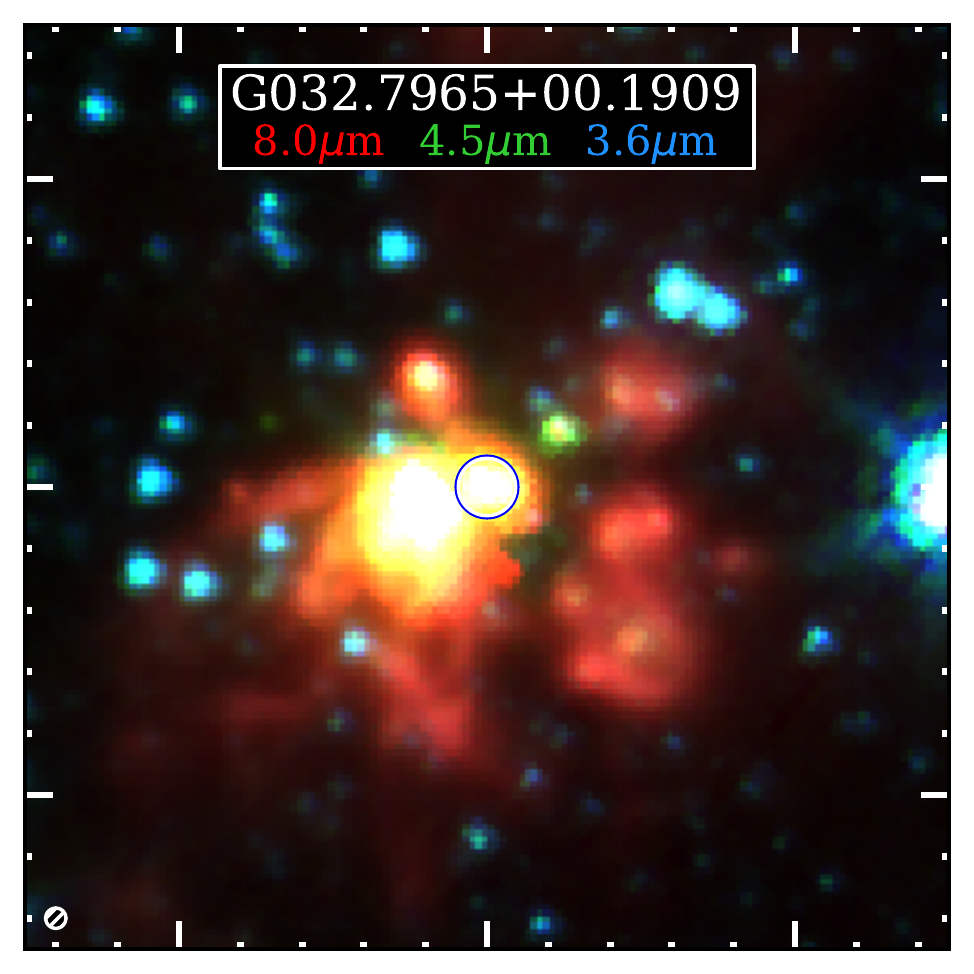}
\includegraphics[width=0.23\textwidth, trim= 0 0 0 0,clip,valign=t]{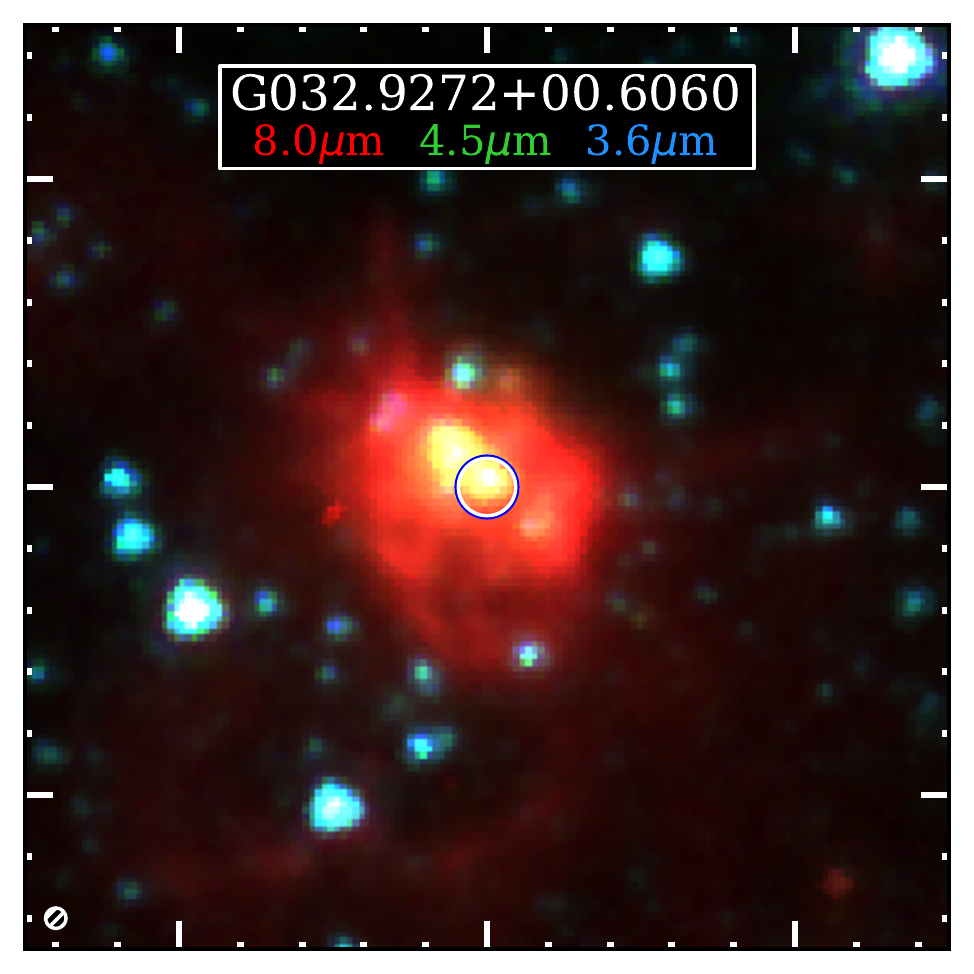}
\includegraphics[width=0.23\textwidth, trim= 0 0 0 0,clip,valign=t]{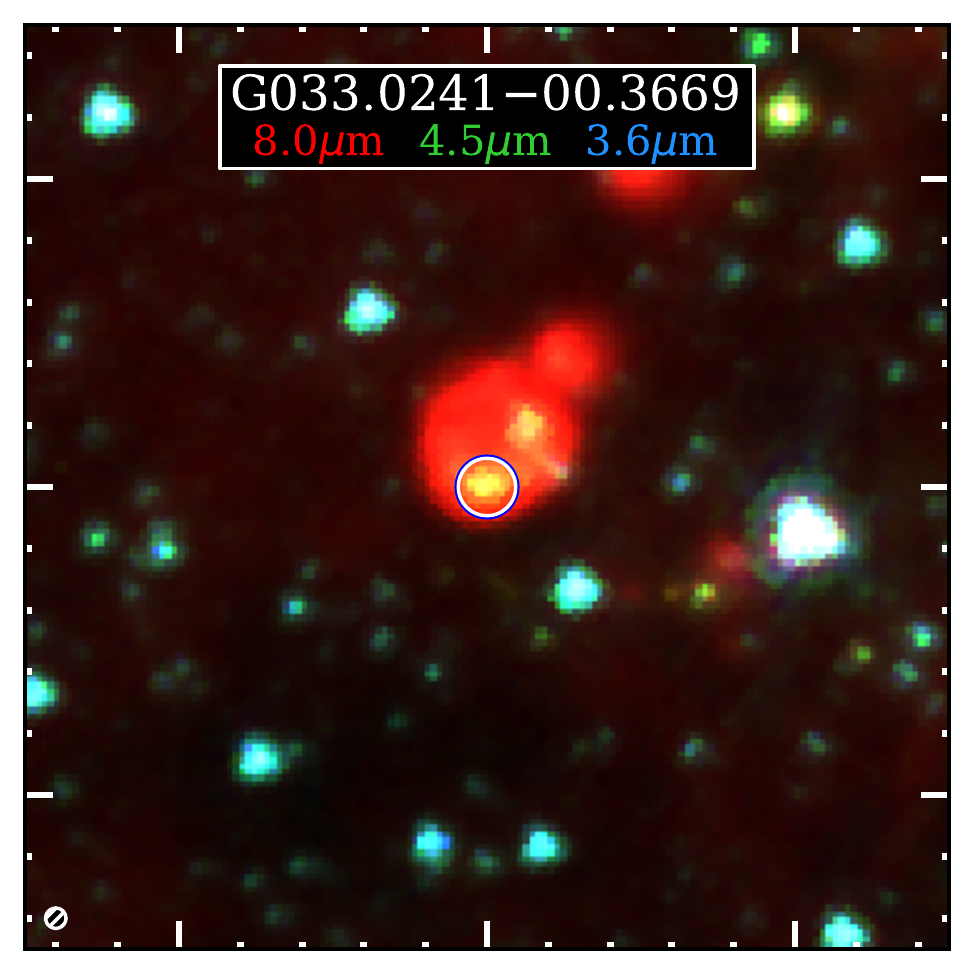}\\    
\caption{Continuation.}
\end{figure*}

\setcounter{figure}{0}\begin{figure*}[!h]
\centering
 \includegraphics[width=0.23\textwidth, trim= 0 0 0 0,clip]{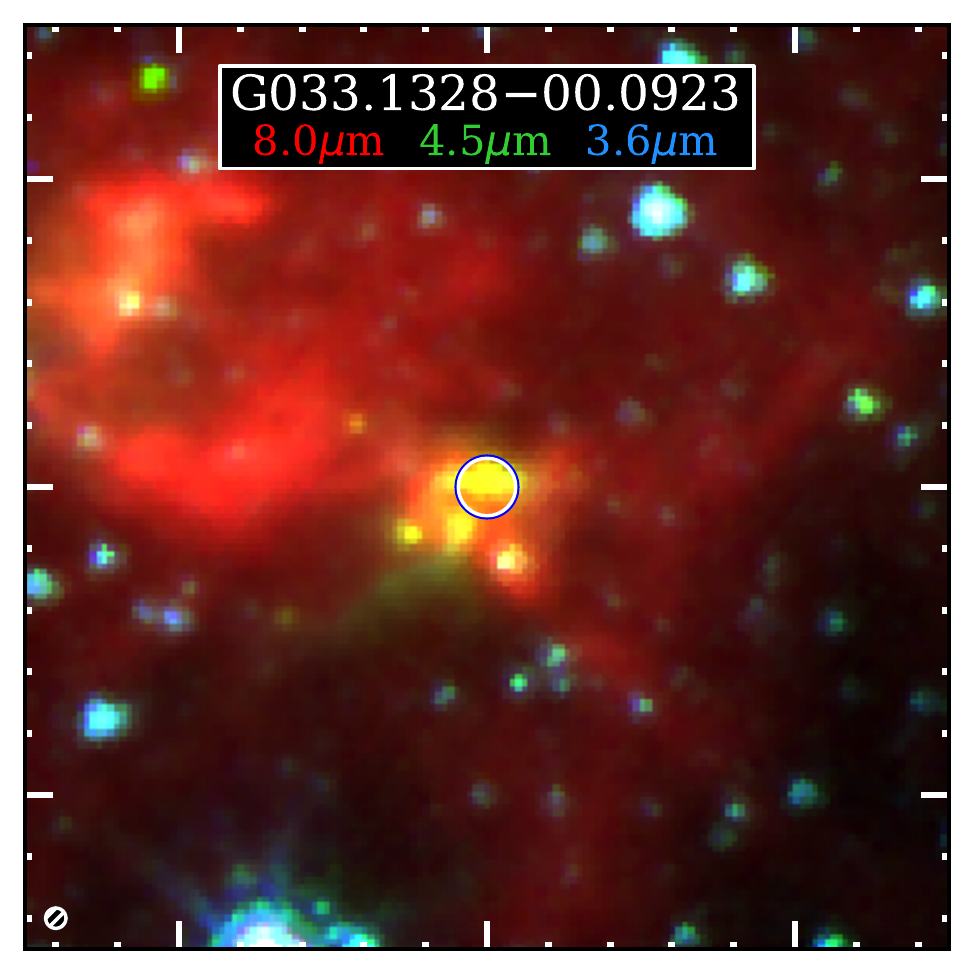}
\includegraphics[width=0.23\textwidth, trim= 0 0 0 0,clip]{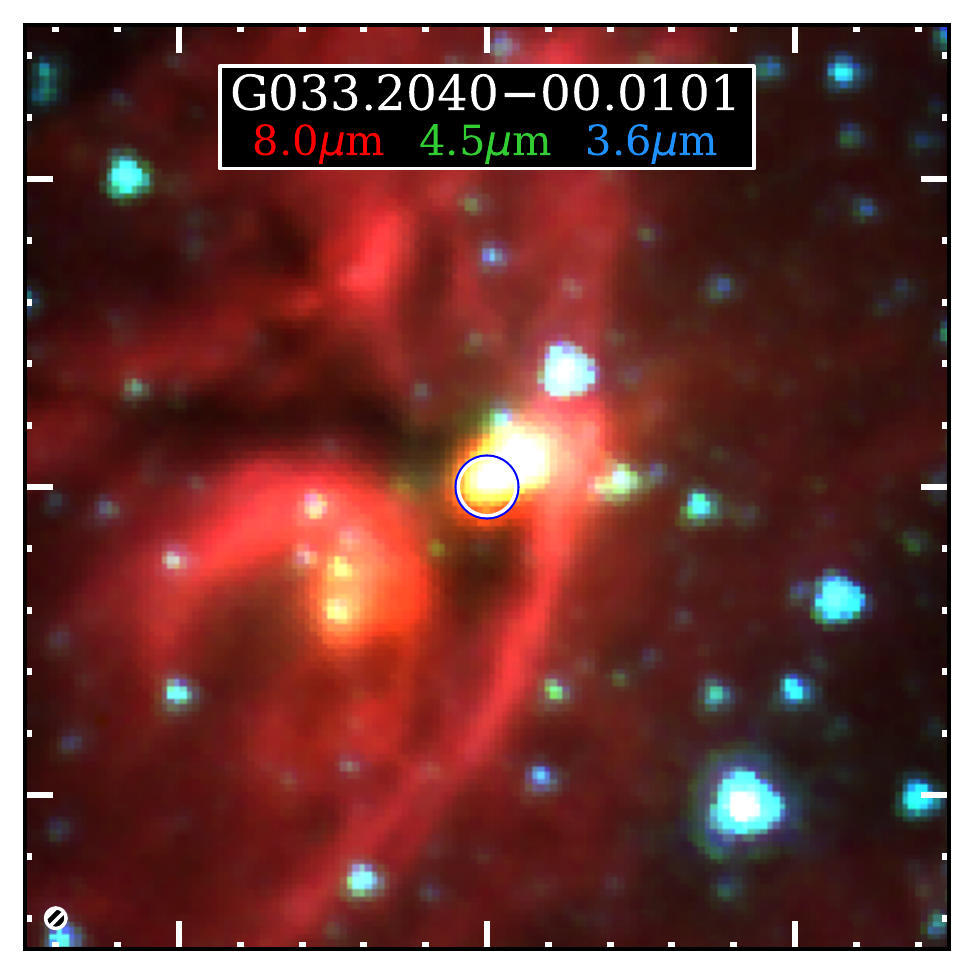}
\includegraphics[width=0.23\textwidth, trim= 0 0 0 0,clip]{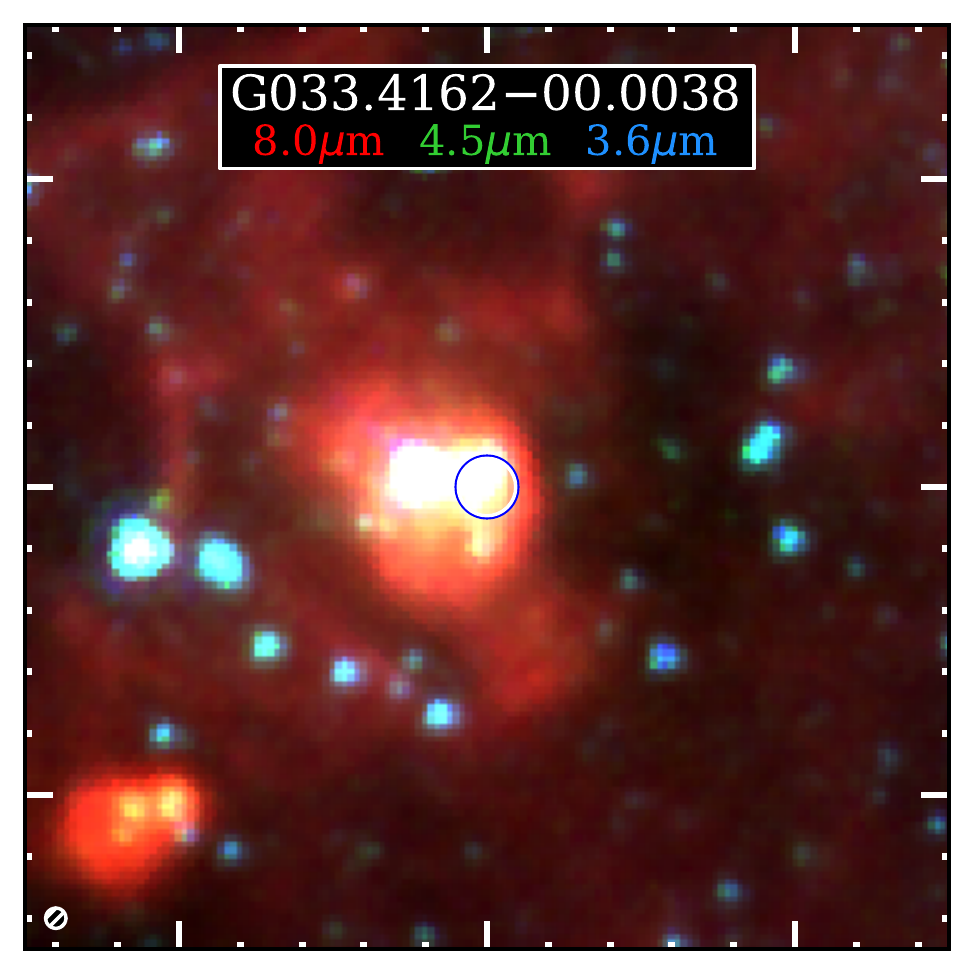}
\includegraphics[width=0.23\textwidth, trim= 0 0 0 0,clip]{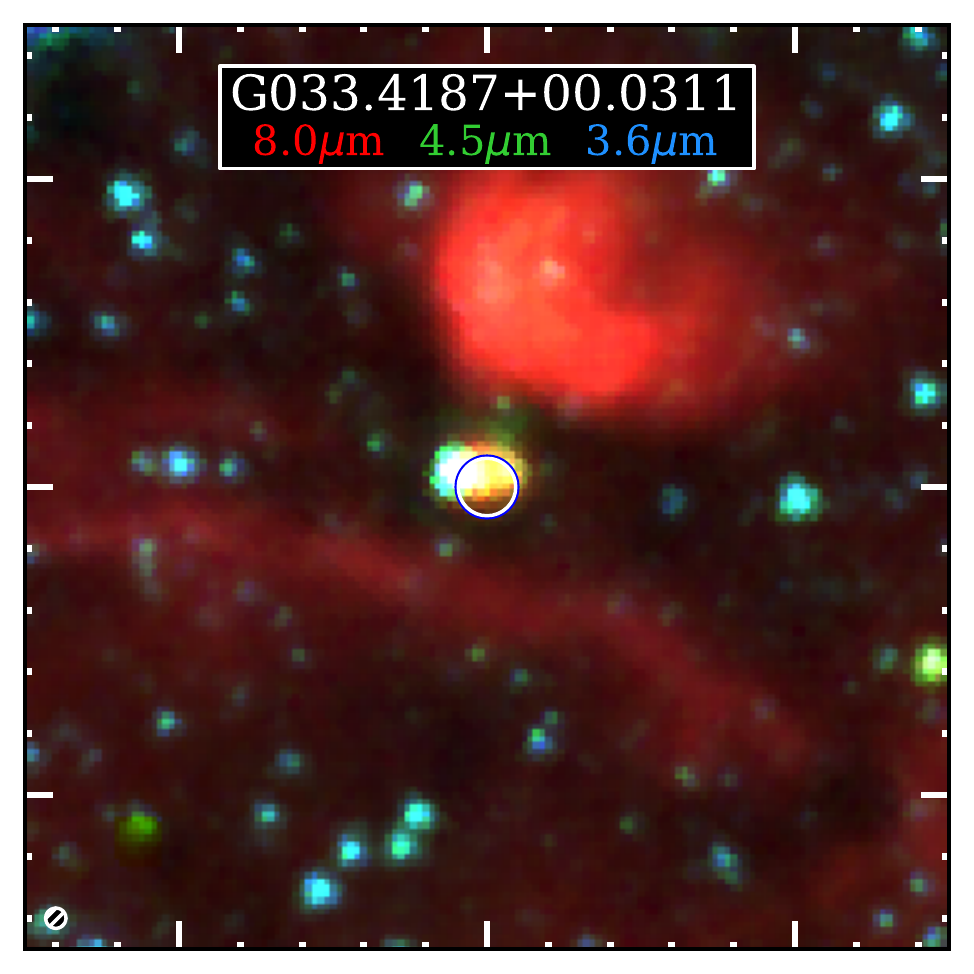}\\ 
\includegraphics[width=0.23\textwidth, trim= 0 0 0 0,clip]{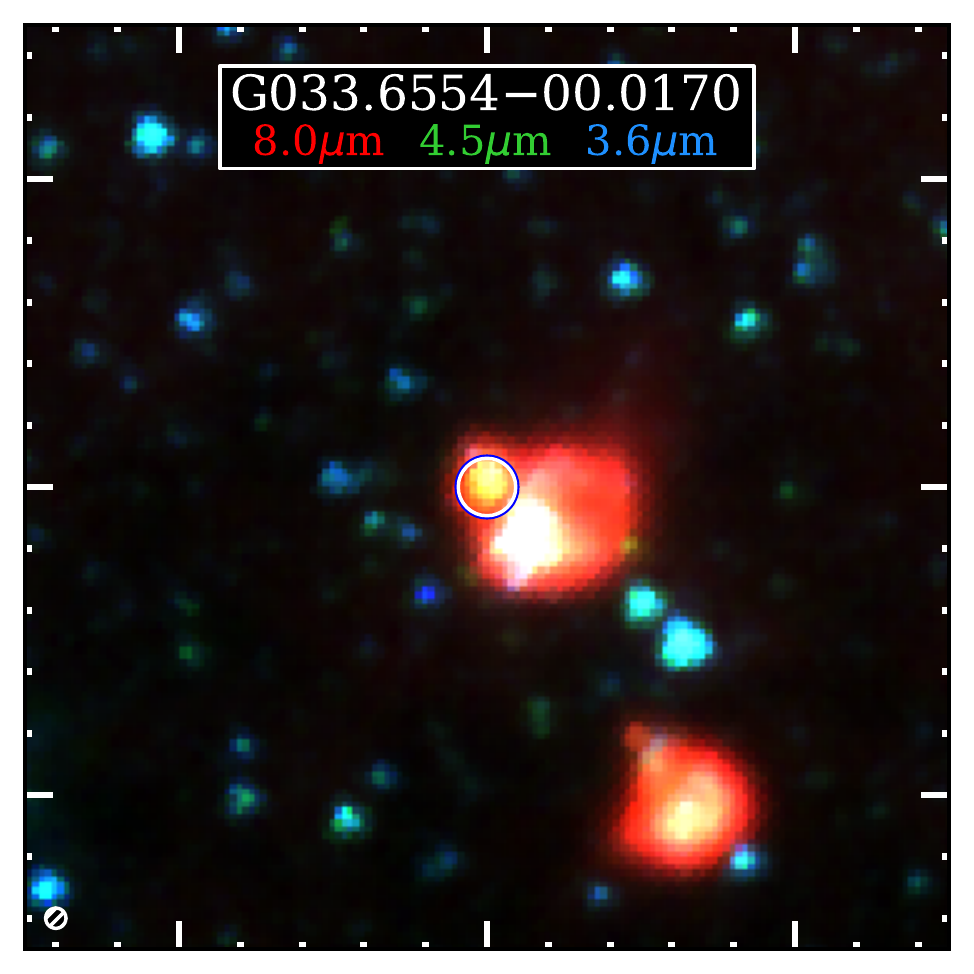}
\includegraphics[width=0.23\textwidth, trim= 0 0 0 0,clip]{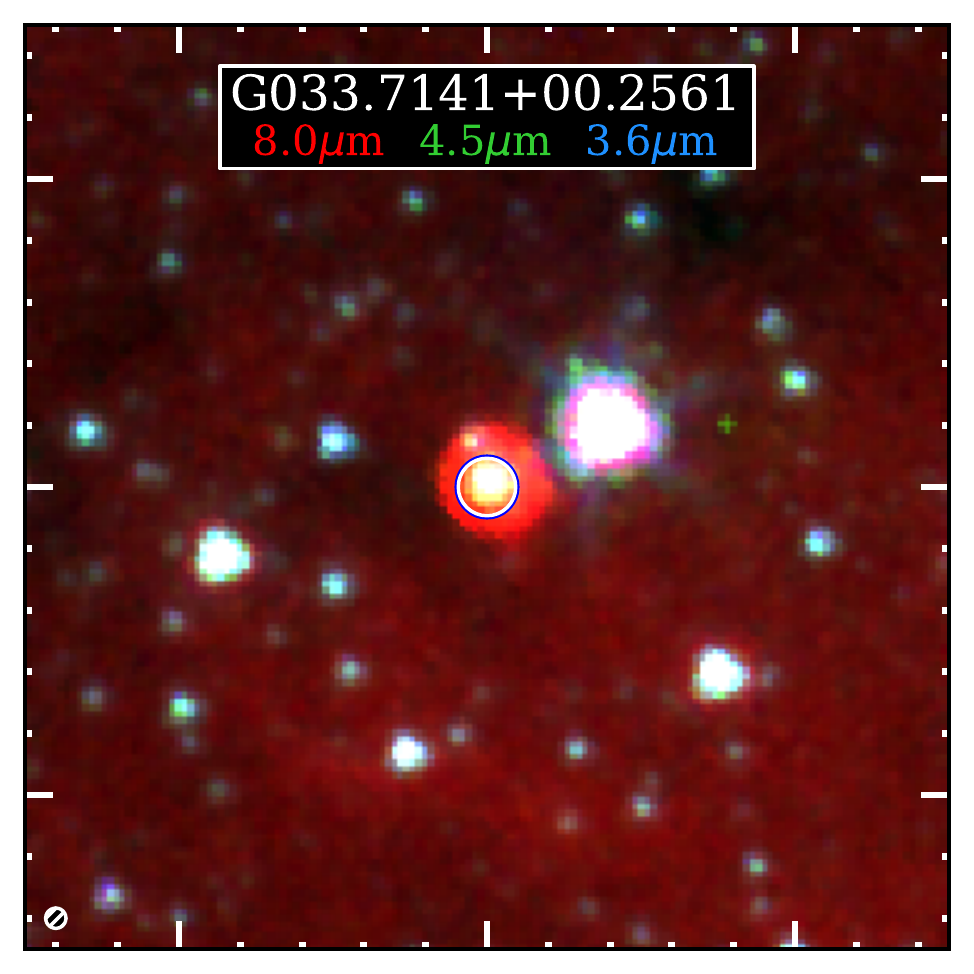}
\includegraphics[width=0.23\textwidth, trim= 0 0 0 0,clip]{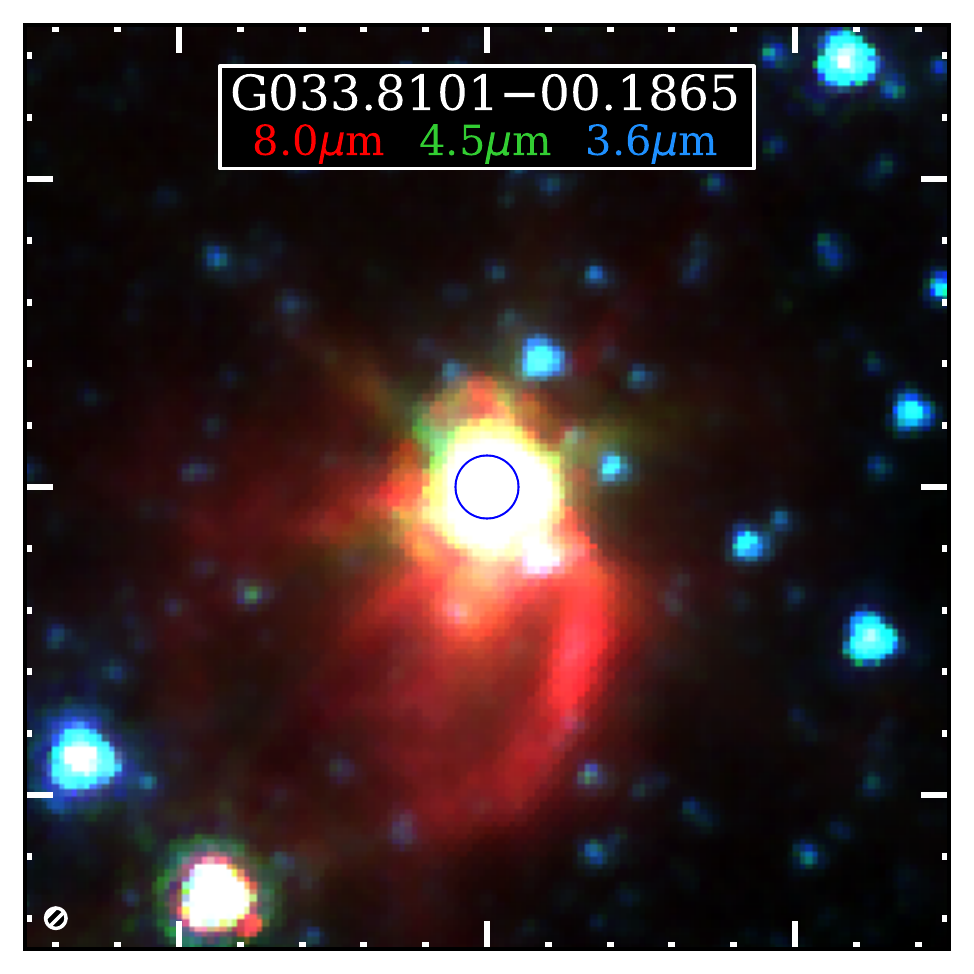}
\includegraphics[width=0.23\textwidth, trim= 0 0 0 0,clip]{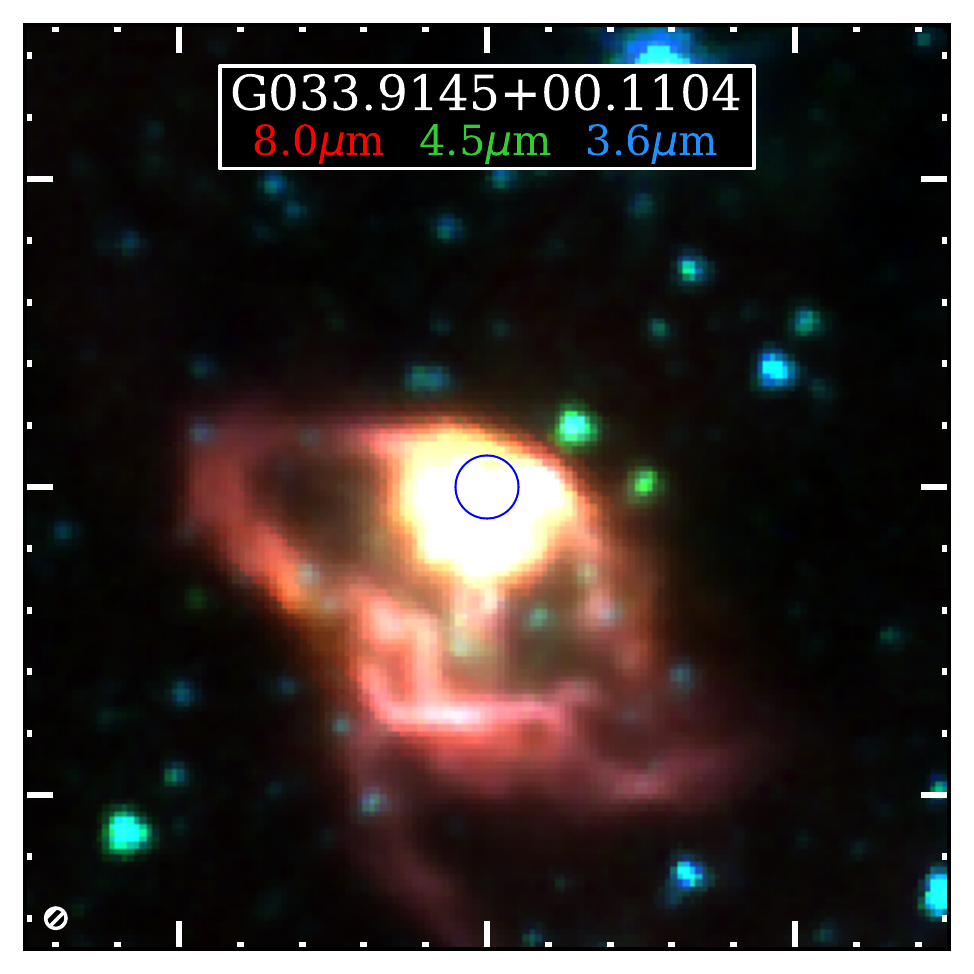}\\ 
\includegraphics[width=0.23\textwidth, trim= 0 0 0 0,clip]{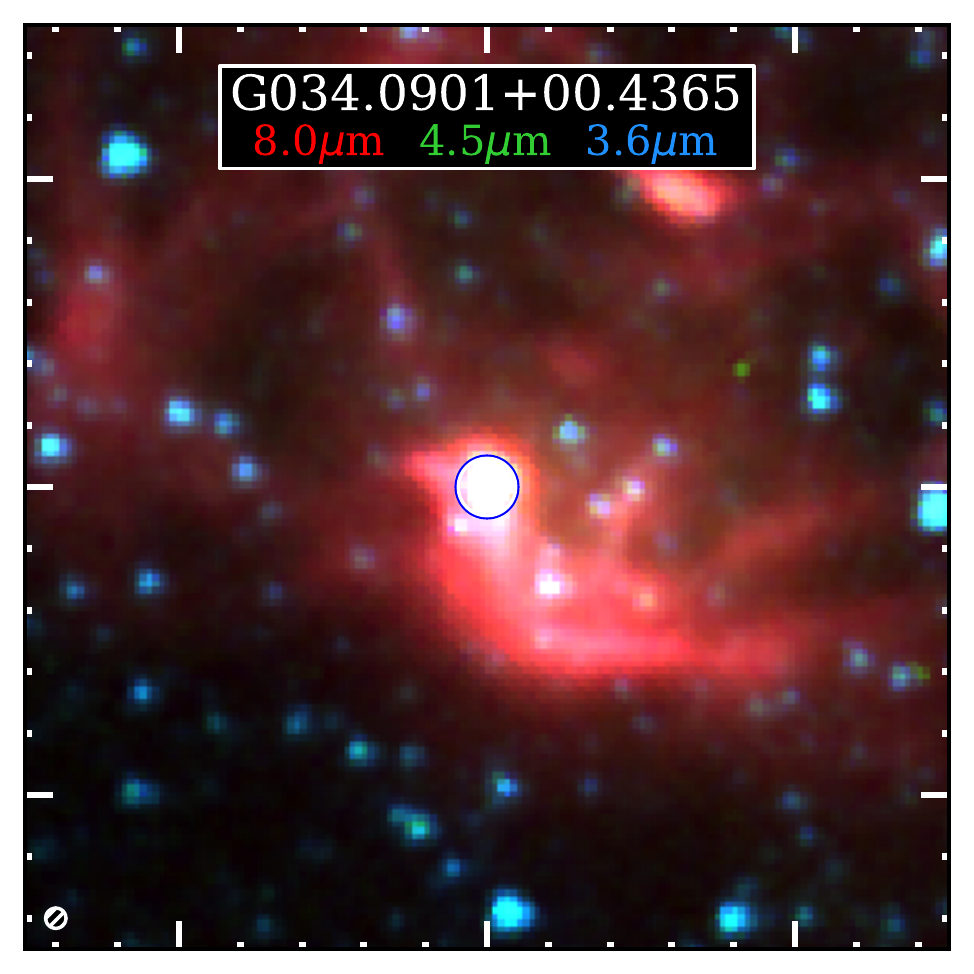}
\includegraphics[width=0.23\textwidth, trim= 0 0 0 0,clip]{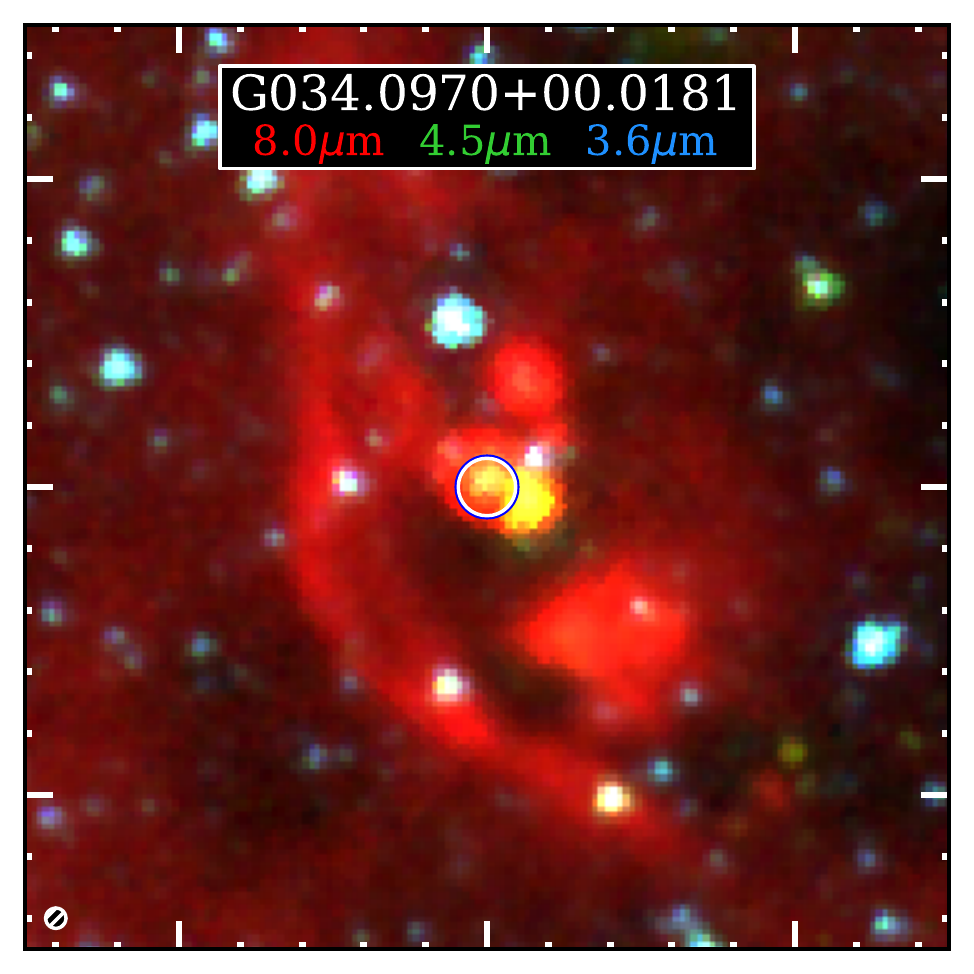}
\includegraphics[width=0.23\textwidth, trim= 0 0 0 0,clip]{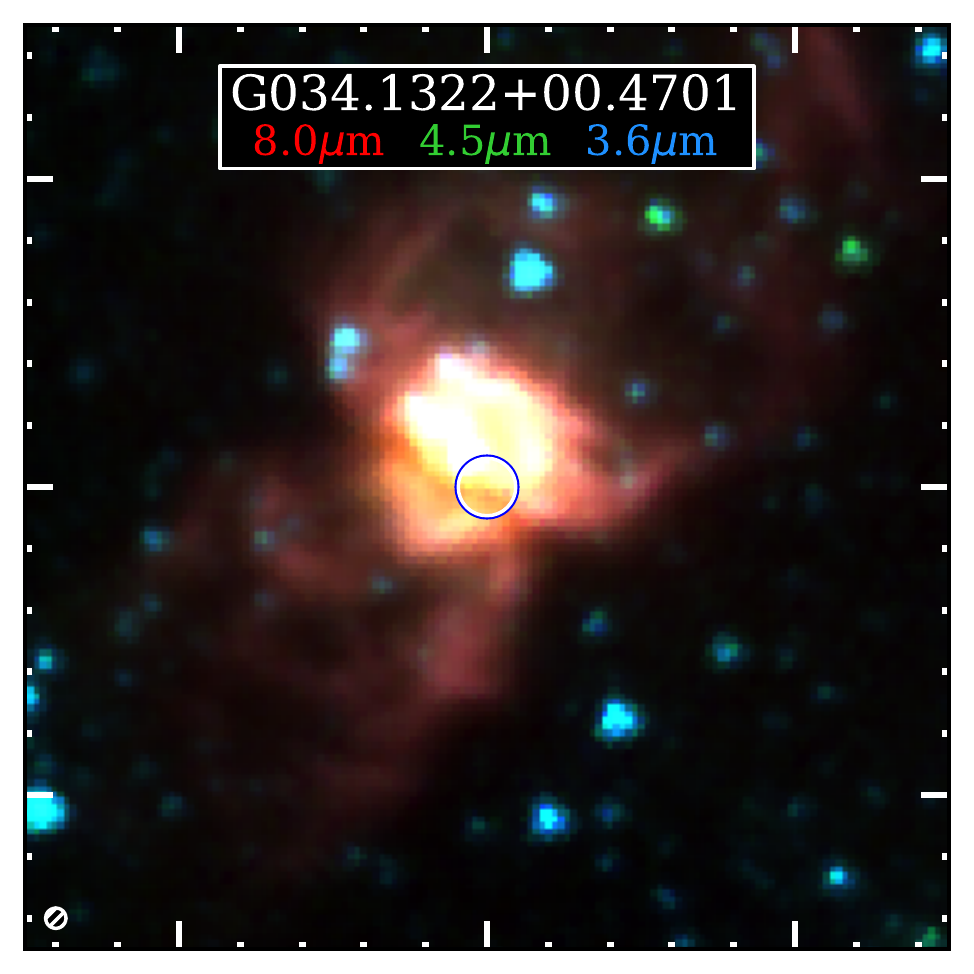}
\includegraphics[width=0.23\textwidth, trim= 0 0 0 0,clip]{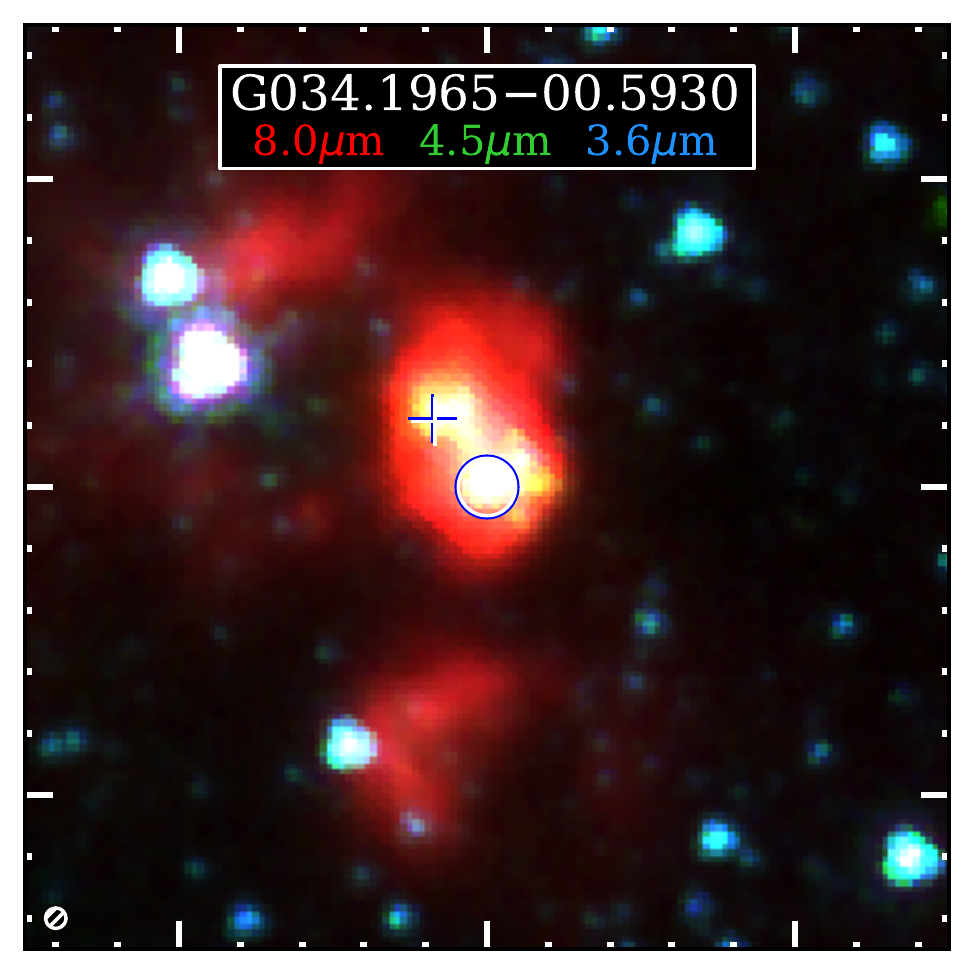}\\ 
\includegraphics[width=0.23\textwidth, trim= 0 0 0 0,clip]{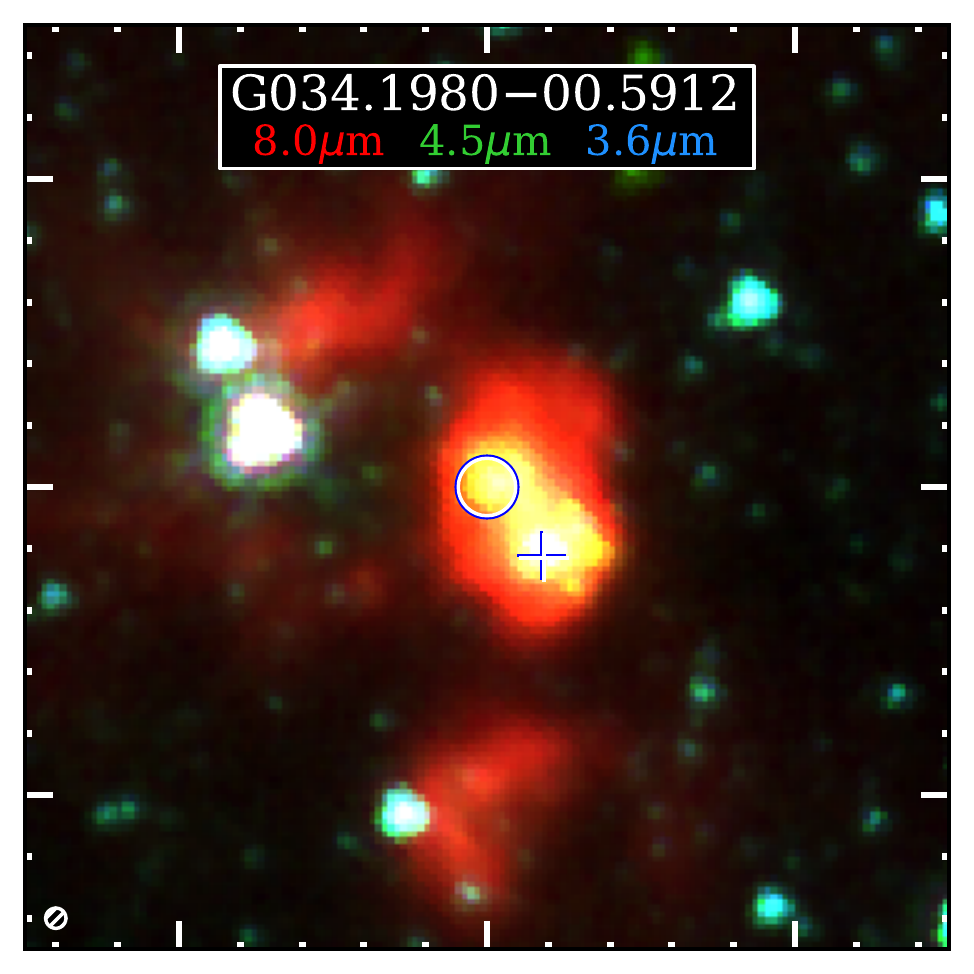}
\includegraphics[width=0.23\textwidth, trim= 0 0 0 0,clip]{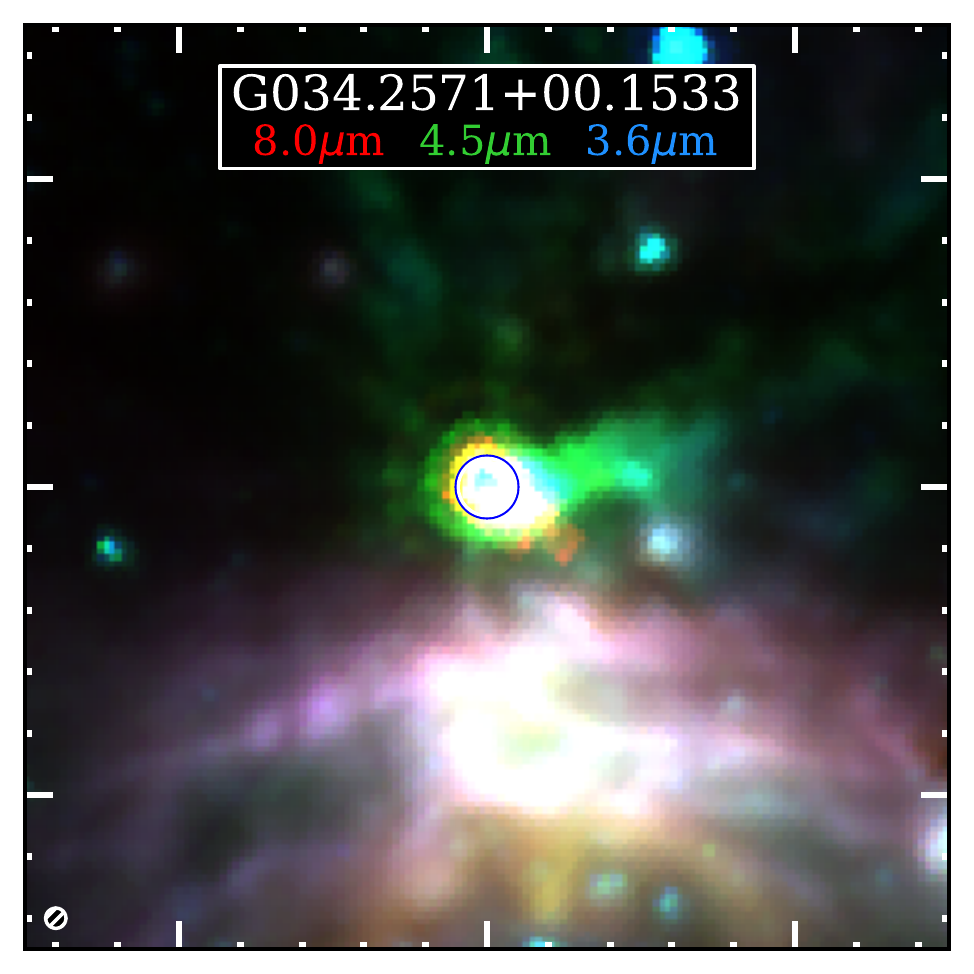}
\includegraphics[width=0.23\textwidth, trim= 0 0 0 0,clip]{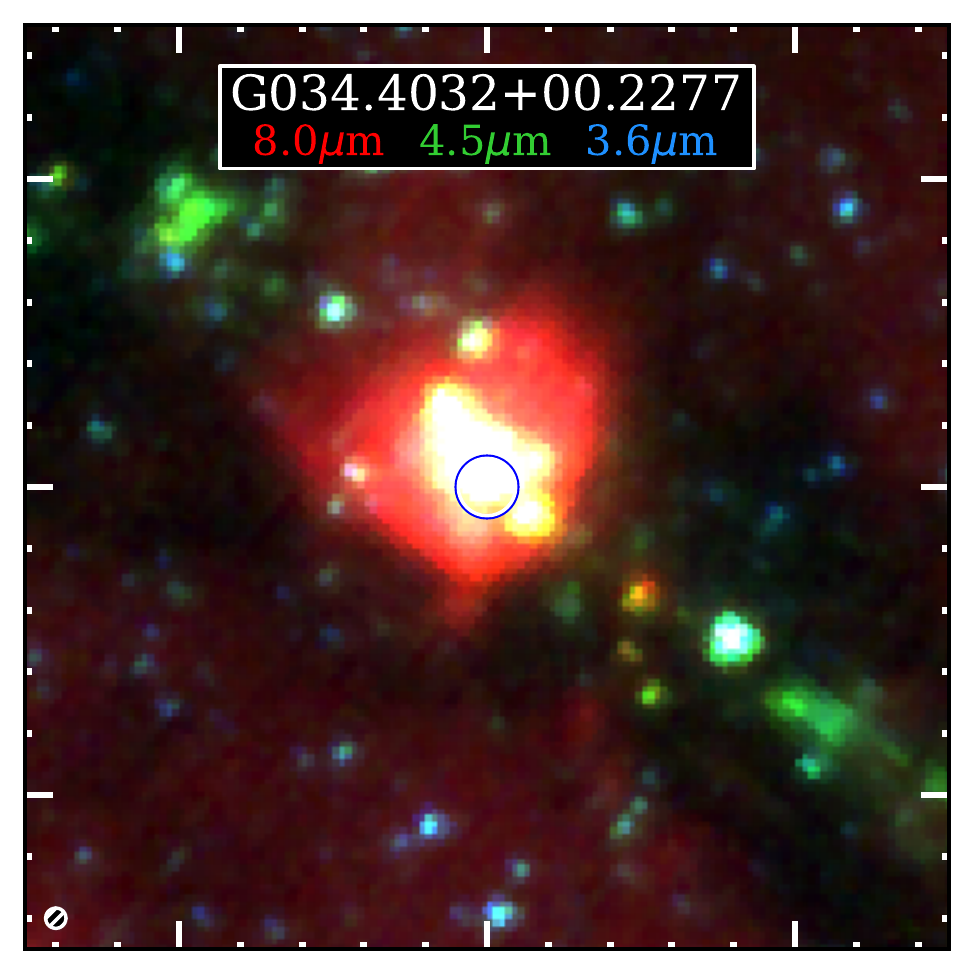}
\includegraphics[width=0.23\textwidth, trim= 0 0 0 0,clip]{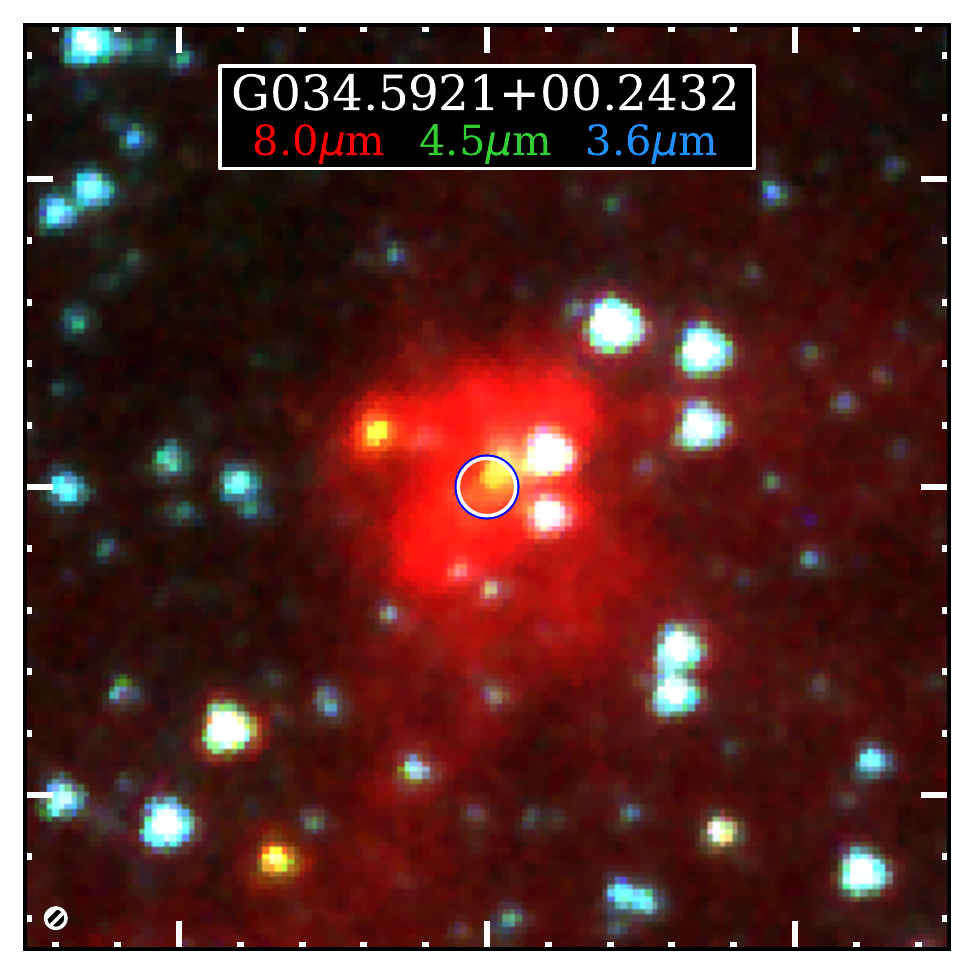}\\ 
\hspace{-0.85cm}\includegraphics[width=0.28\textwidth, trim= 0 0 0 0,clip,valign=t]{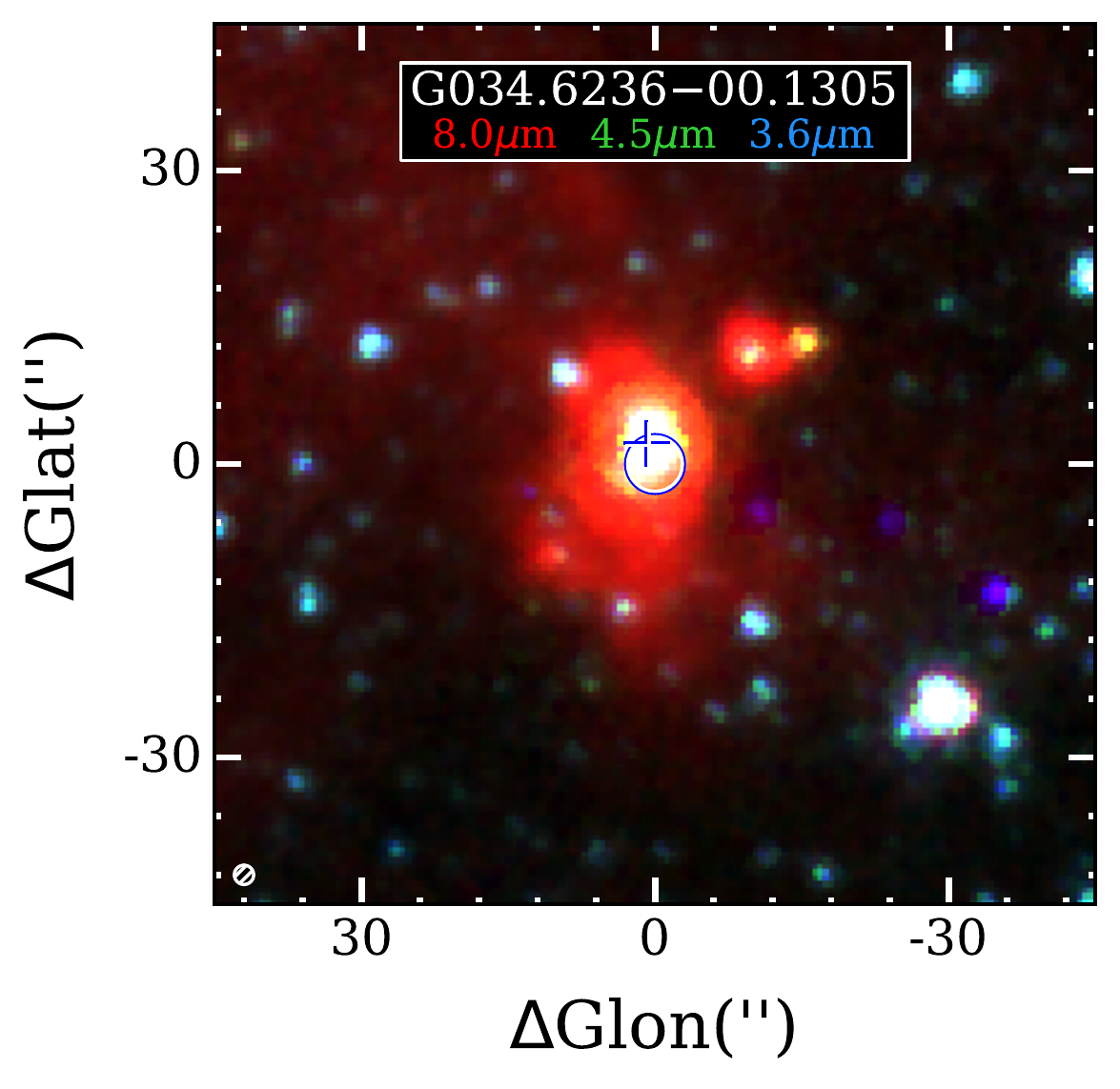}
\includegraphics[width=0.23\textwidth, trim= 0 0 0 0,clip,valign=t]{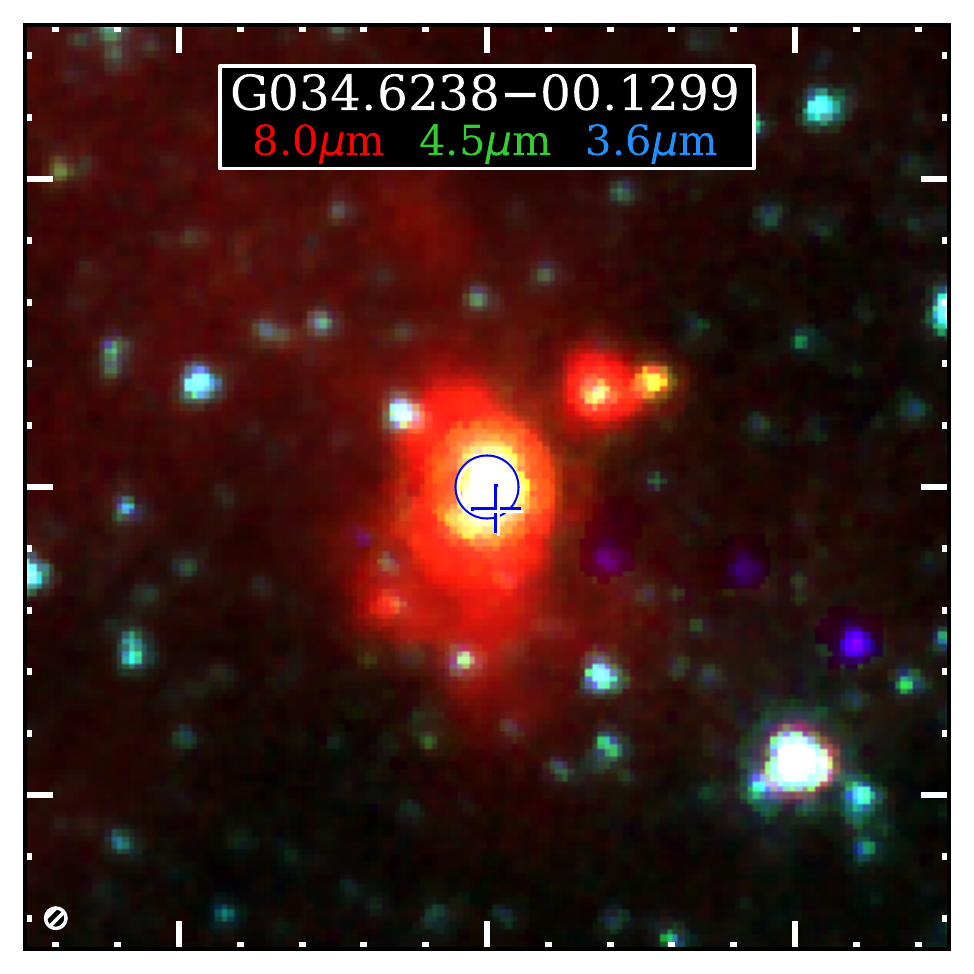}
\includegraphics[width=0.23\textwidth, trim= 0 0 0 0,clip,valign=t]{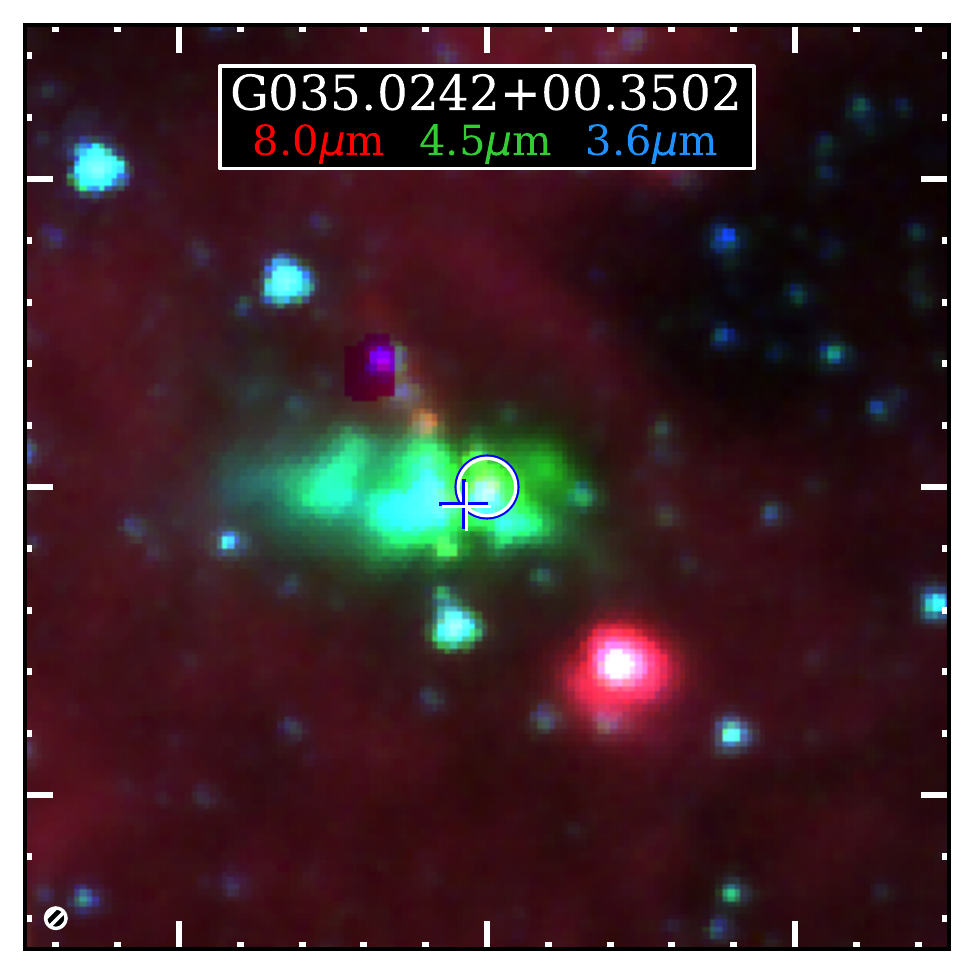}
\includegraphics[width=0.23\textwidth, trim= 0 0 0 0,clip,valign=t]{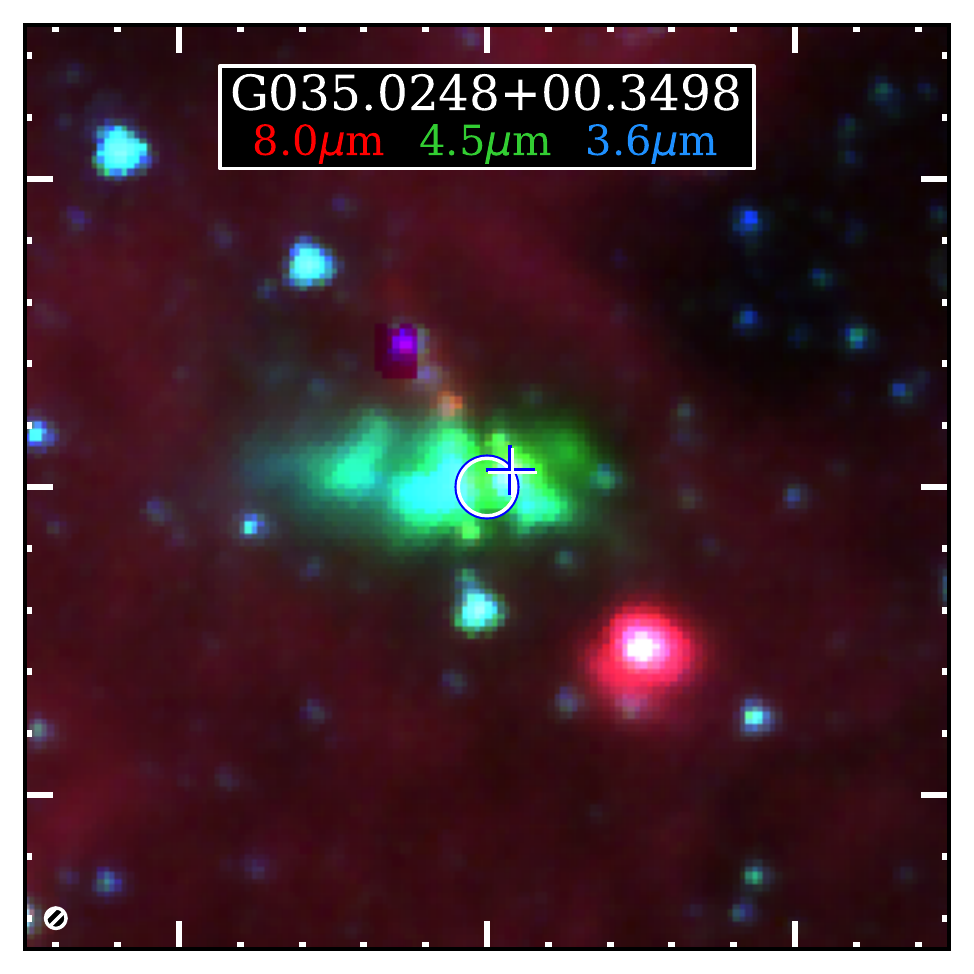}\\    
\caption{Continuation.}
\end{figure*}

\setcounter{figure}{0}\begin{figure*}[!h]
\centering
\includegraphics[width=0.23\textwidth, trim= 0 0 0 0,clip]{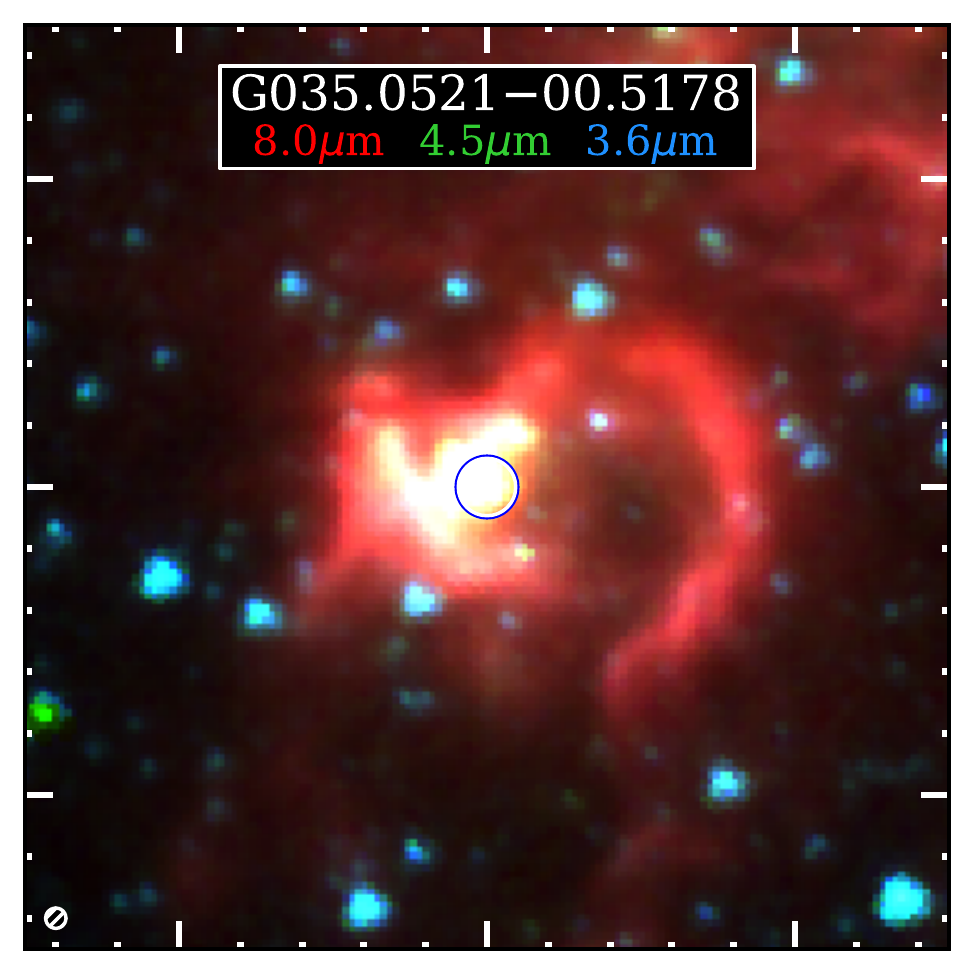}
\includegraphics[width=0.23\textwidth, trim= 0 0 0 0,clip]{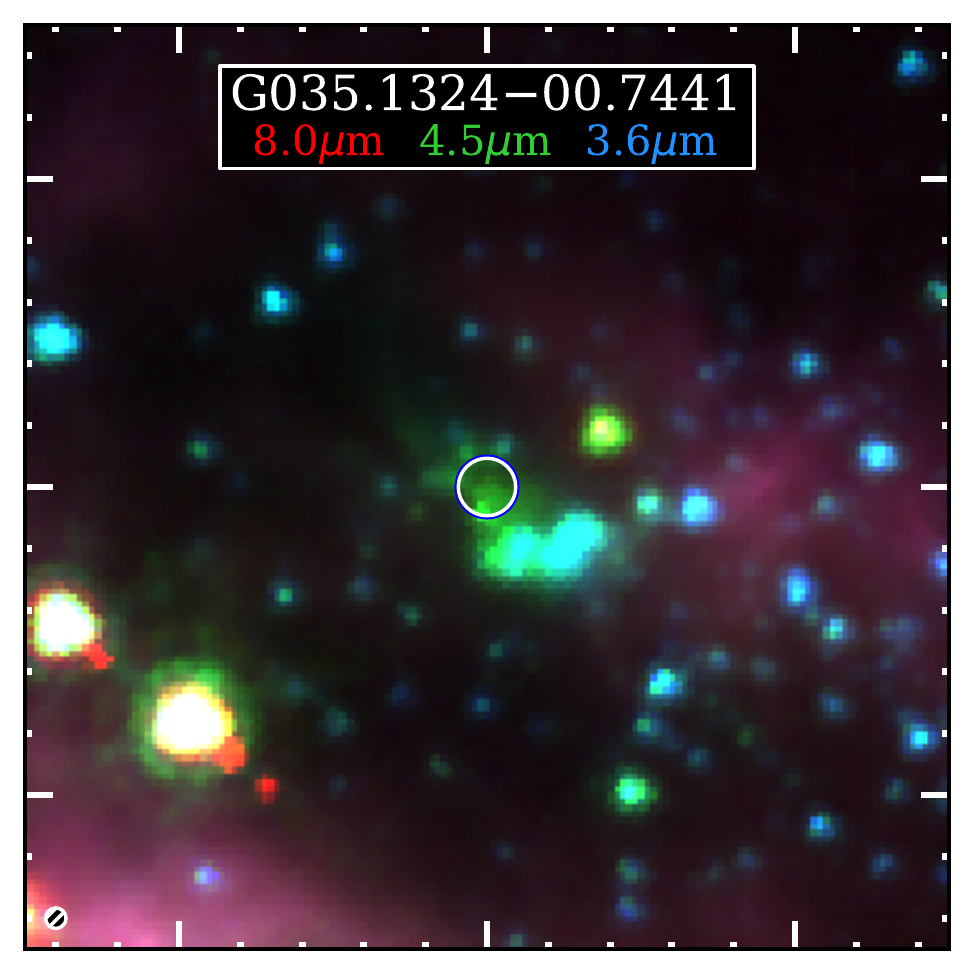}
\includegraphics[width=0.23\textwidth, trim= 0 0 0 0,clip]{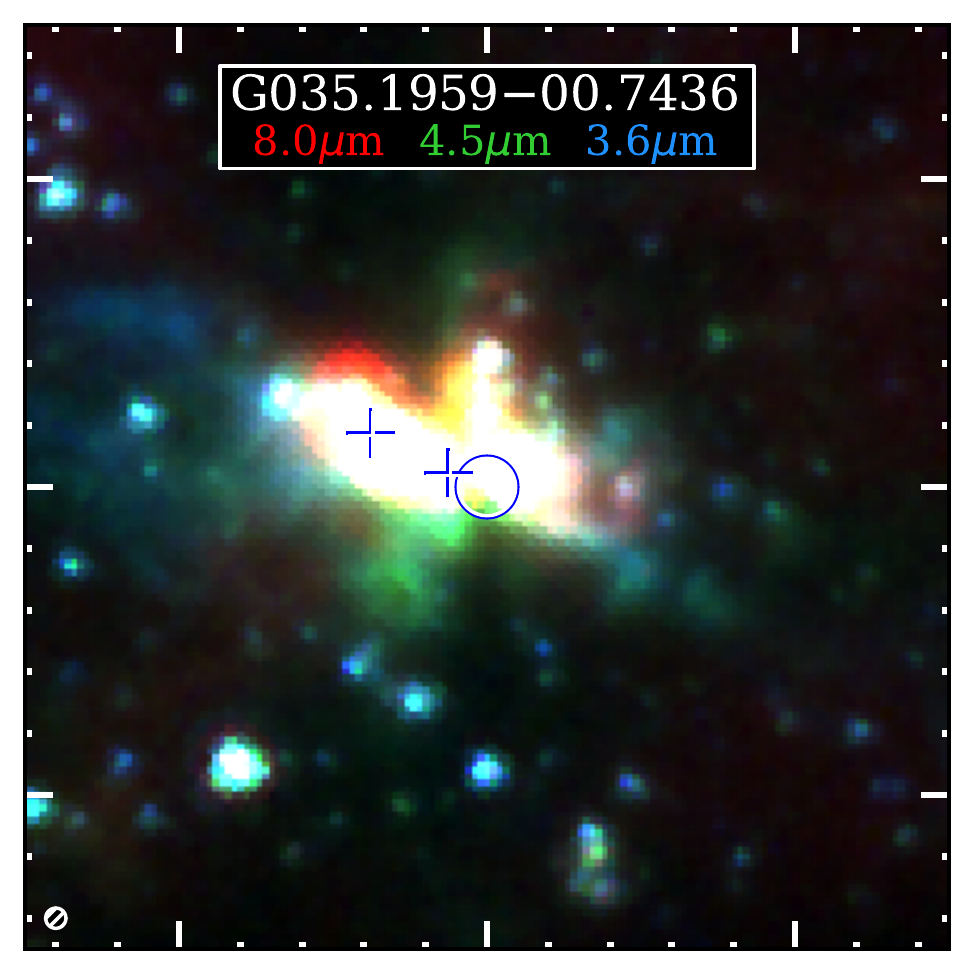}
\includegraphics[width=0.23\textwidth, trim= 0 0 0 0,clip]{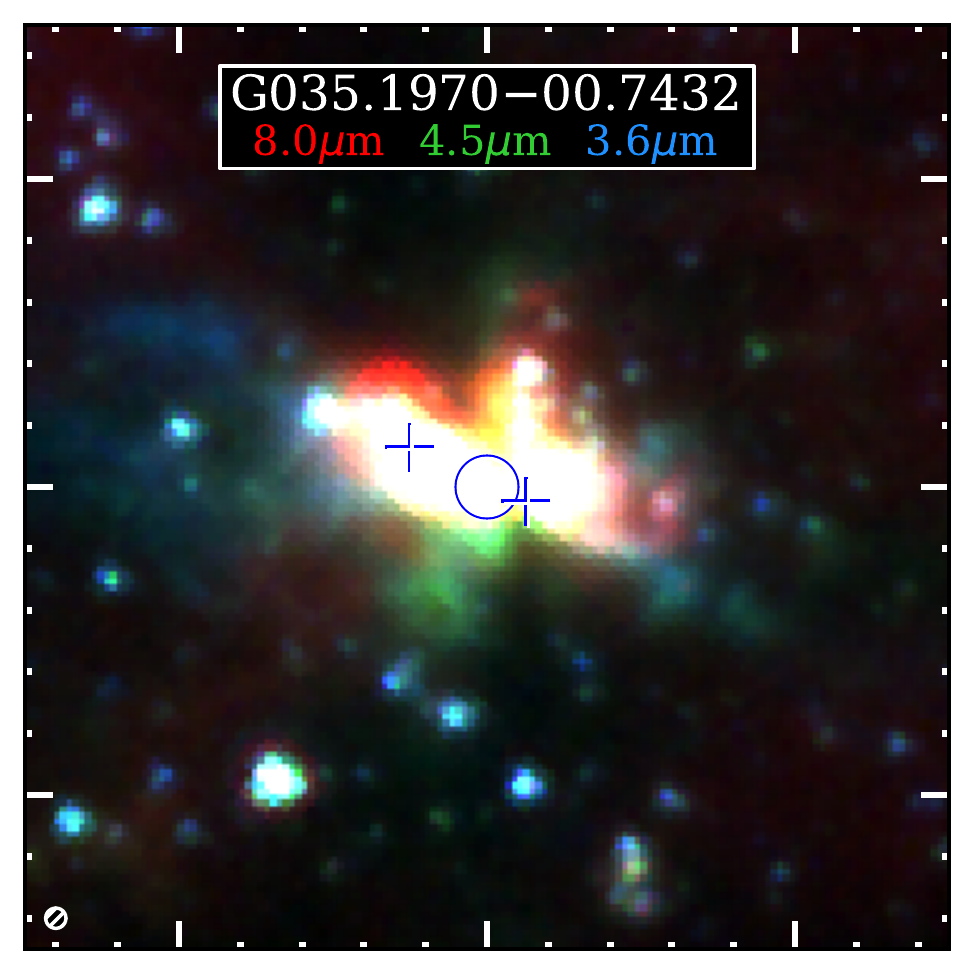}\\ 
\includegraphics[width=0.23\textwidth, trim= 0 0 0 0,clip]{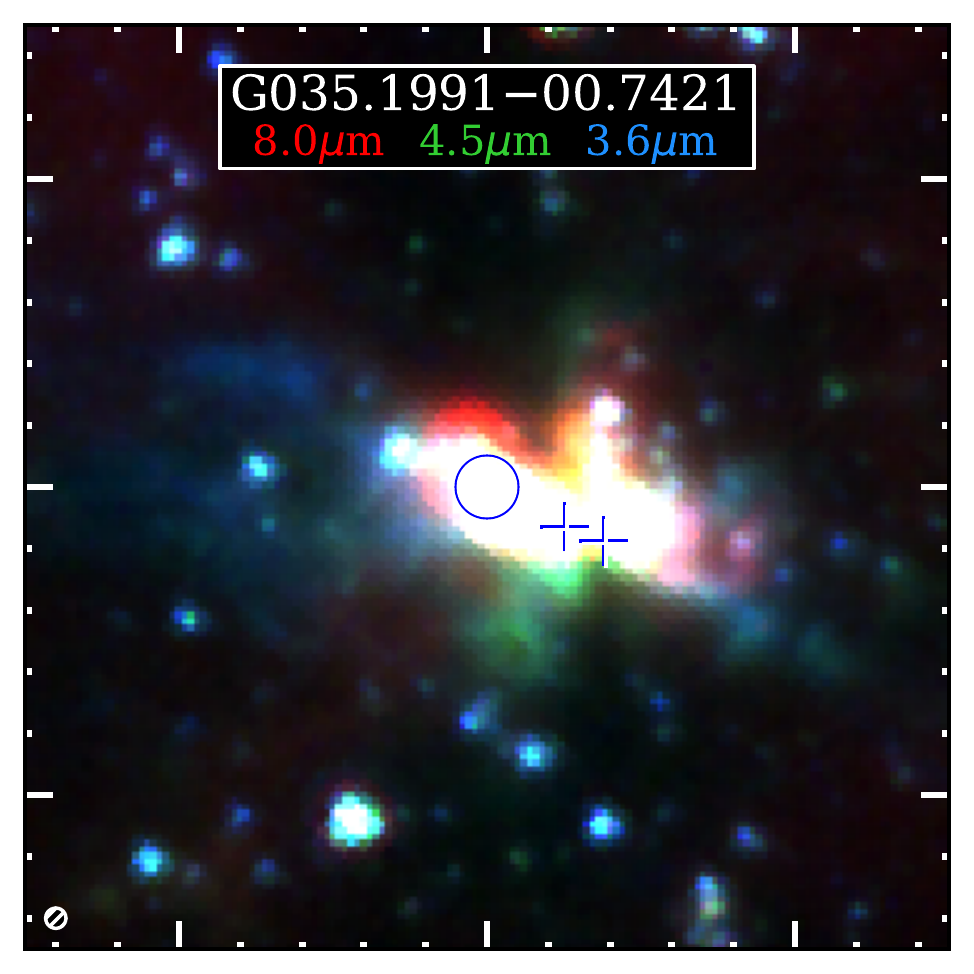}
\includegraphics[width=0.23\textwidth, trim= 0 0 0 0,clip]{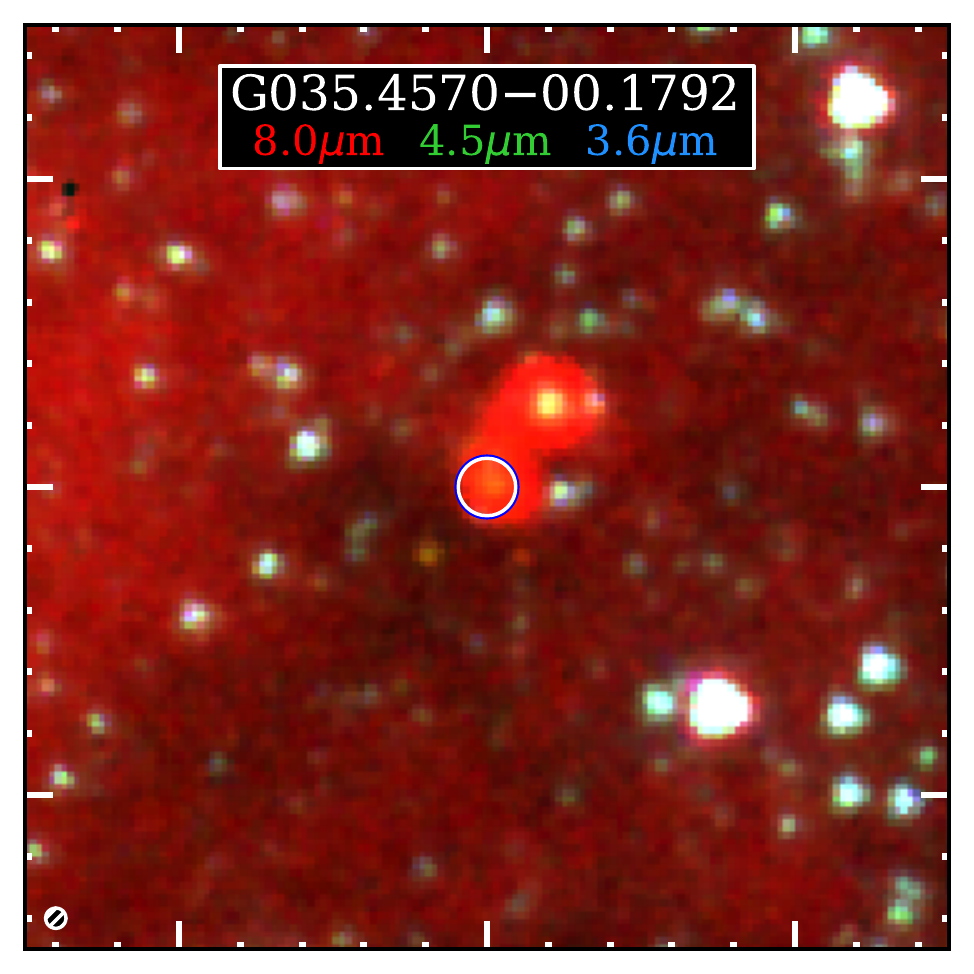}
\includegraphics[width=0.23\textwidth, trim= 0 0 0 0,clip]{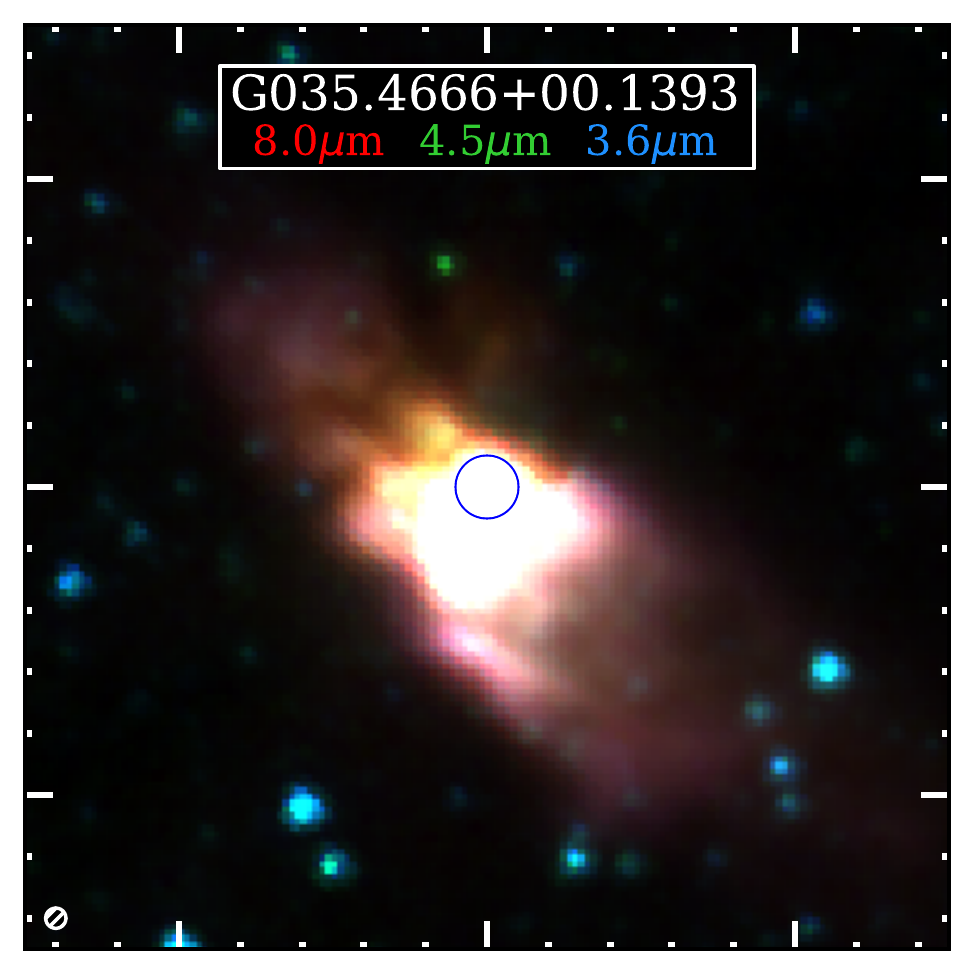}
\includegraphics[width=0.23\textwidth, trim= 0 0 0 0,clip]{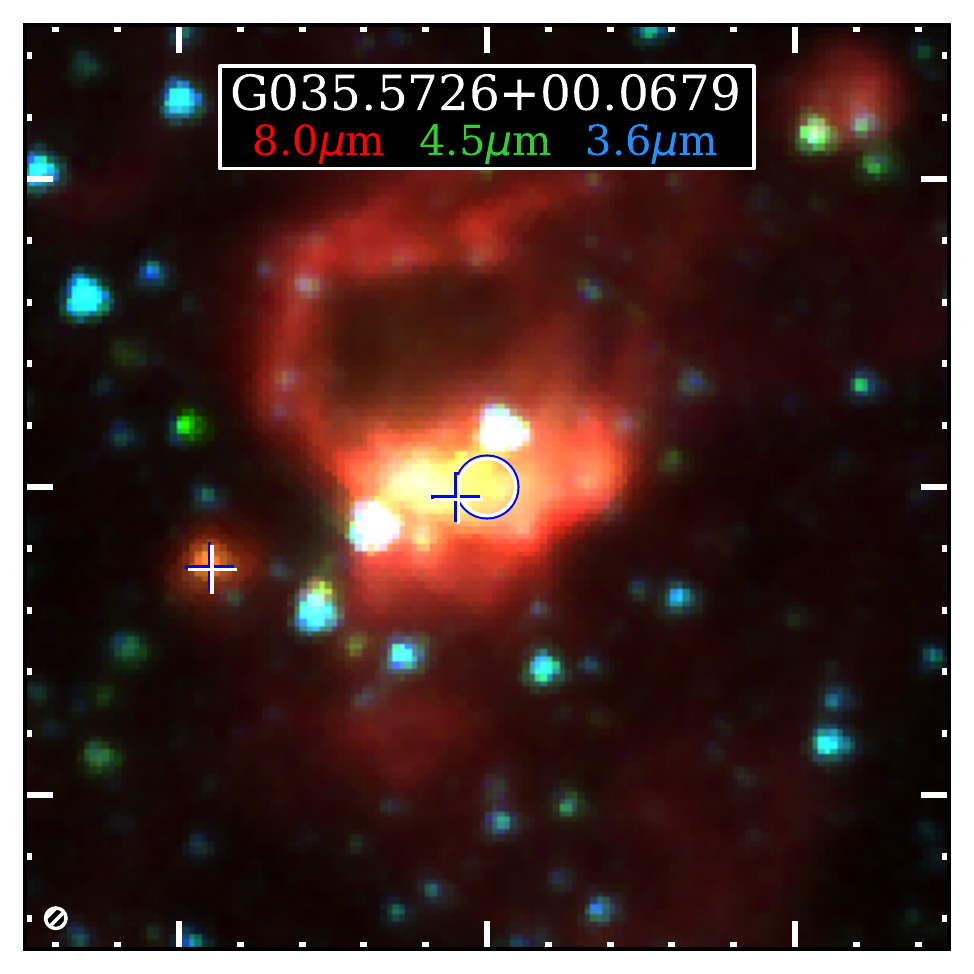}\\ 
\includegraphics[width=0.23\textwidth, trim= 0 0 0 0,clip]{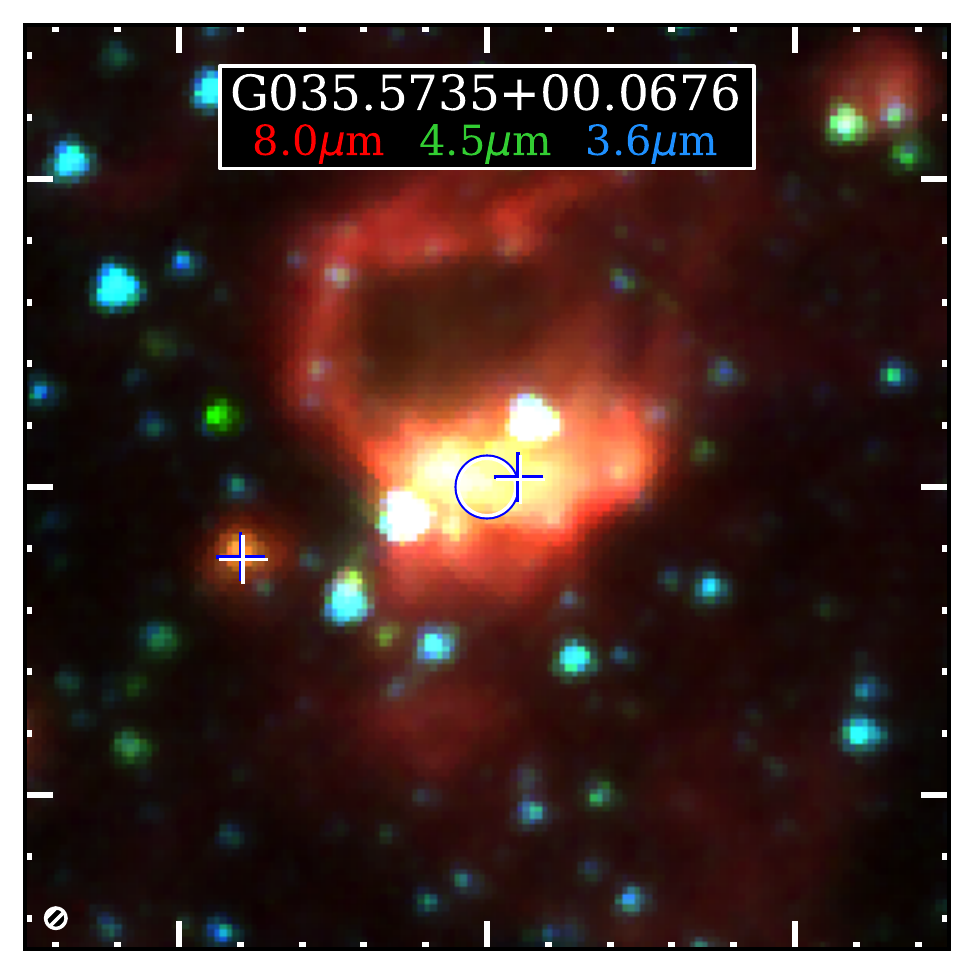}
\includegraphics[width=0.23\textwidth, trim= 0 0 0 0,clip]{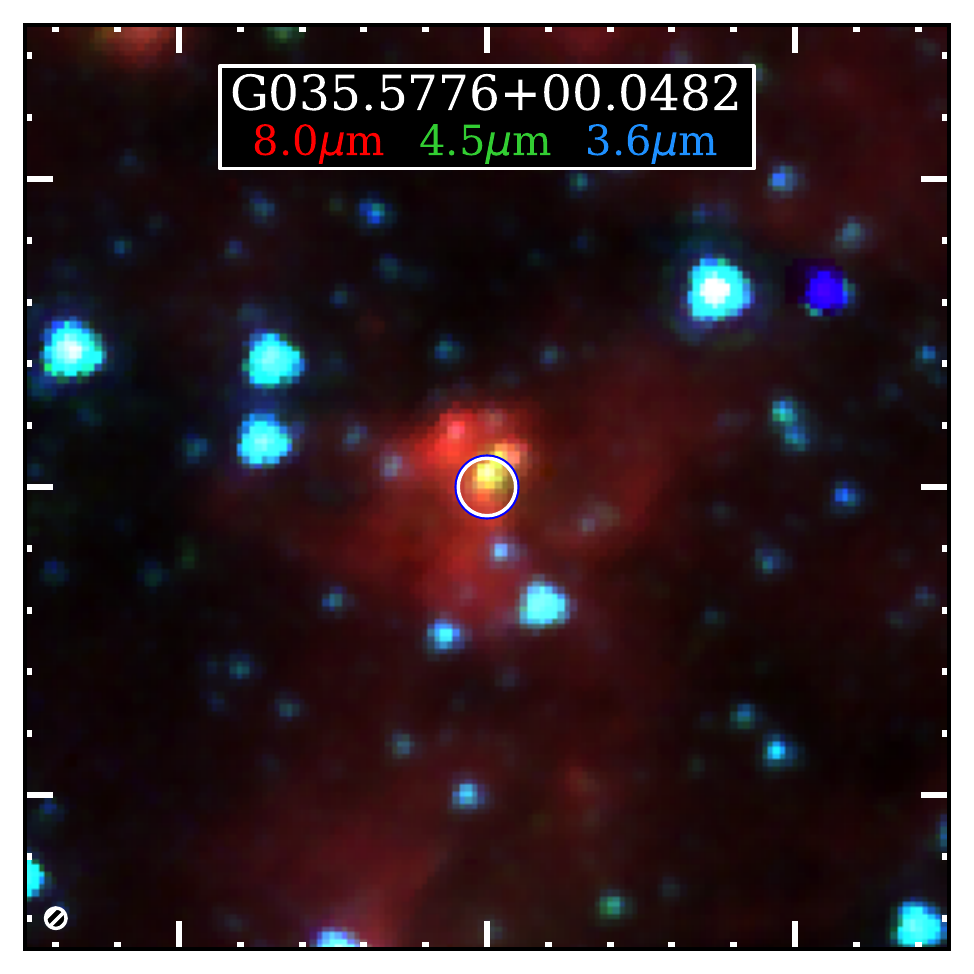}
\includegraphics[width=0.23\textwidth, trim= 0 0 0 0,clip]{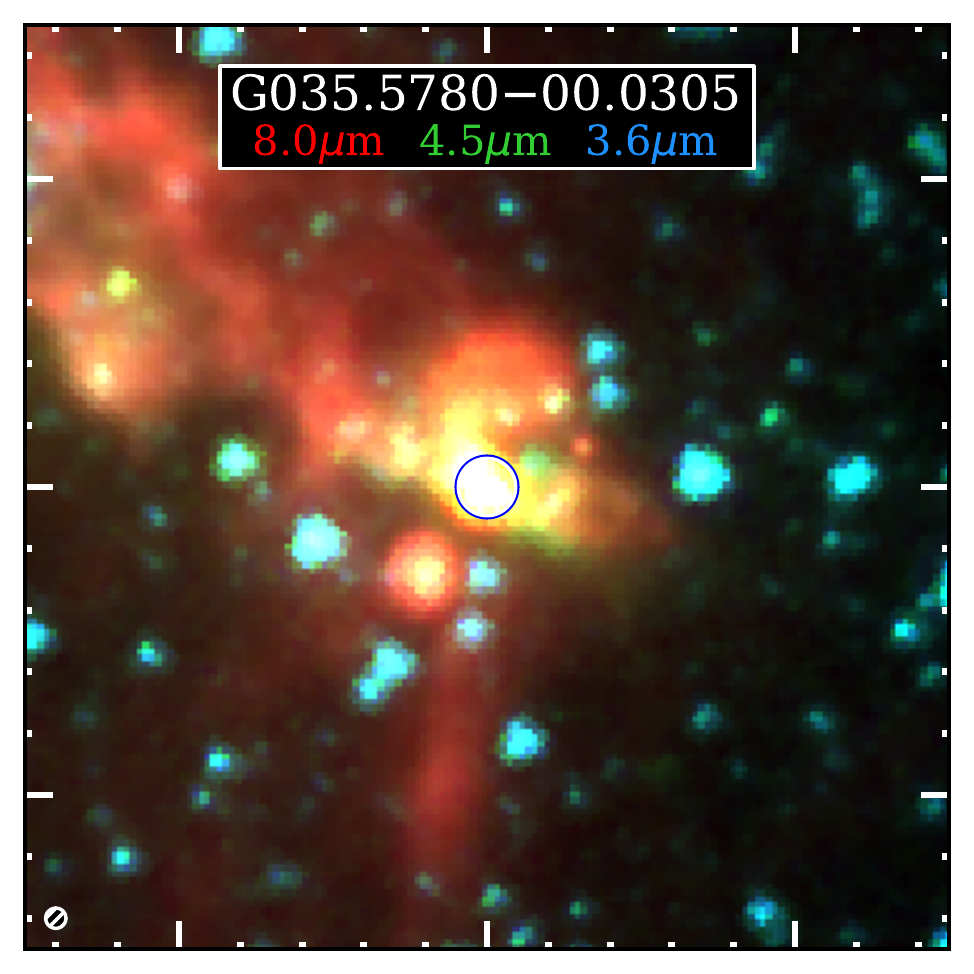}
\includegraphics[width=0.23\textwidth, trim= 0 0 0 0,clip]{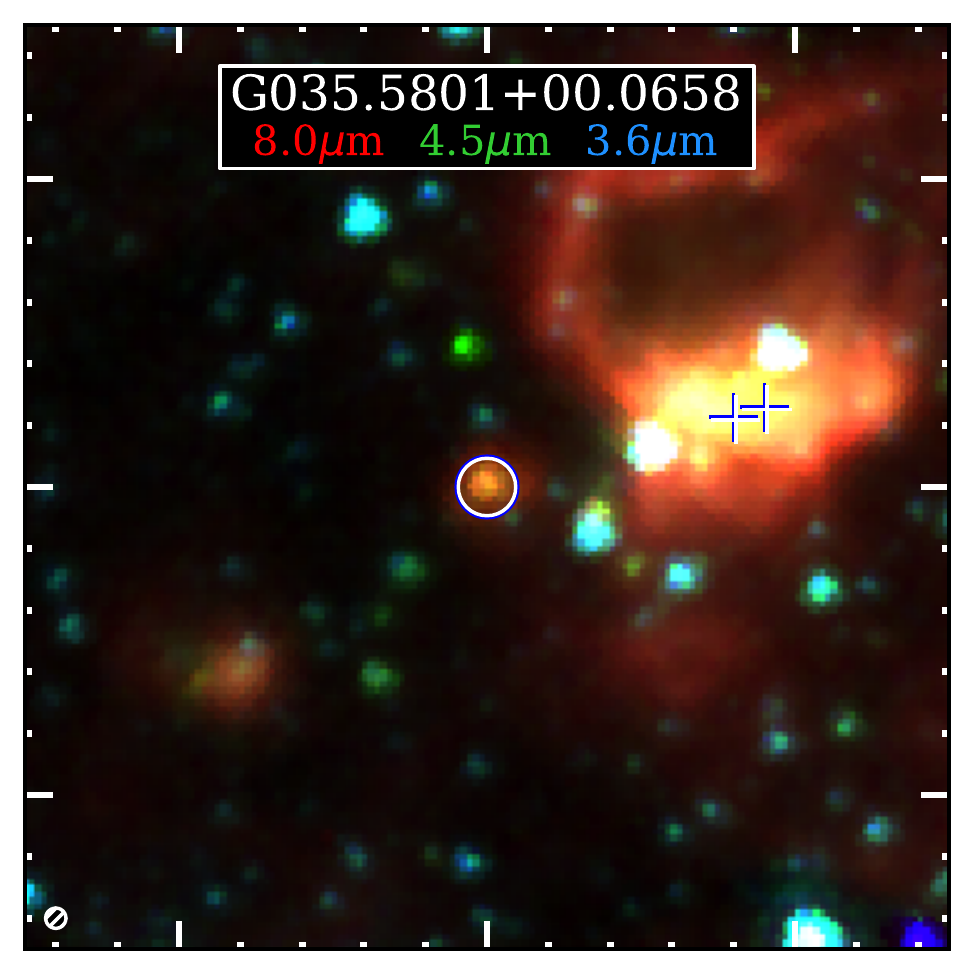}\\ 
\hspace{-13.80cm}\includegraphics[width=0.276\textwidth, trim= 0 0 0 0,clip]{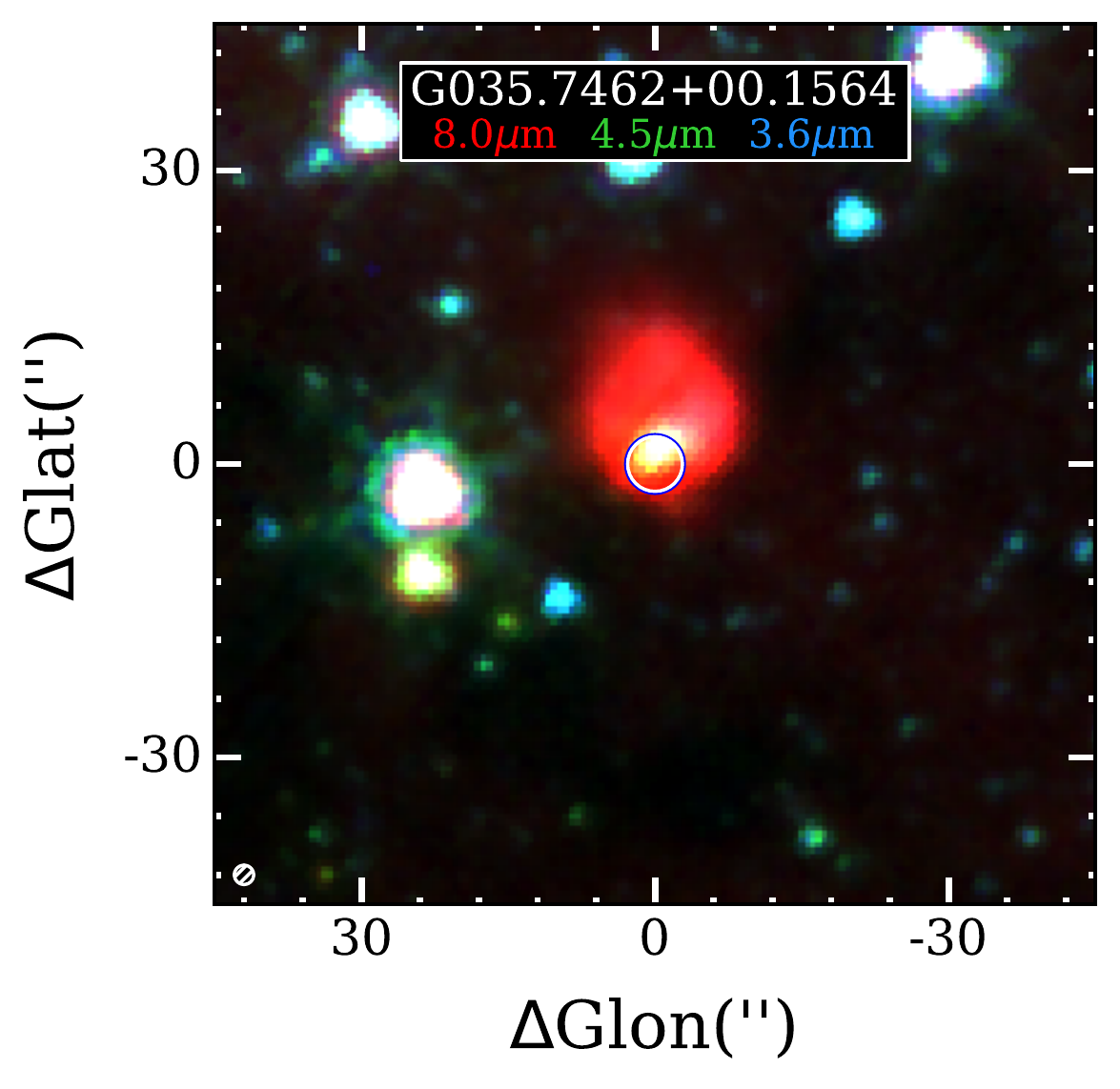}\\
\caption{Continuation.}
\end{figure*}

\end{appendix}

\end{document}